\documentstyle[prb,aps,twocolumn]{revtex}

\begin{document}
\input epsf.tex
\draft
\preprint{Submitted to Physical Review B}
\wideabs{
\title{Magnetic Susceptibilities of Spin-1/2 Antiferromagnetic Heisenberg
Ladders and
\protect\\ Applications to Ladder Oxide Compounds}
\author{D. C. Johnston$^1$, M. Troyer$^{2,3}$, S. Miyahara$^2$, D.
Lidsky$^2$, K. Ueda$^2$, M. Azuma$^4$, Z. Hiroi$^{4,}$\cite{Hiroi}, M.
Takano$^4$,\protect\\ M. Isobe$^2$, Y. Ueda$^2$, M. A. Korotin$^5$, V. I.
Anisimov$^5$, A. V. Mahajan$^{1,}$\cite{Mahajan}, and L. L. Miller$^1$}
\address{$^1$Ames Laboratory and Department of Physics and Astronomy, Iowa
State University, Ames, Iowa 50011}
\address{$^2$Institute for Solid State Physics, University of Tokyo,
Roppongi 7-22-1, Tokyo 106, Japan}
\address{$^3$Theoretische Physik,
Eidgen\"ossische Technische Hochschule-Z\"urich, CH-8093 Z\"urich,
Switzerland}
\address{$^4$Institute for Chemical Research, Kyoto University, Uji, Kyoto
611, Japan}
\address{$^5$Institute of Metal Physics, Ekaterinburg GSP-170, Russia}
\date{January 8, 2000}
\maketitle
\begin{abstract}An extensive theoretical and experimental study is
presented of the magnetic susceptibility versus temperature $\chi(T)$ of
spin $S = 1/2$ two- and three-leg Heisenberg ladders and ladder oxide
compounds.  The $\chi(T)$ of isolated two-leg ladders with spatially
anisotropic antiferromagnetic (AF) Heisenberg exchange was calculated by
quantum Monte Carlo (QMC) simulations, with and without ferromagnetic (FM)
second-neighbor diagonal intraladder coupling, and for two-leg ladders
coupled into a two-dimensional (2D) stacked ladder configuration or a 3D
LaCuO$_{2.5}$-type interladder coupling configuration.  We present accurate
analytical fits and interpolations of these data and of previously reported
related QMC $\chi(T)$ simulation data for the isolated ladder with
spatially isotropic exchange, for the 2D trellis layer configuration and
for isotropic and anisotropic three-leg ladders.  We have also calculated
the one- and two-magnon dispersion relations for the isolated $2 \times 12$
ladder with \mbox{$0.5 \leq J^\prime/J \leq 1$}, where $J$ is the AF
coupling constant along the legs and $J^\prime$ is that along the rungs. 
The exchange constants in the two-leg ladder compound SrCu$_2$O$_3$ are
estimated from LDA+U calculations.  We report the detailed crystal
structure of SrCu$_2$O$_3$ and of the three-leg ladder compound
Sr$_2$Cu$_3$O$_5$.  New experimental $\chi(T)$ data for the two-leg ladder
cuprates SrCu$_2$O$_3$ and LaCuO$_{2.5}$, and for the nominally two-leg
ladder vanadates CaV$_2$O$_5$ and MgV$_2$O$_5$ which are structurally
similar to SrCu$_2$O$_3$, are presented.  These and literature $\chi(T)$
data for these compounds and Sr$_2$Cu$_3$O$_5$ are modeled using the QMC
simulation fits.  For SrCu$_2$O$_3$, we find that $J^\prime/J \approx 0.5$
and $J/k_{\rm B} \approx 1900$\,K, assuming a spectroscopic splitting factor
$g \approx 2.1$, confirming the previous modeling results of D.~C.~Johnston,
Phys.\ Rev.\ B {\bf 54}, 13\,009 (1996).  The interladder coupling
$J^{\prime\prime\prime}/J = 0.01(1)$ perpendicular to the ladder layers is
found to be very weak and on the spin-gapped side of the quantum critical
point (QCP) at $J_{\rm QCP}^{\prime\prime\prime}/J = 0.048(2)$ for
$J^\prime/J = 0.5$.  Sr$_2$Cu$_3$O$_5$ is also found to exhibit strong
intraladder exchange anisotropy, with $J^\prime/J = 0.66(5)$ and $J/k_{\rm
B} \approx 1810(150)$\,K for $g = 2.1(1)$.  The $\chi(T)$ data for
LaCuO$_{2.5}$ are consistent with $J^\prime/J \approx 0.5$ with $J/k_{\rm
B} \approx 1700$\,K, again assuming $g \approx 2.1$, and with a 3D FM
interladder coupling $J^{\rm 3D}/J \approx -0.05$ which is close to and on
the AF ordered side of the QCP at $J_{\rm QCP}^{\rm 3D}/J = -0.036(1)$ for
$J^\prime/J = 0.5$, consistent with the observed AF-ordered ground state. 
The surpisingly strong spatial anisotropy of the bilinear exchange
constants within the cuprate ladders is discussed together with the results
of other experiments sensitive to this anisotropy.  Recent theoretical
predictions are discussed including those which indicate that a four-spin
cyclic exchange interaction within a Cu$_4$ plaquette is important to
determining the magnetic properties.  On the other hand, CaV$_2$O$_5$ is
found to be essentially a dimer compound with AF intradimer coupling
669(3)\,K ($g = 1.96$), in agreement with the results of M.~Onoda and
N.~Nishiguchi, J. Solid State Chem.\ {\bf 127}, 359 (1996).  The leg and
rung exchange constants found for isostructural MgV$_2$O$_5$ are very
different from those in CaV$_2$O$_5$, as predicted  previously from LDA+U
calculations.
\end{abstract}
\pacs{PACS numbers: 71.27.+a, 71.70.Gm, 75.30.Et, 75.50.Ee}
}

\section{Introduction}

Low-dimensional quantum spin systems have attracted much attention over the
past decade mainly due to their possible relevance to the mechanism for the
high superconducting transition temperatures in the layered cuprate
superconductors, which contain Cu$^{+2}$ spin $S = 1/2$
antiferromagnetic (AF) square lattice layers.  One avenue to approach the
physics of the two-dimensional (2D) AF square lattice Heisenberg
antiferromagnet is to study how the magnetic properties of AF
Heisenberg spin ladders evolve with increasing number of legs and/or with
coupling between them.\cite{reviews}  The study of such systems is also
interesting in its own right.  The spin-ladder field has been motivated and
guided by theory.  Odd-leg ladders with AF leg and rung couplings were
predicted to have no energy gap (``spin gap'') from the spin singlet $S =
0$ ground state to the lowest magnetic triplet $S = 1$ excited states (as
in the ``one-leg'' isolated chain), whereas, surprisingly, even-leg ladders
were predicted to have a spin gap for any finite AF rung coupling
$J^\prime$.
\cite{Dagotto1988,Dagotto1992,Strong1992,Barnes1993,Rice1993,Gopalan1994,%
White1994,Poilblanc1994,Bariev1994,Rojo1996,Greven1996,Sierra1996,%
Chakravarty1996,Oitmaa1996,%
Schollwock1996,Shelton1996,Arrigoni1996,DellAringa1997,Schmeltzer1998}
For even-leg ladders in which the ratio of the rung to leg exchange
constants is $J^\prime/J \lesssim 1$, the spin gap decreases exponentially
with increasing number of legs.
\cite{White1994,Poilblanc1994,Greven1996,Sierra1996,Chakravarty1996,%
DellAringa1997} 
A close relationship of these generic spin gap behaviors of $S = 1/2$ even-
and odd-leg AF Heisenberg spin ladders was established with AF integer-spin
and half-integer-spin Heisenberg chains, which are gapful and gapless,
respectively.
\cite{Sierra1996,Chakravarty1996,White1996} A spin gap also occurs for AF
leg coupling if $J^\prime$ is any finite ferromagnetic (FM) value,
although the dependence of the gap on the magnitude of $J^\prime$ is
different from the dependence when $J^\prime$ is AF; a second-order
transition between the two spin-gapped ground states occurs when the spin
gap is zero as the rung coupling passes from AF values through zero to
FM values.\cite{Wang1999}  Subsequent developments in the field involved a
close interaction of theory and experiment, mainly on oxide spin ladder
compounds.  Interest in the properties of oxide spin ladder materials was
stimulated by the stripe picture for the high-$T_{\rm c}$ layered cuprate
superconductors, in which the doped CuO$_2$ layers may be viewed as
containing undoped $S = 1/2$ AF $n$-leg spin ladders separated by domain
walls containing the doped charges.
\cite{Cho1992,Emery1993,Tranquada1995,Moshchalcov1999}  Undoped
non-oxide $S = 1/2$ two-leg AF spin ladders also exist in nature.  The best
studied of these is ${\rm Cu_2(C_5H_{12}N_2)_2Cl_4}$, for which the Cu-Cu
exchange interactions are much weaker than in the layered and spin ladder
cuprate compounds, which in turn has allowed extensive studies to be done
of the low-temperature magnetic field-temperature phase diagram and
associated critical spin dynamics.\cite{Chaboussant1998}  

To provide a basis for further understanding of the properties of both
undoped and doped spin ladders, it is important to determine the values of
the superexchange interactions present in undoped spin ladder compounds. 
The work reported here was motivated by the surprising inference by one of
us in 1996,\cite{Johnston1996} based on several theoretical analyses of the
magnetic susceptibility versus temperature $\chi(T)$ of the two-leg ladder
cuprate compound SrCu$_2$O$_3$, that the exchange interaction $J$ between
nearest-neighbor Cu spins-1/2 along the legs of the ladder is about a
factor of two stronger than the exchange interaction $J^\prime$ across the
rungs.  This conclusion strongly disagrees with $J^\prime/J \gtrsim 1$ as
expected from well-established ``empirical rules'' for superexchange in
oxides, and has wide ramifications for the understanding of superexchange
interactions not only in cuprate spin ladders but also, e.g., in the
high-$T_{\rm c}$ layered cuprate superconductors.

We therefore considered it important to conclusively test the modeling
results of Ref.~\onlinecite{Johnston1996}.  To do so, we carried out
extensive high-accuracy quantum Monte Carlo (QMC) simulations of $\chi(T)$
not only for isolated two-leg spin $S = 1/2$ Heisenberg ladders with
spatially anisotropic exchange as suggested in
Ref.~\onlinecite{Johnston1996}, but also for both isotropic and anisotropic
two-leg ladders coupled together in a two-dimensional (2D) stacked-ladder
configuration and in a 3D ``LaCuO$_{2.5}$-type'' configuration.  These QMC
simulations of $\chi(T)$ are in addition to QMC simulations that have
already been reported in the literature for isolated and spatially
isotropic two- and three-leg ladders and, as several of us reported
recently,\cite{Miyahara1998} for both isotropic and anisotropic intraladder
exchange in the 2D trellis layer coupled-ladder configuration present in
SrCu$_2$O$_3$.  In order to reliably and precisely model experimental
$\chi(T)$ data, we obtained high-accuracy analytic fits to all of these QMC
simulation data, and the fit functions and parameters are reported.  Two
additional calculations were carried out.  First, we computed the
one- and two-magnon dispersion relations for two-leg
$2\times 12$ ladders in the parameter range $0.5 \leq J^\prime/J \leq 1$
for use in determining exchange constants in two-leg ladder compounds from
inelastic neutron scattering data.  Second, for comparison with our
modeling results for SrCu$_2$O$_3$, the exchange constants in this
compound  were estimated using LDA+U calculations.

On the experimental side, we report the detailed crystal structure of
SrCu$_2$O$_3$, required as input to our LDA+U calculations for this
compound, along with that of the three-leg ladder compound
Sr$_2$Cu$_3$O$_5$.  New
$\chi(T)$ data are reported for the two-leg ladder cuprates SrCu$_2$O$_3$
and LaCuO$_{2.5}$, and for the two-leg ladder vanadates CaV$_2$O$_5$ and
MgV$_2$O$_5$ which have a trellis-layer structure similar to that of
SrCu$_2$O$_3$.  The intraladder, and in several cases the interladder,
exchange constants in each of these materials and in Sr$_2$Cu$_3$O$_5$ were
determined by fitting the new as well as previously reported $\chi(T)$ data
for one to three samples of each compound using our fits to the QMC
$\chi(T)$ simulation data.  In each of the three cuprate ladder compounds,
the intraladder exchange constants were found to be strongly anisotropic,
with $J^\prime/J \approx 0.5$--0.7, confirming the results of
Ref.~\onlinecite{Johnston1996}.

In the following sections we discuss in more detail the important previous
developments in theory and experiment on spin ladder oxide compounds
relevant to the present work, which is necessary to better place in
perspective our own comprehensive theoretical and experimental study on
undoped spin ladders, and then give the plan for the remainder of the
paper.  There is an extensive literature on the theory of doped spin ladders
which we will not cite or discuss except in passing.

\subsection{SrCu$_2$O$_3$ and Sr$_2$Cu$_3$O$_5$}
\begin{figure}
\epsfxsize=3in
\centerline{\epsfbox{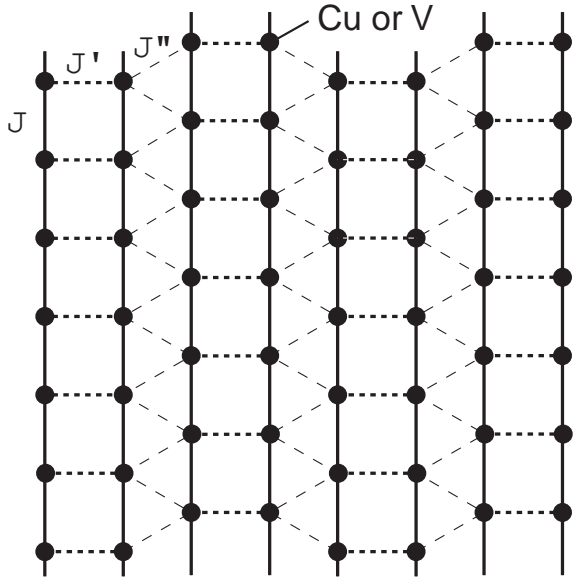}}
\vglue 0.1in
\caption{Sketch showing the basic trellis layer structure of the $M = $ Cu
and V sublattices in the two-leg ladder compound SrCu$_2$O$_3$ and in 
CaV$_2$O$_5$ and MgV$_2$O$_5$, respectively.  The figure shows the
intraladder leg (or chain) ($J$) and rung ($J^\prime$) exchange couplings
and the interladder ($J^{\prime\prime}$) coupling within the
trellis layer.  The intraladder diagonal coupling ($J^{\rm diag}$)
between second-nearest-neighbors within each $M_4$ plaquette of each ladder
and the nearest-neighbor interladder coupling ($J^{\prime\prime\prime}$)
between pairs of  $M$ atoms in adjacent stacked layers are not shown.}
\label{Fig01}
\end{figure}

\vglue0.12in
Rice, Gopalan and Sigrist\cite{Rice1993,Gopalan1994} recognized that a
class of undoped layered strontium cuprates discovered by \mbox{Hiroi} and
Takano, with general formula Sr$_{m-1}$Cu$_{m+1}$O$_{2m}$ ($m = 3$, 5,
$\ldots$),\cite{Hiroi1991,Kazakov1997,Kobayashi1997} may exhibit properties
characteristic of nearly isolated spin $S =1/2$ $n$-leg ladders with $n =
(m+1)/2$, due to geometric frustration between the ladders in the ``trellis
layer'' structure\cite{Hiroi1991} which effectively decouples the ladders
magnetically.  A sketch of the Cu trellis layer substructure of the $n = 2$
two-leg ladder compound SrCu$_2$O$_3$ is shown in
Fig.~\ref{Fig01}.  The above spin-gap 
\noindent predictions were
subsequently verified experimentally from $\chi(T)$ measurements on the
two-leg ladder compound SrCu$_2$O$_3$ which exhibits a spin-gap and for the
three-leg ladder compound Sr$_2$Cu$_3$O$_5$ which does
not.\cite{Azuma1994,Ishida1994,Kojima1995,Ishida1996,Azuma1998}   

Normand {\it et al.}\cite{Normand1997b} have investigated the ground state  
magnetic phase diagram of the trellis layer in exchange parameter space for
antiferromagnetic (AF) spin interactions.  They find spin-gap,
N\'eel-ordered and spiral ordered phases, depending on the relative
strengths of the interactions.  Thermodynamic and other properties of the
trellis layer with spatially isotropic coupling within each ladder have
been calculated using the Hubbard model by Kontani and
Ueda.\cite{Kontani1998}  At half filling,  with a ratio of interladder
($t^{\prime\prime}$) to isotropic intraladder ($t$)  hopping parameters
$t^{\prime\prime}/t = 0.15$ and with an on-site Coulomb repulsion parameter
$U$ given by $U/|t| =  2.9$, they find that a pseudogap opens in the
electronic density of states with decreasing temperature $T$, and that
$d$-wave superconductivity develops in the presence of this pseudogap at
lower $T$, similar to behaviors observed for ``underdoped'' high-$T_{\rm c}$
layered cuprate superconductors.

Experimental work on spin ladder systems has been strongly motivated by such
theoretical predictions that even-leg ladder systems (with spin gaps) may
exhibit a $d$-wave-like superconducting ground state via an electronic
mechanism when appropriately doped.
\cite{Dagotto1992,Rice1993,Kontani1998,Sigrist1994,Tsunetsugu1994,%
Khveshchenko1994,Hayward1995,Yanagisawa1995,Troyer1996,Hayward1996,%
Yoshioka1996,Kuroki1996,Noack1997,Kishine1997b,%
Riera1999,Schulz1998,Kishine1998}  
Theoretical studies suggest that superconductivity may also occur in
doped three-leg ladders which have no spin gap, but with interesting
subtleties.\cite{Kimura1996,Rice1997,Kimura1998,White1998}  Thus far, it has
not proved possible to dope SrCu$_2$O$_3$ or Sr$_2$Cu$_3$O$_5$ into the
superconducing state, although electron doping has been achieved by
substituting limited amounts of Sr by La in SrCu$_2$O$_3$.\cite{Azuma1995} 

Surprisingly, introducing small amounts of disorder in the Cu sublattice by
substituting nonmagnetic isoelectronic Zn$^{+2}$ for Cu$^{+2}$ in
Sr(Cu$_{1-x}$Zn$_x$)$_2$O$_3$ was found to destroy the spin gap and induce
long-range AF ordering at $T_{\rm N} \sim 3$--8\,K for $0.01 \leq x \leq
0.08$.\cite{Azuma1998,Azuma1997,Fujiwara1998,Ohsugi1999}  Specific heat
measurements above $T_{\rm N}$ for $x = 0.02$ and~0.04 indicated an
electronic specific heat coefficient $\gamma \sim 3.5$\,mJ/mol\,K$^2$ and a
gapless ground state; from this $\gamma$ value, an (average) exchange
constant in the ladders $J/k_{\rm B} \sim 1600$\,K was
derived.\cite{Azuma1997}  Many theoretical studies have been carried out on
site-depleted and otherwise disordered two-leg ladders to interpret these
experiments.
\cite{Fukuyama1996,Motome1996,Nagaosa1996,Sigrist1996,Iino1996,Imada1997,%
Fukui1997,Mikeska1997,Miyazaki1998,Laukamp1998,Greven1998,Orignac1998,%
Sachdev1999}   
The essential feature of the theoretical results is that the spin
vacancy induces a localized magnetic moment around it as well as a
static staggered magnetization that enhances the AF correlations between
the spins in the vicinity of the vacancy.  The enhanced staggered
magnetization fields around the respective spin vacancies interfere
constructively, resulting in a quasi-long range AF order along the ladder,
so that even weak interladder couplings are presumably sufficient to induce
3D AF long-range order at finite temperatures.  To our knowledge, no
quantitative calculations have yet been done of the 3D AF ordering
temperature $T_{\rm N}$ in the N\'eel-ordered regime versus interladder
coupling strengths for the 3D stacked trellis layer lattice spin coupling
configuration of the type present in SrCu$_2$O$_3$ either with or without
Zn doping.

\subsection{LaCuO$_{2.5}$}
\vglue0.08in
Another candidate for doping is the two-leg spin-ladder compound
LaCuO$_{2.5}$ (high-pressure
form),\cite{LaPlaca1993,Hiroi1995,Hiroi1996,Khasanova1996} which has the
oxygen-vacancy-ordered CaMnO$_{2.5}$ (Ref.~\onlinecite{Poeppelmeier1982})
structure.  The interladder exchange coupling is evidently stronger than in
SrCu$_2$O$_3$, since long-range AF ordering was observed to occur in
LaCuO$_{2.5}$ from $^{63}$Cu NMR and muon spin rotation/relaxation ($\mu$SR)
measurements at a N\'eel temperature $T_{\rm N} =
110$--125\,K.\cite{Matsumoto1996,Kadono1996}  Metallic hole-doped compounds
La$_{1-x}$Sr$_x$CuO$_{2.5}$ can be
formed,\cite{Hiroi1995,Hiroi1996,Otzschi1993} but superconductivity has not
yet been observed at ambient pressure above 1.8\,K for \mbox{$0 \leq x \leq
0.20$} or at high pressures up to
8\,GPa.\cite{Hiroi1995,Hiroi1996,Hiroi1999a}

\begin{figure}
\epsfxsize=2.5in
\centerline{\epsfbox{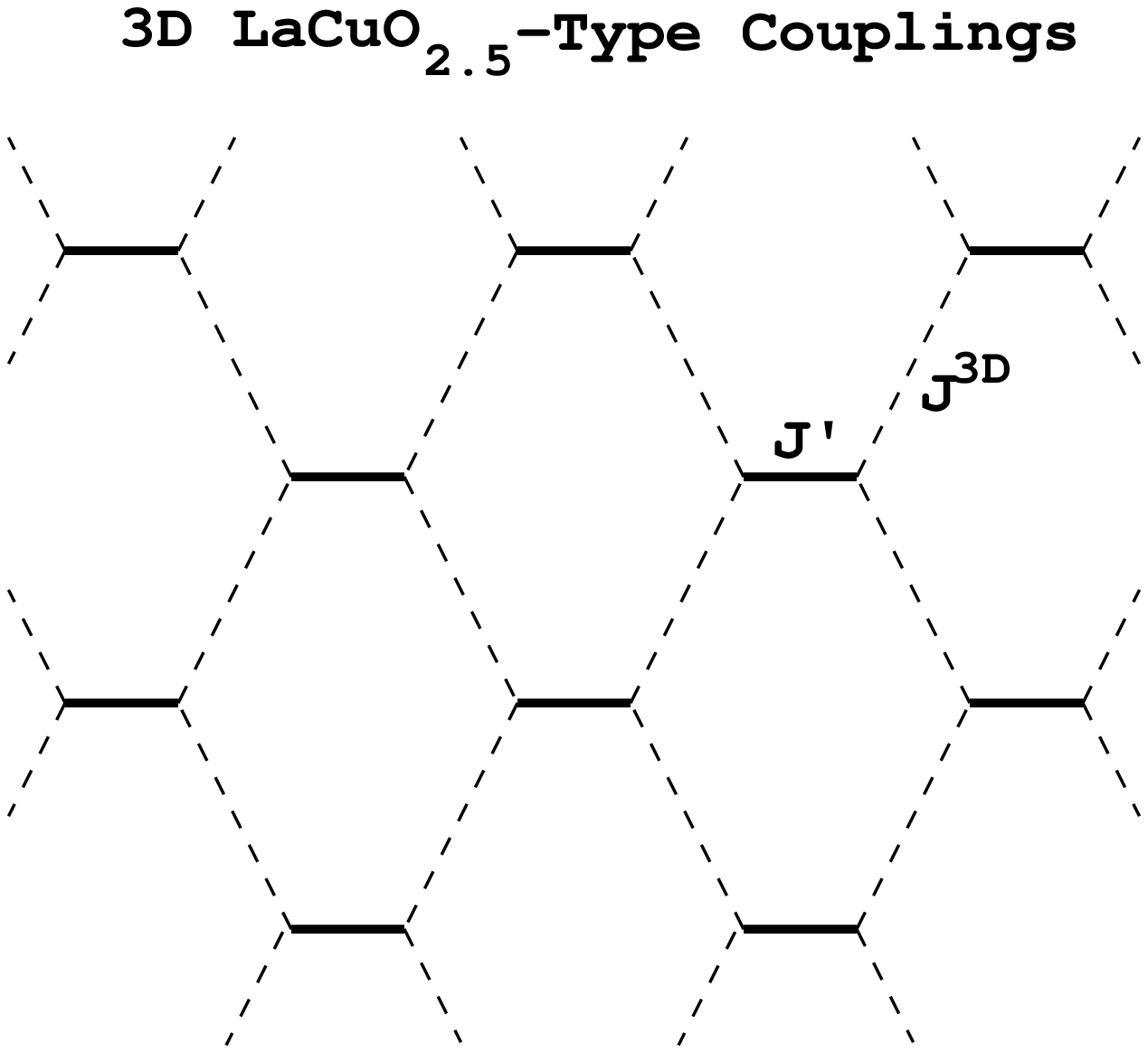}}
\vglue 0.1in
\caption{Exchange coupling topology
proposed\protect\cite{Normand1996,Troyer1997} for the two-leg ladder
compound LaCuO$_{2.5}$.  The ladder legs run perpendicular to the page,
along which the exchange coupling is $J$ (not shown).  The figure shows the
intraladder rung ($J^\prime$) and interladder ($J^{\rm 3D}$) exchange
couplings.}
\label{Fig02}
\end{figure}

Normand and coworkers have carried out detailed analytical calculations for
LaCuO$_{2.5}$.\cite{Normand1996,Normand1997,Normand1998}  The 3D 
exchange coupling topology proposed\cite{Normand1996,Troyer1997} for this
compound is shown in Fig.~\ref{Fig02}.  If the leg coupling $J = 0$, the
spin lattice is a 2D spatially anisotropic honeycomb lattice,
whereas if \,the \,rung \,coupling \ $J^\prime = 0$, one \,has \,a
2D~\,anisotropic 

\noindent square lattice.  From a tight-binding fit to the LDA band
structure,\cite{Mattheiss1996} Normand and Rice inferred that the ratio
of the rung to leg exchange coupling constants is $J^\prime/J \sim 1$, with
an AF interladder superexchange interaction $J^{\rm 3D}/J \approx
0.25$.\cite{Normand1996}  Then using a mean-field analysis of the spin
ground state, they found (for $J^\prime/J = 1$) that with increasing
$J^{\rm 3D}$ the spin gap disappears at a quantum critical point (QCP) at
$J^{\rm 3D}_{\rm QCP}/J = 0.121$ separating the spin-liquid from the
long-range  AF ordered ground state, and thereby inferred that the ground
state of LaCuO$_{2.5}$ is AF ordered ($J^{\rm 3D}/J$ is on the ordered side
of $J^{\rm 3D}_{\rm QCP}/J$).  The critical on-site Coulomb repulsion
parameter $U_{\rm c}$ necessary to induce AF order was found to be given by
$U_{\rm c}/W \approx 0.2$ where $W$ is the bandwidth; the small value of
this ratio is a reflection of the substantial 1D character of the bands,
the interladder coupling notwithstanding.

Troyer, Zhitomirsky and Ueda\cite{Troyer1997} confirmed using large-scale
QMC simulations of $\chi(T)$ for $J^\prime/J = 1$ that $J^{\rm 3D}_{\rm
QCP}/J \approx 0.11$, and confirmed the predicted\cite{Chubukov1994}
behavior $\chi\propto T^2$ (up to logarithmic corrections as three
dimensions is the upper critical dimension) at the 3D QCP\@.  Additional
calculations by Normand and Rice\cite{Normand1997} in the vicinity of the
QCP predicted that $\chi = \chi(0) + b T^2$ with further increases in
$J^{\rm 3D}/J$, where $\chi(0)$ increases with increasing $(J^{\rm
3D}-J^{\rm 3D}_{\rm QCP})/J$, consistent with the
simulations.\cite{Troyer1997}  Comparison of the calculated
$T_{\rm N}(J^{\rm 3D}/J)$ with the experimental results substantiated that 
$J^{\rm 3D}/J$ in LaCuO$_{2.5}$ is only slightly larger than $J^{\rm
3D}_{\rm QCP}/J$.\cite{Normand1997,Troyer1997}  Other calculations, using
as input the results of x-ray photoelectron spectroscopy measurements,
indicate that the interladder coupling is ferromagnetic (FM) with $|J^{\rm
3D}/J| < 0.1$,\cite{Mizokawa1997} rather than AF\@.  However, the same
generic behaviors near the QCP described above are expected regardless of
the sign of $J^{\rm 3D}$, which can be determined from magnetic neutron
Bragg diffraction intensities below $T_{\rm N}$;\cite{Normand1997} these
measurements have not been done yet.

The possibility of superconductivity occurring in the doped system
La$_{1-x}$Sr$_x$CuO$_{2.5}$ with $0.05 < x < 0.2$ was recently investigated
within spin fluctuation theory by Normand, Agterberg and Rice, who found
that
$d$-wave-like superconductivity should occur within this entire doping
range.\cite{Normand1998}  They  suggested that the reason that 
superconductivity has not been observed to date in this system may be
associated with crystalline imperfections and/or with the disruptive
influence of the intrinsic random disorder which occurs upon substituting
La by Sr.  They suggested that future improvements in the materials may
allow superconductivity to occur.

Two conclusions from Refs.~\onlinecite{Normand1997}
and~\onlinecite{Troyer1997} are important to the present experimental
$\chi(T)$ studies and modeling.  First, for intraladder exchange couplings
in the proximity of the QCP, the existence of relatively weak interladder
coupling does not change $\chi(T)$ significantly for $T\gtrsim J^{\rm
3D}/k_{\rm B}$ from that of the isolated ladders [except for the usual
mean-field shift of $\chi$ as in Eq.~(\ref{EqMFT:a}) below].  Second, the
onset of long-range AF ordering has a nearly unobservable effect on the
spherically-averaged $\chi(T)$ of polycrystalline samples, as was also
found\cite{Johnston1997} for the undoped high-$T_{\rm c}$ layered cuprate
parent compounds.  Although the studies of Refs.~\onlinecite{Normand1997}
and~\onlinecite{Troyer1997} were carried out primarily for ladders with
spatially isotropic exchange ($J^\prime/J = 1$), these conclusions are
general and do not depend on the precise value of $J^\prime/J$ within a
ladder.  They explain both why the experimental $\chi(T)$ data for
LaCuO$_{2.5}$ could be fitted assuming a spin gap\cite{Hiroi1995}
[Eq.~(\ref{EqTroyer}) below] even though this compound does not have one,
and why this experimental $\chi(T)$ study did not detect the AF ordering
transition at $T_{\rm N}$. 

\subsection{(Sr,Ca,La)$_{14}$Cu$_{24}$O$_{41}$}

A related class of compounds with general formula $A_{14}$Cu$_{24}$O$_{41}$
has been extensively investigated over the past several years, especially 
since large single crystals have become available.  The structure consists
of Cu$_2$O$_3$ trellis layers, as in SrCu$_2$O$_3$, alternating with
CuO$_2$ chain layers and $A$ layers, where the ladders and chains are both
oriented in the direction of the
$c$-axis.\cite{McCarron1988,Siegrist1988,Ohta1997}  The CuO$_2$ chains
consist of edge-sharing Cu-centered CuO$_4$ squares with an approximately
90$^\circ$ Cu-O-Cu bond angle, so from the Goodenough-Konamori-Anderson
superexchange rules the Cu-Cu superexchange interaction is expected to be
weakly FM.

From $\chi(T)$,\cite{Carter1996,Matsuda1996b} specific heat and polarized
and unpolarized neutron diffraction measurements,\cite{Matsuda1996b} the
undoped compound with $A_{14}$ = La$_6$Ca$_8$ shows long-range magnetic
ordering of the chain Cu spins below $T_{\rm N} = 12.20(5)$\,K, but the
detailed magnetic structure could not be solved.\cite{Matsuda1996b}  Similar
measurements on a single crystal of the slightly doped compound with
$A_{14}$ = La$_5$Ca$_9$ showed long-range commensurate AF ordering below
$T_{\rm N} = 10.5$\,K with FM alignment of the spins within the chains and
AF alignment between nearest-neighbor chains; the ordered Cu moment is
$\sim 0.2\,\mu_{\rm B}$/Cu with an intrachain FM-aligned O moment $\sim
0.02\,\mu_{\rm B}$/O.\cite{Matsuda1998}  Long-range AF ordering of the
chain-Cu spins has also been found in single crystals of the system with
$A_{14}$ = Sr$_{14-x}$La$_x$ for $x = 6$, 5 and~3 at $T_{\rm N} = 16$\,K,
12\,K and~2\,K, respectively,\cite{Kumagai1998} and for $A_{14}$ =
Sr$_{2.5}$Ca$_{11.5}$ at
$T_{\rm N} \approx 2.1$\,K.\cite{Nagata1998}  From Cu NMR and NQR
measurements on a Sr$_{2.5}$Ca$_{11.5}$Cu$_{24}$O$_{41}$ crystal, Ohsugi
{\it et al.}\ found that the Cu spins in the two-leg ladder trellis layers
have an ordered moment of only $\sim 0.02\,\mu_{\rm B}$, whereas the
ordered moment on the magnetic Cu sites in the chains is $\sim
0.56\,\mu_{\rm B}$.\cite{Ohsugi1999b}

The doped-chain compounds Ca$_{0.83}$CuO$_2$, Sr$_{0.73}$CuO$_2$,
Ca$_{0.4}$Y$_{0.4}$CuO$_2$ and Ca$_{0.55}$Y$_{0.25}$CuO$_2$, containing the
same type of edge-sharing CuO$_4$-plaquette CuO$_2$ chains as in the
$A_{14}$Cu$_{24}$O$_{41}$ materials, exhibit long-range AF ordering at
$T_{\rm N}
\approx 12.2$\,K, 10.0\,K, 29\,K and 23\,K,
respectively.\cite{Meijer1998,Matsuda1998b,Meijer1999,Hayashi1998,Hiroi1999b}

The stoichiometric compound Sr$_{14}$Cu$_{24}$O$_{41}$ is self-doped; the
average oxidation state of the Cu is ${+2.25}$, corresponding to a doping
level of 0.25 holes/Cu.  \mbox{McElfresh} {\it et al.}\cite{McElfresh1989}
found that single crystals show highly resistive semiconducting behavior
with an activation energy of 0.18~eV between 125 and 300\,K for conduction
in the direction of the chains and ladders.  Valence bond sum
calculations,\cite{Kato1996b} 
$\chi(T)$ (Refs.~\onlinecite{Carter1996,Matsuda1996}) and
ESR\cite{Matsuda1996} measurements indicated that the localized doped holes
reside primarily on O atoms within the CuO$_2$ chains, with their spins
forming nonmagnetic Zhang-Rice singlets\cite{Zhang1988} with chain Cu
spins.\cite{Carter1996} The remaining chain-Cu spins show a maximum in
$\chi(T)$ at $\sim 60$--80\,K (after subtracting a Curie $C/T$ term due to
$\sim 1.5$\% of isolated Cu defect spins) arising from short-range AF
ordering and the formation of a spin-gap  $\Delta/k_{\rm B} \sim
100$--150\,K;
\cite{Carter1996,McElfresh1989,Matsuda1996,Matsuda1996c} the ladders do not
contribute significantly to $\chi(T)$ below 300\,K due to their larger
spin-gap.  An inelastic neutron scattering investigation on single crystals
by Regnault {\it et al.}\cite{Regnault1998} found that at temperatures
below 150\,K, spin correlations develop within the chain layers.  The data
were modeled as arising from chain dimers with an AF Heisenberg intradimer
interaction $\approx 116$\,K, a FM interdimer intrachain interaction
$-$12.8\,K and an AF interdimer interchain interaction 19.7\,K\@.  Similar
measurements on a single crystal by Matsuda {\it et al.}\ yielded somewhat
different values of these three exchange constants.\cite{Matsuda1999b} 
Specific heat measurements\cite{Shaviv1990} from 5.7 to 347\,K and elastic
constant measurements\cite{Konig1997} on a single crystal from~5 to~110\,K
showed no evidence for any phase transitions.  An elastic constant study to
300\,K indicated broad anamolies in $c_{11}$ and $c_{33}$ at $\sim 110$\,K
and~230\,K, possibly associated with charge-ordering effects in the CuO$_2$
chains.\cite{Schwenk1999}

In the series Sr$_{14-x}$Ca$_x$Cu$_{24}$O$_{41}$, substituting
isoelectronic Ca for Sr up to $x = 8.4$ increases the
conductivity.\cite{Kato1994,Kato1996} At the composition $x = 8$, a
spin-gap $\Delta/k_{\rm B} = 140(20)$\,K was found on the chains which was
modeled as due to AF-coupled dimers comprising about 29\% of the Cu in the
chains,\cite{Carter1996} where $k_{\rm B}$ is Boltzmann's constant.  Osafune
{\it et al.}\cite{Osafune1997} inferred from optical conductivity
measurements that holes are transferred from the CuO$_2$ chains to the
Cu$_2$O$_3$ ladders with increasing $x$; high pressure enhances this
redistribution.\cite{Isobe1998,Motoyama1997}

The spin excitations in the Cu$_2$O$_3$ trellis layers in ${\rm Sr_{14}
Cu_{24}O_{41}}$ single crystals were studied using inelastic neutron
scattering by Eccleston {\it et al.}\cite{Eccleston1998} and \mbox{Regnault
{\it et al.}}\cite{Regnault1999} and in ${\rm
Sr_{2.5}Ca_{11.5}Cu_{24}O_{41}}$ single crystals using the same technique
by Katano {\it et al.}\cite{Katano1999}  The spin gaps for ladder spin
excitations were found to be $\Delta/k_{\rm B} = 377(1)$\,K, 370\,K and
372(35)\,K, respectively.  Essentially the same spin gap (380\,K) was
obtained by Azuma {\it et al.}\cite{Azuma1998} from inelastic neutron
scattering measurements on a polycrystalline sample of SrCu$_2$O$_3$.  The
good agreement among all four spin gap values indicates that the hole-doping
inferred\cite{Osafune1997} to occur in the Cu$_2$O$_3$ trellis layer
ladders in Sr$_{14}{\rm Cu_{24}O_{41}}$ and
${\rm Sr_{2.5}Ca_{11.5}Cu_{24}O_{41}}$ has little influence on the
ladder spin gap, consistent with $^{17}$O NMR results of Imai {\it et
al.}\cite{Imai1998}  Dagotto {\it et al.}\cite{Dagotto1998} recently
predicted theoretically that lightly hole-doped two-leg ladders should
exhibit two branches in the lowest-energy spin excitation spectra from
neutron scattering experiments, with different gaps for each occurring at
wavevector ($\pi,\pi$).  These two branches have evidently not (yet) been
observed or at least distinguished experimentally.

Many Cu NMR and NQR studies of the paramagnetic shifts and spin dynamics in
$A_{14}{\rm Cu_{24}O_{41}}$ compounds have been reported.
\cite{Kumagai1998,Imai1998,Tsuji1996,Kumagai1997,Carretta1997,Magishi1997,Takigawa1998,%
Carretta1998a,Carretta1998b,Magishi1998,Melzi1998}  Melzi and
Carretta,\cite{Melzi1998} Kishine and Fukuyama,\cite{Kishine1997} Ivanov
and Lee\cite{Ivanov1998} and Naef and Wang\cite{Naef1999} have discussed the
spin gaps obtained from these measurements and have presented analyses
which may explain why some of these inferred spin gaps do not agree with
each other and/or with the spin gaps derived independently from other
measurements such as inelastic neutron scattering and $\chi(T)$.

Superconductivity was discovered by Uehara {\it et al.}\cite{Uehara1996}
under high pressure (3--4.5 GPa) in Sr$_{14-x}$Ca$_x$Cu$_{24}$O$_{41}$ for
$x = 13.6$ at temperatures up to $\sim 12$\,K, and subsequently 
confirmed.\cite{Nagata1998,Isobe1998}  According to NMR measurements by
Mayaffre {\it et al.}, a spin gap is absent at high pressure in the
Cu$_2$O$_3$ ladders of the superconducting material,\cite{Mayaffre1998} a
result subsequently  studied theoretically.\cite{Schmeltzer1998b}  Metallic
interladder conduction within the Cu$_2$O$_3$ trellis layers occurs at high
pressure in a superconducting single crystal with $x =
11.5$.\cite{Nagata1998}  These results suggest a picture in which the
superconductivity originates from 2D metallic trellis layers with no spin
gap.  However, the ladder spin gap determined from $^{63}$Cu NMR
measurements by \mbox{Mito {\it et al.}}\ in a crystal with $x = 12$ at
ambient pressure and at 1.7\,GPa, when extrapolated into the
superconducting pressure region, suggested that the spin gap may persist in
the normal state at the pressures at which superconductivity is
found.\cite{Mito1999}  Further, inelastic neutron scattering measurements of
${\rm Sr_{2.5}Ca_{11.5}Cu_{24}O_{41}}$ single crystals under pressures up to
2.1\,GPa, which is somewhat below the pressure at which superconductivity is
induced, suggested that the spin gap does not change significantly with
pressure, although the scattered intensity decreases with increasing
pressure.\cite{Katano1999}  Thus whether the superconductivity occurs in
the presence of a spin gap or not is currently controversial.  

The crystal structure of the $A_{14}$Cu$_{24}$O$_{41}$ compounds can be
considered to be an ordered intergrowth of Cu$_2$O$_3$ spin ladder layers
and CuO$_2$ spin chain layers, and the composition can be written as
$[A_2{\rm Cu_2O_3]_7[CuO_2]_{10}}$.  A different configuration occurs as
$[A_2{\rm Cu_2O_3]_5[CuO_2]_7}$,\cite{Jensen1993} corresponding to the
overall composition  $A_{10}{\rm Cu_{17}O_{29}}$.  Single crystals of this
phase have been grown at ambient pressure with a deficiency ($\sim 10$\%) in
Cu and in which $A_{10}$ is a mixture of Sr, Ca, Bi, Y and Pb, and
sometimes Al, which were found to become superconducting at a temperature
of 80\,K at ambient pressure from both resistivity and magnetization
measurements.\cite{Leonyuk1999}

\subsection{CaV$_2$O$_5$ and MgV$_2$O$_5$}

The $d^1$ vanadium oxide CaV$_2$O$_5$ has a crystal
structure\cite{Onoda1996} containing (puckered) V$_2$O$_3$ trellis layers
with one additional O above or below each V atom, where adjacent two-leg
ladders are displaced to opposite sides of the trellis layer plane. 
CaV$_2$O$_5$ is a member of the $R$V$_2$O$_5$ ($R =$ Li, Na, Cs, Ca, Mg)
family of compounds, each of which exhibits interesting  low-dimensional
quantum magnetic properties.\cite{Ueda1998}   Because the structure is
similar to that of SrCu$_2$O$_3$, and CaV$_2$O$_5$ was found to possess a
spin gap from NMR measurements,\cite{Iwase1996} this vanadium oxide was
suggested to be a possible candidate for a $S = 1/2$ two-leg ladder
compound.\cite{Iwase1996}  However, \mbox{Onoda} and
Nishiguchi\cite{Onoda1996} found that the $\chi(T)$ could be fitted well by
the prediction for isolated dimers, with a spin singlet ground state and an
intradimer exchange constant and spin gap $J^\prime/k_{\rm B} =
\Delta/k_{\rm B} = 660$\,K\@.  On the other hand, Luke {\it et al.}\
concluded from $\mu$SR and magnetization measurements that spin freezing
occurs in the bulk of CaV$_2$O$_5$ below $\sim 50$\,K.\cite{Luke1998}  This
is most likely caused by impurities and/or defects in the samples.  The
spin gap is thus evidently destroyed or converted into a pseudogap upon
even a small amount of doping and/or disorder, as was also seen by Azuma
{\it et al.}\ in Zn-doped SrCu$_{2-x}$Zn$_x$O$_3$ as described above.

For the isostructural compound MgV$_2$O$_5$ which contains the same type of
V$_2$O$_3$ trellis layers as in
CaV$_2$O$_5$,\cite{Bouloux1976,Millet1998b,Onoda1998} Millet {\it et
al.}\cite{Millet1998} found from analysis of $\chi(T)$ measurements using
Eq.~(\ref{EqTroyer}) below that $\Delta/k_{\rm B} = 14.8$\,K, a remarkable
factor of 44 smaller than in CaV$_2$O$_5$.  $\mu$SR measurements did not
show any static magnetic ordering above 2.5\,K, and  $\chi(T)$ and
high-field ($\leq 30$\,T) magnetization measurements  yielded a value
$\Delta/k_{\rm B} \sim 17$\,K,\cite{Isobe1998b} similar to the result by
Millet {\it et al.} Inelastic neutron scattering measurements indicated a
gap $\Delta/k_{\rm B} \sim 20$\,K at a wave vector of
$(\pi,\pi)$,\cite{Isobe1998b}  which however need not be the magnon
dispersion minimum because  strong frustration effects can shift the spin
gap minimum away from this
wavevector.\cite{Miyahara1998,Normand1997b,Lidsky1998} Substituting V by up
to 10\% Ti introduces a strong local moment Curie ($C/T$) contribution to
$\chi(T)$; however, no long-range AF ordering was induced in contrast to
lightly Zn-doped SrCu$_2$O$_3$.\cite{Isobe1998c}  The exchange interactions
in CaV$_2$O$_5$ and MgV$_2$O$_5$ have recently been estimated by three of us
using LDA+U calculations.\cite{Korotin1999}  Additional calculations were
carried out which explain why the spin gaps and exchange interactions are
so different in these two compounds.\cite{Korotin2000} 

\subsection{Exchange Couplings from Fits of the Uniform Susceptibility by 
Theoretical Models}
\label{SecExchCoup}
\vglue0.05in
Of particular interest in this paper are the signs, magnitudes and spatial
anisotropies of the exchange interactions between the transition metal
spins in undoped spin-ladder oxide compounds.  These interactions have a
direct bearing on the electronic (including superconducting) properties
predicted for the doped materials and are of intrinsic interest in their
own right.  The primary experimental tool we employ here is $\chi(T)$
measurements.  The first $\chi(T)$ measurements on SrCu$_2$O$_3$ by
\mbox{Azuma {\it et al.}}\cite{Azuma1994} were modeled by the
low-temperature approximation to $\chi(T)$ of a spin $S = 1/2$ two-leg
ladder derived by Troyer, Tsunetsugu, and  W\"{u}rtz\cite{Troyer1994}
\begin{equation}
\chi(T) = {A\over\sqrt{T}}\,{\rm e}^{-\Delta/(k_{\rm B}T)}~,
\label{EqTroyer}
\end{equation} where a spin-gap $\Delta/k_{\rm B} = 420$\,K was found and
$k_{\rm B}$ is Boltzmann's constant.  If one assumes spatially isotropic
exchange interactions
$J^\prime = J$ within isolated two-leg ladders, then using $\Delta/J
\approx 0.5$ appropriate to this case (see Sec.~\ref{SecSimFits}) yields
$J/k_{\rm B}
\approx 840$\,K, about a factor of two smaller than in the layered high
$T_{\rm c}$ cuprate parent compounds.\cite{Johnston1997}  On the other hand,
one of us inferred from analysis of the value of the prefactor $A$, which
is not an adjustable parameter but instead is a function of $\Delta$ which
in turn is a function of $J$ and $J^\prime$, and from fits to the $\chi(T)$
data by numerical calculations for isolated ladders, all assuming a
$g$-factor $g = 2.1$, that a strong anisotropy exists between the rung
coupling constant $J^\prime$ and the leg coupling constant $J$: $J^\prime/J
\sim 0.5$, $J/k_{\rm B} \sim 2000$\,K, for which the spin-gap is similar to
that cited above.\cite{Johnston1996}  If confirmed, which we in fact do
here for  SrCu$_2$O$_3$ and LaCuO$_{2.5}$ as well as for the three-leg
ladder cuprate Sr$_2$Cu$_3$O$_5$, this suppression of $J^\prime$ with
respect to $J$ is predicted to suppress superconducting correlations in the
doped spin-ladders.\cite{Dagotto1992,Tsunetsugu1994,Noack1997,Riera1999}

The large spatial anisotropy in the exchange interactions and the large
value of $J$ inferred in Ref.~\onlinecite{Johnston1996} for SrCu$_2$O$_3$ 
were very surprising.  The Cu-Cu distance across a rung in this compound is
3.858\,\AA, and that along a leg is 3.934\,\AA\ (see the crystal structure
refinement data in Sec.~\ref{SecSrCu2O3Struct}), so if the nearest-neighbor
Cu-Cu distance were the only criterion for determining the exchange
constants, one would have expected $J^\prime/J > 1$, not $J^\prime/J \ll
1$.  This inference is strengthened when one notes that the Cu-O-Cu bond
angle across a rung is
$180^\circ$, whereas that along a leg is smaller (174.22$^\circ$).  Further,
the Cu-Cu distance in the layered cuprates is $\approx 3.80$\,\AA, shorter
than either the rung or leg Cu-Cu distance in SrCu$_2$O$_3$, with similar
$\approx 180^\circ$ Cu-O-Cu bond angles, and it is well-established that
$J/k_{\rm B} \approx 1500$\,K in the undoped layered cuprate parent
compounds,\cite{Johnston1997} so on this basis one would expect $J$ and
$J^\prime$ in SrCu$_2$O$_3$ to both be smaller than this value, not one of
them much larger.  On the other hand, an O ion in a rung of a ladder in
SrCu$_2$O$_3$, with two Cu nearest neighbors, is not crystallographically
or electronically equivalent to an O ion in a leg, with three Cu nearest
neighbors, so the respective superexchange constants $J$ and $J^\prime$
involving these different types of O ions are not expected to be
identical.  Second, the experimentally inferred spin-gap is approximately
reproduced assuming $J^\prime/J \sim 0.5$ and $J/k_{\rm B} \sim 2000$\,K,
as noted above.  Third, the nearly ideal linear Heisenberg chain compound
Sr$_2$CuO$_3$ with $180^\circ$ Cu-O-Cu bonds has a Cu-Cu exchange constant
estimated from $\chi(T)$ data as $J/k_{\rm B} = 2150^{+150}_{-100}$\,K,
\cite{Johnston1997,Ami1995,Motoyama1996,Johnston1997a} and from optical
measurements as 2850--3000\,K,\cite{Suzuura1996} which are much larger than
in the layered cuprates even though the Cu-Cu distance along the chain is
3.91\,\AA, significantly larger than the 3.80\,\AA\ in the layered
cuprates; this itself is surprising.

Regarding the subject of the present paper, it is important to keep in mind
that fitting experimental $\chi(T)$ data by theoretical predictions for a
given model Hamiltonian can test consistency with the assumed model, but
{\it cannot} prove uniqueness of that model.  A recent example in the
spin-ladder area clearly illustrates this point.  The V$^{+4}\ d^1$
compound vanadyl pyrophosphate, ${\rm (VO)_2P_2O_7}$ (``VOPO''), has an
orthorhombic crystal structure\cite{Gorbunova1979,Johnson1984,Hiroi1999}
which can be viewed crystallographically as containing $S = 1/2$ two-leg
ladders.\cite{Johnston1987}  However, the $\chi(T)$ was initially fitted by
one of us to high precision by the prediction for the $S = 1/2$ AF
alternating-exchange Heisenberg chain;\cite{Johnston1987} a spin-ladder
model fit was not possible at that time (1987) due to lack of theoretical
predictions for $\chi(T)$ of this model.  When such calculations were
eventually done,\cite{Barnes1994} it was found that the same experimental
$\chi(T)$ data set\cite{Johnston1987} could be fitted by the spin ladder
model to the same high precision as for the very different
alternating-exchange chain model.\cite{Barnes1994}  Inelastic neutron
scattering measurements on a polycrystalline sample reportedly confirmed
the spin-ladder model.\cite{Eccleston1994}  However, subsequent inelastic
neutron scattering results on single crystals proved that ${\rm
(VO)_2P_2O_7}$ is not a spin-ladder compound.\cite{Garrett1997a}  The
current evidence again indicates that ${\rm (VO)_2P_2O_7}$ may be an
alternating-exchange chain compound,\cite{Garrett1997a} although the
compound has continued to be studied both experimentally
\cite{Garrett1997a,Garrett1997b,Prokofiev1998,Wolf1999,Grove2000}
and theoretically,\cite{Uhrig1998,Damle1999} and an alternative 2D model has
been proposed.\cite{Weisse1998}  Recent $^{31}$P and $^{51}$V NMR and
high-field magnetization measurements have indicated that there are two
magnetically distinct types of alternating-exchange V chains in ${\rm
(VO)_2P_2O_7}$, interpenetrating with each other, each with its own spin
gap;\cite{Kikuchi1999} this finding has important implications for the
interpretation of the neutron scattering data.  A high-pressure phase of
${\rm (VO)_2P_2O_7}$ was recently discovered by Azuma {\it et al.}\ which
has a simpler structure containing a single type of $S = 1/2$ AF
alternating-exchange Heisenberg chain.\cite{Azuma1999}

\subsection{Plan of the Paper}

Herein we report a combined theoretical and experimental study of the
$\chi(T)$ of $S = 1/2$ spin-ladders and spin-ladder oxides.  Extensive new
quantum Monte Carlo (QMC) simulations of $\chi(T)$ are presented in
Sec.~\ref{SecSimulations} for isolated two-leg ladders with spatially
anisotropic intraladder exchange, including a FM diagonal second-neighbor
intraladder coupling in addition to $J$ and $J^\prime$, and of two-leg
ladders interacting with each other with stacked ladder (for $J^\prime/J =
0.5,\ 1$) and proposed\cite{Normand1996,Troyer1997} 3D LaCuO$_{2.5}$-type
interladder exchange configurations (for $J^\prime/J = 0.5$).  We
discuss the previous QMC simulation data of Frischmuth, Ammon and
Troyer\cite{Frischmuth1996} for the isolated ladder with
$J^\prime/J = 1$ and for three-leg ladders with spatially isotropic and
anisotropic intraladder couplings, of Miyahara {\it et
al.}\cite{Miyahara1998} for anisotropic two-leg ladder trellis layers and
of Troyer, Zhitomirsky and Ueda\cite{Troyer1997} for isotropic ($J^\prime/J
= 1$) two-leg ladders with 3D LaCuO$_{2.5}$-type interladder couplings. 
In this section we also obtain accurate estimates for $J^\prime/J
= 0.5$ and~1 of the values of the interladder exchange interactions at
which quantum critical points occur for the 2D stacked ladder exchange
coupling configuration and for ladders coupled in the 3D
${\rm LaCuO_{2.5}}$-type configuration.

A major part of the present work was obtaining a functional form to
accurately and reliably fit, interpolate and extrapolate the
multidimensional ($T$ and one or two types of exchange constants) QMC
$\chi(T)$ simulation data.  Four sections are devoted to this topic, where
we discuss and incorporate into the fit function some of the physics of
spin ladders.  The spin gap of the isolated two-leg ladder is part of our
general fit function, so in Sec.~\ref{SecSpinGap} we obtain accurate
analytic fits to the reported literature data for the spin gap versus the
intraladder exchange constants $J$ and $J^\prime$.  High temperature series
expansions (HTSEs) of $\chi(T)$ for isolated and coupled isotropic and
anisotropic ladders are considered in Sec.~\ref{SecHTSE}.  The first few
terms of the general HTSE for an arbitrary Heisenberg spin lattice with
spatially anisotropic exchange, which to our knowledge have not been
reported before, are given and are incorporated into the fit function so
that the function can be accurately extrapolated to arbitrarily high
temperatures, and also so that the function can be used to reliably fit QMC
data sets which contain few or no data at high temperatures.  In this
section we also give the HTSE to lowest order in $1/T$ for the magnetic
contribution to the specific heat of $S = 1/2$ spin ladders, and correct an
error in the literature.  The fit function itself that we use for most of
the fits to the QMC $\chi(T)$ data is then presented and discussed in
Sec.~\ref{SecGenFitFcn}.  In special cases where sufficient low-temperature
QMC data are not available, a fit function containing a minimum number
(perhaps only zero, one or two) of fitting parameters must be used, so in
Sec.~\ref{SecMFT} we consider such functions formulated on the basis of the
molecular field approximation.  The Appendix gives the detailed procedures
we used to fit our new QMC $\chi(T)$ simulation results and the previously
reported QMC data cited above.  Tables containing the fitted parameters
obtained from all fifteen of the \mbox{one-,} two- and three-dimensional
fits to the various sets of QMC data are also given in the Appendix.  We
hope that the general fit function and the extensive high-accuracy fits we
have obtained will prove to be generally useful to both theorists and
experimentalists working in the spin ladder and low-dimensional magnetism
fields.

Our calculations of the one- and two-magnon dispersion relations for
isolated two-leg ladders with \mbox{$0.5 \leq J^\prime/J \leq 1$}, in
increments of 0.1, are presented in Sec.~\ref{SecDisprsnRlns}; these extend
the earlier calculations by Barnes and Riera\cite{Barnes1994} for
$J^\prime/J = 0.5$, 1 and~2.  Additionally we discuss the influence of
interladder couplings on these dispersions.  Dynamical spin structure
factor $S(\bbox{q},\omega)$ calculations for $J^\prime/J = 0.5$, the
experimentally relevant exchange constant ratio, are also presented in this
section and compared with previous related work.  These calculations show
that two-magnon excitations should be included in the modeling of
inelastic neutron scattering data when using such data to derive the full
one-magnon triplet dispersion relation including the higher-energy part. 
Our calculations of the intraladder and interladder exchange constants in
SrCu$_2$O$_3$, obtained using the LDA+U method, are presented in
Sec.~\ref{SecLDA+U}.

We begin the experimental part of the paper by presenting in
Sec.~\ref{SecSrCu2O3Struct} a structure refinement of SrCu$_2$O$_3$,
necessary as input to the LDA+U calculations, and also of
Sr$_2$Cu$_3$O$_5$.  In Sec.~\ref{SecExpDatModeling}, we present our new
experimental $\chi(T)$ data for SrCu$_2$O$_3$, LaCuO$_{2.5}$, CaV$_2$O$_5$
and MgV$_2$O$_5$ and estimate the exchange constants in these compounds and
in Sr$_2$Cu$_3$O$_5$ by modeling the respective $\chi(T)$ data using our
fits to the QMC $\chi(T)$ simulation results.

The paper concludes in Sec.~\ref{SecSummaryDisc} with a summary and
discussion of our theoretical and experimental results, their relationships
to previous work and a discussion of how the presence of four-spin cyclic
exchange, as presented in the literature, can affect the magnetic
properties of spin ladders and the exchange constants derived assuming the
presence of only bilinear exchange interactions.  Further implications of
the presence of this cyclic exchange interaction are also discussed.

\section{Quantum Monte Carlo Simulations}
\label{SecSimulations}
\vglue-0.1in
Throughout this paper, the Heisenberg Hamiltonian for bilinear exchange
interactions between spins is assumed,
\begin{equation} {\cal H} = \sum_{<ij>} J_{ij}\,\bbox{S}_i\cdot\bbox{S}_j~,
\label{EqHeisHam}
\end{equation} where $J_{ij}$ is the exchange constant linking spins
$\bbox{S}_i$ and
$\bbox{S}_j$, $J_{ij}$ is positive (negative) for AF (FM) coupling and the
sum is over distinct exchange bonds.  For notational convenience, we define
the reduced spin susceptibility $\chi^*$, reduced temperature $t$ and
reduced spin gap $\Delta^*$ as
\begin{mathletters}
\label{EqRedPars:all}
\begin{equation}
\chi^* \equiv \frac{\chi^{\rm spin} J^{\rm max}}{N g^2\mu_{\rm B}^2}~,
\label{EqRedPars:a}
\end{equation}
\begin{equation} t \equiv \frac{k_{\rm B} T}{J^{\rm max}}~,
\label{EqRedPars:b}
\end{equation}
\begin{equation}
\Delta^*\equiv \frac{\Delta}{J^{\rm max}}~,
\end{equation}
\end{mathletters} where $\chi^{\rm spin}$ is the magnetic spin
susceptibility, $J^{\rm max}$ is the largest (AF) exchange constant in the
system, $N$ is the number of spins,
$g$ is the spectroscopic splitting factor, $\mu_{\rm B}$ is the Bohr
magneton and $k_{\rm B}$ is Boltzmann's constant.  All of the QMC
simulations presented and/or discussed here are for spins $S = 1/2$.  We
summarize below the definitions of the exchange constants to be used
throughout the rest of the paper:
\begin{eqnarray} J~~~~&  {\rm Nearest-neighbor,\ leg}& ~~2\times\nonumber\\
J^\prime~~~~& {\rm Nearest-neighbor,\ rung}& ~~1\times\nonumber\\ J^{\rm
diag}~~~&  {\rm Second-neighbor,}& ~~2\times\nonumber\\  &{\rm diagonal\
intraladder}\label{EqJDefs}\\ J^{\prime\prime}~~~&  {\rm Trellis\ layer,\
interladder}& ~~2\times\nonumber\\ J^{\prime\prime\prime}~~~&  {\rm
Stacked\ ladder,\ interladder}& ~~2\times\nonumber\\ J^{\rm 3D}~~&  {\rm
3D\ interladder,}& ~~2\times\nonumber\\ & {\rm LaCuO_{2.5}-type~.}\nonumber
\end{eqnarray}

The uniform susceptibilities $\chi(T)$ of $S = 1/2$ Heisenberg spin ladder
models were simulated using the continuous time version of the quantum
Monte Carlo (QMC) loop  algorithm.\cite{Loop}  This algorithm uses no
discretization of the imaginary  time direction and the only source of
systematic errors is thus  finite size effects.  The lattice sizes were
chosen large enough so  that these errors are much smaller than the
statistical errors of the  QMC simulations.  The simulations of the trellis
layer suffer from the ``negative sign problem'' caused by the frustrating
interladder interaction $J^{\prime\prime}$. Improved
estimators\cite{Ammon1998} were used to lessen the sign problem in this
case.\cite{Miyahara1998} 

\begin{figure}
\epsfxsize=3.3in
\centerline{\epsfbox{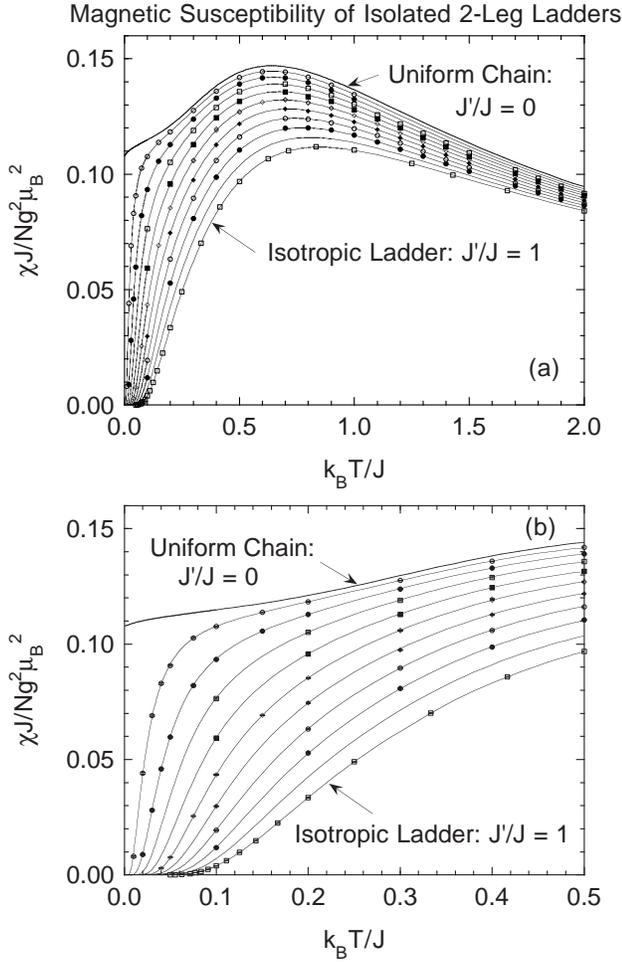}}
\vglue 0.1in
\caption{(a) Quantum Monte Carlo (QMC) $\chi(T)$ simulation data  for
isolated two-leg ladders with $J^\prime/J = 0.1$, 0.2, 0.3, 0.4, 0.5, 0.6,
0.7, 0.8 and 1.0 (symbols, top to bottom).  Additional simulation data for
$J^\prime/J = 0.25,\ 0.35,\ 0.45$ and~0.55 are not shown.  The data for
$J^\prime/J = 1$ are the QMC data of Frischmuth {\it et
al}.\protect\cite{Frischmuth1996}  The set of solid curves through the data
points is a two-dimensional fit to all the QMC data.  Also shown are the fit
for the $S = 1/2$ uniform Heisenberg chain\protect\cite{Johnston1999}
($J^\prime/J = 0$) and an interpolation curve for $J^\prime/J = 0.9$. 
(b)~Expanded plots at low temperatures of the data, now including error
bars, and curves in (a).}
\label{Fig03}
\end{figure}

\subsection{Isolated Ladders}

$\chi^*(t)$ was simulated for isolated two-leg ladders of size $2\times 200$
spins~1/2 with $J^\prime/J = 0.1$, 0.2, 0.25, 0.3, $\ldots$, 0.6, 0.7 and
0.8 with maximum temperature range $t = 0.01$ to~3.0, comprising 348 data
points; here, $J^{\rm max} = J$.  A selection of the results for $t
\leq 2$ in $J^\prime/J$ increments of 0.1 is shown as open and filled
symbols in Fig.~\ref{Fig03}(a) along with the QMC simulation
results of Frischmuth {\it et al}.\cite{Frischmuth1996} for $J^\prime/J =
1$ (30 data points from $t = 0.05$ to 5).  Expanded plots of the data, now
including error bars, are shown in Fig.~\ref{Fig03}(b), where
the error bars are seen to be on the order of or smaller than the size of
the data point symbols.  

\begin{figure}
\epsfxsize=3.3in
\centerline{\epsfbox{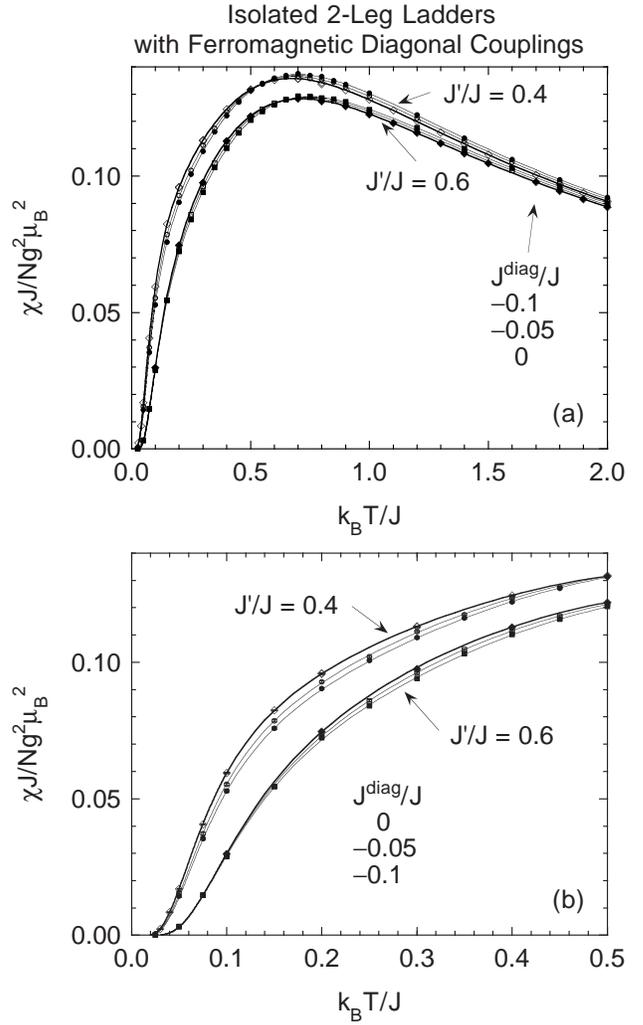}}
\vglue 0.1in
\caption{(a) Quantum Monte Carlo simulations of the magnetic spin
susceptibility $\chi$ vs\ temperature $T$ for isolated $S = 1/2$ two-leg
ladders with $J^\prime/J = 0.4$ and~0.6 and each with ferromagnetic
second-neighbor diagonal couplings $J^{\rm diag}/J = 0,\ -0.05$ and $-0.1$
(data for $J^{\rm diag}/J = -0.111$ are not shown).  The set of solid curves
is a three-dimensional fit to all the data, including those not shown for
$J^\prime/J = 0.45$, 0.5, 0.55 and~0.65, each with $J^{\rm diag}/J = 0,\
-0.05,$ $-0.1$ and $-0.111$.  (b) Expanded plots at low temperatures of the
data and fit in (a), where the error bars on the data points are also
plotted.}
\label{Fig04}
\end{figure}

We have also simulated $\chi^*(t)$ for isolated two-leg ladders with
$J^\prime/J = 0.4$, 0.45, $\ldots$, 0.65, each with ferromagnetic diagonal
coupling $J^{\rm diag}/J = -0.05,\ -0.1$ and~$-0.111$, and with maximum
temperature range $t = 0.025$ to~2, comprising 457 data points.  This FM
sign of the diagonal intraladder coupling was motivated by the LDA+U
results reported below in Sec.~\ref{SecLDA+U}.  The QMC results with error
bars for $J^\prime/J = 0.4$ and~0.6 and $J^{\rm diag}/J = -0.05$ and~$-0.1$
are shown as the symbols in Fig.~\ref{Fig04} along with our
results above for $J^\prime/J = 0.4$ and~0.6 and $J^{\rm diag} = 0$.  The
$\chi^*(t)$ is seen to be only weakly affected by the presence of $J^{\rm
diag}$.  \ Hence one expects that
\begin{figure}
\epsfxsize=3.3in
\centerline{\epsfbox{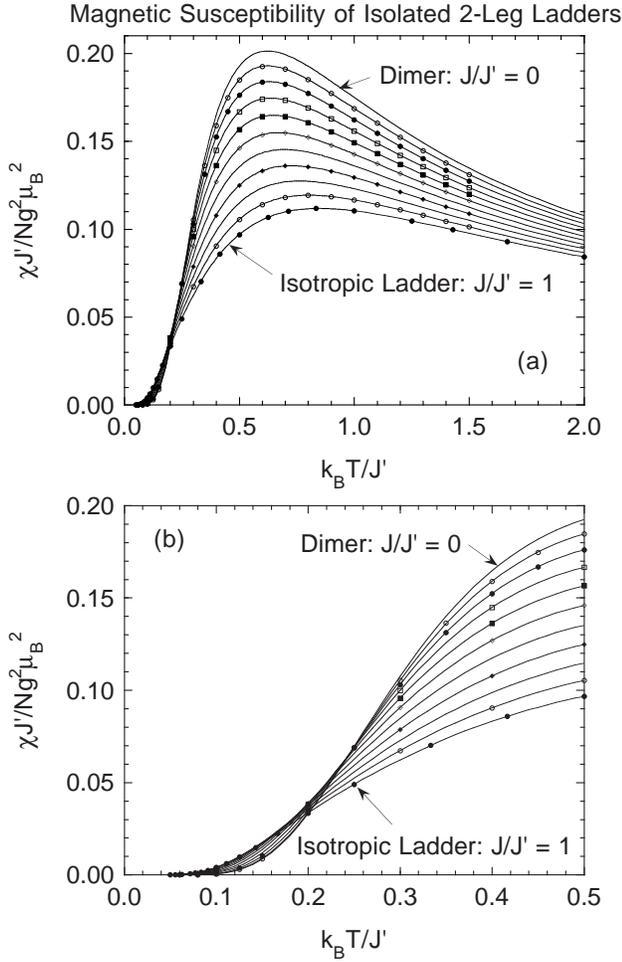}}
\vglue 0.1in
\caption{Quantum Monte Carlo simulations of $\chi(T)$ for isolated ladders
with $J/J^\prime = 0.1$, 0.2, 0.3, 0.4, 0.5, 0.7, 0.9 and 1.0 (symbols, top
to bottom).  The data for $J/J^\prime = 1$ are from Frischmuth {\it et
al}.\protect\cite{Frischmuth1996}  In all cases, the error bars are smaller
than the data symbols.  The set of solid curves is a two-dimensional fit to
all the data.  The solid curve for $J/J^\prime = 0$ is a fit to the exact
theoretical $\chi(T)$ for the isolated dimer.  Interpolation curves for
$J/J^\prime = 0.6$ and~0.8 are also shown.}
\label{Fig05}
\end{figure}
\noindent fits of experimental $\chi(T)$ data for spin ladder compounds by
the simulations will not be capable of determining $J^{\rm diag}$
quantitatively if $|J^{\rm diag}/J| \ll 1$.

For stronger interchain couplings $J^\prime/J > 1$, $\chi^*(t)$ was
simulated for isolated two-leg ladders with $J/J^\prime = 0.1$, 0.2, 0.3,
0.4, 0.5, 0.7 and 0.9 with maximum temperature range $t = 0.06$ to~1.5,
comprising 119 data points; here, in Eqs.~(\ref{EqRedPars:all}) one has
$J^{\rm max} = J^\prime$.  The results with error bars are shown as open
and filled symbols in Fig.~\ref{Fig05}, along with the QMC
simulation results of Frischmuth {\it et al}.\cite{Frischmuth1996} for
$J/J^\prime = 1$.  The susceptibility $\chi^{*,{\rm dimer}}(t)$ of the
isolated antiferromagnetically-coupled
$S = 1/2$ dimer ($J/J^\prime = 0$) is shown for comparison, where
\begin{equation}
\chi^{*,{\rm dimer}}(t) = {1\over t(3 + {\rm e}^{1/t})}~~.
\label{EqChiDimer}
\end{equation}
\begin{figure}
\epsfxsize=3.3in
\centerline{\epsfbox{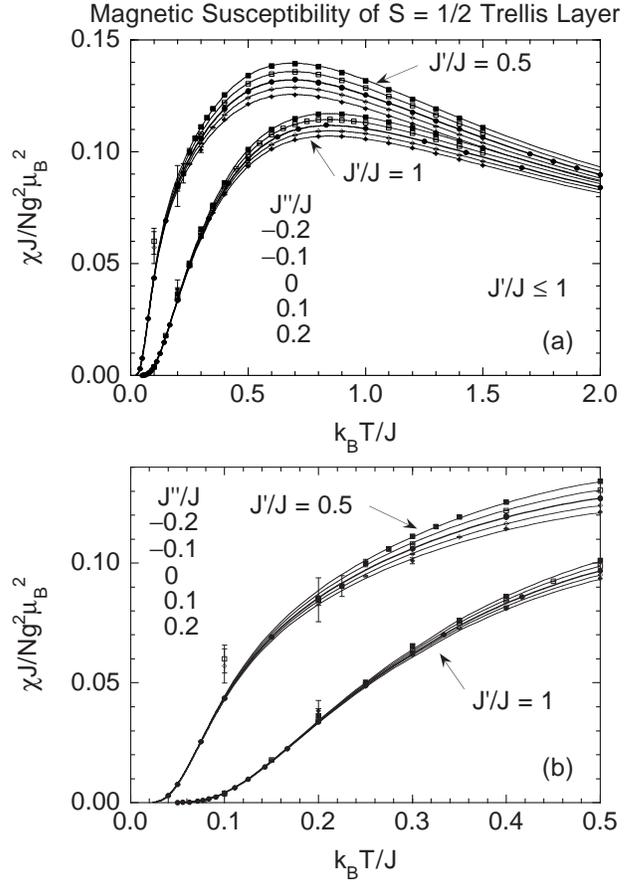}}
\vglue 0.1in
\caption{(a) Quantum Monte Carlo $\chi$ data (symbols) vs $T$ for the
trellis layer with $J^\prime/J = 0.5$ and~1, where the set of solid curves
for $J^{\prime\prime} \neq 0$ is the molecular-field-theory prediction
obtained from the fit for $J^{\prime\prime} = 0$ with no adjustable
parameters.  (b) Expanded plots at low temperatures of the data and fit in
(a).  The error bars for all data points are shown in both~(a) and~(b).}
\label{Fig06}
\end{figure}

The $\chi^*(t)$ for the $S = 1/2$ uniform
Heisenberg chain ($J^\prime/J = 0$) was obtained essentially exactly by
Eggert, Affleck and Takahashi in 1994.\cite{Eggert1994}  A fit to the
recently refined numerical calculations of
Kl\"umper\cite{Klumper1998,Johnston1999} for this chain in the temperature
range $0.01\leq t\leq 5$ is shown in Fig.~\ref{Fig03} for comparison with
the data; this high-accuracy (15\,ppm rms fit deviation) seven-parameter
analytical fit to these calculated data, using the fitting scheme in
Sec.~\ref{SecSimFits} below, was obtained (``Fit~1'') in
Ref.~\onlinecite{Johnston1999}.

\subsection{Coupled Ladders}

\subsubsection{Trellis Layer Interladder Interactions}

Miyahara {\it et al.}\cite{Miyahara1998} carried out QMC simulations of
$\chi^*(t)$ for trellis layers with $J'/J = 0.5$ (64 data points) and~1 (72
data points) over a maximum temperature range $0.1\leq t\leq 1.5$,  with
trellis layer interladder couplings $J^{\prime\prime}/J \,= \,-0.2,\
\,-0.1,\ \,0.1$ \,and~\,0.2 \,for \,each $J'/J$ \,value, 
\begin{figure}
\epsfxsize=3.3in
\centerline{\epsfbox{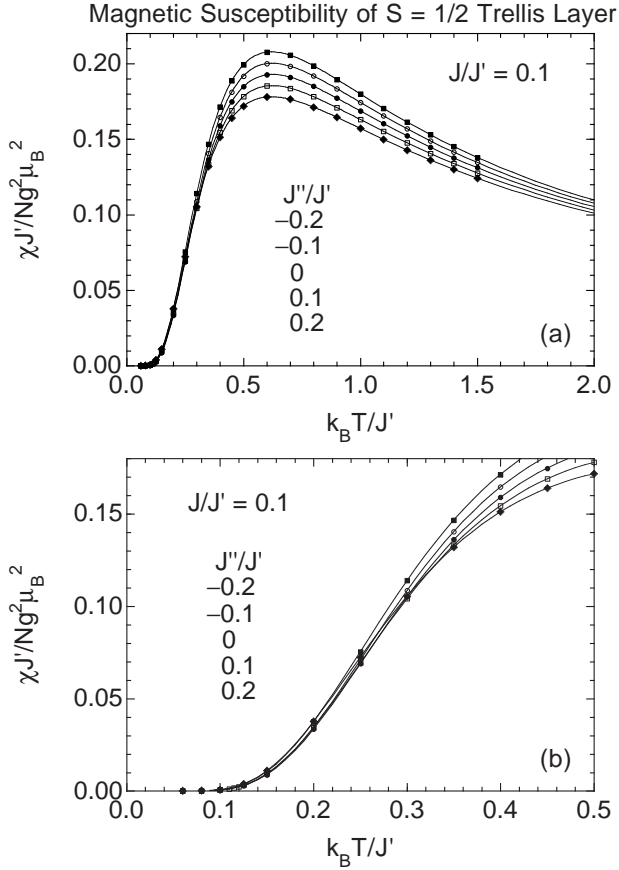}}
\vglue 0.1in
\caption{(a) Quantum Monte Carlo simulations of the magnetic spin
susceptibility
$\chi$ of the $S = 1/2$ trellis layer vs temperature $T$ for intraladder
couplings $J/J^\prime = 0.1$ and interladder couplings
$J^{\prime\prime}/J^\prime = -0.2, -0.1,$ 0, 0.1
and~0.2.\protect\cite{Miyahara1998}  \mbox{(b)~Expanded} plots of the data
in (a) at low temperatures.  In (a) and~(b), the error bars on the data
points are all smaller than the data point symbols.  The set of solid
curves is a three-dimensional fit to all these data together with those in
Fig.~\protect\ref{Fig08}, which total 162 data points for
$J^{\prime\prime}/J^\prime \neq 0$.}
\label{Fig07}
\end{figure}
\vglue0.45in
\noindent and for additional exchange parameters $J^{\prime\prime}/J = \pm
0.5$ which we do not discuss here.  The results are shown in
Fig.~\ref{Fig06}, along with the above isolated ladder results for $J'/J =
0.5$ and~1 and $J^{\prime\prime} = 0$.  We also show their  QMC simulations
for the strong-coupling regime over the maximum $t$ range $0.08
\leq t \leq 1.5$, for the same values of $J^{\prime\prime}/J$ and with
$J/J^\prime = 0.1$ (84 data points, Fig.~\ref{Fig07}) and $J/J^\prime = 0.2$
(78 data points, Fig.~\ref{Fig08}).  The results are seen to be quite
insensitive to the frustrating interladder exchange interaction
$J^{\prime\prime}$.  In addition, for ladders with spatially isotropic
exchange, Gopalan, Rice and Sigrist have deduced that the spin gap is nearly
independent of $J^{\prime\prime}$ for weak coupling.\cite{Gopalan1994}  Thus
one expects that fits of experimental $\chi(T)$ data by our fits to the QMC
data will not be able to establish a quantitative value of
$J^{\prime\prime}$ for trellis layer compounds with weak interladder
interactions.

\begin{figure}
\epsfxsize=3.3in
\centerline{\epsfbox{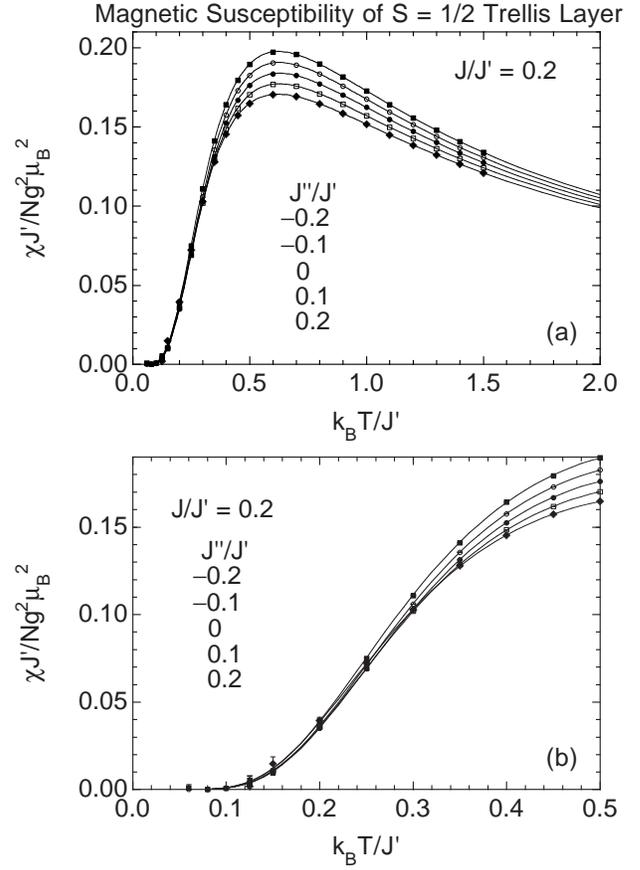}}
\vglue 0.1in
\caption{(a) Quantum Monte Carlo magnetic spin susceptibility $\chi$ data
(symbols) for the $S = 1/2$ trellis layer vs temperature $T$ for intraladder
couplings $J/J^\prime = 0.2$ and interladder couplings
$J^{\prime\prime}/J^\prime = -0.2, -0.1,$ 0, 0.1
and~0.2.\protect\cite{Miyahara1998}  \mbox{(b) Expanded} plots of the data
in (a) at low temperatures, where the error bars are also plotted.  The set
of solid curves is a three-dimensional fit to all these data together with
those in Fig.~\protect\ref{Fig07}.}
\label{Fig08}
\end{figure}

\subsubsection{Stacked Ladder Interladder Interactions}
\begin{figure}
\epsfxsize=3.3in
\centerline{\epsfbox{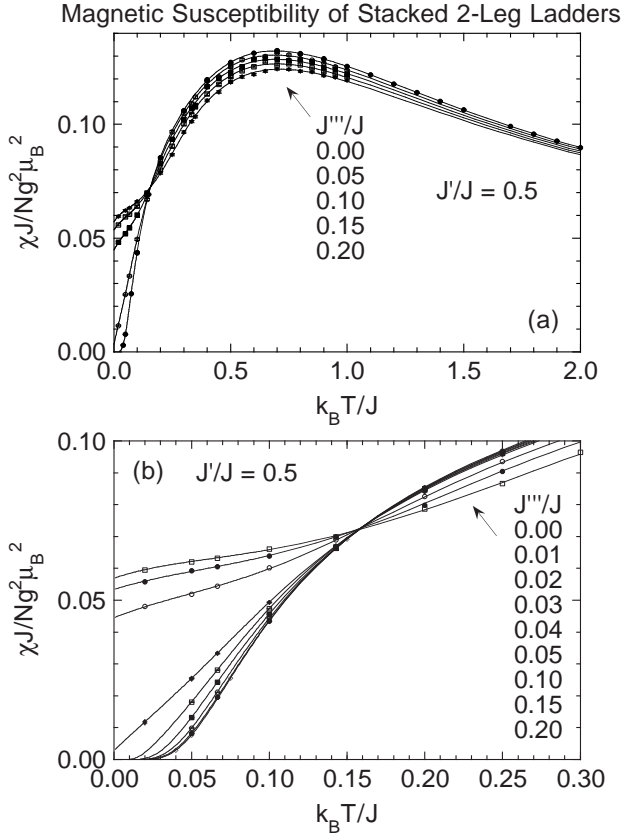}}
\vglue 0.1in
\caption{(a) Quantum Monte Carlo magnetic spin susceptibility $\chi$ data
(symbols) for stacked $S = 1/2$ 2-leg ladders vs temperature $T$ for
intraladder couplings $J^\prime/J = 0.5$ and stacked interladder couplings
$J^{\prime\prime\prime}/J = 0,$ 0.05, 0.10, 0.15 and~0.2.  (b) Expanded
plot of the data in (a), plus additional data for $J^{\prime\prime\prime}/J
= 0.01,$ 0.02, 0.03 and~0.04, at low temperatures.  In (a) and (b), the
error bars are smaller than the data symbols.  The solid curves are two
two-dimensional fits to the data in the gapped regime
($J^{\prime\prime\prime}/J = 0$--0.04) and gapless regime
($J^{\prime\prime\prime}/J = 0.05$--0.2), respectively.  Note the unique
crossing point of all the curves at $k_{\rm B}T/J \approx 0.16$, where
$\chi$ is independent of $J^{\prime\prime\prime}/J$.}
\label{Fig09}
\end{figure}

\begin{figure}
\epsfxsize=3.3in
\centerline{\epsfbox{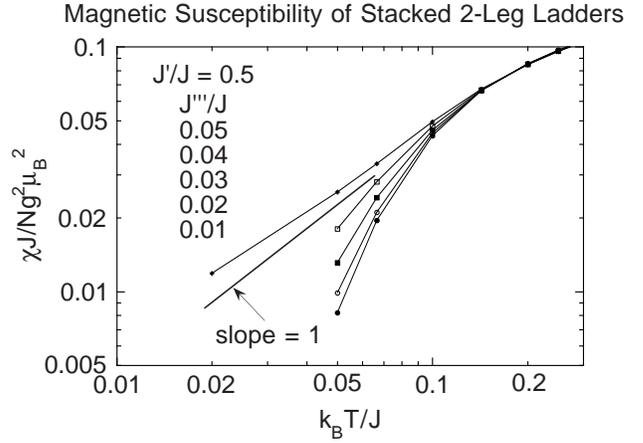}}
\vglue 0.1in
\caption{Log-log plot of the low-temperature QMC magnetic spin
susceptibility $\chi$ data (symbols) for stacked $S = 1/2$ 2-leg ladders vs
temperature $T$ for intraladder couplings $J^\prime/J = 0.5$ and interladder
couplings $J^{\prime\prime\prime}/J = 0.01$ to~0.05 (bottom to top data
sets) from Fig.~\protect\ref{Fig09}.  The error bars (not shown)
are smaller than the data point symbols.  The lines connecting the data
points are guides to the eye.  Comparison of the data with the heavy line
with a slope of~1 indicates that the quantum critical point occurs for
$J^{\prime\prime\prime}/J$ between 0.04 and~0.05.}
\label{Fig10}
\end{figure}

\begin{figure}
\epsfxsize=3.3in
\centerline{\epsfbox{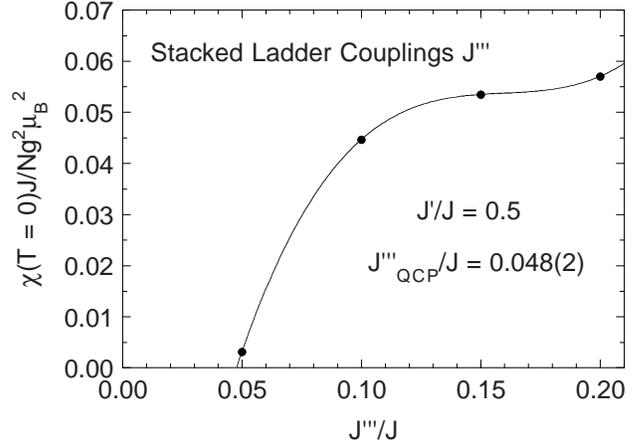}}
\vglue 0.1in
\caption{Magnetic spin susceptibility at zero temperature $\chi(T = 0)$ data
($\bullet$) for gapless stacked $S = 1/2$ 2-leg ladders vs stacked
interladder coupling $J^{\prime\prime\prime}/J$ for intraladder
couplings $J^\prime/J = 0.5$.  The solid curve is an exact third-order
polynomial fit to the data.  The quantum critical point (QCP) occurs when
$\chi(T=0,J^{\prime\prime\prime}/J)\to 0$, from which we find 
$J_{\rm QCP}^{\prime\prime\prime}/J = 0.048(2)$.}
\label{Fig11}
\end{figure}

\vglue-0.06in
Another interladder coupling path, with exchange constant
$J^{\prime\prime\prime}$, is from each spin in a ladder to one spin in each
of two ladders directly above and below the first ladder, a 2D array termed
a ``stacked ladder'' configuration.  We have carried out QMC simulations of
$\chi^*(t)$ for $J^\prime/J = 0.5$ (96 data points) and~1 (94 data points)
over a maximum temperature range $0.02\leq t\leq 1$ with AF stacked ladder
couplings
$J^{\prime\prime\prime}/J = 0.05,\ 0.1,\ 0.15$ and~0.2 for each $J^\prime/J$
value, and also for $J^{\prime\prime\prime}/J = 0.01$, 0.02, 0.03 and~0.04
for
$J^\prime/J = 0.5$ (106 data points).  The results for $J^\prime/J= 0.5$
are shown in Fig.~\ref{Fig09}.  A log-log plot of the
low-$t$ data for
$J^{\prime\prime\prime}/J = 0.01$ to~0.05 from
Fig.~\ref{Fig09} is shown separately in
Fig.~\ref{Fig10}.  According to
theory,\cite{Chubukov1994,Chubukov1993,Sandvik1994,Elstner1998} the quantum
critical point (QCP) separating the spin-gapped phase from the AF ordered
phase in a 2D system is characterized by the behavior $\chi^* \propto t$ at
low $t$.  A comparison of the data in Fig.~\ref{Fig10} with the heavy
solid line with slope~1 indicates 
\noindent that a QCP occurs for
$J^\prime/J = 0.5$ at $0.04 < J^{\prime\prime\prime}_{\rm QCP}/J < 0.05$. 
In order to obtain a more precise estimate of $J^{\prime\prime\prime}/J$ at
the QCP, in Fig.~\ref{Fig11} we plot $\chi^*(t=0)$ vs
$J^{\prime\prime\prime}/J$, where the $\chi^*(t=0)$ values were determined
by fits to the data as described later.  By fitting these
$\chi^*(t=0,J^{\prime\prime\prime}/J)$ data by various polynomials, such as
the third order polynomial shown as the solid curve in the figure, and by
noninteger power laws, we estimate $J_{\rm QCP}^{\prime\prime\prime}/J =
0.048(2)$ with conservative error bars.

\begin{figure}
\epsfxsize=3.3in
\centerline{\epsfbox{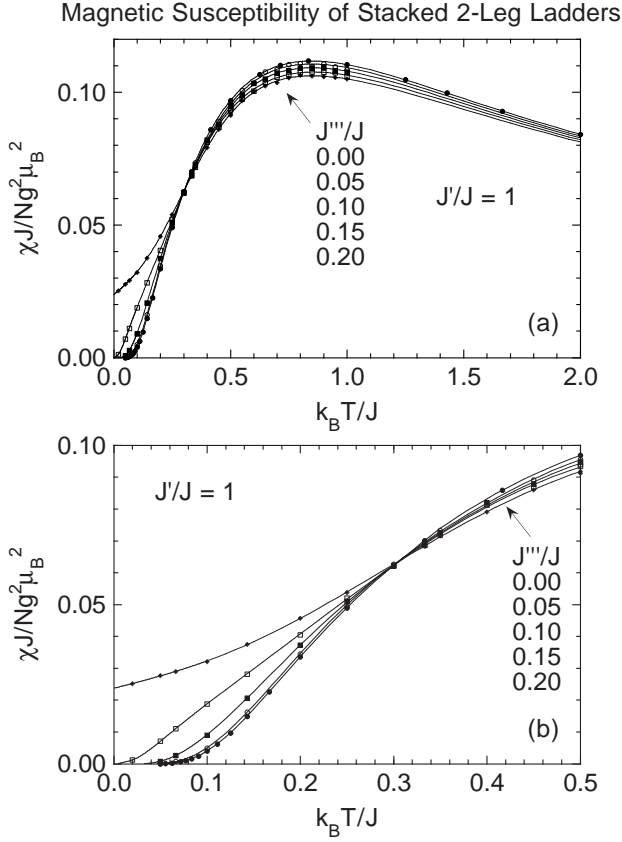}}
\vglue 0.1in
\caption{Quantum Monte Carlo magnetic spin susceptibility $\chi$ data
(symbols) for stacked $S = 1/2$ 2-leg ladders vs temperature $T$ for
intraladder couplings $J^\prime/J = 1$ and interladder couplings
$J^{\prime\prime\prime}/J = 0,$ 0.05, 0.1, 0.15 and~0.2.  The system has a
spin gap for $J^{\prime\prime\prime}/J = 0,$ 0.05, 0.1 and~0.15 but is
gapless for $J^{\prime\prime\prime}/J = 0.2$.  The set of solid curves
through the data for $J^{\prime\prime\prime}/J = 0,$ 0.05, 0.1 and~0.15 is a
two-dimensional fit to  these data.  A separate fit was obtained for
$J^{\prime\prime\prime}/J = 0.2$, shown as the solid curve through those
data.  Note the unique crossing point of all the curves at $k_{\rm B}T/J
\approx 0.31$, where $\chi$ is independent of $J^{\prime\prime\prime}/J$.}
\label{Fig12}
\end{figure}

The $\chi^*(t)$ data for isotropic ($J^\prime/J= 1$) stacked ladders 
are shown in Fig~\ref{Fig12}, along with the above isolated
ladder results for $J'/J = 1$ and $J^{\prime\prime\prime} = 0$.  A QCP is
seen to occur at $J^{\prime\prime\prime}_{\rm QCP}/J \approx 0.16$.  This
value of $J^{\prime\prime\prime}_{\rm QCP}/J$ is much smaller than the
various values 0.32(2)~(Ref.~\onlinecite{Imada1997}),
0.43~(Ref.~\onlinecite{Normand1996}) and 0.30~(Ref.~\onlinecite{Koga1999})
inferred for spatially isotropic ladders arranged in a nonfrustrated flat 2D
array, because the interladder spin coordination number for the 2D stacked
\noindent ladder configuration is two whereas for ladders arranged in 2D
flat layers it is only one.  On the other hand, our
$J^{\prime\prime\prime}_{\rm QCP}/J$ for $J^\prime/J = 1$ is somewhat larger
than that ($\approx 0.11$, Refs.~\onlinecite{Normand1996,Troyer1997}) for
the 3D LaCuO$_{2.5}$-type coupling configuration (see below), even though
the interladder spin coordination number is the same; here the effects of
dimensionality evidently come into play, with fluctuation effects generally
being stronger in lower dimensional sys-
\noindent tems (other factors being equal). 

Another interesting aspect of the QMC data for the stacked ladders is the
striking well-defined crossing points of $\chi^*(t)$ versus
$J^{\prime\prime\prime}/J$ at $t \approx 0.16$ for $J^\prime/J = 0.5$ in
Fig.~\ref{Fig09} and at $t \approx 0.31$ for
$J^\prime/J = 1$ in Fig.~\ref{Fig12}.  Similar crossing
points in the specific heats of strongly correlated electron systems versus
some thermodynamic variable (e.g., pressure, magnetic field, interaction
parameter) have been pointed out by Vollhardt and
coworkers.\cite{Vollhardt1997,Chandra1998}

\begin{figure}
\epsfxsize=3.3in
\centerline{\epsfbox{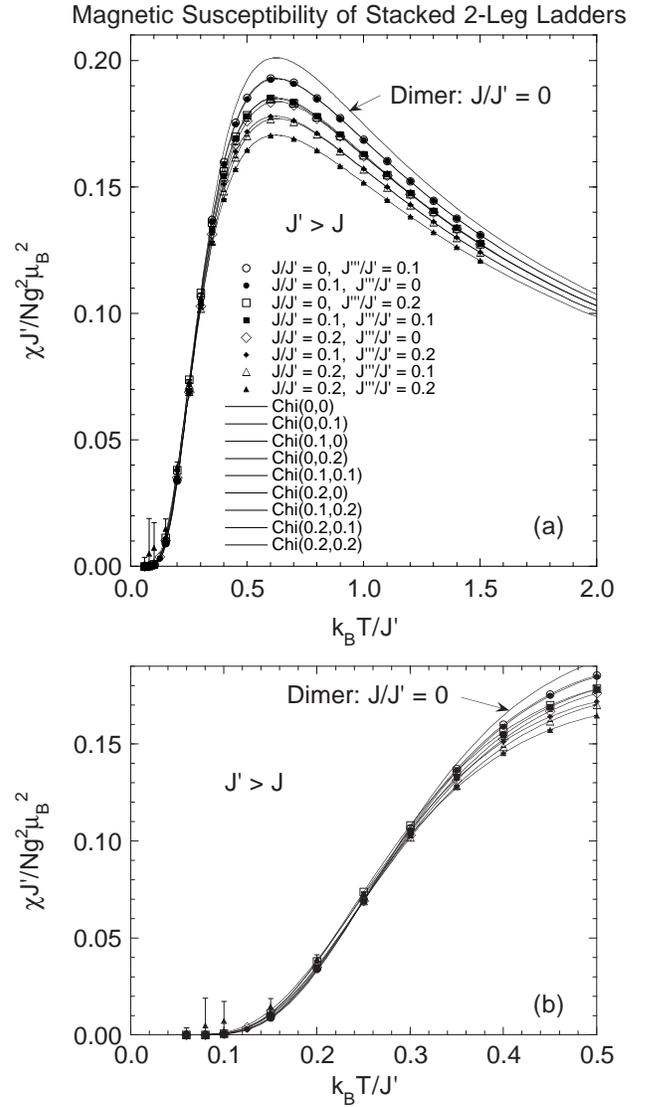}}
\vglue 0.1in
\caption{(a) Quantum Monte Carlo magnetic spin susceptibility $\chi$ data
(symbols) for stacked $S = 1/2$ 2-leg ladders vs temperature $T$ for
intraladder couplings $J/J^\prime = 0,\ 0.1$ and~0.2 and interladder
couplings $J^{\prime\prime\prime}/J^\prime = 0,$ 0.1 and~0.2 for each
$J/J^\prime$ value.  (b)  Expanded plot of the data in (a) at low
temperatures.  The error bars for each data point are shown in both (a)
and~(b).  Note that the pairs of data sets with a fixed value (0.1, 0.2 or
0.3) of $J^{\prime\prime\prime}/J^\prime + J/J^\prime$ are nearly
coincident, and thus all these data closely follow the prediction of
molecular field theory.  The set of solid curves is a three-dimensional fit
to all the data.}
\label{Fig13}
\end{figure}

We have also obtained $\chi^*(t)$ data for strong interchain intraladder
couplings $J/J^\prime = 0$ (42 data points), 0.1 (39 data points) and~0.2
(38 data points) over a maximum temperature range $0.06\leq t\leq 1.5$ with
stacked ladder couplings $J^{\prime\prime\prime}/J^\prime = 0.1$ and~0.2 for
each $J/J^\prime$ value. The results are shown in
Fig.~\ref{Fig13}, along with $\chi^*(t)$ for the isolated
dimer from Eq.~(\ref{EqChiDimer}) and the above isolated ladder results for
$J/J^\prime = 0.1$ and~0.2 and
$J^{\prime\prime\prime} = 0$.  Over these exchange interaction ranges, the
spin gap persists and  a
\vglue0.33in
\noindent QCP is not traversed.  Note that the pairs of data
sets with a fixed value (0.1, 0.2 or 0.3) of
$J^{\prime\prime\prime}/J^\prime + J/J^\prime$ are nearly coincident, and
thus closely follow molecular field theory which predicts (see
Sec.~\ref{SecMFT}) that $\chi^*(t)$ of coupled dimers only depends on the
sum of the exchange interactions between a spin in a dimer and all other
spins outside the dimer.

\begin{figure}
\epsfxsize=3.3in
\centerline{\epsfbox{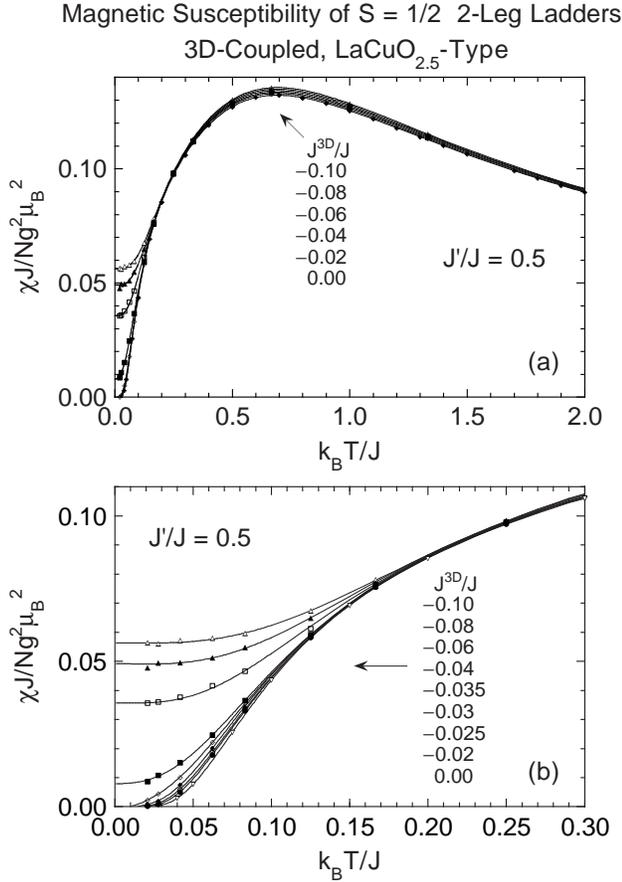}}
\vglue 0.1in
\caption{(a) Quantum Monte Carlo magnetic spin susceptibility $\chi$ data
(symbols) for isolated ($J^{\rm 3D}/J$ = 0) and three-dimensionally
ferromagnetically-coupled ($J^{\rm 3D}/J = -$0.02, $-$0.04, $-$0.06,
$-$0.08 and $-$0.1) two-leg ladders vs reduced temperature $t = k_{\rm
B}T/J$ for intraladder couplings $J^\prime/J = 0.5$.  The 3D interladder
coupling topology is that proposed for LaCuO$_{2.5}$.  (b) Expanded plots
at low temperatures of the data in (a) plus additional data for $J^{\rm
3D}/J = -$0.025, $-$0.03 and $-$0.035.  In~(a) and~(b), the error bars on
the data are smaller than the data symbols.  The two sets of solid curves
are two fits to the data with and without spin gaps, respectively.  A spin
gap occurs for $J^{\rm 3D}/J = 0,$ $-$0.02, $-$0.025, $-$0.03 and $-$0.035,
and the set of solid curves through these data is a two-dimensional ($t,\
J^{\rm 3D}/J$) fit at fixed $J^\prime/J = 0.5$.  For $J^{\rm 3D}/J = -$0.04,
$-$0.06, $-$0.08 and $-$0.1, the magnetic excitation spectra are gapless,
and the set of solid curves through these data is a  three-dimensional
($t,\ J^\prime/J,\ J^{\rm 3D}/J$) fit to these data together with the data
for $J^\prime/J = 0.6$, 0.7, 0.8, 0.9 and 1.0 in
Figs.~\protect\ref{Fig16} and~\protect\ref{Fig18} with
$J^{\rm 3D}/J$ values in the gapless regime given in
Sec.~\protect\ref{SecLaCuO2.5NoGap}.  Note that $\chi$ is nearly independent
of $J^{\rm 3D}/J$ at $t \approx 0.21$, even though the different data sets
do not cross.}
\label{Fig14}
\end{figure}

\subsubsection{3D LaCuO$_{2.5}$-Type Interladder Interactions}
\vglue0.1in
Additional simulations of $\chi^*(t)$ were performed which incorporated the
nonfrustrated 3D interladder coupling 
\begin{figure}
\epsfxsize=3.3in
\centerline{\epsfbox{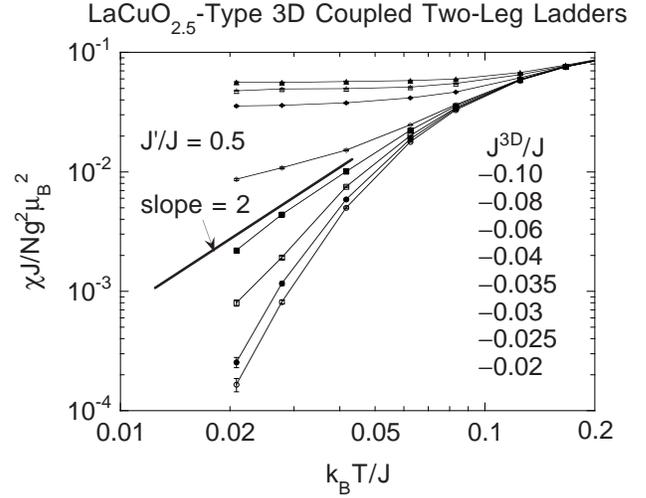}}
\vglue 0.1in
\caption{Quantum Monte Carlo magnetic spin susceptibility $\chi$ data
(symbols) from Fig.~\protect\ref{Fig14}(b), replotted on log-log scales. 
The light lines are guides to the eye.  At the quantum critical point
(QCP), $\chi \propto T^2$ at low $T$ as indicated by the heavy line with
slope~2.  A comparison of this line with the data indicates that the QCP
occurs in the range $0.035 < |J^{\rm 3D}/J| < 0.040$.}
\label{Fig15}
\end{figure}
\noindent configuration
proposed\cite{Normand1996,Troyer1997} for the two-leg ladder compound
LaCuO$_{2.5}$, in which each spin is coupled by exchange constant
$J^{\rm 3D}$ to one nearest-neighbor spin in each of two adjacent ladders
diagonally above and below the first ladder in adjacent layers,
respectively.  Simulations are reported here for $J^\prime/J = 0.5$ and
$J^{\rm 3D}/J =  0.2$, 0.1, 0.05,  $-0.02$, $-$0.025, $-$0.03, $-$0.035,
$-$0.04, $-$0.06, $-$0.08 and $-$0.1 over the maximum temperature range
$0.021 \leq t \leq 3$ (204 data points), for a 3D lattice of size $6 \times
6$\,ladder$^2 \times 40$\,spins.  The simulations for ferromagnetic (FM,
negative) interladder couplings were motivated by the recent findings of
Mizokawa {\it et  al.}\cite{Mizokawa1997} mentioned in the Introduction. 
The data for FM couplings at $t \leq 2$, plotted in Fig.~\ref{Fig14},
indicate a loss of the spin gap for $0.035 < |J^{\rm 3D}/J| < 0.040$.  At
the QCP in a 3D system, $\chi^*$ is predicted theoretically to be
proportional to $t^2$.\cite{Normand1997,Troyer1997}  We have plotted our
$\chi^*(t)$ simulation data on double logarithmic axes in
Fig.~\ref{Fig15}, where by comparison of the data with the
heavy line with slope~2, the quantum critical point for $J^\prime/J = 0.5$
is indeed seen to be in this range.

\begin{figure}
\epsfxsize=3.3in
\centerline{\epsfbox{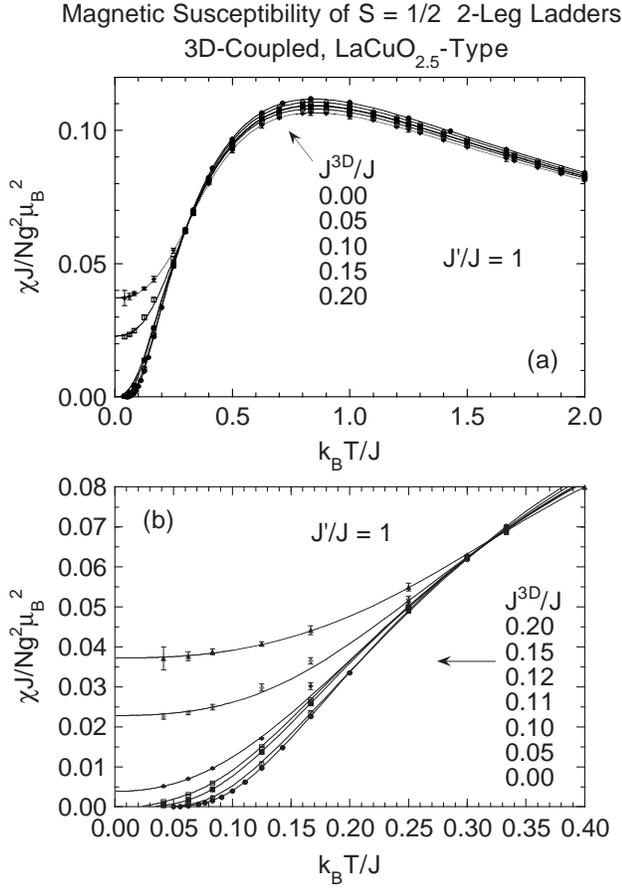}}
\vglue 0.1in
\caption{(a) Quantum Monte Carlo magnetic spin susceptibility $\chi$ data
(symbols with error bars) for isolated ($J^{\rm 3D}/J = 0$) and
three-dimensionally antiferromagnetically-coupled ($J^{\rm 3D}/J = 0.05$,
0.1, 0.15, 0.2) two-leg ladders vs reduced temperature $t = k_{\rm B}T/J$
for intraladder couplings $J^\prime/J = 1$.\protect\cite{Troyer1997}  The
interladder coupling topology is that proposed for LaCuO$_{2.5}$.  (b)
Expanded plots of the the data in~(a) and additional data for $J^{\rm 3D}/J
= 0.11$ and~0.12.  In (a) and~(b), the data for $J^{\rm 3D}/J = 0$, 0.05,
0.1 and~0.11 are in the gapped regime, whereas those for $J^{\rm 3D}/J =
0.12$, 0.15 and~0.2 are in the gapless  regime.  The set of solid curves
for the gapped regime is a two-dimensional ($t,\ J^{\rm 3D}/J$) fit to
these data, whereas the set of solid curves for the gapless regime is a
three-dimensional ($t,\ J^\prime/J,\ J^{\rm 3D}/J$) fit to these data
together with the data for $J^\prime/J = 0.5$, 0.6, 0.7, 0.8 and 0.9 in
Figs.~\protect\ref{Fig14} and~\protect\ref{Fig18} in the gapless
regime with $J^{\rm 3D}/J$ values given in
Sec.~\protect\ref{SecLaCuO2.5NoGap}.  Note the crossing point of all the
curves at $t \approx 0.32$, where $\chi$ is nearly independent of
$J^{\rm 3D}/J$.}
\label{Fig16}
\end{figure}

The previously reported\cite{Troyer1997} $\chi^*(t)$ simulation data for
$J'/J = 1$ and $J^{\rm 3D}/J = 0.05$, 0.1, 0.11, 0.12, 0.15 and~0.2 (a
total of 169 data points), which show a quantum critical point at $J^{\rm
3D}_{\rm QCP}/J \approx 0.11$,\cite{Troyer1997} are shown in
Fig.~\ref{Fig16}.

\begin{figure}
\epsfxsize=3.3in
\centerline{\epsfbox{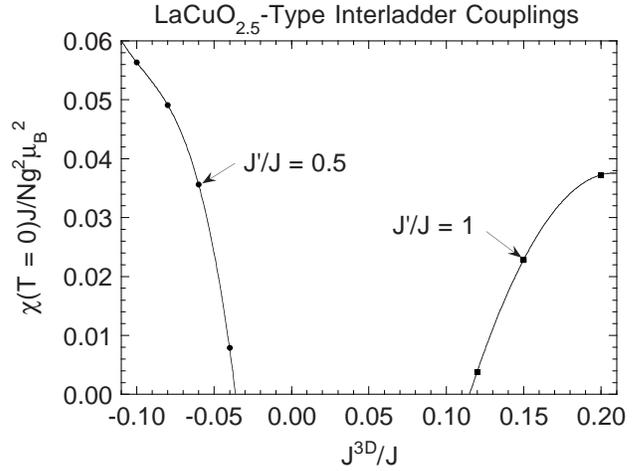}}
\vglue 0.1in
\caption{Quantum Monte Carlo magnetic spin susceptibility data extrapolated
to zero temperature $\chi(T = 0)$ (symbols) for gapless two-leg ladders vs
interladder coupling $J^{\rm 3D}/J$ and intraladder couplings
$J^\prime/J = 0.5$ and~1.  The interladder coupling topology is that
proposed for LaCuO$_{2.5}$.  The solid curves are exact polynomial fits to
the respective sets of data.  The quantum critical point (QCP) is
determined by the condition $\chi(T=0)=0$, yielding $J^{\rm 3D}_{\rm QCP}/J
= -0.036(1)$ for $J^\prime/J = 0.5$ and $J^{\rm 3D}_{\rm QCP}/J = 0.115(1)$
for $J^\prime/J = 1$.  There is a QCP for $J^{\rm 3D}/J$ of the
opposite respective sign for each value of $J^\prime/J$.}
\label{Fig17}
\end{figure}

Since one expects that upon approaching the QCP from the AF-ordered side
that
$\chi^*(t = 0) \to 0$, to determine more precisely the QCPs we have plotted
$\chi^*(t = 0)$ vs $J^{\rm 3D}/J$ for each of $J^\prime/J = 0.5$ and~1 in
Fig.~\ref{Fig17}.  The extrapolated $\chi^*(t = 0)$ values were
determined by fits to the $\chi^*(t)$ data described in
Sec.~\ref{SecSimFits} below and in the Appendix.  From exact polynomial
fits to the data in Fig.~\ref{Fig17}, we find $J^{\rm 3D}_{\rm
QCP}/J = -0.036(1)$ for $J^\prime/J = 0.5$ and $J^{\rm 3D}_{\rm QCP}/J$  
\vglue0.3in
\noindent $= 0.115(1)$ for $J^\prime/J = 1$.  There
also exist QCPs for the opposite sign of $J^{\rm 3D}/J$ in each case,
respectively.

We report here additional $\chi^*(t)$ simulation data for intermediate
values of
$J^\prime/J = 0.6$, 0.7, 0.8 and~0.9, each with $J^{\rm 3D}/J = 0.1$, 0.15
and~0.2 at temperatures $0.3 \leq t \leq 3$ (a total of 325 data points). 
A  selection of these data for $J^{\rm 3D}/J = 0.1$ and~0.2 are plotted in
Figs.~\ref{Fig18}(a) and~\ref{Fig18}(b), respectively.

\subsection{Three-Leg $\bbox{S = 1/2}$ Ladders}

The $\chi^*(t)$ for $S = 1/2$ $n$-leg ladders with $n = 1,$ \mbox{2,
$\ldots,\ 6$} and isotropic exchange ($J^\prime/J = 1$) and for $n = 3$ with
spatially anisotropic exchange was computed using QMC simulations for $0.02
\lesssim t \leq 5$ by \mbox{Frischmuth {\it et al.}}\cite{Frischmuth1996} 
As noted in the Introduction, the even-leg ladders exhibit a spin-gap but
the odd-leg ladders do not.  In this paper we will be fitting experimental
$\chi(T)$ data for the three-leg ladder compound ${\rm Sr_2Cu_3O_5}$.  The
QMC $\chi^*(t)$ data for three-leg ladders with both spatially isotropic and
anistropic exchange\cite{Frischmuth1996} are shown in
Fig.~\ref{Fig19}.  The $\chi^*(t)$ is seen to be sensitive to the
value of $J^\prime/J$.  As a consequence, we expect to be able to obtain an
accurate value of $J^\prime/J$ for ${\rm Sr_2Cu_3O_5}$ from fits to the
experimental $\chi(T)$ data.  In contrast, Kim {\it et al.}\ have found from
QMC simulations that $\chi^*(t)$ is very insensitive to the {\it
interladder} coupling between isotropic three-leg ladders arranged in a
layer,\cite{Kim1999} although it should be noted that they calculated
$\chi^*(t)$ for nonfrustrated interladder couplings and not for the
frustrated trellis layer interladder coupling configuration present in ${\rm
Sr_2Cu_3O_5}$.

\section{Fits to the QMC $\bbox{\chi^*(\lowercase{t})}$ Simulation Data}
\label{SecSimFits}

In order to precisely fit experimental $\chi(T)$ data by the QMC $\chi^*(t)$
simulations, it is essential to first obtain accu-
\begin{figure}
\epsfxsize=3.3in
\centerline{\epsfbox{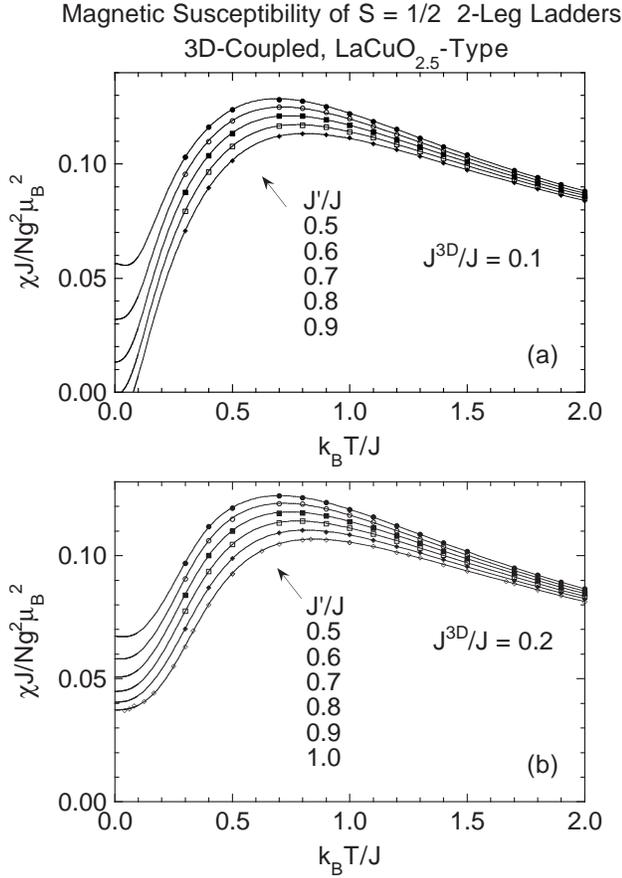}}
\vglue 0.1in
\caption{Quantum Monte Carlo magnetic spin susceptibility $\chi$ vs
temperature $T$ data (symbols) for three-dimensionally
antiferromagnetically-coupled gapless two-leg ladders with intraladder
couplings $J^\prime/J = 0.5$ to~1 and interladder couplings $J^{\rm 3D}/J =
0.1$~(a) and~0.2~(b).  The interladder coupling topology is that proposed
for LaCuO$_{2.5}$.  The set of solid curves in (a) and~(b) is a single
three-dimensional fit to these data together with additional data (not
shown) for $J^{\rm 3D}/J = 0.05$ and $J^\prime/J = 0.6$, $J^{\rm 3D}/J =
0.15$ and $J^\prime/J = 0.6$--0.9, and the additional data for $J^\prime/J =
0.5$ and~1.0 in Figs.~\protect\ref{Fig14} and~\protect\ref{Fig16} in the
gapless regime with $J^{\rm 3D}/J$ values given in the respective
captions.  Extrapolations of the fit to low temperatures are also shown. 
Note that the extrapolations for $J^{\rm 3D}/J = 0.1$ and $J^\prime/J =
0.8$ and~0.9 are negative (unphysical) at the lowest temperatures.}
\label{Fig18}
\end{figure}
\noindent rate analytical fits to the
simulation data.  As part of our QMC data fit function for two-leg ladders,
we first obtain fits to the known dependence of the spin gap $\Delta$ on
$J$ and $J^\prime$ for isolated ladders.  We then discuss the first few
terms of the high-temperature series expansion (HTSE) for the magnetic
susceptibility, which will also be incorporated into the fit function so
that the function can be accurately extrapolated to arbitrarily high
temperatures.  The fit function itself is then presented and discussed. 
For some sets of exchange parameters (for the frustrated trellis layer) it
was not possible to obtain extensive QMC $\chi^*(t)$ data at low
temperatures due to the ``negative sign problem''.  For such cases, it is
sometimes 
\begin{figure}
\epsfxsize=3.3in
\centerline{\epsfbox{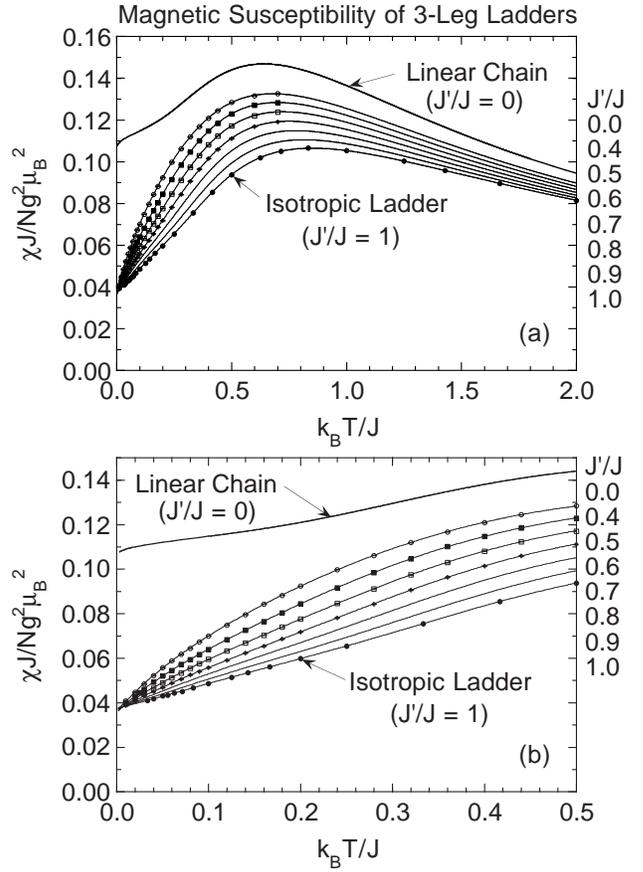}}
\vglue 0.1in
\caption{(a) Quantum Monte Carlo (QMC) magnetic spin susceptibility $\chi$
vs temperature $T$ data (symbols) for three-leg ladders with spatially
anisotropic intraladder exchange.\protect\cite{Frischmuth1996}  The set of
solid curves through the data points is a global two-dimensional fit.  Also
shown are extrapolations of the QMC data fits for $0.4\leq J^\prime/J\leq
0.7$ by a factor of about three to higher $T$, and two interpolation curves
for $J^\prime/J = 0.8$ and~0.9.  The data for the linear chain were
calculated by Eggert, Affleck and Takahashi.\protect\cite{Eggert1994}  (b)
Expanded plots of the data and fit at low temperatures.}
\label{Fig19}
\end{figure}
\noindent necessary to use a fit
function, containing a minimum number (even zero) of fitting parameters,
derived from the molecular field theory for coupled subsystems.  We
therefore also discuss such fit functions.

\subsection{Spin Gap for Isolated Two-Leg Ladders}
\label{SecSpinGap}
\vglue0.28in
For any finite $J^\prime/J > 0$ an energy gap (``spin gap'')  $\Delta$
exists in the magnetic excitation spectrum between the singlet ground state
and the lowest triplet excited states of $S = 1/2$ two-leg Heisenberg
ladders.\cite{Barnes1993}  The numerical $\Delta$ values of Barnes {\it et
al}.\cite{Barnes1993} in the range $0 \leq J^\prime/J\leq 1$, obtained by
extrapolating exact diagonalization results for short two-leg ladders to
the bulk limit, were previously fitted by one of us with the
expression\cite{Johnston1996}
\begin{figure}
\epsfxsize=3.3in
\centerline{\epsfbox{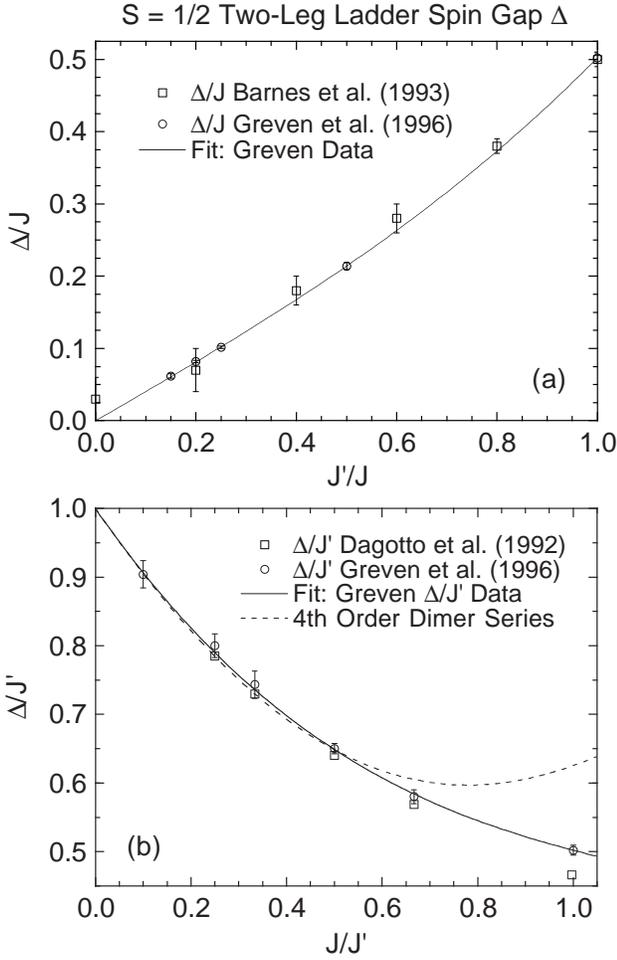}}
\vglue 0.1in
\caption{Spin gap $\Delta$ of isolated two-leg ladders versus (a) $J'/J$
for $0 \leq J'/J \leq 1$ from numerical calculations of Barnes {\it et
al}.\protect\cite{Barnes1993} and of Greven {\it et
al}.\protect\cite{Greven1996} and (b) $J/J'$ for $0 \leq J/J' \leq 1$ from
numerical calculations of Dagotto {\it et al}.\protect\cite{Dagotto1992}
and of Greven {\it et al}.\protect\cite{Greven1996}  The dashed curve
in~(b) is the exact dimer series expansion to fourth order in $J/J^\prime$
in Eq.~(\protect\ref{EqDelta2b}).  In (a) and (b), the solid curves are
weighted fits to the data of Greven {\it et al}.\ by
Eqs.~(\protect\ref{EqDelta1}) and~(\protect\ref{EqDelta2:all}),
respectively.}
\label{Fig20}
\end{figure}
\vglue0.05in
\begin{equation}
\Delta^* \equiv {\Delta\over J} = 0.4\bigg({J^\prime\over J}\bigg) +
0.1\bigg({J^\prime\over J}\bigg)^2~.
\label{EqDelta0}
\end{equation}
Higher accuracy $\Delta$ values were subsequently obtained numerically from
Monte Carlo simulations by Greven, \mbox{Birgeneau} and
Wiese\cite{Greven1996} as shown in Fig.~\ref{Fig20}(a) where the
previous data of Barnes {\it et al}.\ are shown for comparison.  We find
that the data of Greven {\it et al}.\ for $J'/J \leq 1$ in
Fig.~\ref{Fig20}(a) are fitted better by
\begin{equation} {\Delta_0\over J}  =  0.4030\bigg({J^\prime\over J}\bigg) +
0.0989\bigg({J^\prime\over J}\bigg)^3~,
\label{EqDelta1}
\end{equation} 
with a statistical $\chi^2/{\rm DOF} = 0.18$ (for the
definition, see the Appendix), as shown by the solid curve in the figure. 
For isotropic ladders ($J^\prime/J = 1$), the fitted $\Delta/J = 0.5019$ is
in good agreement with the values 0.50(1) \mbox{\,of} \mbox{ Barnes}
\begin{figure}
\epsfxsize=3.3in
\centerline{\epsfbox{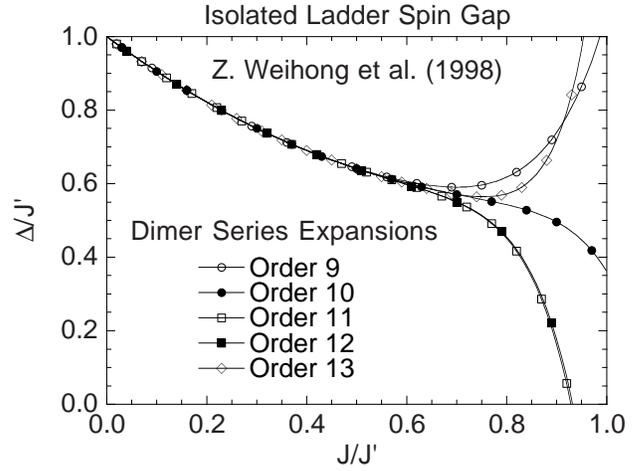}}
\vglue 0.1in
\caption{Dimer series expansions from 9th to 13th order in $J/J^\prime$ of
the spin gap $\Delta/J^\prime$ of isolated two-leg ladders versus
$J/J^\prime$ (solid curves) as computed by Weihong, Kotov and
Oitmaa.\protect\cite{Weihong1998}  The symbols identify the curves.}
\label{Fig21}
\end{figure}
\noindent  {\it et
al.},\cite{Barnes1993} 0.504(7) of Oitmaa, Singh and
Weihong,\cite{Oitmaa1996} 0.504 of White, Noack and
Scalapino\cite{White1994} and 0.5028(8) of \mbox{Weihong}, Kotov and
Oitmaa.\cite{Weihong1998b}  Using exact diagonalizations for ladders of size
up to $2\times 15$ spins, Flocke obtained an extrapolated value 0.49\,996
for the bulk limit, and conjectured that the exact value is~${1\over
2}$.\cite{Flocke1997}  Equation~(\ref{EqDelta1}) predicts values of
$\Delta_0/J$ which are systematically larger than the extrapolated
bulk-limit values of Flocke at smaller $J^\prime/J$, e.g.\ by 0.003 at
$J^\prime/J = 0.8$ and by 0.02 at $J^\prime/J = 0.2$.  The fitted initial
slope in Eq.~(\ref{EqDelta1}) agrees with the estimates 0.41(1) of Greven
{\it et al.}\cite{Greven1996} and 0.405(15) of Weihong, Kotov and
Oitmaa.\cite{Weihong1998b}

For the strong interchain coupling regime $0 \leq J/J^\prime \leq 1$ of the
two-leg ladder, the exact dimer series expansion for the spin gap to seventh
order in $J/J^\prime$ is
\cite{Barnes1993,Weihong1998b,Reigrotzki1994,Cabra1997,Piekarewicz1998}
\begin{eqnarray}
{\Delta\over J^\prime}  =  1 &-& \bigg({J\over J^\prime}\bigg) + {1\over
2}\bigg({J\over J^\prime}\bigg)^2  + {1\over 4}\bigg({J\over
J^\prime}\bigg)^3 - {1\over 8}\bigg({J\over J^\prime}\bigg)^4\nonumber\\
&-& {35\over 128}\bigg({J\over J^\prime}\bigg)^5 - {157\over
1024}\bigg({J\over J^\prime}\bigg)^6 + {503\over
2048}\bigg({J\over J^\prime}\bigg)^7~.
\label{EqDelta2b}
\end{eqnarray} 
The fourth-order series is plotted as the dashed curve in
Fig.~\ref{Fig20}(b).  Comparison of this prediction with the
numerical data of Dagotto, \mbox{Riera} and Scalapino\cite{Dagotto1992} and
of Greven {\it et al}.\cite{Greven1996} in the figure shows that the
fourth-order series is a poor description of the data for $J^\prime/J
\gtrsim 0.5$.  The dimer series expansion has been computed to 13th order by
Weihong, Kotov and Oitmaa.\cite{Weihong1998}  Plots of their 9th to 13th
order series are shown in Fig.~\ref{Fig21}.  The series is seen to
converge very slowly with increasing order for $J/J^\prime \gtrsim 0.5$. 
In fact, Piekarewicz and Shepard concluded, on the basis of dimer series
expansions of the ground state energy per site at 50th order in
perturbation theory for \mbox{4-,} 6- and 8-rung ladders, that the radius of
convergence of the dimer series is only $\approx
0.7$--0.8.\cite{Piekarewicz1998}  Therefore, for both of these reasons, to
obtain an expression for $\Delta(J/J^\prime)$ to use in our QMC $\chi^*(t)$
data fit function for the entire strong-coupling range $0 \leq J/J^\prime
\leq 1$, we carried out a weighted fit of the data of Greven {\it et
al.}\cite{Greven1996} for $0.1 \leq J/J^\prime\leq 1$ in Fig.~\ref{Fig20}(b)
by the simple two-parameter third-order polynomial
\begin{mathletters}
\label{EqDelta2:all}
\begin{equation} {\Delta_0\over J^\prime}  =  1 - \bigg({J\over
J^\prime}\bigg) + a\bigg({J\over J^\prime}\bigg)^2 + b\bigg({J\over
J^\prime}\bigg)^3~,
\label{EqDelta2:a}
\end{equation} yielding the parameters
\begin{equation} a = 0.6878,~~~b = -0.1861~.
\end{equation}
\end{mathletters} The first two terms in Eq.~(\ref{EqDelta2:a}) were set to
be the same as the corresponding exact terms in Eq.~(\ref{EqDelta2b}).  The
fit in Eqs.~(\ref{EqDelta2:all}) is shown as the solid curve in
Fig.~\ref{Fig20}(b).  The high precision of the fit is characterized
by the small $\chi^2/{\rm DOF} = 0.16$.  The value of $\Delta/J^\prime$ at
$J/J^\prime = 1$ is 0.5017, which matches very well the value 0.5019 of
the fit for \mbox{$0\leq J^\prime/J \leq 1$} in Eq.~(\ref{EqDelta1}) at this
isotropic-ladder crossover point between the two fits.

\subsection{High-Temperature Series Expansions for $\bbox{\chi^*(t)}$
\protect\\ and the Magnetic Specific Heat $\bbox{C_{\rm mag}(T)}$}
\label{SecHTSE}

As the second component of our fit function for the QMC $\chi^*(t)$
simulation data, we next consider the high temperature series expansion
(HTSE) of $\chi^*(t)$ for a general Heisenberg spin lattice containing
magnetically equivalent spins.  Spins are magnetically
equivalent if they have identical magnetic coordination spheres.  Note
that the HTSEs we discuss here, and HTSEs in general, are not restricted to
AF couplings (with a positive sign as defined in this paper); the
expansions are equally valid if the couplings are all FM (negative) or if
they are a mixture of AF and FM couplings.
 
\mbox{HTSEs} for $\chi^*(t)$ are calculated as, and the results are
normally expressed directly as, a power series in $1/t$.  However, as
mentioned by \mbox{Rushbrooke} and Wood\cite{Rushbrooke1958} and discussed
in Ref.~\onlinecite{Johnston1997}, the expressions for the expansion
coefficients for a general spin lattice containing magnetically equivalent
spins considerably simplify if the HTSE for $3\chi^*t/[S(S+1)]$ in powers
of $1/t$ is inverted (the underlying physics of this is unclear).  Indeed,
Rushbrooke and Wood\cite{Rushbrooke1958} presented their calculated
expansion coefficients in precisely this form.  In fact
for any Heisenberg spin lattice (in any dimension) containing magnetically
equivalent spins interacting with spatially isotropic nearest-neighbor AF
or FM Heisenberg exchange, a simple universal HTSE of $\chi^*(t)t$ exists
up to second order in $1/t$, and for geometrically nonfrustrated lattices
to third order,\cite{Johnston1997} which we write for $S = 1/2$ as
\begin{eqnarray} 
{4 k_{\rm B}T\chi(T)\over N g^2\mu_{\rm B}^2} = \bigg[1 & 
+ & {z J\over 4 k_{\rm B}T} + {z J^2\over 8(k_{\rm B}T)^2}\nonumber\\ 
&+&{z J^3\over 24(k_{\rm B}T)^3} + \cdots\bigg]^{-1}~,
\label{EqGenHTS}
\end{eqnarray} 
where $z$ is the coordination number of a spin by other
spins and $J$ is the unique exchange constant in the system.  The same form
of the HTSE of $\chi(T)T$ is valid for any spin $S$, but where of course the
numerical coefficients in Eq.~(\ref{EqGenHTS}) depend on $S$.  Each term
listed on the right-hand-side (but not the  higher-order terms) depends
only on $z$ (and $S$) and not on any other feature of the spin lattice or
magnetic behavior; as noted above, however, additional term(s) are added to
the numerator of the last term if geometric frustration is present or if
second-neighbor interactions are present (see below).  Hence, one can
generalize Eq.~(\ref{EqGenHTS}) to systems containing equivalent spins but
unequal exchange constants $J_{ij}$ by the replacement $z J^n\to\sum_j
J_{ij}^n$, yielding using Eqs.~(\ref{EqRedPars:all})
\begin{eqnarray}
\chi^*(t) = {1\over 4 t}\bigg[1 & + & {\sum_j J_{ij}/J^{\rm max}\over 4 t}
+ {\sum_j J_{ij}^2/{J^{\rm max}}^2\over 8 t^2}\nonumber\\ 
 & + &{\sum_j J_{ij}^3/{J^{\rm max}}^3\over 24 t^3} +
\cdots\bigg]^{-1}~,\label{EqGenHTS2}
\end{eqnarray} 
which we write as
\begin{mathletters}
\label{EqGenHTS3}
\begin{equation}
\chi^*(t)  \equiv  {1\over 4 t}\bigg[1 + {d_1\over t} + {d_2\over t^2} +
{d_3\over t^3} + \cdots\bigg]^{-1}~~.
\label{EqGenHTS3:a}
\end{equation} 
with
\begin{equation} d_1 = {1\over 4 J^{\rm max}}\sum_j J_{ij}~,~~~d_2 =
{1\over {8 J^{\rm max}}^2}\sum_j J_{ij}^2~,
\label{EqGenHTS3:b}
\end{equation}
\begin{equation} d_3 = {1\over {24 J^{\rm max}}^3}\sum_j J_{ij}^3~.
\label{EqGenHTS3:c}
\end{equation}
\end{mathletters} 
Including only the first term on the right-hand-side of
Eq.~(\ref{EqGenHTS3:a}) gives the Curie law $\chi^*(t) = C^*/t$ with reduced
Curie constant $C^* \equiv C/(N g^2\mu_{\rm B}^2) = S(S+1)/3 = 1/4$, whereas
the first and second terms together yield the Curie-Weiss law
$\chi^*(t) = C^*/(t - \theta^*)$ with reduced Weiss temperature $\theta^*
\equiv k_{\rm B}\theta/J^{\rm max} = -d_1 = -(1/4)\sum_j J_{ij}/J^{\rm max}$
(see Sec.~\ref{SecMFT} below).

A geometrically frustrated spin lattice is one in which there exist closed
exchange path loops containing an odd number of bonds.  Usually, the
exchange path loops are triangles containing three bonds (such as in the 2D
triangular lattice), where at least two nearest neighbors of a given spin
are nearest neighbors of each other, although e.g.\ spin rings with any odd
number of spins (and therefore an odd number of bonds) are also
geometrically frustrated.  Another example of a system containing
triangular exchange path loops is the dimer system ${\rm SrCu_2(BO_3)_2}$ 
(intradimer interaction $J_1\equiv J^{\rm max}$) with a partially
frustrating interdimer interaction $J_2$.  One can show that
Eq.~(\ref{EqGenHTS2}) agrees exactly to ${\cal O}(1/t^2)$ with the
HTSE\cite{Weihong1998,Miyahara1998b} for $\chi(T)T$ of this system.  The
frustration first becomes apparent in the HTSE as an additional additive
term [$(-15/4)(J_2/J_1)^2$ in this case] in the numerator of the $1/t^3$
coefficient in the square brackets in Eq.~(\ref{EqGenHTS2}).  In the
context of the present discussion, a second- or further-nearest-neighbor
interaction is equivalent to a nearest-neighbor one in a system with
geometric frustration, and hence the general  expansion~(\ref{EqGenHTS2})
for $\chi^* t$ is still exact to
${\cal O}(1/t^2)$ for such systems, provided again that all spins are
magnetically equivalent.

For our isolated and coupled ladder QMC simulation fits, the three $d_n$
HTSE coefficients in Eqs.~(\ref{EqGenHTS3}) are
\begin{eqnarray} d_1 = {1\over 4 J^{\rm max}}\Big[2 J + J^\prime &+& 2
{J^{\rm diag}} + 2 J^{\prime\prime}\nonumber\\
 &+& 2 J^{\prime\prime\prime} + 2 J^{\rm 3D} \Big]~,\nonumber\\  d_2 =
{1\over 8 {J^{\rm max}}^2}\Big[2 J^2 + {J^\prime}^2 &+& 2 {J^{\rm diag}}^2
+ 2 {J^{\prime\prime}}^2\nonumber\\
 &+& 2 {J^{\prime\prime\prime}}^2 + 2 {J^{\rm 3D}}^2\Big]~,\nonumber\\  d_3
= {1\over 24 {J^{\rm max}}^3}\Big[2 J^3 + {J^\prime}^3 &+& 2
{J^{\prime\prime}}^3 + 2 {J^{\prime\prime\prime}}^3\nonumber\\
 &+& 2 {J^{\rm 3D}}^3 - 9 J{J^{\prime\prime}}^2/4\Big]~,
\label{Eqsd}
\end{eqnarray} where the last term in $d_3$, given in the HTSE in
Ref.~\onlinecite{Miyahara1998} for $\chi^*(t)$ of the trellis layer, arises
due to the geometric frustration in the trellis layer interladder
coupling.  The $d_n$ in Eqs.~(\ref{Eqsd}) are the correct HTSE coefficients
in Eq.~(\ref{EqGenHTS3:a}), except for $d_3$ in the case of diagonal
second-neighbor intraladder couplings $J^{\rm diag}$; in this latter case
we will not use $d_3$ in the fit function.

Weihong, Singh and Oitmaa have computed the HTSE for the product $\chi(T)T$
of the isolated $S = 1/2$ two-leg Heisenberg ladder with spatially
anisotropic exchange in the leg and rung to 9th order in $1/T$, which
contains a total of 54 nonzero coefficients in powers of $J$ and/or
$J^\prime$.\cite{Weihong1997}  As anticipated above, the series simplifies
if it is inverted.  In addition, this inversion allowed us to easily
estimate the rational fractions approximated by the ten-significant-figure
decimal coefficients given by \mbox{Weihong {\it et al}}.  Our result for
the inverted ninth-order series, containing 42 nonzero terms, is
\newpage
\begin{eqnarray} {Ng^2\mu_{\rm B}^2\over 4 \chi T} = 1 &+& \Big(2 J + 
{J^\prime}\Big) {x\over 2}  + \Big(2 J^2 +  {J^\prime}^2\Big) {x^2\over
2}\nonumber\\ &+& \Big(2 J^3 +  {J^\prime}^3\Big) {x^3\over 3} + 
  \Big(3 J^4 + 4  {J^\prime}^4\Big) {x^4\over 24}\nonumber\\ &-& \Big(116
J^5 + 99 J  {J^\prime}^4 - 32{J^\prime}^5\Big)  {x^5\over 480}\nonumber\\
&-& \Big(317 J^6 - 111 J^4{J^\prime}^2 - 96 J^3{J^\prime}^3 \nonumber\\ &+&
642 J^2{J^\prime}^4 + 297 J{J^\prime}^5 - 32{J^\prime}^6\Big)  {x^6\over
1440}\nonumber\\  &+& \Big(792 J^7 + 3444 J^5  {J^\prime}^2 + 5068 J^4 
{J^\prime}^3 \nonumber\\ &-& 22932 J^3  {J^\prime}^4 - 10332 J^2 
{J^\prime}^5 \nonumber\\ &-& 10395 J  {J^\prime}^6 + 256{J^\prime}^7\Big)
{x^7\over 40320} \nonumber\\ &+& \Big(6165 J^8 - 411 J^6  {J^\prime}^2 +
604 J^5  {J^\prime}^3 \nonumber\\ &-& 26477 J^4  {J^\prime}^4 - 8220 J^3 
{J^\prime}^5 \nonumber\\ &-& 9580 J^2 {J^\prime}^6 - 9702 J{J^\prime}^7 +
64{J^\prime}^8\Big) {x^8\over 40320} \nonumber\\ &+& \Big(23674 J^9 - 29916
J^7  {J^\prime}^2 - 46269 J^6  {J^\prime}^3 \nonumber\\ &-& 228168 J^5 
{J^\prime}^4 - 65340 J^4  {J^\prime}^5 \nonumber\\ &-& 78516 J^3 
{J^\prime}^6 - 51840 J^2{J^\prime}^7 \nonumber\\ &-& 68607 J  {J^\prime}^8
+ 128{J^\prime}^9\Big){x^9\over 362880}~,
\label{EqWeihongHTSEInv}
\end{eqnarray} where
\[ x \equiv {1\over 2 k_{\rm B}T}~.
\] Upon inverting the series in Eq.~(\ref{EqWeihongHTSEInv}) and then
converting each resulting rational fraction coefficient to a
ten-significant-figure decimal value to compare with the HTSE of Weihong
{\it et al.}, each of the 54 coefficients is found to be identical to the
corresponding ten-significant-figure coefficient given by Weihong {\it et
al.}\cite{Weihong1997} The HTSE in Eq.~(\ref{EqWeihongHTSEInv}) is
identical to order $1/T^3$ with the HTSE for
$(\chi T)^{-1}$ in Eq.~(\ref{EqGenHTS3:a}) in which the $d_1,$ $d_2$ and
$d_3$ coefficients are given for the general $S = 1/2$ two-leg ladder by
Eq.~(\ref{Eqsd}), but where in the present case only $J$ and $J^\prime$ are
nonzero.  An interesting aspect of the HTSE in Eq.~(\ref{EqWeihongHTSEInv})
is that in the expression for the coefficient of each $x^n$ term shown, the
coefficient of the $J^{n-1}J^\prime$ term vanishes.

Gu, Yu and Shen have derived an analytic expression for the magnetic field-
and temperature-dependent free energy of the two-leg ladder for strong
interchain couplings \mbox{$J/J^\prime \ll 1$} using perturbation theory to
third order in $J/J^\prime$.\cite{Gu1999}  Our HTSE of $1/[4\chi^*(t)t]$
obtained from their free energy expression is identical to order $1/T^3$
with Eq.~(\ref{EqWeihongHTSEInv}).  As expected, the coefficients of the
fourth order and higher order terms of the HTSE do not agree with the
corresponding correct coefficients in Eq.~(\ref{EqWeihongHTSEInv}).

Just as there is a universal expression for the first three to four HTSE
terms for
$\chi(T)T$ of a Heisenberg spin lattice containing magnetically equivalent
spins as discussed above, a universal HTSE for the magnetic specific heat
$C_{\rm mag}(T)$ of such a spin lattice exists to order $1/T^2$ to $1/T^3$
and is given for $S = 1/2$ by
\begin{equation} {C_{\rm mag}(T)\over N k_{\rm B}} = {3\over
32}\bigg[\frac{\sum_j J_{ij}^2}{(k_{\rm B}T)^2} + \frac{\sum_j J_{ij}^3}{2
(k_{\rm B}T)^3} + {\cal O}\Big({1\over T^4}\Big)\bigg]~.
\label{EqCmHTSE}
\end{equation} The sums are over all exchange bonds from any given spin
$\bbox{S}_i$ to magnetic nearest-neighbor spins $\bbox{S}_j$.  The first
term holds for any spin lattice containing magnetically equivalent spins,
but the second term holds only for geometrically nonfrustrated spin
lattices in which the crystallographic and magnetic nearest-neighbors of
any given spin are the same.  Higher order terms all depend on the
structure and dimensionality of the spin lattice.  The HTSE  for $C_{\rm
mag}(T)$ to (lowest) order $1/T^2$ is the specific heat analogue of the
Curie-Weiss law for the magnetic susceptibility, i.e., they can both be
derived from the same lowest (first) order term in $1/T$ of the
magnetic-nearest-neighbor instantaneous two-spin correlation
function.\cite{Johnston1997}  Equation~(\ref{EqCmHTSE}) is therefore
accurate in the same high-temperature region in which the Curie-Weiss law
for the magnetic susceptibility is accurate.  Physically, the reason that
the lowest-order HTSE terms of
$C_{\rm mag}(T)$ are of the form $J_{ij}^2/T^2$ is that $C_{\rm mag}(T)$
cannot be negative, regardless of the sign(s) of the $J_{ij}$.  

For the two-leg spin ladder couplings considered in this paper, to lowest
order in
$1/T$ Eq.~(\ref{EqCmHTSE}) yields
\begin{eqnarray} {C_{\rm mag}(T)\over N k_{\rm B}} = {3\over 32(k_{\rm
B}T)^2}\Big(2 J^2 &+& {J^\prime}^2 + 2 {J^{\rm diag}}^2 +
2{J^{\prime\prime}}^2 \nonumber\\ &+& 2 {J^{\prime\prime\prime}}^2 + 2
{J^{\rm 3D}}^2\Big)~.
\label{EqCmHTSELadds}
\end{eqnarray} From comparison with Eq.~(\ref{EqCmHTSELadds}), the HTSE for
$C_{\rm mag}(T)$ of isolated $S = 1/2$ spatially anistropic two-leg
Heisenberg ladders ($J,J^\prime\neq 0$) given to lowest order ($1/T^2$) in
Ref.~\onlinecite{Troyer1994} is found to be incorrect.

\subsection{General $\bbox{\chi^*(t)}$ Fit Function}
\label{SecGenFitFcn}

The following fit function incorporating the above considerations, and
containing the Pad\'e approximant ${\cal P}^{(p)}_{(q)}(t)$, was found
capable of fitting the QMC $\chi^*(t)$ data for a given exchange parameter
set to within the accuracy of those data (i.e., to within a $\chi^2/{\rm
DOF}\sim 1$)
\begin{mathletters}
\label{EqChiFit:all}
\begin{equation}
\chi^*(t) = \frac{{\rm e}^{-\Delta^*_{\rm fit}/t}}{4t}\, {\cal
P}^{(p)}_{(q)}(t)~,
\label{EqChiFit:a}
\end{equation}
\begin{equation} {\cal P}^{(p)}_{(q)}(t) = \frac{1+ \sum_{n=1}^p N_n/t^n}{1
+ \sum_{n=1}^q D_n/t^n}~,
\label{EqChiFit:b}
\end{equation}
\end{mathletters} 
\newpage
\noindent which satisfies the Curie law at high temperatures, where
$\Delta^*_{\rm fit}$ is not necessarily the same as the true spin gap.  In
order to further constrain the fit and also to produce a fit which can be 
accurately extrapolated to high temperatures, we require that a HTSE of
Eqs.~(\ref{EqChiFit:all}) reproduce the HTSE in
Eqs.~(\ref{EqGenHTS2})--(\ref{Eqsd}), which in turn yields the constraints
\begin{mathletters}
\label{EqD:all}
\begin{eqnarray} D_1 = (d_1 &+& N_1) - \Delta^*_{\rm fit}~,
\label{EqD:a}\\ D_2 = (d_2 &+& d_1 N_1 + N_2) - \Delta^*_{\rm fit}(d_1 +
N_1) + {{\Delta^*_{\rm fit}}^2\over 2}~,
\label{EqD:b}\\ D_3 = (d_3 &+& d_2 N_1 + d_1 N_2 + N_3)  - \Delta^*_{\rm
fit}(d_2 + d_1 N_1 + N_2)\nonumber\\
 &+& {{\Delta^*_{\rm fit}}^2\over 2}(d_1 + N_1) - {{\Delta^*_{\rm
fit}}^3\over 6}~,
\label{EqD:c}
\end{eqnarray}
\end{mathletters} In general, one has
\begin{equation} D_n = \sum_{p=0}^n \frac{(-{\Delta^*_{\rm
fit}})^p}{p!}\sum_{m=0}^{n-p} d_m N_{n-p-m}~.
\label{EqD:d}
\end{equation} Unless otherwise explicitly noted for a specific fit, $D_1,\
D_2$ and $D_3$ are not independent fitting parameters but are rather
determined from the fitting parameters $N_1,\ N_2,\ N_3$ and $\Delta^*_{\rm
fit}$ in Eqs.~(\ref{EqD:all}), where $\Delta^*_{\rm fit}$ can also be a
fitting parameter.  To obtain a fit to a QMC $\chi^*(t)$ data set for a
specific set of exchange constants to within the accuracy of the data,
i.e.\ which yielded $\chi^2/{\rm DOF} \sim 1$, typically required a total
of 6--9 independent fitting parameters, which was essentially independent
of the number of data points in the data set.

Finally, we reformulated the fit function into a two- or three-dimensional
one so that it could not only interpolate and extrapolate $\chi^*$ versus
$t$ for a given set of exchange constants but could also interpolate
$\chi^*(t)$ for a range of exchange constants.  To do this, we expressed
the parameters
$N_n,\ D_n$ and sometimes $\Delta^*_{\rm fit}$ in Eqs.~(\ref{EqChiFit:all})
as power series in the exchange constants; this also considerably reduced
the total number of fitting parameters required to obtain a global fit to
$\chi^*(t)$ data for a given range of exchange constants.  This scheme was
successfully used except for exchange constant ranges traversing a QCP, for
which two piecewise continuous interpolation fits were required for the two
exchange constant ranges on opposite sides of the QCP, respectively.  The
resulting fits to the QMC $\chi^*(t)$ simulation data and several exchange
parameter interpolations are shown as the sets of solid curves in the above
QMC data figures, as described in the captions.  For most of the QMC
simulations, a $\chi^2/{\rm DOF}\sim 1$ was obtained.  The high quality of
the fits may therefore perhaps be appreciated from the small errors
estimated for the QMC data, especially at the higher temperatures, which
varied from
$\sim 1$--10\% for $0.01\lesssim t \lesssim 0.1$ to 0.03--0.1\% for
$t\gtrsim 0.5$.  The details of the fitting procedures and tables of fitted
parameters are given in the Appendix.

\newpage

\subsection{Fit Functions for $\bbox{\chi^*(t)}$ Derived from \protect\\ 
Molecular Field Theory}
\label{SecMFT}

For Heisenberg spin lattices consisting of identical spin subsystems which
are weakly coupled to each other, it is sometimes necessary to use a fit
function for theoretical $\chi^*(t)$ data in the paramagnetic phase which
contains a minimum number (perhaps only zero, one or two) of fitting
parameters, and which still provides a reasonably good fit to the data. 
Such fit functions can be provided by molecular field theory (MFT) and its
extensions as described in this section.  Each isolated spin subsystem is
assumed to have a known susceptibility $\chi^*_0(t)$.  It can be easily
shown that if each spin in the entire system is magnetically equivalent to
every other spin, with spins in adjacent subsystems coupled by Heisenberg
exchange, then the reduced susceptibility $\chi^*(t)$ in the paramagnetic
state of the system is given by MFT as
\begin{mathletters}
\label{EqMFT:all}
\begin{equation}
\chi^*(t) = \frac{\chi^*_0(t)}{1 + \lambda\,\chi^*_0(t)}~,
\label{EqMFT:a}
\end{equation} or equivalently
\begin{equation} {1\over \chi^*(t)} = {1\over \chi^*_0(t)} + \lambda~,
\label{EqMFT:b}
\end{equation} where the MFT coupling constant $\lambda$ is given by
\begin{equation}
\lambda = \sum_j^{\ \ \ \ \ \ \prime} {J_{ij}\over J^{\rm max}}~,
\label{EqMFT:c}
\end{equation}
\end{mathletters} the prime on the sum over $j$ signifies that the sum is
only taken over exchange bonds $J_{ij}$ from a given spin $\bbox{S}_i$ to
spins $\bbox{S}_j$ not in the same spin subsystem, and $J^{\rm max}$ is the
exchange constant in the system with the largest magnitude.  By definition,
the expression for
$\chi^*_0(t)$ does not contain any of these $J_{ij}$ interactions which are
external to a subsystem.  Within MFT, Eqs.~(\ref{EqMFT:all}) are correct at
each temperature in the paramagnetic state not only for bipartite AF spin
systems, but also for any system containing subsystems coupled together by
any set of FM and/or AF Heisenberg exchange interactions.  The only
restriction, as noted above, is that each spin in the system is
magnetically equivalent to every other spin in the system.  Thus, our fit
functions derived in this section could have been used to fit
$\chi^*(t)$ data for any  of the coupled two-leg ladder spin lattices
discussed in this paper, although in general to much lower accuracy than
obtained in the above section and the Appendix.  However, they do not apply,
e.g., to trellis layers containing three-leg ladders, because in this case
the spins are not all magnetically equivalent since the magnetic
environment of a spin in the central leg of such a ladder is different from
that of a spin in the outer two legs of the ladder.

Before proceeding further, we first make contact with the familiar case in
which a subsystem consists of a single spin.  Then $\chi_0(T)$ is the Curie
law,
\begin{mathletters}
\label{EqCurieLaw:all}
\begin{equation}
\chi_0(T) = {C\over T}~,
\label{EqCurieLaw:a}
\end{equation} where the Curie constant is
\begin{equation} C = {N g^2 \mu_{\rm B}^2 S(S+1)\over 3 k_{\rm B}}~.
\label{EqCurieLaw:b}
\end{equation} In reduced units, the Curie law reads
\begin{equation}
\chi^*_0(t) \equiv {\chi_0(T)J^{\rm max}\over Ng^2\mu_{\rm b}^2} = {C^*\over
t}~,
\label{EqCurieLaw:c}
\end{equation} where the reduced Curie constant $C^*$ and reduced
temperature $t$ are defined as
\begin{equation} C^* \equiv {C\over Ng^2\mu_{\rm B}^2} = {S(S+1)\over 3}~.
\label{EqCurieLaw:d}
\end{equation}
\begin{equation} t \equiv {k_{\rm B}T\over J^{\rm max}}~.
\end{equation}
\end{mathletters} Then our general expression~(\ref{EqMFT:a}) incorporating
interactions between the spins yields the Curie-Weiss law $\chi(T) = C/(T -
\theta)$ in reduced form as
\begin{mathletters}
\label{EqCurie-Weiss:all}
\begin{equation}
\chi^*(t) = {C^*\over t - \theta^*}
\label{EqCurie-Weiss:a}
\end{equation} with reduced Weiss temperature $\theta^*$ given by
\begin{equation}
\theta^* \equiv -{k_{\rm B}\theta\over J^{\rm max}} = -{S(S+1)\over 3}\sum_j
{J_{ij}\over J^{\rm max}}~,
\label{EqCurie-Weiss:b}
\end{equation}
\end{mathletters} where we have removed the prime from the sum because in
this case all exchange interactions in the system are external to a
subsystem which consists only of a single spin $\bbox{S}_i$.

Equation~(\ref{EqMFT:a}) can be used as a fit function containing no
adjustable parameters to parametrize numerical $\chi^*(t)$ data for spin
systems with weak intersubsystem coupling.  In the following, we extend the
MFT framework to provide latitude for including adjustable fitting
parameters to improve the quality of the fit.  To obtain a general form for
the fit function for $S
 = 1/2$ Heisenberg spin systems we first rewrite the HTSE for $\chi^*(t)$ in
Eqs.~(\ref{EqGenHTS3}), absorbing the HTSE terms for a subsystem back into
the exact $\chi^*_0(t)$ for the subsystem which already implicitly contains
the correct HTSE for the subsystem, leaving only the external interactions
explicit, yielding the modified HTSE
\begin{mathletters}
\label{EqMFTHTSE:all}
\begin{equation} {1\over \chi^*(t)} = {1\over \chi^*_0(t)} + 4d_1^\prime +
{4d_2^\prime\over t} + {4d_3^\prime\over t^2} + \cdots~,
\label{EqMFTHTSE:a}
\end{equation} with
\begin{equation} 4d_1^\prime = {1\over J^{\rm max}}\sum_j^{\ \ \ \ \ \prime}
J_{ij}~,~~~4d_2^\prime = {1\over {2 J^{\rm max}}^2}\sum_j^{\ \ \ \ \ \prime}
J_{ij}^2~,
\label{EqMFTHTSE:b}
\end{equation}
\begin{equation} 4d_3^\prime = {1\over {6 J^{\rm max}}^3}\sum_j^{\ \ \ \ \
\prime} J_{ij}^3~,
\label{EqMFTHTSE:c}
\end{equation}
\end{mathletters} where the prime on the sums has the same meaning as in
Eq.~(\ref{EqMFT:c}).  Again, $d_3^\prime$ has additional terms if geometric
frustration and/or second-neighbor interactions are present in the
intersubsystem couplings.  Comparison of Eqs.~(\ref{EqMFTHTSE:all})
and~(\ref{EqMFT:all}) shows explicitly that MFT exactly yields the first
expansion term ($\lambda = 4d_1^\prime$) of the quantum mechanical HTSE for
$\chi^*(t)$ in terms of the exchange constants external to a subsystem. 
Note that the Weiss temperature in the Curie-Weiss law is always given by
Eq.~(\ref{EqCurie-Weiss:b}), where the sum over $j$ is over all magnetic
nearest neighbors of a given spin and not just over those external to a
subsystem.

We now rewrite the HTSE in Eqs.~(\ref{EqMFTHTSE:all}) in the form
\begin{mathletters}
\label{EqMFTGenf:all}
\begin{equation}
\chi^*(t) = {\chi^*_0(t)\over 1 + f(J_{ij},t)\chi^*_0(t)}~,
\label{EqMFTGenf:a}
\end{equation} with
\begin{equation} f(J_{ij},t) = 4d_1^\prime + {4d_2^\prime\over t} +
{4d_3^\prime\over t^2} + \cdots~,
\label{EqMFTGenf:b}
\end{equation}
\end{mathletters} where the $4d_n^\prime$ parameters for $n=1$--3 are the
same as given in Eqs.~(\ref{EqMFTHTSE:b}) and~(\ref{EqMFTHTSE:c}). 
Equation~(\ref{EqMFTGenf:a}), which is an extension of the MFT prediction in
Eq.~(\ref{EqMFT:a}), can be used as a function to fit
$\chi^*(t)$ data for coupled two-leg ladders, where $\chi^*_0(t)$ is then
the susceptibility of isolated ladders.  The function $f(J_{ij},t)$ in
Eq.~(\ref{EqMFTGenf:b}), which contains (apart from $J^{\rm max}$) only the
intersubsystem exchange constants $J_{ij}$ coupling the ladders to each
other and is expected to be most accurate at high temperatures, can be
modified to provide for the introduction of adjustable fitting parameters
as will be further discussed in the Appendix where we obtain \mbox{ fit }
\mbox{functions } for our
\widetext
\begin{table}
\caption{One-magnon dispersion relations ($k_xa/\pi = 1$) and the lower
boundary of the two-magnon continua ($k_xa/\pi = 0$) for $S = 1/2$ $2\times
12$ two-leg Heisenberg ladders with $J^\prime/J = 0.5$, 0.6, 0.7, 0.8, 0.9
and 1.0, calculated by exact diagonalization using the Lanczos algorithm. 
The spin-gap between the singlet ground state and the lowest triplet
excited state is $\Delta = E(k_ya/\pi = 1)$ for the one-magnon dispersion
relation.}
\label{TabDisprsnRlnDat}
\begin{tabular}{cccccccc} &&&&$k_ya/\pi$\\ & 0	& 1/6	& 1/3	& 1/2	& 2/3	&
5/6	& 1\\
\hline
$J^\prime/J$&&&&One-Magnon $E/J$\\ 0.5 & 1.13926	& 1.25058	& 1.62977	&
1.79736	& 1.52857	& 0.908547 & 0.289208 \\ 0.6	& 1.28127	& 1.38810	&
1.70428	& 1.82438 & 1.52561 & 0.902179 & 0.319648 \\ 0.7	& 1.42954	&
1.52908	& 1.78656 & 1.85470 & 1.52828 & 0.904685 & 0.358123 \\ 0.8	&
1.58158	& 1.67096 & 1.87366 & 1.88871 & 1.53729 & 0.915965 & 0.403962 \\
0.9 & 1.73510 & 1.81177 & 1.96301 & 1.92667 & 1.55290 & 0.935748 & 0.456448
\\ 1.0 & 1.88782 & 1.94993 & 2.05281 & 1.96863 & 1.57507 & 0.963567 &
0.514784 \\
\hline
$J^\prime/J$&&&&Two-Magnon $E/J$\\ 0.5 && 0.960355	& 1.56230 & 1.86742	&
1.78996 & 1.43543 & 1.15334 \\ 0.6 & & 1.01850 & 1.61171	& 1.92615 &
1.87887	& 1.57866 & 1.35375 \\ 0.7 & & 1.08763	& 1.67135 & 1.99573	&
1.98118 & 1.73555	& 1.55798 \\ 0.8 & & 1.16790	& 1.74149	& 2.07589 &
2.09499	& 1.90220 & 1.76404 \\ 0.9 & & 1.25911	& 1.82206	& 2.16602	&
2.21834 & 2.07507	& 1.96985 \\ 1.0 & & 1.36070	& 1.91271	& 2.26527	&
2.34932	& 2.25116 & 2.17341 
\end{tabular}
\end{table}
\narrowtext
\noindent  QMC $\chi^*(t)$ data for the two-leg ladder trellis layer.  In
addition, especially when the intersubsystem interactions significantly
change the spin gap, terms which include the $J_{ij}$ interactions external
to a subsystem and additional fitting parameters could be included in the
$\chi^*_0(t)$ function itself.\section{Dispersion Relations}
\label{SecDisprsnRlns}

\begin{figure}
\epsfxsize=3.3in
\centerline{\epsfbox{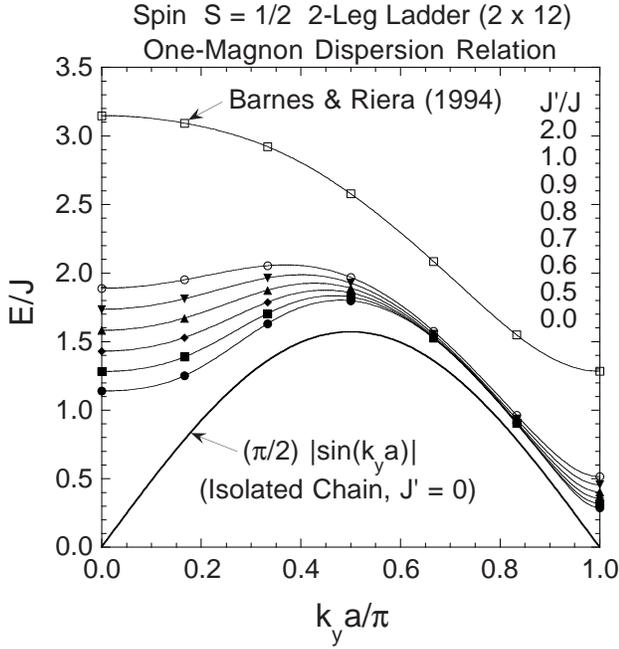}}
\vglue 0.1in
\caption{One-magnon dispersion relations for the isolated two-leg $2\times
12$ ladder calculated for $J^\prime/J = 0.5,\ 0.6,\ \ldots, 1.0$ using the
Lanczos algorithm.  The wavevector $k_y$ is parallel to the legs of the
ladder and $a$ is the nearest-neighbor spin-spin distance ($k_xa/\pi = 1$). 
The data for $J^\prime/J = 2$ calculated for the same lattice using the same
algorithm by Barnes and Riera\protect\cite{Barnes1994} and the dispersion
relation for the isolated chain\protect\cite{desCloizeaux1962} are shown for
comparison.  The solid  curves for $J^\prime/J > 0$ are exact fits to the
data by the square root of an even seven-term Fourier series with period
$\Delta k_ya/\pi = 2$.}
\label{Fig22}
\end{figure}

The one- and two-magnon dispersion relations $E(k_y)$ were computed for $S
={1\over 2}$ $2\times 12$ ladders by exact diagonalization using the Lanczos
algorithm for $J^\prime/J = 0.5,\ 0.6,\ \ldots$, 1.0 and in each case for
wavevectors $k_ya/\pi = 0,\ 1/6,\ 2/6,\ \ldots,\ 1$, where $k_y$ is the
wavevector in the ladder leg direction and $a$ is the nearest-neighbor
spin-spin distance; in the discussion below we set $a = 1$.  The results are
given in Table~\ref{TabDisprsnRlnDat}.  Our value of the spin-gap for
$J^\prime/J = 1$, $\Delta/J = 0.514\,784$, is identical to the six
significant figures with that calculated for the same spin lattice by Yang
and Haxton.\cite{Yang1998}  This value is about 1.8\% 
higher than for the
$2\times 16$ ladder (0.505\,460\,384, Ref.~\onlinecite{Dagotto1998}) and for
the bulk limit discussed in Sec.~\ref{SecSimFits}.  The one-magnon
dispersion relation data are shown as the symbols in
Fig.~\ref{Fig22}.  In the limit $J^\prime/J\to 0$, the exact
dispersion relation calculated for the $S = 1/2$ AF uniform Heisenberg
chain by des Cloizeaux and Pearson in 1962 is
$E(k_y) = (\pi J/2)|\sin(k_y)| =  (\pi J/2)[{1\over 2} - {1\over 2}\cos(2
k_y)]^{1/2}$,\cite{desCloizeaux1962} also shown in
Fig.~\ref{Fig22}, whereas for $J^\prime/J \gg 1$ one has
$E(k_y) = J^\prime + J\cos(k_y)$.\cite{Barnes1993,Reigrotzki1994}  Since our
dispersion relations are for exchange constant ratios $0.5 \leq J^\prime/J
\leq 1$ closer to the former limit, we obtained exact fits to the data by
the {\em square root} of an even seven-term Fourier
series,\cite{Johnston1996,Barnes1994} shown as the solid curves in
Fig.~\ref{Fig22}.  The spin-gap $\Delta$ for each
$J^\prime/J$ occurs at wavevector $k_y = \pi/a$.
\noindent Also included in Fig.~\ref{Fig22} are the earlier results of
Barnes and Riera\cite{Barnes1994} for $J^\prime/J = 2$ computed using the
same algorithm on the same spin lattice.  Their data for $J^\prime/J = 1$
and~0.5 (not shown) are in good agreement with our data for these
$J^\prime/J$ values.

\begin{figure}
\epsfxsize=3.3in
\centerline{\epsfbox{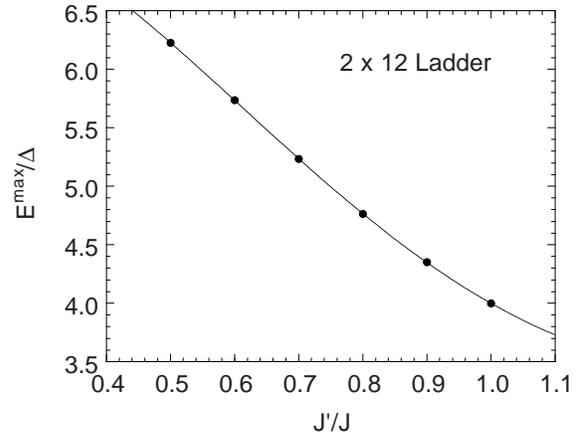}}
\vglue 0.1in
\caption{Ratio of the maximum to the minimum one-magnon excitation energy,
$E^{\rm max}/\Delta$ ($\bullet$), versus $J^\prime/J$ for the $2\times 12$
two-leg ladder, for magnons with wavevector parallel to the legs of the
ladder, from the fits to the data in Fig.~\protect\ref{Fig22}.  The solid
curve is a polynomial fit to the data.}
\label{Fig23}
\end{figure}

A notable feature of the data in Fig.~\ref{Fig22} is that the
ratio
$E^{\rm max}/\Delta$ of the maximum to the minimum energy of each dispersion
relation estimated using the above exact fits to the data is a strong
function of $J^\prime/J$, as shown in Fig.~\ref{Fig23}, which
facilitates obtaining highly precise estimates of $J^\prime/J$  from neutron
scattering data.  This observation, previously made by us on the basis of
the earlier dispersion relations of Barnes and Riera,\cite{Barnes1994} was
used by Eccleston~{\it et al.}\ to estimate $J^\prime/J$ for the two-leg
ladders in ${\rm Sr_{14}Cu_{24}O_{41}}$ from their neutron scattering data
on single crystals of this compound.\cite{Eccleston1998}  Their data, in
turn, motivated us in the present work to compute the dispersion relations
on a finer grid of $J^\prime/J$ values than existed previously.  We will
discuss these calculations and experimental data further in
Sec.~\ref{SecSummaryDisc}.

\begin{figure}
\epsfxsize=3.3in
\centerline{\epsfbox{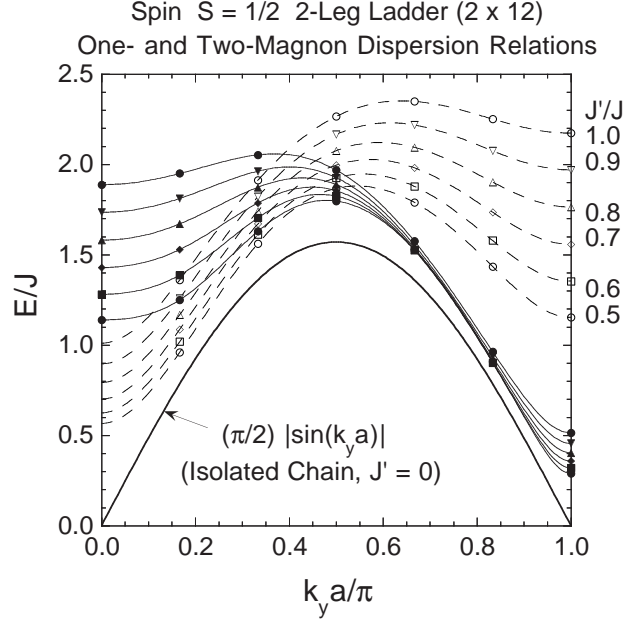}}
\vglue 0.1in
\caption{The lower boundary of the two-magnon continuum ($k_xa/\pi = 0$,
open symbols) for the isolated two-leg $2\times 12$ ladder calculated for
$J^\prime/J = 0.5,\ 0.6,\ \ldots, 1.0$ using the \mbox{Lanczos} algorithm. 
The wavevector $k_y$ is parallel to the legs of the ladder and $a$ is the
nearest-neighbor spin-spin distance.  For comparison, the dispersion
relations for one-magnon excitations ($k_xa/\pi = 0$, filled symbols, from
Fig.~\protect\ref{Fig22}) are included. The dashed curves for
$J^\prime/J > 0$ are exact fits to the two-magnon data by the square root of
an even six-term Fourier series with period $\Delta k_ya/\pi = 2$.}
\label{Fig24}
\end{figure}

The lower boundary of the two-magnon continuum ($k_x = 0$) is shown in
Fig.~\ref{Fig24} for $J^\prime/J \leq 1$ over much of the $k_y$ range. 
From our data on the finite-size ladder, we cannot clearly distinguish
between the two-magnon scattering \,states \,and \,bound \,states lying near
the lower
\noindent boundary of the two-magnon continuum for $k_y \sim
\pi$.\cite{Sushkov1998,Jurecka1999}  Each dashed curve is an exact fit to
the respective two-magnon data by the square root (see above) of a six-term
Fourier series.  The two-magnon excitations 
are degenerate with the
one-magnon spectra over much of the Brillouin zone.

Interladder coupling within the trellis-layer ($J''$) has virtually no 
influence on the one-magnon dispersion and on the spin structure factor
close to the dispersion minimum, as the contributions due to this coupling 
interfere destructively at \mbox{$k_{y}a=\pi$}.  A coupling $J'''$ in the 
third ($z$) dimension, perpendicular to the trellis layer, will however
give an additional dispersion $J'''\cos(k_{z})$ that has to be taken into
account.  Away from the minimum the magnon band disperses also due to the
trellis layer interladder coupling $J''$, most strongly close to the
dispersion maximum around $k_{y}a\approx\pi/2$.  As was shown by Lidsky and
Troyer,\cite{Lidsky1998} the band center is not moved substantially by
$J''$, and averaging over all momenta perpendicular to the ladders, as done
by Eccleston~{\it et al.},\cite{Eccleston1998} essentially recovers the
uncoupled ladder dispersion. The neutron scattering function depends of course on the relative
contributions of all the magnetic excitations.  Our calculations of the
dynamical spin structure factor \mbox{ $S(\bbox{q},\omega)$ } for
\widetext
\begin{figure}
\epsfxsize=6.6in
\centerline{\epsfbox{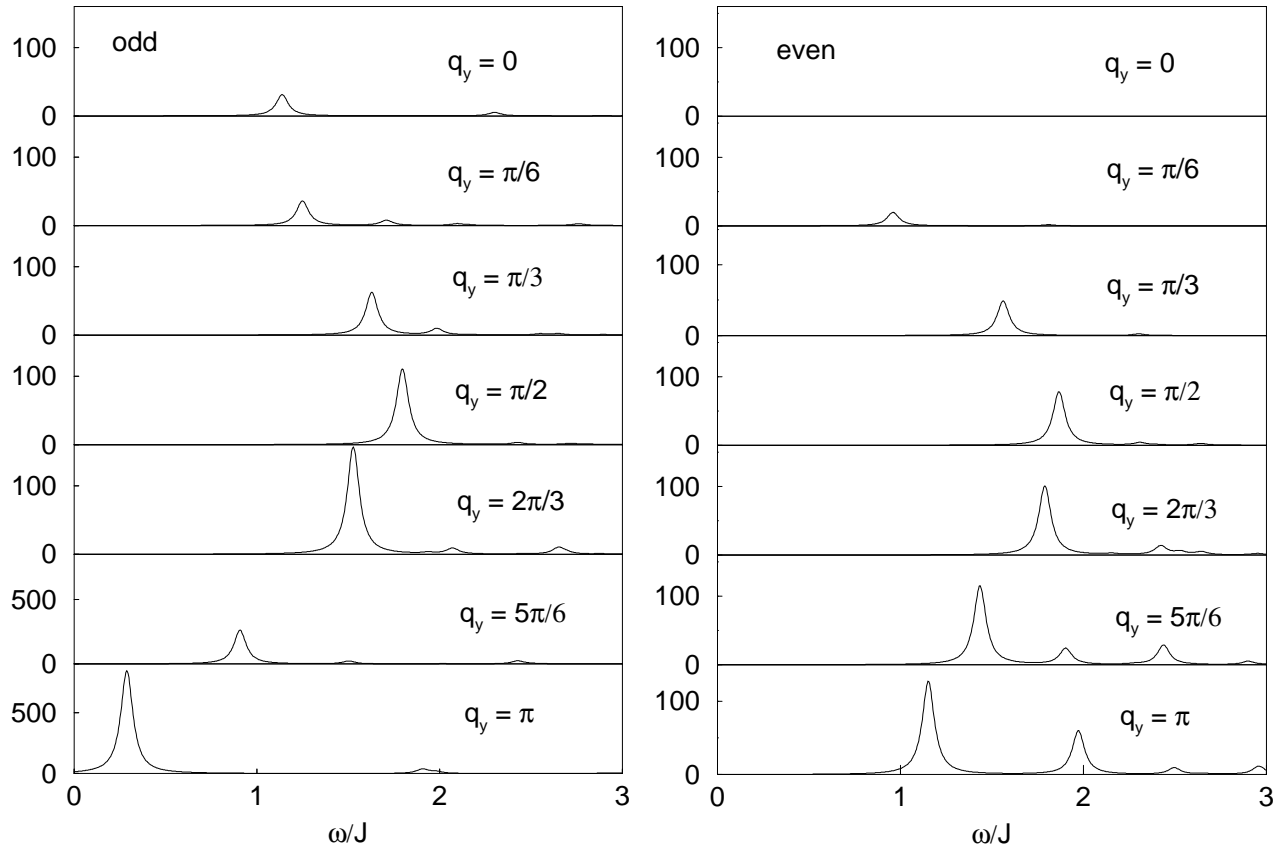}}
\vglue 0.1in
\caption{Dynamical spin structure factor $S(q_{y},\omega)$ (in arbitrary
units) for the isolated $S = 1/2$ AF two-leg Heisenberg ladder with $L=12$
rungs for the physically relevant coupling $J^\prime/J = 1/2$ and for both
the odd number of magnons (left panel) and even number of magnons (right
panel) sectors.  As in the text, the $y$ direction is defined to be along
the ladder legs and the $x$ direction along the ladder rungs.  In the
abscissa label we have set $\hbar = 1$.  Note that the ordinate scale is
different for the odd number of magnons sector at $q_{y} = 5\pi/6$ and
$q_{y} = \pi$ than for the other plots. It is apparent that for momenta
along the ladder $q_{y}$ near the top of the one-magnon band, the two-magnon
continuum (lowest peaks in the even sector) has energy and spectral weight
comparable to those of the single magnon collective mode.}
\label{Fig25}
\end{figure}
\narrowtext
\noindent 
a $2\times 12$ ladder and for the experimentally relevant (see
Sec.~\ref{SecExpDatModeling}) intraladder coupling ratio $J^\prime/J =
0.5$, shown in Fig.~\ref{Fig25},  demonstrate that around the top of the
one-magnon band the two-magnon states have energy and weight comparable to
those of the single magnon band.  Therefore the two-magnon states should be
taken into account when fitting inelastic neutron scattering data to obtain
the part of the  one-magnon dispersion relation at the higher energies.

Related results for $S(\bbox{q},\omega)$ of $S = 1/2$ two-leg Heisenberg
ladders have been obtained previously.  A study of the $2\times 12$ ladder
with spatially anisotropic exchange by Yang and Haxton indicated that the
contribution of the lowest one-magnon triplet states to the response
function at wavevector ($\pi,\pi$) increased from 91.7\% to 96.7\% of the
total response at that wavevector as $J^\prime/J$ increased from 0.4 to 1;
the total response itself at this wavevector peaked at $J^\prime/J \sim
0.5$.\cite{Yang1998}  Calculations of the odd number of magnons
sector of $S(\bbox{q},\omega)$ for the $2 \times 16$ ladder
with $J^\prime/J\ =\ 1$\ \ were\ \ reported \ by \ \mbox{Dagotto {\it et
al.}}\cite{Dagotto1998} \ \,Their 
\newpage
\noindent results showed that the one-magnon contribution to
$S(\bbox{q},\omega)$ continuously decreases as the wavevector decreases
from ($\pi,\pi$) to ($\pi$,0),\cite{Dagotto1998} which is qualitatively the
same as we have found for $J^\prime/J = 0.5$.

\section{LDA+U Calculations of Exchange Constants in 
S\lowercase{r}C\lowercase{u}$_2$O$_3$}
\label{SecLDA+U}

An {\it ab-initio} calculation using the LDA+U method\cite{ci1} was
enlisted to compute the electronic structure of SrCu$_2$O$_3$ and to extract
from it the exchange couplings.  The atomic coordinates used are those
given in Sec.~\ref{SecSrCu2O3Struct} below.  The LDA+U method has been
shown to give good results for insulating transition metal oxides with a
partially filled
$d$-shell.\cite{review}  The exchange interaction parameters can be
calculated using a procedure based on the Greens function method which was
developed by  A.~I.~Lichtenstein.\cite{lichtexchange,lichtan}  This method
was successfully applied for calculation of the exchange couplings in 
KCuF$_3$ (Ref.~\onlinecite{lichtan}) and in layered vanadates
CaV$_n$O$_{2n+1}$.\cite{Korotin1999}

The LDA+U method\cite{ci1,review} is essentially the Local Density
Approximation (LDA) modified  by a potential correction restoring a proper
description of the Coulomb interaction between localized $d$-electrons of
transition metal ions.  This is written in the form of a projection operator
\begin{equation}
\widehat{H}=\widehat{H}_{\rm LSDA}+
\sum_{mm^{\prime }}\mid inlm\sigma \rangle V_{mm^{\prime }}^\sigma \langle
inlm^{\prime }\sigma \mid~,
\label{hamilt}
\end{equation}
\begin{eqnarray} V_{mm^{\prime }}^\sigma &=&\sum_{\{m\}}
\{\langle m,m^{\prime \prime }\mid V_{ee} \mid m^{\prime },m^{\prime \prime
\prime }
\rangle n_{m^{\prime \prime } m^{\prime \prime \prime }}^{-\sigma }
\nonumber \\ &&+(\langle m,m^{\prime \prime }\mid V_{ee}
\mid m^{\prime },m^{\prime \prime \prime }\rangle \label{Pot}
 \\ &&-\langle m,m^{\prime \prime }
\mid V_{ee}\mid m^{\prime \prime \prime }, m^{\prime }\rangle )
n_{m^{\prime \prime }m^{\prime \prime
\prime }}^\sigma \} \nonumber \\ &&-U\Big(N-\frac
12\Big)+J\Big(N^{\sigma}-\frac 12\Big)~,\nonumber 
\end{eqnarray} where $\mid inlm\sigma \rangle $ ($i$ denotes the site, $n$
the main  quantum number, $l$- orbital quantum number, $m$- magnetic number
and 
$\sigma$- spin index) are $d$-orbitals of transition metal ions.  The 
density matrix is defined by
\begin{equation} n_{mm^{\prime }}^\sigma =-\frac 1\pi \int^{E_F}{\rm Im}
G_{inlm,inlm^ {\prime}}^\sigma(E)\,dE~,
\label{Occ}
\end{equation} where $G_{inlm,inlm^{^{\prime }}}^\sigma (E)= \langle
inlm\sigma \mid  (E-\widehat{H})^{-1}\mid inlm^{^{\prime }}\sigma \rangle $
are the  elements of the Green function matrix, $N^\sigma 
=Tr(n_{mm^{\prime}}^\sigma )$, and $N=N^{\uparrow }+N^{\downarrow }$.
$U$ and $J$ are screened Coulomb and exchange parameters calculated  via
the so-called ``supercell'' procedure \cite{superlsda} and found to be
$U=7.79$\,eV and $J=0.92$\,eV, respectively.  The calculation scheme was
realized in the framework of the Linear Muffin-Tin Orbitals (LMTO) method
\cite{lmto} based on the Stuttgart TBLMTO-47 computer code.

\newpage

The inter-site exchange couplings were calculated with a formula which  was
derived using the Green function method as the second derivative of  the
ground state energy with respect to the magnetic moment rotation  angle
\cite{lichtexchange,lichtan}
\begin{equation}
\label{exchange} J_{ij}=\sum_{\{m\}}I_{mm^{\prime }}^i
\chi _{mm^{\prime }m^{\prime \prime} m^{\prime \prime \prime }}^{ij}
I_{m^{\prime \prime }m^{\prime \prime \prime }}^j~,
\end{equation} where the spin-dependent potentials $I$ are expressed in
terms of the potentials of Eq.~(\ref{Pot}) as
\begin{equation}  
\label{magpot} I_{mm^{\prime }}^i=V_{mm^{\prime }}^{i\uparrow }-
V_{mm^{\prime}}^{i\downarrow }~.
\end{equation} The effective inter-sublattice susceptibilities are defined
in terms of  the LDA+U eigenfunctions $\psi$ as
\begin{equation}
\label{suscep}
\chi _{mm^{\prime }m^{\prime \prime }m^{\prime \prime \prime }}^{ij}=\sum_{
{\bf k}nn^{\prime }}\frac{n_{n{\bf k\uparrow }}-n_{n^{\prime }{\bf
k\downarrow }}}{\epsilon _{n{\bf k\uparrow }}-\epsilon _{n^{\prime }{\bf
k\downarrow }}}\psi _{n{\bf k\uparrow }}^{ilm^{*}}\psi _{n{\bf k\uparrow }
}^{jlm^{\prime \prime }}\psi _{n^{\prime }{\bf k\downarrow }}^{ilm^{\prime
}}\psi _{n^{\prime }{\bf k\downarrow }}^{jlm^{\prime \prime \prime *}}.
\end{equation}

Equation (\ref{exchange}) was derived as a second derivative of the  total
energy with respect to the angle between spin directions of the  LDA+U
solution.  The LDA+U method is the analogue of the Hartree-Fock  (HF,
mean-field) approximation for a degenerate Hubbard model.\cite{review} 
While in the multi-orbital case a mean-field  approximation gives reasonably
good estimates for the total energy, for the non-degenerate Hubbard model
it is known to underestimate the triplet-singlet energy difference (and
thus the value of effective  exchange parameter $J_{ij}$) by a factor of
two for a two-site problem ($E_{\rm HF}=\frac{2t^2}{U}$ and $E_{\rm exact}
=\frac{4t^2}{U}$, where $t\ll  U$ is the inter-site hopping parameter).
Thus the
$J$ value calculated by expression (\ref{exchange}) was multiplied by a
factor of two to correct the Hartree-Fock value.  The calculated results are
presented in Table~\ref{tab:calcJ}.
\vglue 0.35in
\begin{table}
\caption{Exchange constants in SrCu$_2$O$_3$, calculated using the LDA+U
method, along with those obtained for CaV$_2$O$_5$ and MgV$_2$O$_5$ by
Korotin {\it et al.}\protect\cite{Korotin1999,Korotin2000} using the same
method.  For the definitions of the exchange constants, see
Eq.~(\protect\ref{EqJDefs}).  Antiferromagnetic (AF) and ferromagnetic (FM)
exchange constants are positive and negative, respectively.  $J^{\rm max}$
is the largest (AF) exchange constant in the system.}
\begin{tabular}{lccc} & SrCu$_2$O$_3$ & CaV$_2$O$_5$ & MgV$_2$O$_5$ \\
\hline
$J/k_{\rm B}\,$(K) & 1795 & 122 & 144 \\
$J^\prime/k_{\rm B}\,$(K) & 809 & 608 & 92 \\
$J^{\rm diag}/k_{\rm B}\,$(K) & $-$200 & 20 & 19 \\
$J^{\prime\prime}/k_{\rm B}\,$(K) & 4 & $-$28 & 60 \\
$J^{\prime\prime\prime}/k_{\rm B}\,$(K) & 3 \\
\hline
$J^{\rm max}$ & $J$ & $J^\prime$ & $J$ \\
$J/J^{\rm max}$ & 1 & 0.201 & 1 \\
$J^\prime/J^{\rm max}$ & 0.451 & 1 & 0.64 \\
$J^{\rm diag}/J^{\rm max}$ & $-$0.111 & 0.033 & 0.13 \\
$J^{\prime\prime}/J^{\rm max}$ & 0.002 & $-$0.046 & 0.42 \\
$J^{\prime\prime\prime}/J^{\rm max}$ & 0.002 \\
\end{tabular}
\label{tab:calcJ}
\end{table}

\section{Structure Refinements of S\lowercase{r}C\lowercase{u}$_2$O$_3$ and
S\lowercase{r}$_2$C\lowercase{u}$_3$O$_5$}
\label{SecSrCu2O3Struct}

Up to now the detailed crystal structure of the two-leg ladder compound
SrCu$_2$O$_3$ has not been reported.  This is necessary as input to our
LDA+U calculations in the previous section.  Here we give our refinement
results for this compound as well as those for the three-leg ladder compound
Sr$_2$Cu$_3$O$_5$.  Powder X-ray diffraction data were collected with a
Rigaku RAD-C diffractometer equipped with a graphite crystal monochrometer
(CuK$\alpha$ radiation, 30\,kV, 100\,mA).  Data were collected from 20 to
120\,$^\circ$ with a step width of 0.02\,$^\circ$.  Lattice parameters and
the atomic positions were refined by a Rietveld technique using Rietan
software.\cite{Izumi1993}

Our crystal data for SrCu$_2$O$_3$ and Sr$_2$Cu$_3$O$_5$ are given in
Table~\ref{TabSrCu2O3XtalDat}.  The atomic positions for Sr$_2$Cu$_3$O$_5$
are in good agreement with those of Kazakov {\it et al.}\cite{Kazakov1997}
determined from Rietveld refinement of powder neutron diffraction data, but
our lattice parameters are significantly smaller, indicating a possible
homogeneity range and/or variable density of defects in this compound. 
Selected interatomic distances and bond angles are listed for both
compounds in Table~\ref{TabBonds}.
\vglue 1.0in
\begin{table}
\caption{Crystal structure refinement data for SrCu$_2$O$_3$ and
Sr$_2$Cu$_3$O$_5$.  Sample color: black; radiation: CuK$\alpha$;
temperature: 290\,K; monochromator: graphite; 2$\theta$ range: 20
to~120$^\circ$; step width: 0.02$^\circ$; crystal system: orthorhombic;
space group: {\it Cmmm}.  SrCu$_2$O$_3$: lattice parameters (\AA): $a$ =
3.9300(1),
$b$ = 11.561(4), $c$ = 3.4925(1); density: 5.497\,g/cm$^3$;
$R_{wp} = 4.51$\%; $R_{p} = 2.78$\%; $R_{e} = 2.76$\%; $R_{I} = 2.76$\%;
$R_{F} = 3.97$\%.  Sr$_2$Cu$_3$O$_5$: lattice parameters (\AA): $a$ =
3.9292(1), $b$ = 19.396(5), $c$ = 3.4600(1); density: 5.614\,g/cm$^3$;
$R_{wp} = 5.03$\%; $R_{p} = 3.35$\%; $R_{e} = 3.00$\%; $R_{I} = 6.92$\%;
$R_{F} = 6.28$\%.}
\label{TabSrCu2O3XtalDat}
\begin{tabular}{lcccc} atom & site & $x$ & $y$ & $z$ \\
\hline &&SrCu$_2$O$_3$ \\
\hline Sr & 2$d$ & 0 & 0 & 1/2 \\ Cu & 4$i$ & 0 & 0.3348(5) & 0 \\ O(1) &
2$b$ & 1/2 & 0 & 0 \\ O(2) & 4$i$ & 0 & 0.1728(24) & 0 \\
\hline &&Sr$_2$Cu$_3$O$_5$ \\
\hline Sr   & 4$j$ & 0 & 0.3987(5) & 1/2 \\ Cu(1) & 2$a$ & 0 & 0 & 0 \\
Cu(2) & 4$i$ & 0 & 0.1996(5) & 0 \\ O(1)  & 2$b$ & 1/2 & 0 & 0 \\ O(2)  &
4$i$ & 0 & 0.1028(24) & 0 \\ O(3)  & 4$i$ & 0 & 0.3001(24) & 0 
\end{tabular}
\end{table}

\section{Experimental Magnetic Susceptibilies and Modeling}
\label{SecExpDatModeling}

\subsection{Introduction}
\label{SecExpDatModIntro}

For the three cuprate spin ladder compounds for which experimental magnetic
susceptibility $\chi(T)\equiv M/H$ data are presented below, where $M$ is
the magnetization and $H = 0.1$ or 1\,T is the applied magnetic field, we
fitted the $\chi(T)$ data per mole of Cu by the expression
\begin{mathletters}
\label{EqChiExp:all}
\begin{equation}
\chi(T) = \chi_0 + \frac{C_{\rm imp}}{T - \theta} + \chi^{\rm spin}(T)~,
\label{EqChiExp:a}
\end{equation} where
\begin{equation}
\chi_0 = \chi^{\rm core} + \chi^{\rm VV}~,
\label{EqChiExp:b}
\end{equation}
\begin{eqnarray}
\chi^{\rm spin}(T) &=& {N_{\rm A}g^2\mu_{\rm B}^2\over J^{\rm
max}}\,\chi^*(t)\nonumber\\  &=& \Big(0.3751\,\frac{\rm cm^3\,K}{\rm
mol}\Big) {g^2\over J^{\rm max}/k_{\rm B}}\chi^*\Big({k_{\rm B}T\over
J^{\rm max}}\Big)
\label{EqChiExp:c}
\end{eqnarray}
\end{mathletters} 
\noindent and $N_{\rm A}$ is Avogadro's number.  The first term
$\chi_0$ in Eq.~(\ref{EqChiExp:a}), according to Eq.~(\ref{EqChiExp:b}), is
the sum of the orbital diamagnetic core contribution $\chi^{\rm core}$ and
paramagnetic Van Vleck contribution $\chi^{\rm VV}$ which are normally
essentially independent of $T$; these contributions, especially $\chi^{\rm
VV}$, are very difficult to estimate accurately {\it a priori}.  The second
term in  Eq.~(\ref{EqChiExp:a}) is the extrinsic impurity Curie-Weiss term
with impurity Curie constant $C_{\rm imp}$ and Weiss temperature $\theta$
which gives a low-temperature 
\begin{table}
\caption{Selected bond lengths\,(\AA) and angles\,($^\circ$) for
SrCu$_2$O$_3$ and Sr$_2$Cu$_3$O$_5$.}
\label{TabBonds}
\begin{tabular}{lcd} & SrCu$_2$O$_3$ \\
\hline Sr-O(1) & & 2.629(0) \\ Sr-O(2) & & 2.654(20) \\ Cu-O(1) & rung &
1.910(6) \\ Cu-O(2) & leg & 1.967(13) \\
\\ Cu-O(1)-Cu & rung & 180.00 \\ Cu-O(2)-Cu & leg  & 174.22 \\ Cu-O(2)-Cu &
interladder & 92.56 \\
\hline & Sr$_2$Cu$_3$O$_5$ \\
\hline Sr-O(1) & & 2.617(7) \\ Sr-O(2) & & 2.618(1) \\ Sr-O(3) & &
2.580(36) \\ Cu(1)-O(1) & center leg & 1.965(1) \\ Cu(1)-O(2) & rung &
1.995(55) \\ Cu(2)-O(2) & rung & 1.877(53) \\ Cu(2)-O(3) & side leg &
1.965(0) \\ Cu(2)-O(3) & interladder & 1.949(2) \\
\\ Cu(1)-O(1)-Cu(1) & center leg & 180.00 \\ Cu(1)-O(2)-Cu(2) & rung &
180.00 \\ Cu(2)-O(3)-Cu(2) & side leg & 180.00 \\ Cu(2)-O(3)-Cu(2) &
interladder & 90.17 \\
\end{tabular}
\end{table}
\noindent upturn in $\chi(T)$ not predicted by theory
for the intrinsic spin susceptibility $\chi^{\rm spin}(T)$ and is assumed
to arise from paramagnetic impurities and/or defects.  The fitted
$C_{\rm imp}$ values are typically $\sim 10^{-3}\,{\rm cm^3\,K/mol\,Cu}$,
corresponding to a few tenths of an atomic percent with respect to Cu of
paramagnetic species with $S = 1/2$ and $g = 2$.  The $\theta$ is typically
$\sim -2$\,K which may indicate AF interactions between the impurity
magnetic moments, the occurrence of single-impurity-ion crystal field
effects, and/or compensate for paramagnetic saturation of the magnetic
impurities at low temperatures in the fixed field of the measurements.  For
a given sample
$C_{\rm imp}$ and $\theta$ are nearly independent of the model and
parameters for $\chi^{\rm spin}(T)$, for which the QMC data were presented
and fitted in previous sections.

For a given experimental $\chi(T)$ data set, there are potentially at least
six fitting parameters: $\chi_0$, $C_{\rm imp}$, $\theta$, $g$, $J^{\rm
max}$ and at least one additional exchange parameter.  In addition, from
Eq.~(\ref{EqChiExp:c}), the fitted $g$ and $J^{\rm max}$ are intrinsically
strongly correlated and thus even minor inaccuracies in the experimental
data can cause the fitted $g$ and $J^{\rm max}$ parameters for a given type
of fit to vary significantly ($\pm 20$\% or more) from sample to sample of
a given compound.  Therefore it is important to constrain the
$g$-value to lie within a physically reasonable range.\cite{Johnston1996} 
Unfortunately, ESR measurements of the intrinsic (bulk) Cu$^{+2}$
$g$-values in SrCu$_2$O$_3$, Sr$_2$Cu$_3$O$_5$ and LaCuO$_{2.5}$ are not
available, although $g = 2.14$ was reported for Cu defects in both
SrCu$_2$O$_3$ and Sr$_2$Cu$_3$O$_5$.\cite{Schwenk1997}  Fortunately, in all
Cu$^{+2}$-containing oxide compounds for which ESR data are available of
which we are aware, the powder-average $g$-value is always in the narrow
approximate range 2.10(5), as illustrated for representative compounds in
Table~\ref{TabCuGs}.
\cite{Matsuda1996,Konig1997,Monod1998,Mingmei1994,Guskos1995,Kojima1987,%
Sreedhar1988,%
Masuda1991,Honda1996,Ohta1992,Sichelschmidt1995,Mehran1988,Shaltiel1989,%
Dolinsek1998} We will also be investigating the $\chi(T)$ of CaV$_2$O$_5$
and MgV$_2$O$_5$ below.  The $g$-values for $S = 1/2$ V$^{+4}$ compounds
are observed to be in a narrow range about $g = 1.96$, the sign and
magnitude of the deviation from 2 being respectively opposite and smaller
than for Cu$^{+2}$ due to the opposite (positive) sign and smaller
magnitude of the spin-orbit coupling constant for V compared to that of
Cu.  Representative $g$-values observed for V$^{+4}$ species in several
materials are given in Table~\ref{TabCuGs}.
\cite{Onoda1996,Onoda1998,Prokofiev1998,Schwenk1996,Vasilev1997,Schmidt1998,%
Lohmann1997}  In our fits, we will use the fixed value $g = 1.96$
determined for powder samples of CaV$_2$O$_5$ and MgV$_2$O$_5$ by Onoda and
coworkers using ESR.\cite{Onoda1996,Onoda1998}

Before proceeding to presentation and modeling of the experimental $\chi(T)$
data, we comment briefly on units.  The popular commercial Quantum Design
SQUID magnetometer, used also here, reads out the magnetic moment of a
sample in units of ``emu''.  Unfortunately this ``unit'' is useless for unit
conversions, and authors use this same ``unit'' variously for magnetic
moment and magnetic susceptibility, which of course are not the same
quantities.  Here we use cgs units with one exception (T).  The unit for
magnetic moment (an ``emu'') is ${\rm G\,cm^3 \equiv erg/G}$ (e.g.,
1\,$\mu_{\rm B} = 9.274\times 10^{-21}$\,G\,cm$^3$), for molar
magnetization G\,cm$^3$/mol, for magnetic field $H$ and magnetic induction
$B$ G = Oe, for susceptibility of a sample cm$^3$ and for molar
susceptibility cm$^3$/mol.  For convenience, we occasionally quote applied
magnetic fields using the SI magnetic field unit T\,$\equiv 10^4$\,G\@.

\subsection{S\lowercase{r}C\lowercase{u}$_2$O$_3$}

\begin{figure}
\epsfxsize=3.3in
\centerline{\epsfbox{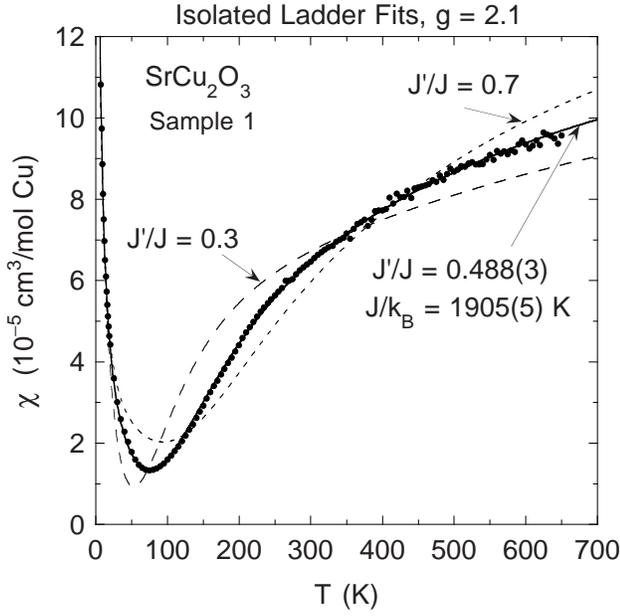}}
\vglue 0.1in
\caption{Magnetic susceptibility $\chi$ versus temperature $T$ for
SrCu$_2$O$_3$ sample~1.\protect\cite{Azuma1994}  The curves are theoretical
fits to the data by Eqs.~(\protect\ref{EqChiExp:all}) assuming that
$\chi^{\rm spin}(T)$ is given by the theory for isolated ladders with $g =
2.1$.  The solid curve shows the best fit to the data, where $J^\prime/J$
was allowed to vary, for which $J^\prime/J = 0.488(3)$ and $J/k_{\rm B} =
1905(5)$\,K were obtained.  For comparison, the long- and short-dashed
curves are the best fits with fixed $J^\prime/J = 0.3$ and~0.7,
respectively.}
\label{Fig26}
\end{figure}

$\chi(T)$ data for three polycrystalline samples of SrCu$_2$O$_3$ were
fitted by the above QMC $\chi^*(t)$ simulations.  Data for sample~1 in the
temperature range from 4 to 650\,K, shown in
Fig.~\ref{Fig26}, have been previously
reported.\cite{Azuma1994}  Samples~2 and~3 are new; data were obtained for
these samples from~4 to 400\,K, as shown in Fig.~\ref{Fig30}
below.  $\chi(T)$ for each of the samples increases monotonically with $T$
from $\sim 70$\,K up to our high-$T$ measurement limit.  At lower $T$, an
upturn in $\chi(T)$ is seen for each sample which we attribute to
paramagnetic impurities and/or defects.

\begin{table}
\caption{$g$-factors parallel ($g_{||}$) and perpendicular ($g_{\bot}$) to
the principal local crystal field and/or crystal structure axis and the
powder-averaged value \big[$\langle g\rangle = \sqrt{(g_1^2 + g_2^2 +
g_3^2)/3}$\,\big] for bulk Cu$^{+2}$ and V$^{+4}$ $S = 1/2$ species in
several representative copper and vanadium oxide compounds, and for defects
in compounds for which a bulk ESR signal has not been observed.  Samples are
polycrystalline unless otherwise noted.  The literature references are
given in the last column.}
\label{TabCuGs}
\begin{tabular}{lcccr} Compound & $g_{||}$ & $g_{\bot}$ & $\langle
g\rangle$ & Ref.\\
\hline CuO           &       &       & 2.125(5) & \onlinecite{Monod1998}\\
~crystal ($b$-axis) & 2.185(5) & & & \onlinecite{Monod1998}\\ BaCuO$_{2+x}$
& 2.23  & 2.06  & 2.12  & \onlinecite{Mingmei1994} \\
              & 2.223 & 2.041, 2.103 & 2.124 & \onlinecite{Guskos1995} \\
              & 2.21  & 2.057, 2.12 & 2.13  & \onlinecite{Guskos1995} \\
Y$_2$BaCuO$_5$ & 2.24  & 2.06 & 2.12 & \onlinecite{Kojima1987} \\
               & 2.22  & 2.08 & 2.13 & \onlinecite{Sreedhar1988} \\
La$_2$BaCuO$_5$ & 2.229 & 2.037 & 2.103 & \onlinecite{Masuda1991} \\
CuGeO$_3$ (crystal) & 2.338 & 2.064 & 2.159 & \onlinecite{Honda1996} \\
${\rm Sr_{14}Cu_{24}O_{41}}$ (crystal) & 2.26 & 2.05, 2.04& 2.12 & 
\onlinecite{Matsuda1996}\\ ~powder& 2.30 & 2.05 & 2.14 &
\onlinecite{Konig1997} \\
${\rm Ca_{0.85}CuO_2}$ & & & 2.0796 & \onlinecite{Dolinsek1998} \\
\hline Cu$^{+2}$ defects in:\\
\hline SrCuO$_2$ & 2.25 & 2.05 & 2.12 & \onlinecite{Ohta1992} \\
${\rm YBa_2Cu_3O_{6.7-6.9}}$ & 2.28 & 2.03 & 2.12 &
\onlinecite{Sichelschmidt1995} \\ ~(crystal)\\
${\rm YBa_2Cu_3}$O$_x$ &  &  & 2.08 & \onlinecite{Mehran1988}\\ ~(crystal,
$T_{\rm c} = 40$\,K) \\
${\rm YBa_2Cu_3}$O$_x$ & 2.295 & 2.042 & 2.130 & \onlinecite{Shaltiel1989}\\
~(crystal, $T_{\rm c} = 30$\,K)\\
\hline Compound & $g_{||}$ & $g_{\bot}$ & $\langle g\rangle$ & Ref.\\
\hline
${\rm CaV_2O_5}$ &&& 1.957(1) & \onlinecite{Onoda1996} \\ MgV$_2$O$_5$
&      &      & 1.96 & \onlinecite{Onoda1998}\\
${\rm (VO)_2P_2O_7}$& 1.94 & 1.98 & 1.97 & \onlinecite{Schwenk1996}\\
~~crystal   & 1.937 & 1.984 & 1.969 & \onlinecite{Prokofiev1998}\\
NaV$_2$O$_5$ (crystal) & 1.938(2) & 1.972(2) & 1.961(2) &
\onlinecite{Vasilev1997}\\ ~~crystal & 1.936(2) &  &  &
\onlinecite{Schmidt1998}\\ ~~crystal & 1.95 & 1.97 & 1.96 &
\onlinecite{Lohmann1997}\\ ~~crystal & 1.936 & 1.974, 1.977 & 1.962 &
\onlinecite{Onoda1999} \\
\end{tabular}
\end{table}

\begin{figure}
\epsfxsize=3.3in
\centerline{\epsfbox{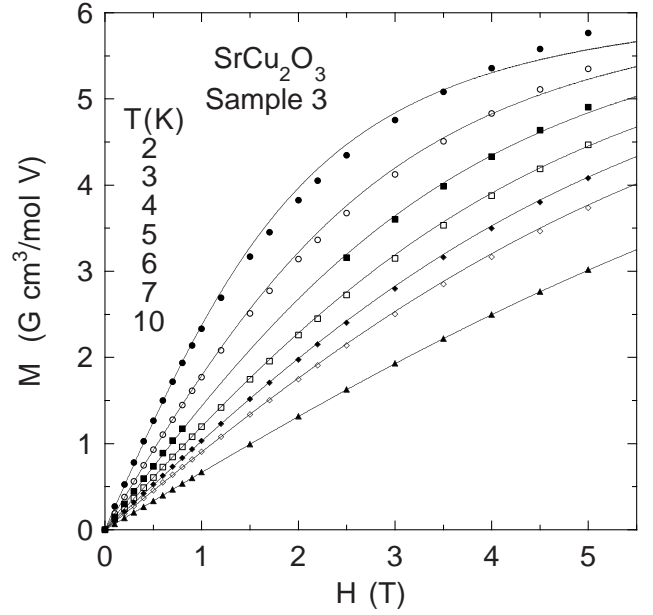}}
\vglue 0.1in
\caption{Magnetization $M$ versus applied magnetic field $H$ for
SrCu$_2$O$_3$ sample~3 at low temperatures $T$.  The set of solid  curves
is a global theoretical fit to all the data by
Eqs.~(\protect\ref{EqMfit:all}), containing a modified Brillouin function.}
\label{Fig27}
\end{figure}

We measured the low-$T$ magnetization $M$ versus applied magnetic field
$H$ in detail for sample~3, as shown in Fig.~\ref{Fig27}. 
For 25\,K $\leq T\leq 300$\,K, $M$ is found to be proportional to $H$ for
this sample to within our precision.  Significant negative curvature arises
at 10\,K and below, attributed to saturation of paramagnetic impurities
and/or defects.  The slope of the lowest-$H$ (1--3\,kG) molar $M(H)$ data
from 2~to 10\,K yielded the true $\chi(T)$ (i.e., the low-field limit),
which was fitted very well by a constant term plus a Curie-Weiss term
\begin{mathletters}
\label{EqCWFit:all}
\begin{equation}
\chi = \chi_0 + \frac{C_{\rm imp}}{T - \theta}\label{EqCWFit:a}
\label{EqCW}
\end{equation} with parameters
\begin{equation}
\chi_0 = 0.52(1)\times 10^{-5}\,{\rm {cm^3\over mol\,Cu}}~,
\label{EqCWFit:b}
\end{equation}
\begin{eqnarray} C_{\rm imp} &=& \frac{f_{\rm imp}N_{\rm A}g_{\rm
imp}^2\mu_{\rm B}^2S_{\rm imp}(S_{\rm imp}+1)}{3 k_{\rm B}}\nonumber\\ &=&
0.000654(9)\,{\rm {cm^3\,K\over mol\,Cu}}~,\label{EqCWFit:c}
\end{eqnarray}
\begin{equation}
\theta = -0.61(3)\,{\rm K}~,\label{EqCWFit:d}
\end{equation}
\end{mathletters} where $f_{\rm imp}$ is the fraction of impurities with
respect to Cu with spin $S_{\rm imp}$ and $g$-factor $g_{\rm imp}$.

We have performed fits to the $M(H)$ data in
Fig.~\ref{Fig27} by
\begin{mathletters}
\label{EqMfit:all}
\begin{equation} M = \chi_0 H + f_{\rm imp} N g_{\rm imp} S_{\rm imp}
\mu_{\rm B} B_{S_{\rm imp}}(x)~,\label{EqMfit:a}
\end{equation}
\begin{equation} x = \frac{g_{\rm imp}S_{\rm imp}\mu_{\rm B}H}{k_{\rm
B}(T-\theta)}~,\label{EqMfit:b}
\end{equation}
\end{mathletters} where $\chi_0$ and $\theta$ were fixed to the observed
values in Eqs.~(\ref{EqCWFit:all}), $B_{S_{\rm imp}}(x)$ is the Brillouin
function\cite{Kittel1971} describing the magnetization vs field of the
impurity spins, and the parameter $x$ has been modified from the usual
form\cite{Kittel1971} (without $\theta$) so that the expansion of
Eqs.~(\ref{EqMfit:all}) for $H\to 0$ gives the correct observed behavior in
Eq.~(\ref{EqCW}).  The impurity Weiss temperature $\theta$ can arise from
interactions between the impurity spins and/or from single-ion effects
associated with splitting of the impurity spin energy levels.  The $f_{\rm
imp}$ is fixed uniquely by the observed $C_{\rm imp}$ and by $S_{\rm imp}$
and 
$g_{\rm imp}$ in Eq.~(\ref{EqCWFit:c}).  Fitting the $M(H)$ data at 10\,K
showed that $g_{\rm imp}\approx 2$ and $S_{\rm imp}\approx 3/2$; lower spin
values cannot give the strong negative curvature in $M(H)$ extending up to
$\sim 10$\,K\@.  Then fixing $S_{\rm imp} = 3/2$, the only remaining
adjustable parameter is $g_{\rm imp}$, and from a two-dimensional global
fit to all the
$M(H)$ data in Fig.~\ref{Fig27} we obtained  $g_{\rm imp} =
2.093(7)$; the fit is shown by the set of solid curves in
Fig.~\ref{Fig27}.  This $g$-value is within the range
expected for Cu$^{+2}$ as discussed above, but is slightly smaller than
that (2.14) found by ESR for magnetic defects in
SrCu$_2$O$_3$.\cite{Schwenk1997}  Hence the impurity spin may consist of 
ferromagnetically coupled Cu$_3$ clusters.  The fit is very good from 5~to
10\,K, but deteriorates progressively for $T = 4$, 3 and 2\,K\@.  Thus
including $\theta$ in Eq.~(\ref{EqMfit:b}), which is a high-$T$ mean-field
like approximation, is not nor was expected to be accurate at the lowest
temperatures.  We have carried out fits of the exact expression (without
$\theta$) for $M(H,T)$ of a spin $S = 3/2$ impurity with the $S_z$ levels
split by a single-ion interaction $D S_z^2$, for which the high-$T$
approximation gives $\theta = -4D/5$.  The $S_z = \pm 1/2$ levels were
indeed found to be the ground levels ($D > 0$), with $D$ close to that
predicted from the observed $\theta$ in Eq.~(\ref{EqCWFit:d}).  This
treatment improved the fit to the low-$T$ data at large $H$ at the expense
of a poorer fit at low
$H$, but with an improvement in the overall fit.  We will not present or
further discuss such detailed fits here.

\subsubsection{Isolated Ladder Fits}

Figure~\ref{Fig26} shows the $\chi(T)$ for sample~1 along
with the best fit (solid curve) by Eqs.~(\ref{EqChiExp:all}), where $J^{\rm
max} = J$, $g \equiv 2.1$ and $\chi^*(t)$ is our fit to the QMC data for
isolated ladders ($J^{\rm diag} = 0$) with spatially anisotropic exchange,
for which the parameters are
\[
\chi_0 = -0.27(4) \times 10^{-5}\,{\rm {cm^3\over mol\,Cu}}~,
\]
\[ C_{\rm imp} = 0.00105(2)\,{\rm {cm^3\,K\over mol\,Cu}}~,~~\theta =
-2.4(1)\,{\rm K},
\]
\begin{equation} {J\over k_{\rm B}} = 1905(5)\,{\rm K}~,~~{J^\prime\over J}
= 0.488(3)~.
\label{EqSCOIsoLadPars}
\end{equation} We will not continue to quote $\chi_0$, $C_{\rm imp}$ and
$\theta$ values from the fits since these were essentially the same for all
the fits to be described.  Also shown as the dashed curves in
Fig.~\ref{Fig26} are the best fits obtained by setting
$J^\prime/J$ at the fixed values of 0.3 and 0.7; for these very poor fits,
$J/k_{\rm B} = 1870(100)$\,K and~1827(21)\,K were obtained, respectively.

\begin{figure}
\epsfxsize=3.3in
\centerline{\epsfbox{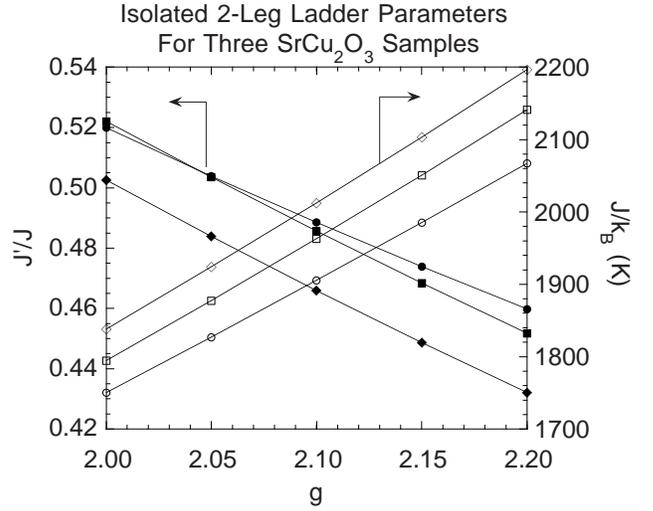}}
\vglue 0.1in
\caption{Exchange constants $J^\prime/J$ (filled symbols, left scale) and
$J/k_{\rm B}$ (open symbols, right scale) versus fixed $g$-value assumed in
the fits by Eqs.~(\protect\ref{EqChiExp:all}) to the data for SrCu$_2$O$_3$
samples~1 (circles), 2 (squares) and 3 (diamonds).}
\label{Fig28}
\end{figure}

Next, we fixed $g$ at 2.0 to 2.2 in 0.05 increments and for each $g$-value
determined the best-fit parameters for each of the three samples for the
isolated ladder model, which are plotted versus $g$ in
Fig.~\ref{Fig28}.  Over the physically most reasonable $g$-value
range $2.10(5)$ discussed above, from the fit parameters for all three
samples taken together we estimate that $J^\prime/J = 0.48(3)$ and
$J/k_{\rm B} = 1970(150)$\,K\@.

\begin{figure}
\epsfxsize=3.3in
\centerline{\epsfbox{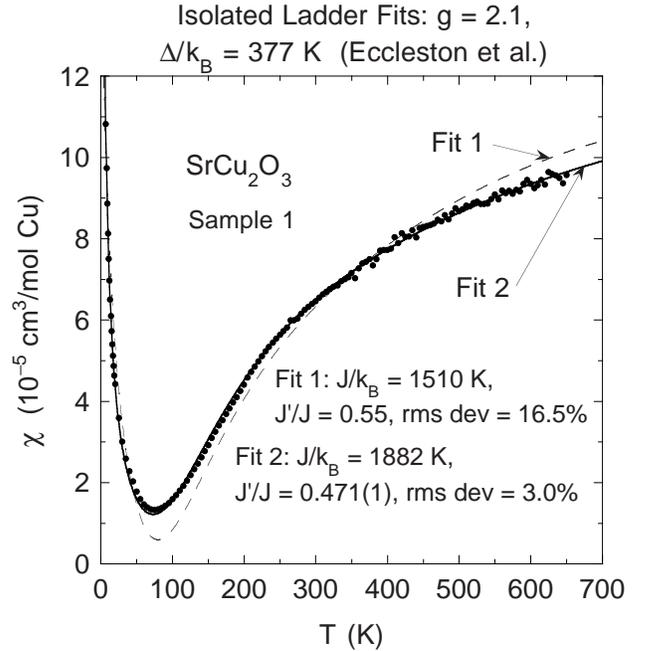}}
\vglue 0.1in
\caption{Magnetic susceptibility $\chi$ versus temperature $T$ for
SrCu$_2$O$_3$ sample~1,\protect\cite{Azuma1994} showing theoretical fits to
the data by Eqs.~(\protect\ref{EqChiExp:all}) assuming $g = 2.1$ and a spin
gap  $\Delta/k_{\rm B} = 377$\,K which was measured using neutron scattering
for the Cu$_2$O$_3$ two-leg ladders in Sr$_{14}$Cu$_{24}$O$_{41}$
crystals.\protect\cite{Eccleston1998}  Fit~1 is for the fixed $J$ and
$J^\prime/J$ values deduced by Eccleston {\it et
al.}\protect\cite{Eccleston1998} whereas for Fit~2 these parameters were
allowed to vary, subject to the constraint that the spin gap is given by the
neutron scattering result $\Delta/k_{\rm B} = 377$\,K\@.}
\label{Fig29}
\end{figure}

Eccleston {\it et al.}\cite{Eccleston1998} and Azuma {\it et
al.}\cite{Azuma1998} have found from inelastic neutron scattering
measurements on the Cu$_2$O$_3$ two-leg ladders in
Sr$_{14}$Cu$_{24}$O$_{41}$ and SrCu$_{2}$O$_{3}$ that $\Delta/k_{\rm B} =
377(1)$\,K and $\approx 380$\,K, respectively.  As discussed in
Sec.~\ref{SecSummaryDisc} below,  Eccleston {\it et al.}\ also infer that
$J^\prime/J = 0.55$ and $J/k_{\rm B} = 1510$\,K\@.  Using the parameters of
Eccleston {\it et al.}, we fitted the data for our SrCu$_2$O$_3$ samples
after setting $g = 2.1$.  The fit for sample~1 is shown as ``Fit~1''
(dashed curve) in Fig.~\ref{Fig29}; similarly bad fits were
obtained for samples~2 and~3.  We then allowed 
\vglue0.54in
\newpage 
\begin{figure}
\epsfxsize=3.3in
\centerline{\epsfbox{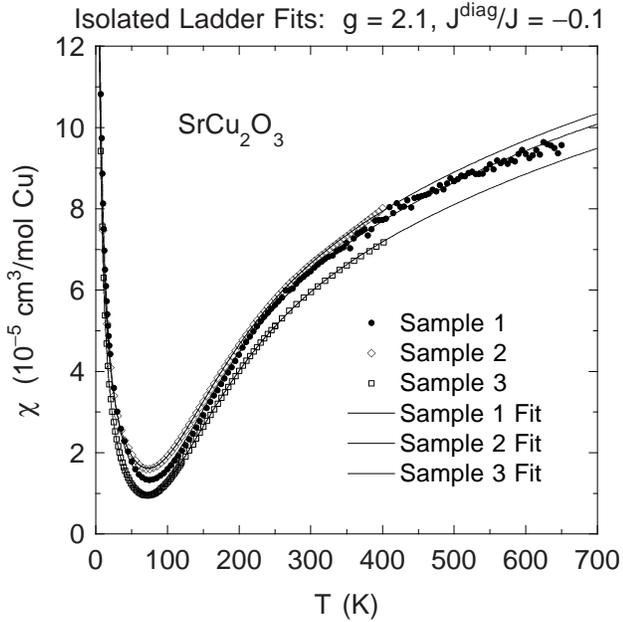}}
\vglue 0.1in
\caption{Magnetic susceptibility $\chi$ versus temperature $T$ for
SrCu$_2$O$_3$ samples~1,\protect\cite{Azuma1994} 2 and~3.  The curves are
theoretical fits to the data by Eqs.~(\protect\ref{EqChiExp:all}) assuming
$g = 2.1$ and $J^{\rm diag}/J = -0.1$, and extrapolations up to 700\,K are
shown.  The fit parameters are given in the text.}
\label{Fig30}
\end{figure}
\vglue0.1in
\noindent $J^\prime/J$ and $J/k_{\rm B}$ to vary, subject in
Eq.~(\ref{EqDelta1}) to the constraint that the spin gap is given by the
accurate neutron scattering result $\Delta/k_{\rm B} = 377$\,K; the result
is ``Fit~2'' in Fig.~\ref{Fig29} (the fit parameters are given in the
figure), which is obviously a much better fit.

\begin{figure}
\epsfxsize=3.3in
\centerline{\epsfbox{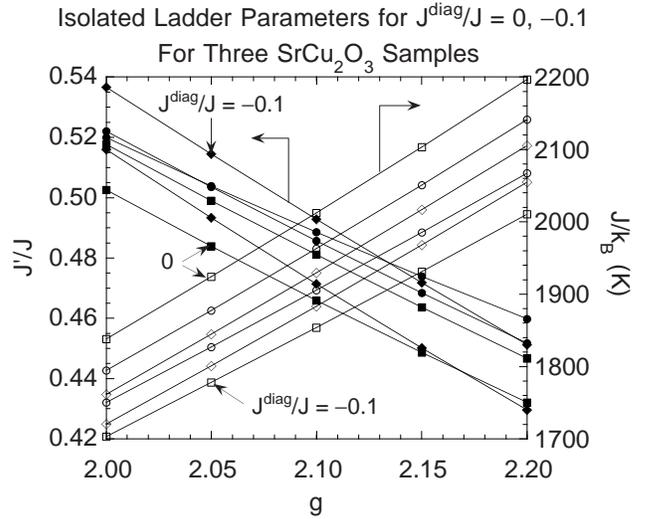}}
\vglue 0.1in
\caption{Exchange constants $J^\prime/J$ (open symbols, left scale) and
$J/k_{\rm B}$ (filled symbols, right scale) versus fixed $g$-value assumed
in the fits by Eqs.~(\protect\ref{EqChiExp:all}) to the data for
SrCu$_2$O$_3$ samples~1~(circles), 2~(squares) and 3~(diamonds); the fit
parameters obtained for both fixed $J^{\rm diag}/J = -0.1$ and~0 are shown.}
\label{Fig31}
\end{figure}

Continuing with the isolated ladder model, we now include the FM
nonfrustrating diagonal intraladder coupling $J^{\rm diag}$ in the fits. 
As discussed in Sec.~\ref{SecLDA+U} above, our LDA+U calculations predict
that $J^{\rm diag}/J
\approx -0.1$.  Shown in Fig.~\ref{Fig30} is the best fit for
each of the three samples assuming fixed $g = 2.1$ and $J^{\rm diag}/J =
-0.1$.  The fits yielded $J^\prime/J = 0.481(4)$ and $J/k_{\rm B} =
1854(6)$\,K for sample~1 with a relative rms deviation $\sigma_{\rm rms} =
1.98$\%, $J^\prime/J = 0.493(2)$, $J/k_{\rm B} = 1883(3)$\,K and
$\sigma_{\rm rms} = 0.54$\% for sample~2, and $J^\prime/J = 0.471(2)$, 
$J/k_{\rm B} = 1930(4)$\,K and
$\sigma_{\rm rms} = 0.71$\% for sample~3.  These $J^\prime/J$ and $J/k_{\rm
B}$ parameters are well within the respective ranges determined above for
$J^{\rm diag} = 0$.  To determine the sensitivity of these parameters to
the assumed
$g$-value in the presence of a $J^{\rm diag} = -0.1$, we again fixed $g$ at
2.0 to 2.2 in 0.05 increments and for each $g$-value determined the best-fit
parameters for each of the three samples, which are plotted versus $g$ in
Fig.~\ref{Fig31}.  For the fixed parameter ranges $g = 2.10(5)$
and
$J^{\rm diag} = -0.05(5)$, from the fit parameters for all three samples
taken together we estimate that $J^\prime/J = 0.48(4)$ and $J/k_{\rm B} =
1950(170)$\,K, nearly the same as for $J^{\rm diag} = 0$ above.  Thus the
fit parameters are not very sensitive to the precise value of $J^{\rm
diag}/J$, at least if its magnitude is much less than unity.

\subsubsection{Coupled Ladder Fits}

\begin{figure}
\epsfxsize=3.3in
\centerline{\epsfbox{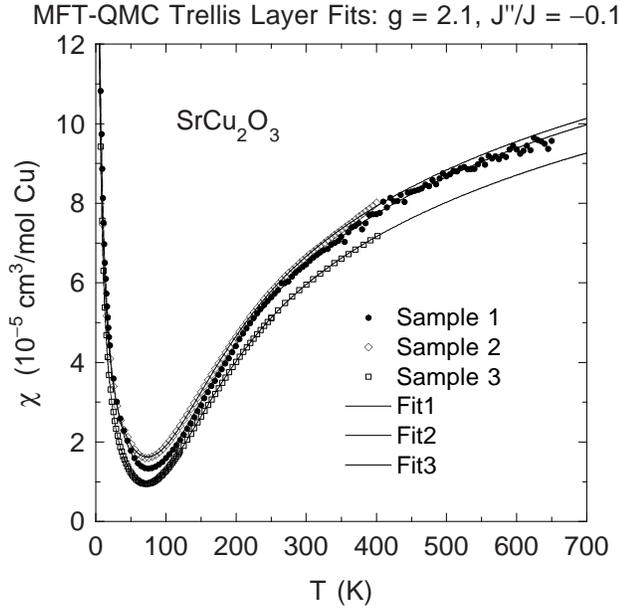}}
\vglue 0.1in
\caption{Magnetic susceptibility $\chi$ versus temperature $T$ for
SrCu$_2$O$_3$ samples~1,\protect\cite{Azuma1994} 2 and~3.  The curves are
theoretical fits for the trellis layer to the data by
Eqs.~(\protect\ref{EqChiExp:all}) assuming $g = 2.1$ and trellis layer
interladder coupling $J^{\prime\prime}/J = -0.1$.  The fit parameters are
given in the text.}
\label{Fig32}
\end{figure}

It has been suggested that the frustrating trellis layer interladder
coupling should be ferromagnetic because it involves Hund's rule coupling
through $90^\circ$ Cu-O-Cu bonds.\cite{Rice1993} Its magnitude was
estimated to be about an order of magnitude smaller than the intraladder
coupling, i.e.\ $J^{\prime\prime}/J \sim -0.1$. Our LDA+U calculations in
Sec.~\ref{SecLDA+U} indicate an even smaller magnitude.  We carried out
fits to the $\chi(T)$ data for all three ${\rm SrCu_2O_3}$ samples by this
model with fixed $g = 2.1$ but allowing all three parameters $J/k_{\rm B},\
J^\prime/J$ and
$J^{\prime\prime}/J$ to vary.  The fitted values of $J^{\prime\prime}/J$
were in the range $-0.7 \leq J^{\prime\prime}/J \leq 0.9$ (two of the three
fitted values are outside the range of validity of the fit), showing that
$J^{\prime\prime}/J$ is too strongly correlated with the other two
parameters to allow all three exchange constants to be simultaneously
varied.  Shown in Fig.~\ref{Fig32} are the best fits to the
data for samples~1, 2 and~3 assuming the fixed values $g = 2.1$ and
$J^{\prime\prime}/J = -0.1$.  The fit parameters are $J/k_{\rm B} =
1944(5)$\,K, $J^\prime/J = 0.476(2)$ and
$\sigma_{\rm rms} = 1.46$\% for sample~1, $J/k_{\rm B} = 2000(3)$\,K,
$J^\prime/J = 0.474(2)$ and
$\sigma_{\rm rms} = 0.65$\% for sample~2 and $J/k_{\rm B} = 2051(3)$\,K,
$J^\prime/J = 0.455(2)$ and $\sigma_{\rm rms} = 0.79$\% for sample~3.

\begin{figure}
\epsfxsize=3.3in
\centerline{\epsfbox{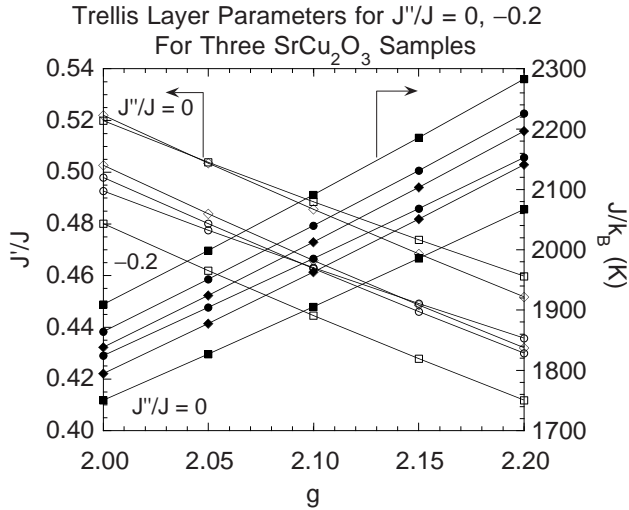}}
\vglue 0.1in
\caption{Intraladder exchange constants $J^\prime/J$ (open symbols, left
scale) and $J/k_{\rm B}$ (filled symbols, right scale) versus fixed
$g$-value assumed in the fits by Eqs.~(\protect\ref{EqChiExp:all}) to the
data for SrCu$_2$O$_3$ samples~1 (circles), 2~(squares) and~3~(diamonds),
for fixed trellis layer interladder couplings $J^{\prime\prime}/J = -0.2$
and~0 as indicated.}
\label{Fig33}
\end{figure}

In Fig.~\ref{Fig33} we compare the parameters obtained for
fixed
$J^{\prime\prime}/J = -0.2$ and~0 and for fixed $g$-values from 2 to~2.2. 
From this figure we infer from the fit parameters for all three samples
taken together that for the ranges $g = 2.10(5)$ and $J^{\prime\prime}/J =
-0.1(1)$, the intraladder exchange constants are $J^\prime/J = 0.465(40)$
and $J/k_{\rm B} = 2000(180)$\,K\@.


The quality of stacked-ladder fits to the data is very sensitive to the
value of $J^\prime/J$ and of the interladder coupling 
$J^{\prime\prime\prime}/J$ perpendicular to the plane of the ladders,
because the shape of the theoretically predicted spin susceptibility
$\chi^*(t)$ depends strongly on these parameters due to the \,proximity \,to
\,a \,QCP\@.  \,Thus \,when \,we fit the 
\begin{figure}
\epsfxsize=3.3in
\centerline{\epsfbox{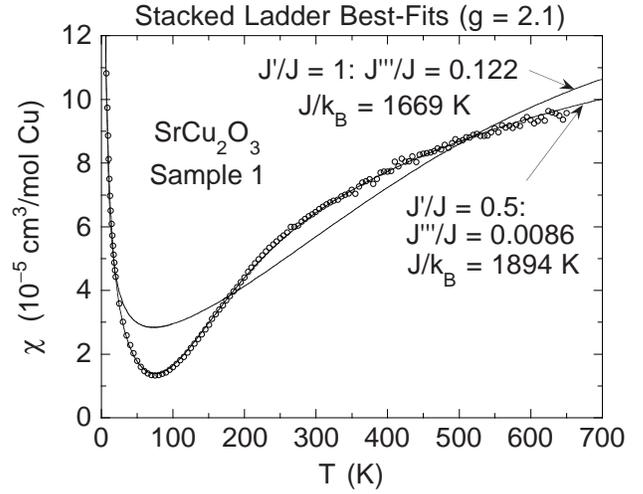}}
\vglue 0.1in
\caption{Magnetic susceptibility $\chi$ versus temperature $T$ for
SrCu$_2$O$_3$ sample~1.\protect\cite{Azuma1994}  The two solid curves are
fits to the data by Eqs.~(\protect\ref{EqChiExp:all}) assuming $g = 2.1$ and
$J^{\prime}/J = 0.5$ and~1, respectively.  The other exchange constants
determined from the respective fits are listed.}
\label{Fig34}
\end{figure}
\vglue-0.05in
\begin{figure}
\epsfxsize=3.3in
\centerline{\epsfbox{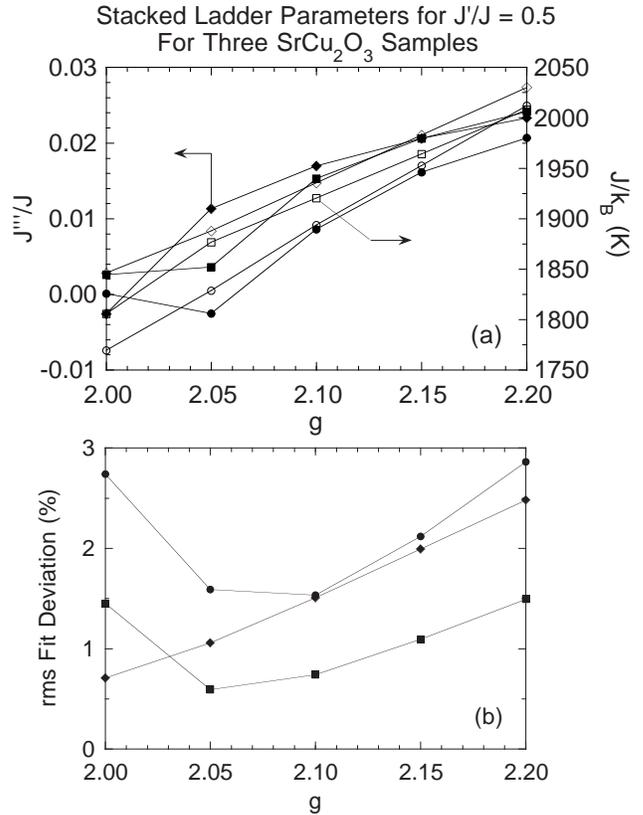}}
\vglue 0.1in
\caption{(a) Exchange constants $J^{\prime\prime\prime}/J$ (filled symbols,
left scale) and $J/k_{\rm B}$ (open symbols, right scale) versus fixed
$g$-value assumed in the stacked-ladder fits by
Eqs.~(\protect\ref{EqChiExp:all}) to the data for SrCu$_2$O$_3$ samples~1
(circles), 2 (squares) and 3 (diamonds). (b) The relative rms fit error for
the fits shown in (a).  The lines connecting the parameter data points in
both~(a) and~(b) are guides to the eye.}
\label{Fig35}
\end{figure}
\noindent experimental data by the predictions only
small values of
$J^{\prime\prime\prime}/J$ can fit the data.  We find that of the two
possibilities
$J^\prime/J = 0.5$ and~1 for which we carried out QMC $\chi^*(t)$
simulations, fits with $J^\prime/J = 1$ are very poor, in contrast to the
excellent fits obtained with
$J^\prime/J = 0.5$.  Shown in Fig.~\ref{Fig34} are the
respective exchange parameters and best fits to the $\chi(T)$ data for
SrCu$_2$O$_3$ sample~1 assuming $g = 2.1$.  For either assumed $J^\prime/J$
value, the fitted $J^{\prime\prime\prime}/J$ values are consistent with
SrCu$_2$O$_3$ being in the gapped part of the phase diagram.  Concentrating
now on fits with $J^\prime/J = 0.5$, the sensitivities of the fitted 
$J^{\prime\prime\prime}/J$ and $J/k_{\rm B}$ values to the assumed
$g$-value are shown in Fig.~\ref{Fig35}(a) for fixed
$g$-values from 2.0 to~2.2.  In contrast to other fits discussed above, the
rms deviation of a fit from the data depends rather strongly on the assumed
$g$-value, as illustrated in Fig.~\ref{Fig35}(b), where
the optimum $g$-values for the fits to the data for two of the three samples
are seen to be consistent with the range 2.10(5) we have assumed when
quoting the exchange constants derived from the other fits above.  From
Fig.~\ref{Fig35}(a), the exchange constants for the 
$g$-value range 2.10(5) are $J^{\prime\prime\prime}/J = 0.01(1)$ and
$J/k_{\rm B} = 1920(70)$\,K\@.  These parameters and the fit quality are
essentially identical with those determined for the optimum isolated ladder
fit in Fig.~\ref{Fig26}, so we will not plot the fit for
the present case.

\subsection{$\bbox{\rm Sr_2Cu_3O_5}$}

\begin{figure}
\epsfxsize=3.3in
\centerline{\epsfbox{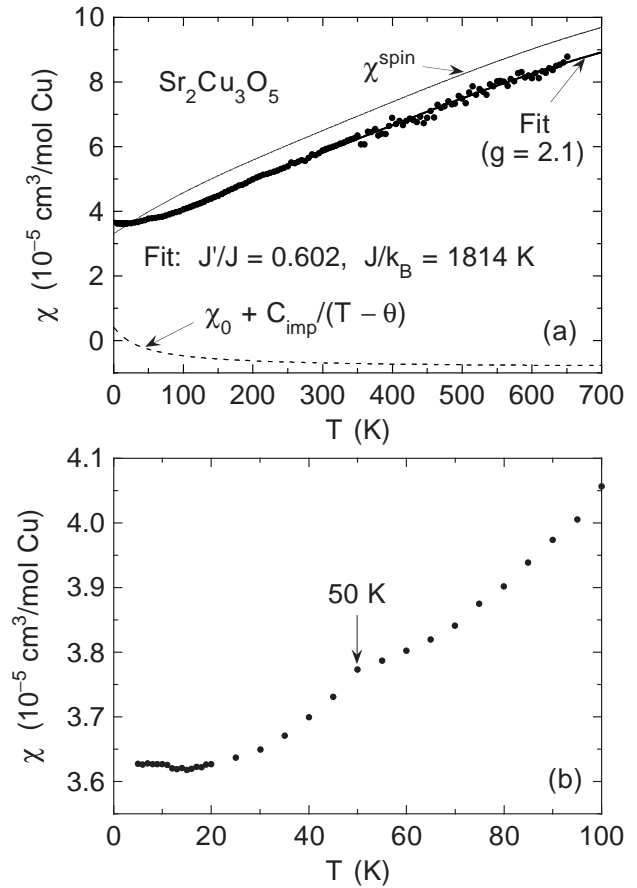}}
\vglue 0.1in
\caption{(a) Magnetic susceptibility $\chi$ versus temperature $T$ for the
three-leg ladder compound Sr$_2$Cu$_3$O$_5$
($\bullet$).\protect\cite{Azuma1994}  The heavy solid curve is a  fit to
the data by theory for isolated three-leg ladders using
Eqs.~(\protect\ref{EqChiExp:all}), yielding the intraladder exchange
constants $J^\prime/J = 0.602$ and $J/k_{\rm B} = 1814$\,K\@.  The other fit
parameters are given in the text.  The light solid curve is the spin
susceptibility $\chi^{\rm spin}$ contribution, and the dashed curve the
contributions $\chi_0 + C_{\rm imp}/(T - \theta)$.  \mbox{(b)~Expanded} plot
of the data in (a) at low temperatures.  A small cusp occurs at $\sim
50$\,K as indicated, probably associated with the short-range AF ordering
seen at $\approx 52$\,K in $\mu$SR measurements.\protect\cite{Kojima1995}}
\label{Fig36}
\end{figure}

The $\chi(T)$ measured in $H = 1$\,T by Azuma {\it et al.}\cite{Azuma1994}
for the three-leg ladder trellis layer compound ${\rm Sr_2Cu_3O_5}$ from~5
to 650\,K is shown in Fig.~\ref{Fig36}(a).  The expanded plot of
the low-$T$ data in Fig.~\ref{Fig36}(b) exhibits a cusp at $\sim
50$\,K, evidently associated with the short-range spin-glass-type ordering
observed for this compound at $\approx 52$\,K from $\mu$SR
measurements.\cite{Kojima1995}  We fitted the data in
Fig.~\ref{Fig36}(a) by Eqs.~(\ref{EqChiExp:all}), where $\chi^*(t)$
is our global fit to the QMC simulation data for the three-leg $S = 1/2$
ladder with spatially anisotropic exchange.  The best fit for $g = 2.1$ is
shown as the heavy solid curve in Fig.~\ref{Fig36}(a); the spin
susceptibility contribution is shown as the light solid curve and the
contributions $\chi_0 + C_{\rm imp}/(T -
\theta)$ as the dashed curve.  The parameters of the fit are
\begin{mathletters}
\label{EqSr2Cu3O5Pars:all}
\begin{equation}
\chi_0 = -0.8(2)\times 10^{-5}\,{\rm {cm^3\over mol\,Cu}}~,
\label{EqSr2Cu3O5Pars:a}
\end{equation}
\begin{equation} C_{\rm imp} = 0.00053(26)\,{\rm {cm^3\,K\over
mol\,Cu}}~,~~~\theta = -41(15)\,{\rm K}~,\label{EqSr2Cu3O5Pars:b}
\end{equation}
\begin{equation} {J^\prime\over J}	= 0.60(4)~,~~~{J\over k_{\rm B}} =
1814(22)\,{\rm K}~.\label{EqSr2Cu3O5Pars:c}
\end{equation}
\end{mathletters} 
\vglue0.1in
The relative rms fit deviation (1.3\%) is found to be
nearly independent of the assumed value of $g$.  For an allowed $g$-value
range 2.1(1), the fitted parameter ranges become
\begin{mathletters}
\label{EqSr2Cu3O5Pars2:all}
\begin{equation}
\chi_0 = -0.8(5)\times 10^{-5}\,{\rm {cm^3\over mol\,Cu}}~,
\label{EqSr2Cu3O5Pars2:a}
\end{equation}
\begin{equation} C_{\rm imp} = 0.0005(4)\,{\rm {cm^3\,K\over
mol\,Cu}}~,~~~\theta = -41(20)\,{\rm K}~,\label{EqSr2Cu3O5Pars2:b}
\end{equation}
\begin{equation} {J^\prime\over J}	= 0.60(7)~,~~~{J\over k_{\rm B}} =
1810(130)\,{\rm K}~.\label{EqSr2Cu3O5Pars2:c}
\end{equation}
\end{mathletters}

On the other hand, the lowest-$T$ data in Fig.~\ref{Fig36}(a) do
not show any direct evidence for the existence of an impurity Curie-Weiss
contribution.  In addition, the expanded plot in
Fig.~\ref{Fig36}(b) shows evidence that the
reported\cite{Kojima1995} spin-glass transition at $\sim 50$\,K affects
$\chi(T)$ and also suggests that there are pretransitional effects.  We
therefore refitted the data only above 100\,K in
Fig.~\ref{Fig36}(a) assuming $C_{\rm imp} = 0$.  The resulting
fitting parameters for the range $g = 2.1(1)$ were
\begin{mathletters}
\label{EqSr2Cu3O5Pars3:all}
\begin{equation}
\chi_0 = -0.4(2)\times 10^{-5}\,{\rm {cm^3\over mol\,Cu}}~,
\label{EqSr2Cu3O5Pars3:a}
\end{equation}
\begin{equation} {J^\prime\over J}	= 0.66(5)~,~~~{J\over k_{\rm B}} =
1810(150)\,{\rm K}~.\label{EqSr2Cu3O5Pars3:b}
\end{equation}
\end{mathletters} For the reasons mentioned, these parameters are
considered to be more reliable than those in
Eqs.~(\ref{EqSr2Cu3O5Pars:all}) and~(\ref{EqSr2Cu3O5Pars2:all}).  The ratio
$J^\prime/J$ thus  appears to be somewhat larger and the value of $J$ a
little smaller than the respective values in SrCu$_2$O$_3$.

With regard to the exchange constants obtained in this section, it should be
kept in mind that we have implicitly assumed that the exchange constant
along the central leg of the three-leg ladder is the same as that along the
outer two legs.  This is not necessarily the case (see the  discussion in
Sec.~\ref{SecSummaryDisc}).

\subsection{LaCuO$_{2.5}$}

\begin{figure}
\epsfxsize=3.3in
\centerline{\epsfbox{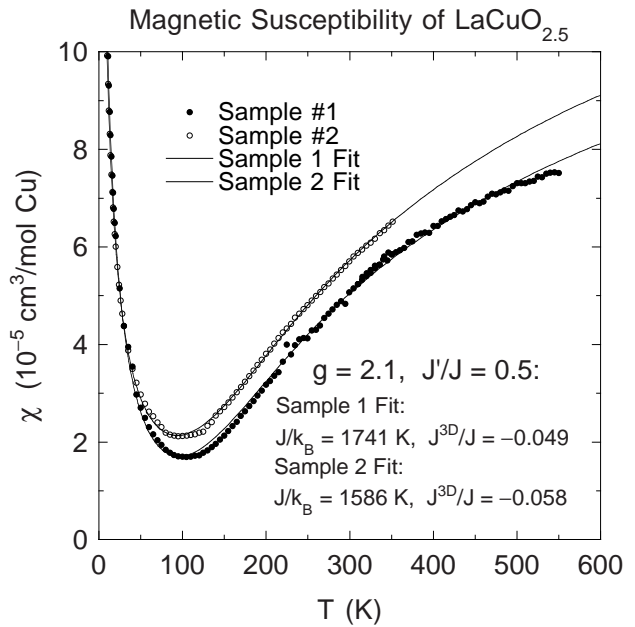}}
\vglue 0.1in
\caption{Magnetic susceptibility $\chi$ versus temperature $T$ for
LaCuO$_{2.5}$ samples~1 ($\bullet$, Ref.~\protect\onlinecite{Hiroi1996})
and~2 ($\circ$).  The curves are theoretical fits for LaCuO$_{2.5}$-type 3D
coupled ladders to the respective data using
Eqs.~(\protect\ref{EqChiExp:all}) and assuming $g = 2.1$ and $J^{\prime}/J
= 0.5$, yielding the intraladder $J$ and interladder $J^{\rm 3D}$ exchange
constants listed for each sample.  The other fit parameters are given in
the text.}
\label{Fig37}
\end{figure}

\begin{figure}
\epsfxsize=3.3in
\centerline{\epsfbox{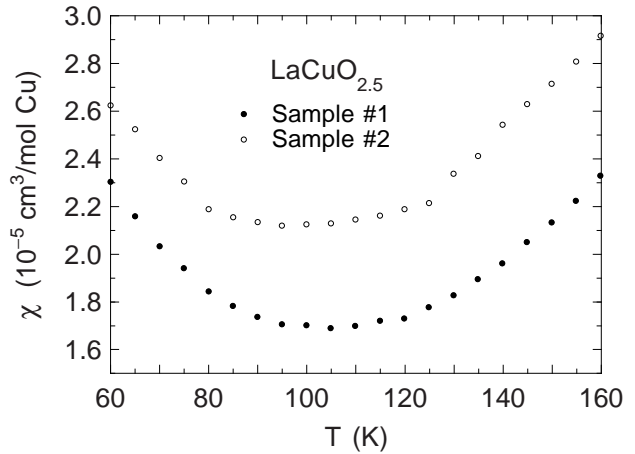}}
\vglue 0.1in
\caption{Expanded plots of the magnetic susceptibility $\chi$ versus
temperature $T$ for LaCuO$_{2.5}$ samples~1
(Ref.~\protect\onlinecite{Hiroi1996}) and~2 from Fig.~\protect\ref{Fig37}.}
\label{Fig38}
\end{figure}

$\chi(T)$ data for two polycrystalline samples of LaCuO$_{2.5}$ were fitted
by our QMC $\chi^*(t)$ simulations.  Data for sample~1 in the temperature
range from 4 to 550\,K have been previously reported.\cite{Hiroi1996} 
Sample~2 is new; data were obtained for this sample from~4 to 350\,K\@. The
data for both samples are shown as filled and open circles in
Fig.~\ref{Fig37}, respectively.  Long-range AF ordering has been
found from NMR and $\mu$SR measurements at $T_{\rm N}\sim
110$--125\,K,\cite{Matsumoto1996,Kadono1996} as noted in the Introduction. 
Shown in Fig.~\ref{Fig38} are expanded plots of the measured
$\chi(T)$ data for the two samples from~60 to~160\,K\@.  Although one could
perhaps infer the occurrence of an anomaly in each set of data in the range
between~115 and~130\,K, we conclude that there is no clearly defined
magnetic ordering anomaly in the $\chi(T)$ data for either of our two
samples in this $T$ range.

We fitted the $\chi(T)$ data for each of the two samples by
Eqs.~(\ref{EqChiExp:all}), where $\chi^*(t)$ is our fit to our QMC
simulations for LaCuO$_{2.5}$-type 3D coupled ladders with no spin gap. 
Because the $\chi^*(t)$ fit function is three-dimensional, we could vary
the fitting parameters $J,\ J^{\rm 3D}/J$ and $J^\prime/J$ simultaneously to
obtain the best fit.  We found that with any reasonable $g\approx 2$, the
possibility $J^\prime/J\approx 1$ could be ruled out by the bad quality of
the fits to the data.  Good fits were obtained for $J^\prime/J \approx
0.5$.  Setting $J^\prime/J = 0.5$ and $g = 2.1$ yielded a fit to the data
for each sample with parameters
\vglue0.5in
\newpage
\begin{mathletters}
\label{EqLCOPars:all}
\begin{eqnarray} {\rm Sample\ 1:}\nonumber\\
\chi_0 &=& -3.0(1)\times 10^{-5}\,{\rm {cm^3\over mol}}~,\nonumber\\ C_{\rm
imp} &=& 0.00187(3)\,{\rm {cm^3\,K\over mol}}~,~~~\theta = -6.4(2)\,{\rm
K}~,\nonumber\\ {J\over k_{\rm B}} &=& 1741(16)\,{\rm K}~,~~~{J^{\rm
3D}\over J} = -0.049(1)~,\\
\nonumber\\ {\rm Sample\ 2:}\nonumber\\
\chi_0 &=& -3.5(1)\times 10^{-5}\,{\rm {cm^3\over mol}}~,\nonumber\\ C_{\rm
imp} &=& 0.00153(1)\,{\rm {cm^3\,K\over mol}}~,~~~\theta = -5.52(9)\,{\rm
K}~,\nonumber\\ {J\over k_{\rm B}} &=& 1586(14)\,{\rm K}~,~~~{J^{\rm
3D}\over J} = -0.058(1)~,
\end{eqnarray}
\end{mathletters} as shown by the solid curves in
Fig.~\ref{Fig37}.  Allowing the parameter $J^\prime/J$ to vary
during the fits yielded equivalent quality fits with exchange constants
$J/k_{\rm B}$ = 2690(560)\,K, $J^\prime/J$ = 0.562(6) and $J^{\rm 3D}/J =
-0.038(10)$ for sample~1 and $J/k_{\rm B}$ = 1505(39)\,K, $J^\prime/J$ =
0.484(8) and $J^{\rm 3D}/J = -0.056(2)$ for sample~2, which are similar to
those in Eqs.~(\ref{EqLCOPars:all}) assuming $J^\prime/J = 0.5$.  The
values of $J^{\rm 3D}/J$ are close to the value $J^{\rm 3D}_{\rm QCP}/J =
-0.036(1)$ at the QCP for FM $J^{\rm 3D}/J$ values and $J^\prime/J = 0.5$,
and are on the ordered side of the QCP as expected from the observed AF
ground state.

We also carried out fits to the $\chi(T)$ data in which we allowed $g$ to
vary along with the exchange constants, but due to the strong correlation
especially between $g$ and $J$, the estimated standard deviations on the
parameters were very large and the fitted $g$ and exchange constant
parameters are therefore uninformative, but are consistent within the
errors with those given above.  Finally, we also carried out fits to the
data assuming that $J^\prime/J = 0.5$ or~1 and that a spin gap exists in
LaCuO$_{2.5}$, but the fits for each
$J^\prime/J$ yielded $J^{\rm 3D}/J$ values outside the ranges of validity
of the respective QMC data fit functions, indicating that the assumption of
the existence of a spin gap is incorrect, consistent with the fitting
results obtained above assuming a gapless excitation spectrum.

\subsection{CaV$_2$O$_5$}
\label{SecCaVO}

\begin{figure}
\epsfxsize=3.3in
\centerline{\epsfbox{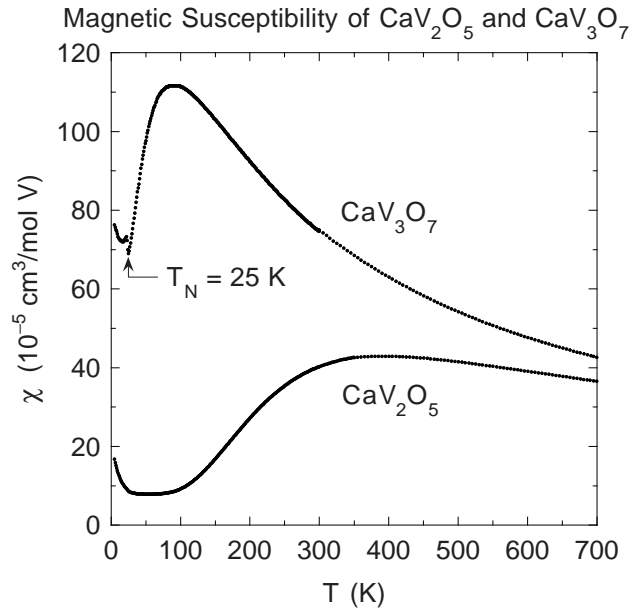}}
\vglue 0.1in
\caption{Magnetic susceptibility $\chi$ versus temperature $T$ for
CaV$_2$O$_5$ (sample~1) and CaV$_3$O$_7$.}
\label{Fig39}
\end{figure}

\begin{figure}
\epsfxsize=3.3in
\centerline{\epsfbox{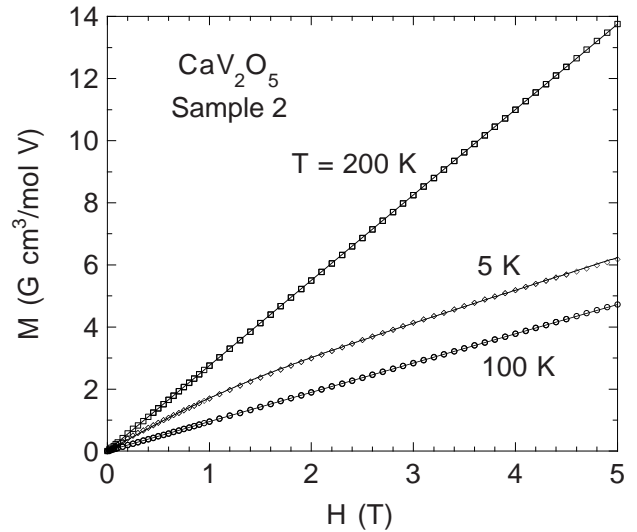}}
\vglue 0.1in
\caption{Magnetization $M$ versus applied magnetic field $H$ for
CaV$_2$O$_5$ sample~2 at 5, 100 and 200\,K\@.  Proportional fits to the
data at 100 and 200\,K are shown.  The solid curve through the data at 5\,K
is a modified-Brillouin function fit described in the text.}
\label{Fig40}
\end{figure}

The $\chi(T)$ data measured for CaV$_2$O$_5$ and CaV$_3$O$_7$ up to 700\,K
are shown in Fig.~\ref{Fig39}; we include data for the latter
compound, which exhibits long-range AF ordering below $T_{\rm N} \sim
23$\,K according to Ref.~\onlinecite{Harashina1996} (we find $T_{\rm N} =
25$\,K), because the former compound contains the latter as an impurity
phase which must be corrected for. 
$M(H)$ isotherms at 5, 100 and 200\,K are shown for CaV$_2$O$_5$ in
Fig.~\ref{Fig40}.  To within our precision, $M\propto H$ at
100 and 200\,K, but pronounced negative curvature is apparent at 5\,K for $H
\lesssim 1$\,T\@.  We were able to fit the data at 5\,K very well by
Eqs.~(\ref{EqMfit:all}) assuming
$g_{\rm imp} = 1.96$ and $S_{\rm imp} = 1/2$, 
\noindent as shown by the solid curve
in the figure.  The fitting parameters were 
\begin{mathletters}
\label{EqCaV2O5MHPars:all}
\begin{equation}
\chi_0 = 10.47\times 10^{-5}\,{\rm {cm^3\over mol\,V}}~,
\label{EqCaV2O5MHPars:a}
\end{equation}
\begin{equation} f_{\rm imp} = 0.0181{\rm \%}~,~~~\theta = +4.17\,{\rm K\
(ferromagnetic)}~,
\label{EqCaV2O5MHPars:b}
\end{equation}
\end{mathletters} with a variance of 0.000420\,(G\,cm$^3$/mol\,V)$^2$.  

\begin{figure}
\epsfxsize=3.3in
\centerline{\epsfbox{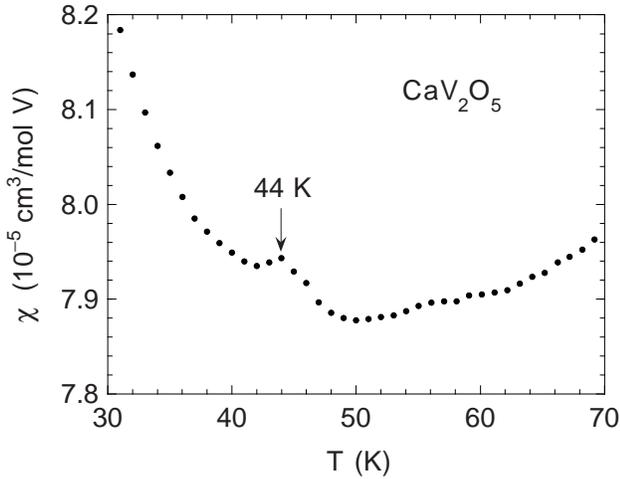}}
\vglue 0.1in
\caption{Expanded plot of the magnetic susceptibility $\chi$ versus
temperature $T$ for CaV$_2$O$_5$ sample~1 from Fig.~\protect\ref{Fig39}.}
\label{Fig41}
\end{figure}

In view of the observation of Luke {\it et al.}\ of an anomaly in $\chi(T)$ 
and of spin-freezing by $\mu$SR at $\sim 50$\,K in
CaV$_2$O$_5$,\cite{Luke1998} we looked carefully at our $\chi(T)$ data for
this compound near this temperature, and indeed found a small but clearly
defined cusp at 44\,K in the measured data, as shown in
Fig.~\ref{Fig41}.  This anomaly presumably cannot arise from the
CaV$_3$O$_7$ impurity phase in our sample, since the AF ordering transition
in pure CaV$_3$O$_7$ occurs at $\approx 25$\,K\@.  In the following, we
limit our theoretical fits to the data for CaV$_2$O$_5$ in the temperature
range $T > 50$\,K\@.

We first fitted the data for CaV$_2$O$_5$ from 50 to 700\,K by
\begin{mathletters}
\label{EqChiCoupDimers:all}
\begin{eqnarray}
\chi(T) = \chi_0 &+& {C_{\rm imp}\over T} + f \chi^{\rm
CaV_3O_7}(T)\nonumber\\
 &+&(1-f)\Big(0.3751\,\frac{\rm cm^3\,K}{\rm mol\,V}\Big) {g^2\over
J^\prime/k_{\rm B}}\,\chi^*(t)~,\label{EqChiCoupDimers:a}
\end{eqnarray}
\begin{equation} 
\chi^*(t) = \frac{\chi^{*,{\rm dimer}}(t)}{1 + \lambda\chi^{*,{\rm
dimer}}(t)}~,\label{EqChiCoupDimers:b}
\end{equation}
\begin{equation} t \equiv {k_{\rm B}T\over J^\prime}~,~~~\lambda \equiv
{\sum^\prime_j J_{ij}\over J^\prime}~,\label{EqChiCoupDimers:c}
\end{equation}
\end{mathletters} 
where $f$ is the molar fraction of the sample with respect to V consisting
of the CaV$_3$O$_7$ impurity phase, $\chi^{\rm CaV_3O_7}(T)$ is a fit to
the measured susceptibility of this impurity phase per mole of~V,
$\chi^*(t)$ is the reduced spin susceptibility of a coupled dimer system
according to the molecular field theory (MFT) in Sec.~\ref{SecMFT}, with
molecular field coupling constant $\lambda$ as defined in
Eq.~(\ref{EqChiCoupDimers:c}) where the sum is over all exchange coupling
constants $J_{ij}$ from a given spin $S_i$ to all other spins $S_j$ outside
its own dimer.  The $\chi^{*,{\rm dimer}}(t)$ of the isolated dimer was
given previously in Eq.~(\ref{EqChiDimer}) and the exchange constant within
a dimer is here denoted by $J^\prime$.   

\begin{figure}
\epsfxsize=3.3in
\centerline{\epsfbox{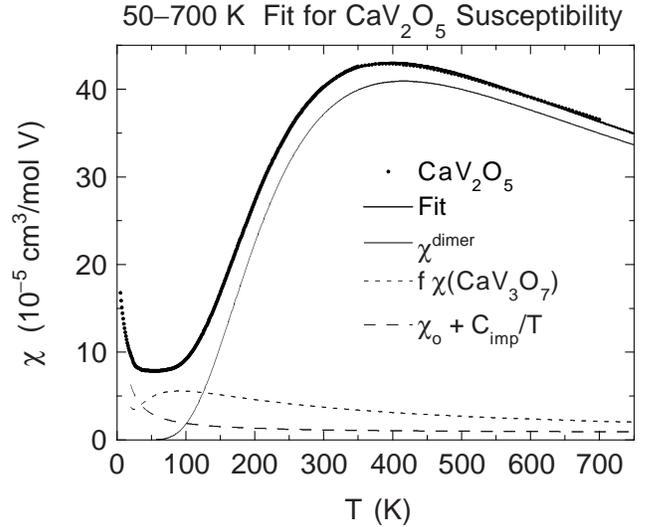}}
\vglue 0.1in
\caption{Magnetic susceptibility $\chi$ versus temperature $T$ for
CaV$_2$O$_5$ sample~1 ($\bullet$) and a fit by
Eqs.~(\protect\ref{EqChiCoupDimers:all}), assuming $g = 1.96$, for AF
quantum $S = 1/2$ dimers which are coupled to each other according to the
molecular field approximation.  The heavy solid curve (not visible over
most of the $T$ range due to overlap with the data points) is the fit, the
light solid curve the contribution from the coupled dimers, the
short-dashed curve the contribution from the CaV$_3$O$_7$ impurity phase
and the long-dashed curve the contributions from the $T$-independent
susceptibility and impurity Curie term. The fit and contribution curves are
extrapolated down to 20\,K and up to 750\,K\@. The fit parameters are given
in the text.}
\label{Fig42}
\end{figure}

Our fit of Eqs.~(\ref{EqChiCoupDimers:all}) to the data, assuming a fixed
$g = 1.96$, is shown as the heavy solid curve in
Fig.~\ref{Fig42}, where 
\noindent the contributions from the various
terms in Eq.~(\ref{EqChiCoupDimers:a}) are also plotted as indicated in the
figure caption.  The parameters of the fit are
\begin{eqnarray}
\chi_0 &=& 0.78\times 10^{-5}\,{\rm {cm^3\over mol\,V}}~,~~~ C_{\rm imp} =
0.0011\,{\rm {cm^3\,K\over mol\,V}}~,\nonumber\\ f &=& 5.0\%~,~~~
{J^\prime\over k_{\rm B}} = 667\,{\rm K}~,~~~\lambda = 0.31~.
\label{EqCVOPars}
\end{eqnarray} The fraction $f$ of the sample consisting of CaV$_3$O$_7$
impurity phase is close to the value of $\approx 4\%$ that we estimate from
our x-ray diffraction measurements on the same CaV$_2$O$_5$ sample.  The
magnitude of $J^\prime$ is surprisingly large for a $d^1$ vanadate.  If we
assume the same types of nearest-neighbor V-V exchange interactions as
discussed above for the Cu-Cu exchange constants in SrCu$_2$O$_3$, then in
that notation we have from Eqs.~(\ref{EqChiCoupDimers:c})
and~(\ref{EqCVOPars})
\begin{equation}
\lambda = \frac{2(J + J^{\prime\prime} + J^{\prime\prime\prime})}{J^\prime}
= 0.31~.
\label{EqCVOExCnsts}
\end{equation} Since we obtained an excellent fit to the data using MFT
which cannot be significantly improved upon, it is not in general possible
to establish from the experimental $\chi(T)$ data alone, without further
theoretical and/or experimental input, which is the V-V dimer bond and to
which V spin(s) outside a dimer a V atom is most strongly coupled.

\begin{figure}
\epsfxsize=3.3in
\centerline{\epsfbox{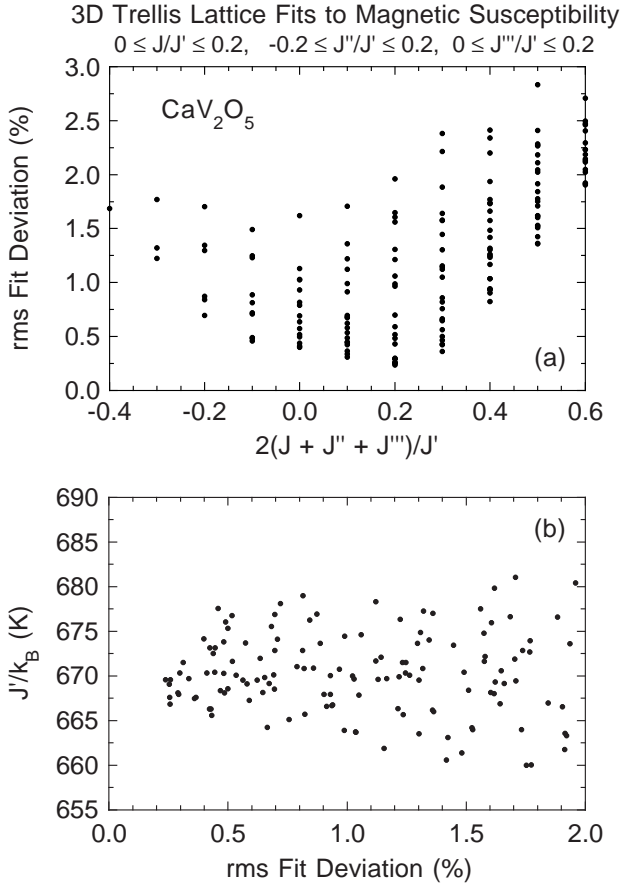}}
\vglue 0.1in
\caption{(a) Scatter plot of the relative rms fit deviation versus twice
the sum of the V-V exchange coupling constants, excluding the rung coupling
constant $J^\prime$, obtained from fits to the magnetic susceptibility of
CaV$_2$O$_5$ by Eq.~(\protect\ref{EqChiExp:a}) assuming $g = 1.96$ and the
3D  trellis lattice model.  (b) Scatter plot of the fitted rung coupling
constant
$J^\prime$ versus rms fit deviation for the fits in (a).}
\label{Fig43}
\end{figure}

We assume now that the strongest V-V exchange bond in the system, the dimer
bond, is across the rungs of the two-leg ladders in the structure.  We then
fitted the experimental data by Eqs.~(\ref{EqChiCoupDimers:all}) using a
fixed
$g = 1.96$, but where the $\chi^*(t)$ is now given by our fits to our QMC
simulations for the trellis layer and stacked ladders.  We carried out 225
fits to the experimental data for the parameter ranges $0 \leq J/J^\prime
\leq 0.2$,
$-0.2 \leq J^{\prime\prime}/J^\prime \leq 0.2$ and $0
\leq J^{\prime\prime\prime}/J^\prime \leq 0.2$, in increments of 0.05 for
each parameter.  A scatter plot of the rms fit deviation versus $\lambda$ in
Eq.~(\ref{EqCVOExCnsts}) is shown in Fig.~\ref{Fig43}(a) for
the fits in which the fraction $f$ was physical (positive), where the
minimum in the deviation is seen to occur for $\lambda \sim 0.2$, in
approximate agreement with the estimate from MFT  in
Eq.~(\ref{EqCVOExCnsts}).  A scatter plot of the $J^\prime$ values from
these fits versus rms fit deviation is shown in
Fig.~\ref{Fig43}(b); the eight best fits with relative rms
deviations below 0.3\% give $J^\prime/k_{\rm B} = 669(3)$\,K, in agreement
with Eq.~(\ref{EqCVOPars}).  For these eight fits, $f = 5.3$--5.7\%,
$J/J^\prime$ and $J^{\prime\prime\prime}/J^\prime$ were either 0, 0.05 \,or
\,0.1 \,and \,$J^{\prime\prime}/J^\prime$ \,was \,either \,$-0.05$, \,0,
\,0.05\,or\,0.1, 
\begin{figure}
\epsfxsize=3.3in
\centerline{\epsfbox{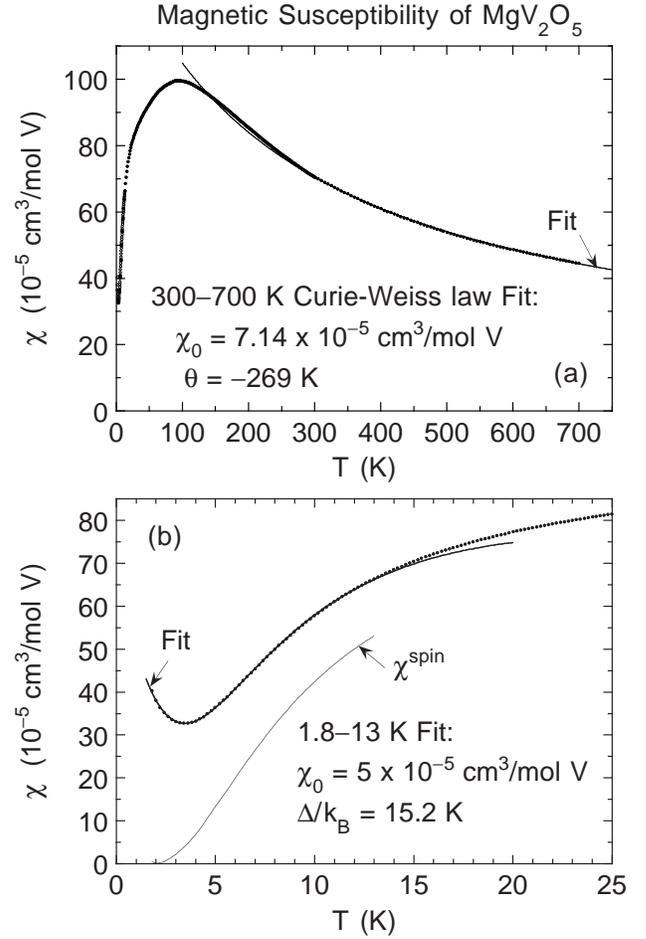}}
\vglue 0.1in
\caption{(a) Magnetic susceptibility $\chi$ versus temperature $T$ for
MgV$_2$O$_5$ ($\bullet$, Ref.~\protect\onlinecite{Isobe1998b} and the
present work).  \mbox{(b) Expanded} plot of the data in~(a) at low
temperatures.  The solid curve in (a) is a 300--700\,K fit to the data by
Eq.~(\protect\ref{EqMVOHiTChiFit}), which is the sum of a constant term and
a Curie-Weiss term; extrapolations down to 100\,K and up to 750\,K are also
shown.  In (b), the heavy solid curve is a 1.8--13\,K fit to the data by
Eqs.~(\protect\ref{EqTroyer}) and~(\protect\ref{EqChiExp:all}), where the
former expression is the low-temperature approximation for the spin
susceptibility of a two-leg ladder with spin gap $\Delta$; an extrapolation
up to 20\,K is also shown.  The spin susceptibility $\chi^{\rm spin}(T)$
from the fit in (b) is plotted as the light solid curve.  The fits in (a)
and (b) both assume
$g = 1.96$.}
\label{Fig44}
\end{figure}
\noindent  subject to the observed constraint that $(J + 
J^{\prime\prime} + J^{\prime\prime\prime})/J^\prime = 0.1$ for all eight
fits.  The corresponding ranges for $\chi_0$ and $C_{\rm imp}$ were 0.02 to
$0.2\times 10^{-5}$\,cm$^3$/mol\,V and 0.0011 to 0.0013\,cm$^3$\,K/mol\,V,
respectively, very similar to the values in Eq.~(\ref{EqCVOPars}).

\subsection{MgV$_2$O$_5$}
\label{SecMgV2O5}
\vglue-0.1in
The $\chi(T)$ of MgV$_2$O$_5$ sample~1 measured by \mbox{Isobe {\it et
al.}}\cite{Isobe1998b} in a field $H = 0.1$\,T below 300\,K and extended
here up to 700\,K is shown in Fig.~\ref{Fig44}(a); an expanded
plot of the data at low temperatures is given in
Fig.~\ref{Fig44}(b).  A broad peak is seen at
$T^{\rm max} \sim 100$\,K, symptomatic of dynamical short-range AF ordering
and of a dominant AF interaction in the compound. 
$T^{\rm max}$ is about a factor of four smaller than that for CaV$_2$O$_5$
in Fig.~\ref{Fig39}, and we thus expect that the largest AF
exchange constant in MgV$_2$O$_5$ is roughly a factor of four smaller than
in CaV$_2$O$_5$, i.e.\ $\sim 170$\,K\@.  Of all the QMC simulations we have
presented in Sec.~\ref{SecSimulations}, the shape of $\chi(T)$ for
MgV$_2$O$_5$ most closely resembles that for the isolated ladder with
$J^\prime/J \sim 0.2$ in Fig.~\ref{Fig03}.  On the other hand,
the exchange constants in CaV$_2$O$_5$ found above are evidently in the
opposite limit $J^\prime/J \gg 1$.  We begin our analysis with the $\chi(T)$
data at high temperatures $T \gtrsim T^{\rm max}$ in
Fig.~\ref{Fig44}(a).

We fitted the $\chi(T)$ data for MgV$_2$O$_5$ from 300 to 700\,K in
Fig.~\ref{Fig44}(a) by the sum of a constant term and a
Curie-Weiss term
\begin{equation}
\chi(T) = \chi_0 + {C\over T - \theta}~,
\label{EqMVOHiTChiFit}
\end{equation} where the Curie constant $C$ is given by
Eq.~(\ref{EqCurieLaw:b}) assuming $g = 1.96$.   The fit is shown by the
solid curve in Fig.~\ref{Fig44}(a), where extrapolations down to
100\,K and up to 750\,K are also shown.  From the fit, we obtained the
parameters
\begin{mathletters}
\label{EqMgV2O5CWPars:all}
\begin{equation}
\chi_0 = 7.144\times 10^{-5}\,{\rm {cm^3\over
mol\,V}}~,\label{EqMgV2O5CWPars:a}
\end{equation}
\begin{equation}
\theta = -268.6\,{\rm K}~,~~~{1\over k_{\rm B}} \sum_j J_{ij} = -4\,\theta =
1074\,{\rm K}~,\label{EqMgV2O5CWPars:b}
\end{equation}
\end{mathletters} where the relationship between $\theta$ and the $J_{ij}$
exchange constants was given in Eq.~(\ref{EqCurie-Weiss:b}).  The relative
rms deviation for this fit is 0.20\%.  If the data are fitted from 200 to
700\,K, the fit parameters change slightly to $\chi_0 = 6.41\times
10^{-5}$\,cm$^3$/mol\,V and
$\theta = -258$\,K, with a larger rms fit deviation of 0.42\%.  The fact
that the extrapolated fit in Fig.~\ref{Fig44}(a) describes the
data well almost down to $T^{\rm max}$ indicates that geometric frustration
may be an important consideration in this compound,\cite{Johnston1997}
consistent with the sizable AF $J^{\rm diag}$ and $J^{\prime\prime}$
couplings (see Table~\ref{tab:calcJ}) calculated using the LDA+U method by
Korotin {\it et al.}\ for MgV$_2$O$_5$.\cite{Korotin1999}   On the other
hand, the value
$\sum_j J_{ij} = 1074\,{\rm K}$ in Eq.~(\ref{EqMgV2O5CWPars:b}) is about a
factor of two larger than calculated using the exchange constants of Korotin
{\it et al.}\ in Table~\ref{tab:calcJ}.  We believe that this discrepancy
arises because the fitted data are not at sufficiently high temperatures to
be in the temperature range where the Curie-Weiss law holds accurately,
which is typically at temperatures $T \gtrsim 10|\theta|$.  We will see
below that a reasonable fit to all the data from 2\,K to 700\,K can be
obtained using a model of coupled two-leg ladders.

Various high-$T$ ($T\ge 200$\,K) fits of the $\chi(T)$ data in
Fig.~\ref{Fig44}(a) by Eq.~(\ref{EqMVOHiTChiFit}), including the two
discussed above, all yielded $\chi_0\sim 4$--8$\times
10^{-5}$\,cm$^3$/mol\,V\@.  On the other hand, Isobe~{\it et
al.}\cite{Isobe1998b} \,concluded \,from \,analysis \,of $M(H)$
\begin{figure}
\epsfxsize=3.3in
\centerline{\epsfbox{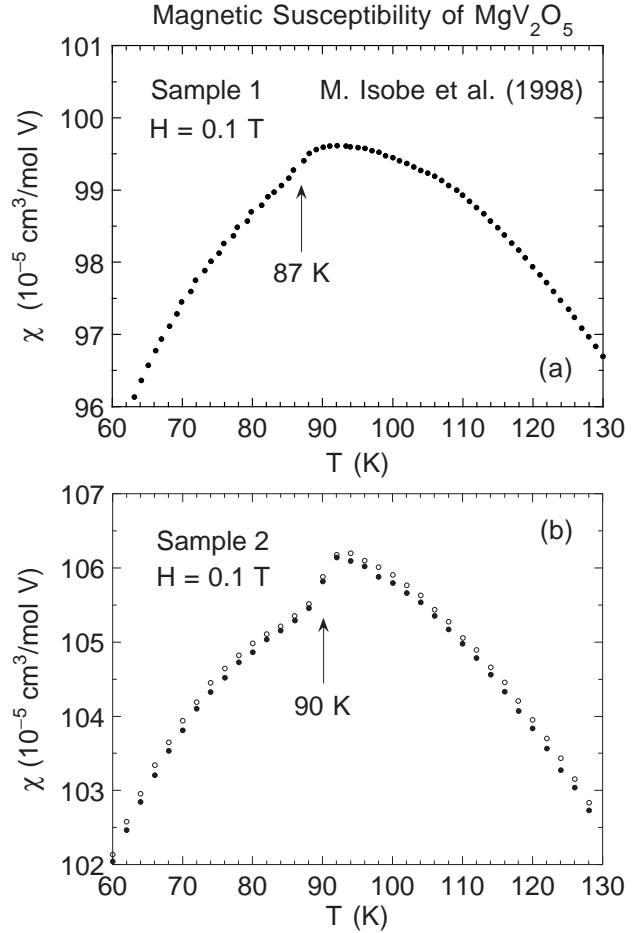}}
\vglue 0.1in
\caption{Expanded plots near 90\,K of the magnetic susceptibility $\chi$ in
an applied magnetic field $H = 0.1$\,T versus temperature $T$ for two
samples of MgV$_2$O$_5$.  Data from Fig.~\protect\ref{Fig44} (sample~1) are
shown in (a) and those for a different sample~2 are shown in~(b).  Both
samples show evidence for some type of phase transition at $\sim 90$\,K,
most likely associated with an impurity phase.}
\label{Fig45}
\end{figure}
\noindent at 2\,K ($0 < H \leq 5$\,T) that $\chi_0 \approx 22\times
10^{-5}$\,cm$^3$/mol\,V\@.  Since other measurements indicated a finite
spin-gap $\Delta$ for which $\chi^{\rm spin}(T = 0) = 0$, this $\chi_0
\equiv \chi(T\to 0)$ was attributed to a large Van Vleck susceptibility. 
We find here that $\chi(T)$ at low $T
\lesssim 15$\,K can be fitted very well by Eqs.~(\ref{EqChiExp:all}), where
the spin susceptibility $\chi^*(t)$ is given by the low-$t$ approximation
for the 2-leg ladder in Eq.~(\ref{EqTroyer}), using a $\chi_0$ consistent
with our range found from the high-$T$ fits to $\chi(T)$.   Our inferred
$\Delta$ is similar to the range of values found by Isobe {\it et al.}\
from different measurements as discussed in the Introduction.  For example,
shown in Fig.~\ref{Fig44}(b) is a 1.8--13\,K fit assuming $\chi_0 = 5\times
10^{-5}$\,cm$^3$/mol\,V and
$g = 1.96$, for which the parameters in Eqs.~(\ref{EqTroyer})
and~(\ref{EqChiExp:all}) are
\begin{mathletters}
\label{EqLoTLadderFitPars:all}
\begin{equation} C = 0.00122\,{\rm {cm^3\,K\over mol\,V}}~,~~~\theta =
-1.72\,{\rm K}~,\label{EqLoTLadderFitPars:a}
\end{equation}
\begin{equation} A = 0.00614\,{\rm {cm^3\,K^{1/2}\over
mol\,V}}~,~~~{\Delta\over k_{\rm B}} = 15.2\,{\rm
K}~.\label{EqLoTLadderFitPars:b}
\end{equation}
\end{mathletters}

There is evidence from expanded plots of $\chi(T)$ near the peak, as shown
in Fig.~\ref{Fig45}, that some type of phase transition occurs
near 90\,K, most likely due to an impurity phase.  This might cause an
increase in $\chi_0$ at low $T$ with respect to that at high $T$.

\begin{figure}
\epsfxsize=6.6in
\centerline{\epsfbox{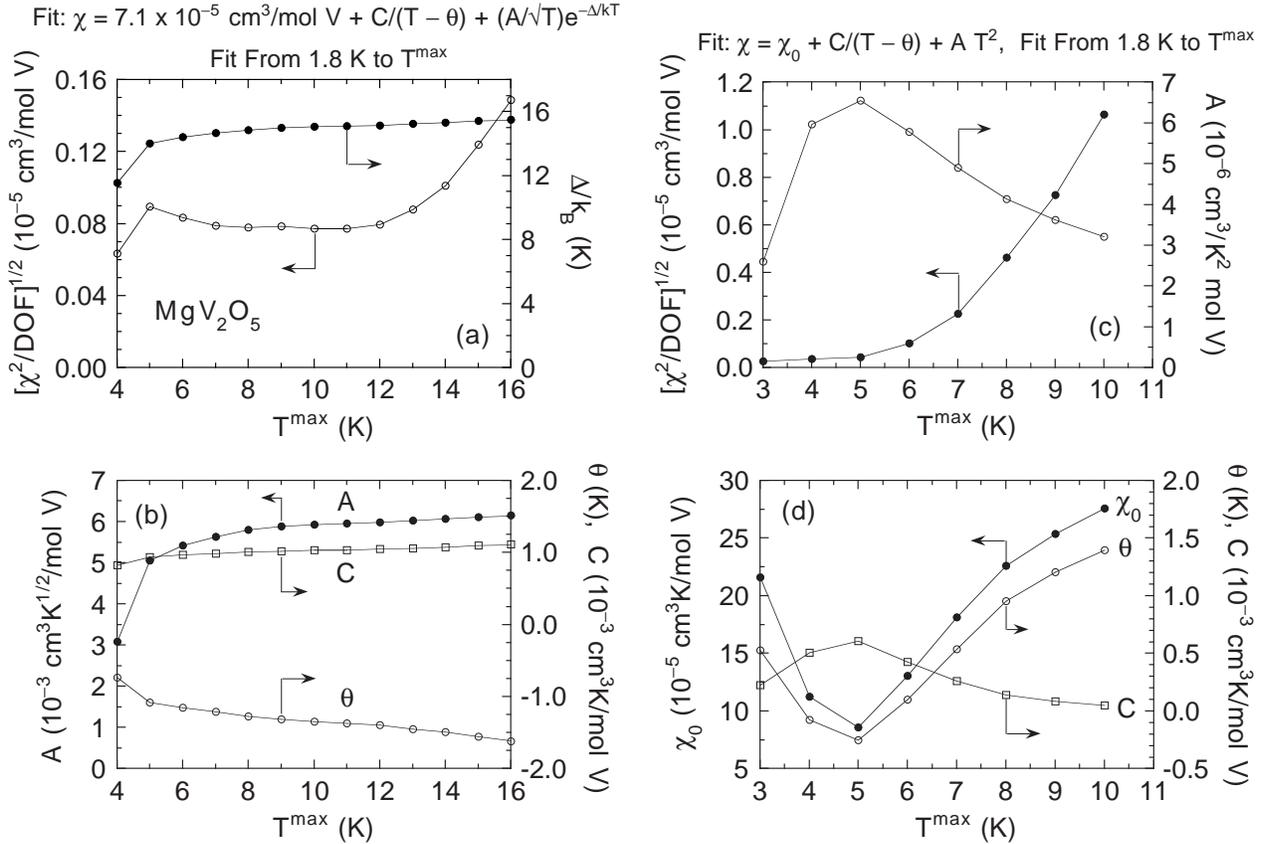}}
\vglue 0.1in
\caption{Fit parameters for the low temperature (1.8\,K to $T^{\rm max}$)
magnetic susceptibility $\chi$ versus temperature $T$ for MgV$_2$O$_5$. 
The fit parameters in (a) and (b) are for fits assuming a spin gap and
those in (c) and (d) are for fits assuming a gapless magnetic excitation
spectrum.}
\label{Fig46}
\end{figure}

On the other hand, one could argue that because the $\chi_0$ we derived at
high $T$ is much smaller than inferred by Isobe {\it et al.}\ from analysis
of low-$T$ $M(H,T=2$\,K) data, this difference might indicate the absence
of a spin gap.  One might then expect MgV$_2$O$_5$ to be close to a 3D QCP;
lack of N\'eel order would then be attributed to disorder effects.  In that
case at low $T$ one would expect $\chi(T) = \chi_0 + A T^2$ plus a
Curie-Weiss impurity term.  For a fit to be valid, one expects that the fit
parameters should not be very sensitive to the temperature range of the
fit.  Therefore, to differentiate the quality and applicability of the
gapped versus gapless fits, we determined the parameters of the two types
of fits to the experimental $\chi(T)$ data from 1.8\,K to a maximum
temperature $T^{\rm max}$.  The fit parameters for both types of fits are
plotted vs $T^{\rm max}$ in Figs.~\ref{Fig46}(a,b) and (c,d),
respectively, where the specific expressions fitted to the data are shown at
the top of the two sets of figures, respectively.  We see that for the QCP
\widetext
\narrowtext
\noindent fit, the $\chi^2$/DOF diverges and the fit parameters change
strongly as $T^{\rm max}$ increases above 5\,K, whereas  $\chi^2$/DOF for
the gapped fit remains small and the parameters  are essentially constant
for \mbox{5\,K $\leq T^{\rm max} \leq 13$}\,K\@.  These results appear to
rule out the gapless 3D QCP scenario and rather indicate that MgV$_2$O$_5$
has a spin-gap $\Delta/k_{\rm B} = 15.0(6)$\,K\@.  

\begin{figure}
\epsfxsize=3.3in
\centerline{\epsfbox{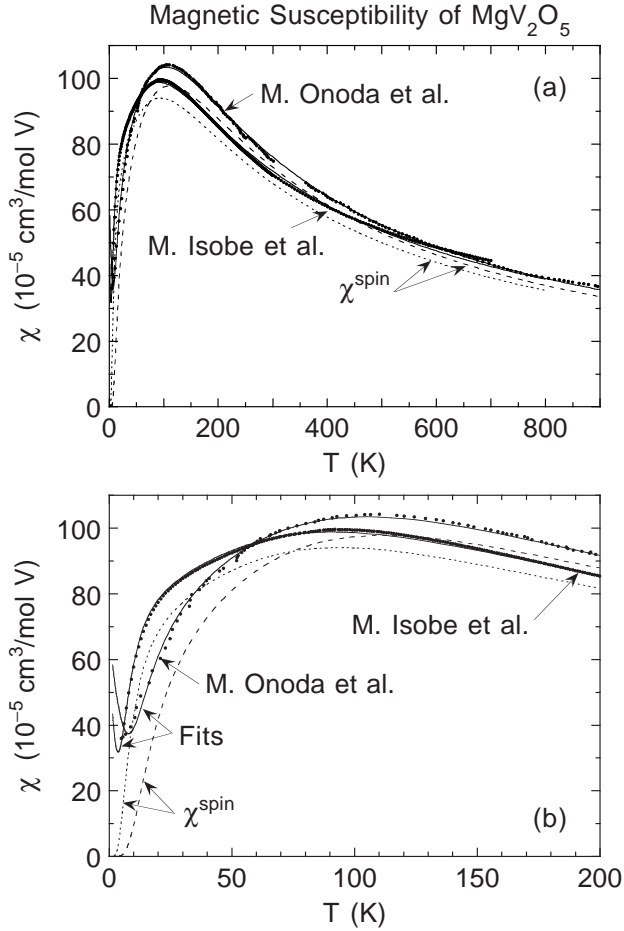}}
\vglue 0.1in
\caption{(a) Magnetic susceptibility $\chi$ versus temperature $T$ for
MgV$_2$O$_5$ ($\bullet$) as measured up to 300\,K by Isobe~{\it et al.}\
(Ref.~\protect\onlinecite{Isobe1998b} and the present work) and for a
different sample by \mbox{Onoda~{\it et al.}},\protect\cite{Onoda1998} as
indicated.  Each solid curve is a fit to the respective data set by
Eqs.~(\protect\ref{EqChiExp:all}).  A model for the spin susceptibility
$\chi^{\rm spin}(T)$ is assumed in which two-leg ladders with a spin gap are
coupled using the molecular field approximation, with fit parameters given
in the text.  The $\chi^{\rm spin}(T)$ for each sample was derived by
subtracting the respective fitted constant term and Curie-Weiss impurity
term from the respective fit, and is shown for the data of
\mbox{Isobe~{\it et al.}}\ by a dotted curve and for the data of
\mbox{Onoda~{\it et al.}}\ by a dashed curve.  (b) Expanded plots below
200\,K of the data and fits in~(a).}
\label{Fig47}
\end{figure}

We therefore carried out a fit by Eqs.~(\ref{EqChiExp:all}) to our
$\chi(T)$ data assuming that the spin susceptibility $\chi^*(t)$ is given
by that for coupled two-leg ladders with spatially anisotropic exchange for
which there is a spin gap.  We assumed a MFT coupling between the ladders,
so the spin susceptibility is given by Eqs.~(\ref{EqMFT:all}), where
$\chi_0^*(t,J^\prime/J)$ is  our 2D fit to our QMC simulation data for
isolated two-leg ladders.  The fit is shown as the solid curve through the
data set labeled ``Isobe {\it et al.}''\ in Fig.~\ref{Fig47}
together with the derived $\chi^{\rm spin}(T)$.  Also shown in
Fig.~\ref{Fig47} are the $\chi(T)$ data up to 900\,K obtained for a
different sample of MgV$_2$O$_5$ by \mbox{Onoda~{\it et
al.}},\cite{Onoda1998} which by comparison with the data for the first
sample illustrates the rather strong variability in $\chi(T)$ that can occur
between different samples.   Our fit to the data of \mbox{Onoda~{\it et
al.}}\ and the derived $\chi^{\rm spin}(T)$ are also shown in the figure. 
The parameters of the two fits are
\newpage

\begin{mathletters} Isobe {\it et al.}:
\label{EqMgV2O5ChiPars:all}
\[
\chi_0 = 0.000032(2)\,{\rm cm^3\over mol\,V},~~~C_{\rm imp} =
0.0015(2)\,{\rm cm^3\,K\over mol\,V}~,
\]
\[
\theta = -2.3(6)\,{\rm K}~,~~~{J\over k_{\rm B}} = 141.8(8)\,{\rm
K}~,~~~{J^\prime\over J} = 0.333(6)~,
\]
\[
\lambda = 3.56(6)~,~~~\chi^2/{\rm DOF = 6.1\times
10^{-11}\,(cm^3/mol\,V)^2}~,
\]
\begin{equation} {\rm relative\ rms\ deviation = 1.31\%}~;
\label{EqMgV2O5ChiPars:b}
\end{equation}

Onoda {\it et al.}:
\label{EqMgV2O5ChiPars:all}
\[
\chi_0 = 0.000016(2)\,{\rm cm^3\over mol\,V},~~~C_{\rm imp} = 0.005(1)\,{\rm
cm^3\,K\over mol\,V}~,
\]
\[
\theta = -7(4)\,{\rm K}~,~~~{J\over k_{\rm B}} = 158(2)\,{\rm
K}~,~~~{J^\prime\over J} = 0.57(2)~,
\]
\[
\lambda = 1.62(6)~,~~~\chi^2/{\rm DOF = 7.7\times
10^{-11}\,(cm^3/mol\,V)^2}~,
\]
\begin{equation} {\rm relative\ rms\ deviation = 1.65\%}~.
\label{EqMgV2O5ChiPars:b}
\end{equation}
\end{mathletters} Both fitted values of $J/k_{\rm B}$ and one of the fitted
values of $J^\prime/J$ are (fortuitously) close to the respective values
144\,K and 0.64 in Table~\ref{tab:calcJ} predicted by \mbox{Korotin {\it et
al.}}\cite{Korotin1999,Korotin2000} from LDA+U calculations, lending
support to the present working hypothesis that MgV$_2$O$_5$ has a spin
gap.  The MFT coupling constant is given by  Eq.~(\ref{EqMFT:c}) to be
$\lambda = 2 J^{\prime\prime}/J$, which from the exchange constants of
\mbox{Korotin {\it et al.}}\ in Table~\ref{tab:calcJ} predicts $\lambda =
0.84$; this value is significantly smaller than both of our fitted
$\lambda$ values.  Of course, the MFT-based fit is only expected to give
accurate values of the exchange constants and $\lambda$ if
$|\lambda|\ll 1$.  In addition, there is no way to include the frustrating
AF diagonal second-neighbor intraladder exchange coupling $J^{\rm diag}$,
such as in Table~\ref{tab:calcJ}, in the present modeling framework and
this coupling has an unknown influence on the values of our fitted
parameters.  Finally, in view of the rather strong variation of the
measured $\chi(T)$ for different polycrystalline samples of MgV$_2$O$_5$
and of possible impurity effects, a definitive evaluation of the exchange
constants from $\chi(T)$ data will probably only be possible using data for
single crystals when such data become available.

\section{Summary and Discussion}
\label{SecSummaryDisc}
\vglue0.15in
We have carried out extensive QMC simulations of $\chi^*(t)$ for a large
range of exchange parameter combinations for both isolated and coupled $S =
1/2$ two-leg Heisenberg ladders, and fitted these and previously published
QMC data accurately by interpolating functions.  Quantum critical points
were determined for both 2D AF stacked-ladder and 3D FM LaCuO$_{2.5}$-type
interladder interactions between AF two-leg ladders.  For each of these
two interladder coupling configurations, but not for the frustrated
trellis layer interladder couplings, and for each of $J^\prime/J = 0.5$
and~1, there is a temperature at which $\chi^*(t)$ is independent of the
strength of the interladder coupling.  It would be interesting to know if
there is any fundamental significance to this result.  The dispersion
relation for the lowest energy one-magnon excitations and for the lower
boundary of the two-magnon continuum were calculated in the range $0.5\leq
J^\prime/J \leq 1$, the regime relevant to cuprate two-leg ladder
compounds.  LDA+U calculations of the exchange constants in SrCu$_2$O$_3$
were carried out.

\begin{figure}
\epsfxsize=3.3in
\centerline{\epsfbox{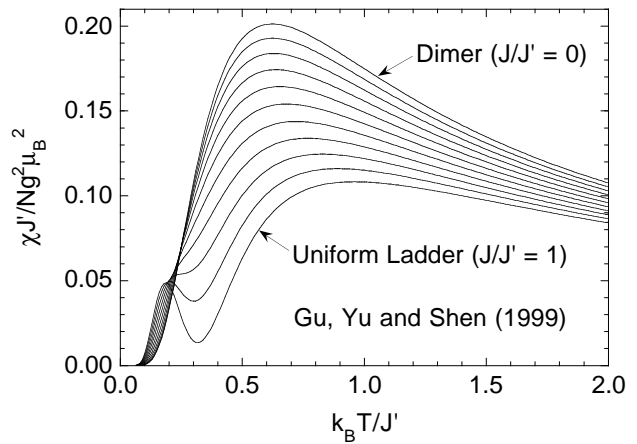}}
\vglue 0.1in
\caption{Magnetic susceptibility $\chi$ versus temperature $T$ for the $S =
1/2$ AF two-leg Heisenberg ladder, obtained by us from the free energy
calculated for $J/J^\prime \ll 1$ from third order perturbation theory by
Gu, Yu and Shen.\protect\cite{Gu1999}  Curves are shown for $J/J^\prime = 0$
to~1 in 0.1 increments (top to bottom).}
\label{Fig48}
\end{figure}

Gu, Yu and Shen have derived an expansion for the temperature- and magnetic
field-dependent free energy of the $S = 1/2$ AF two-leg Heisenberg ladder
for $|J/J^\prime| \ll 1$ using perturbation theory to third order in
$J/J^\prime$.\cite{Gu1999}  To compare our QMC $\chi^*(t)$ data fit in this
parameter regime with the prediction of this analytic theory, we derived
the zero-field $\chi^*(t)$ from their expression for the free energy, and
the result is
\begin{eqnarray}
\chi^*(t) &=&{1\over t}\Bigg\{{1\over 3 + {\rm e}^{\beta}} -
{J\over J^\prime}\bigg[{2\beta\over (3 + {\rm e}^{\beta})^2}\bigg]
\nonumber\\  &-& \Big({J\over J^\prime}\Big)^2\bigg[{3\beta({\rm
e}^{2\beta}-1) - \beta^2(5 + {\rm e}^{2\beta})\over 4(3 + {\rm
e}^{\beta})^3}\bigg]\nonumber\\ 
&-&\Big({J\over J^\prime}\Big)^3\bigg[{3 \beta({\rm
e}^{2\beta}-1)\over 8(3 + {\rm e}^{\beta})^3}\nonumber\\  &-& {9 \beta^2
{\rm e}^{\beta}(1 + 3{\rm e}^{\beta}) - \beta^3(7{\rm e}^{2 \beta} - 9{\rm
e}^{\beta} - 12)\over 12(3 + {\rm e}^{\beta})^4}\bigg]\Bigg\}~,
\label{EqGuYuShen}
\end{eqnarray}
where
\[
\beta \equiv {1\over t} = {J^\prime\over k_{\rm B}T}~.
\]
The first term in Eq.~(\ref{EqGuYuShen}), corresponding to $J = 0$, gives
the exact  susceptibility of the isolated dimer in Eq.~(\ref{EqChiDimer}),
as it should.  An overview of $\chi^*(t)$ for $0 \leq J/J^\prime \leq 1$ is
shown in Fig.~\ref{Fig48}, where for the larger $J/J^\prime$ values the
theory is not expected to apply.  In fact, a pronounced unphysical hump in
$\chi^*$ is seen to develop at $t\approx 0.2$ for $J/J^\prime \gtrsim 0.6$,
so this value of $J/J^\prime$ is an approximate upper limit of the
$J/J^\prime$ range for which Eq.~(\ref{EqGuYuShen}) can be useful for all
temperatures.  Of course, since the HTSE of this $\chi^*(t)$ is accurate to
order $1/T^3$ as we discussed in Sec.~\ref{SecHTSE}, Eq.~(\ref{EqGuYuShen})
can be used for any ratio of $J/J^\prime$ at sufficiently high
temperatures; however, in this case it is easier and simpler just to use
the HTSE itself.

\begin{figure}
\epsfxsize=3.3in
\centerline{\epsfbox{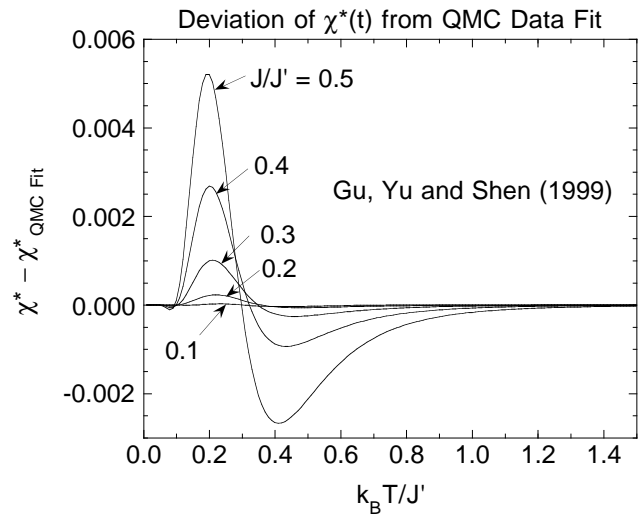}}
\vglue 0.1in
\caption{Deviation of the magnetic susceptibility $\chi^*$ versus
temperature $T$ in Fig.~\protect\ref{Fig48} from our fit to our QMC
$\chi^*(t)$ data.}
\label{Fig49}
\end{figure}

Shown in Fig.~\ref{Fig49} is the deviation versus temperature of the
$\chi^*(t)$ prediction in Eq.~(\ref{EqGuYuShen}) for $J/J^\prime = 0.1$,
0.2, 0.3, 0.4 and~0.5 from our accurate global two-dimensional fit to our
QMC $\chi^*(t)$ data in the range $0 \leq J/J^\prime \leq 1$.  This figure
shows, for example, that for Eq.~(\ref{EqGuYuShen}) to maintain an accuracy
of 1\% or better of the maximum value of $\chi^*(t)$, the value of
$J/J^\prime$ should not exceed about 0.3.  The comparison we have done here
illustrates that our QMC $\chi^*(t)$ data fit functions can be quite useful
for easily and quantitatively comparing with, and evaluating the accuracy
of, other theoretical calculations of the susceptibility of spin ladders,
in addition to their nominal use for modeling experimental data as we have
done extensively in this paper.

Our QMC $\chi^*(t)$ simulations for $-0.111 \leq J^{\rm diag}/J \leq 0$ and
$0.4 \leq J^\prime/J \leq 0.65$ (see Fig.~\ref{Fig04}) suggest
that the spin gap \,does \,not \,change \,significantly for a given
$J^\prime/J$ over 
\vglue0.04in
\noindent this range of $J^{\rm diag}/J$.  We are not aware of any
quantitative theoretical calculations of how the spin gap of isolated
ladders is affected by FM $J^{\rm diag}$ couplings.  For AF (positive)
$J^{\rm diag}$ interactions (which are geometrically frustrating), Wang has
found from density-matrix renormalization-group calculations (for $T = 0$)
that the spin gap is very nearly independent of $J^{\rm diag}/J$ for $0 \leq
J^{\rm diag}/J\lesssim 0.4$ and $J^\prime/J = 1$, whereas, e.g., for
$J^\prime/J = 0.5$ the spin gap strongly decreases with increasing $J^{\rm
diag}/J$.\cite{Wang1999,Wang1996}  This difference in behavior arises from
the presence of a phase boundary in the $J^{\rm diag}/J$ versus $J^\prime/J$
ground state phase diagram between the dimer singlet phase at small $J^{\rm
diag}/J$ and a Haldane phase at large $J^{\rm diag}/J$.  A similar
behavior of the spin gap vs $J^{\rm diag}/J$ for $J^\prime/J = 1$ was
obtained by \mbox{Nakamura~{\it et al.}}\cite{Nakamura1997} for the railroad
trestle lattice, which topologically is a two-leg ladder with
one instead of two diagonal couplings per four-spin \mbox{plaquette}. 
Interestingly, Nakamura and Okamoto have found that $\chi(T)$ calculated
using QMC simulations for the railroad trestle model, the
alternating-exchange chain model and the anisotropic two-leg ladder model,
can all accurately fit the experimental $\chi(T)$ data for KCuCl$_3$ (with
of course different exchange constants for each model),\cite{Nakamura1998}
which  illustrates that fits to $\chi(T)$ data can establish consistency
of a given Hamiltonian for a given spin system, but not the uniqueness of
that Hamiltonian, as discussed in the Introduction.  The natures of the
ground states and low-lying spin excitations of geometrically frustrated $S
= 1/2$ two-leg Heisenberg ladders have been extensively studied.
\cite{Nakamura1997,Xian1995,Allen1999,Kim1999,Nedelcu1999}

Turning now to the experimental part of the paper, the crystal structure of
SrCu$_2$O$_3$, required as input for our LDA+U calculations, was reported
here together with the structure of Sr$_2$Cu$_3$O$_5$.  Experimental
$\chi(T)$ data for SrCu$_2$O$_3$, Sr$_2$Cu$_3$O$_5$, LaCuO$_{2.5}$,
CaV$_2$O$_5$ and MgV$_2$O$_5$ were fitted by our analytic fits to the QMC
simulations to obtain estimates of the superexchange interactions between
the spins-1/2 in these two- and three-leg ladder compounds.  A summary of
the exchange constants determined for the cuprate spin ladder materials is
given in Table~\ref{TabJSumm}.

As shown in Table~\ref{TabJSumm}, our results confirm the preliminary
conclusion of Ref.~\onlinecite{Johnston1996} for SrCu$_2$O$_3$ based on
analyses of $\chi(T)$ data that $J^\prime/J \sim 0.5$ and $J/k_{\rm B} \sim
2000$\,K, assuming that the spherically-averaged $g$-value is in the
vicinity of 2.1.  Due to the insensitivities of the calculated $\chi^*(t)$
to the presence of either a weak FM diagonal intraladder coupling $J^{\rm
diag}$ or a FM trellis layer interladder coupling $J^{\prime\prime}$, we
could not determine either of these parameters from fits to the experimental
$\chi(T)$ data.  Setting $g = 2.1$ 
\widetext
\begin{table}
\caption{Summary of the exchange constants for cuprate spin ladder compounds
obtained by fitting the experimental magnetic susceptibility data by QMC
simulations for the $S = 1/2$ Heisenberg model for various combinations of
the different types of exchange interactions.  A parameter with an asterisk
($\ast$) was fixed to the listed value or range during the fit or fits,
respectively.  The error, or fixed fitting range, in the last digit of a
quantity is given in parentheses.  In the third set of exchange constants
for
${\rm SrCu_2O_3}$, $J$ and $J^\prime$ were constrained by the requirement
that the spin gap $\Delta/k_{\rm B} = 377$\,K as found from inelastic
neutron scattering measurements.  Note that for the three-leg ladder
compound ${\rm Sr_2Cu_3O_5}$ we have assumed that the nearest-neighbor
exchange constant along the central leg is the same as along the outer two
legs, which is not necessarily the case.}
\begin{tabular}{ldcddddd} Compound & g & $J/k_{\rm B}$\,(K) &  $J^\prime/J$
& $J^{\rm diag}/J$ & $J^{\prime\prime}/J$ &
$J^{\prime\prime\prime}/J$ & $J^{\rm 3D}/J$ \\
\hline
${\rm SrCu_2O_3}$ & 2.1$^*$ & 1905(5) & 0.488(3) \\
                  & 2.10(5)$^*$ & 1970(150) & 0.48(3) \\
                  & 2.1$^*$ & 1882 & 0.471(1) \\
                  & 2.1$^*$ & 1890(40) & 0.482(13) & $-$0.1$^*$ \\
                  & 2.10(5)$^*$ & 1950(170) & 0.48(4) & $-$0.05(5)$^*$ \\
                  & 2.1$^*$ & 2000(60) & 0.465(13) &           & $-$0.1$^*$
\\
                  & 2.10(5)$^*$ & 2000(180) & 0.465(40) &   & $-$0.1(1)$^*$
\\
                  & 2.1$^*$ & 1894(8) & 0.5$^*$ &     &           &
0.009(4) \\
                  & 2.10(5) & 1920(70)  & 0.5$^*$ &     &      & 0.01(1) \\
${\rm Sr_2Cu_3O_5}$& 2.1(1)$^*$ & 1810(150) & 0.66(5) \\
${\rm LaCuO_{2.5}}$ & 2.1$^*$ & 1665(95) & 0.5$^*$ &      &  &   &
$-$0.054(6) \\
 & 2.1$^*$ & 2400(900) & 0.53(5) &      &  &   & $-$0.043(15) \\
\end{tabular}
\label{TabJSumm}
\end{table}
\narrowtext
\noindent  and the exchange constant ratio $J^{\rm
diag}/J = -0.1$, which is the value in \mbox{Table~\ref{tab:calcJ}}
predicted by our LDA+U calculations, we obtained a good fit to the data for
$J$ and $J^\prime$ values which are within about 5\% of  the LDA+U
predictions.  The interladder $J^{\prime\prime}$ and stacked-ladder
$J^{\prime\prime\prime}$ couplings determined from our LDA+U calculations
and the $J^{\prime\prime\prime}$ from fits to our experimental $\chi(T)$
data are consistently found to be very small.  Our theoretical esimate
$J^{\prime\prime}/J = 0.002$ and our fitted intraladder exchange constants
$J$ and $J^\prime$ place SrCu$_2$O$_3$ within the spin-gap region of the
phase diagram in Fig.~5 of Normand {\it et al.}\cite{Normand1997b} for the
trellis layer, consistent with the occurrence of an experimentally observed
spin gap.  We note, however, that this phase diagram in exchange-parameter
space assumes that the intraladder diagonal coupling $J^{\rm diag} = 0$,
whereas our LDA+U calculations indicate a moderately strong
FM diagonal coupling in SrCu$_2$O$_3$.  

It is possible to obtain nearly as good a fit to the experimental $\chi(T)$
data for SrCu$_2$O$_3$ by the isolated ladder model ($J,\ J^\prime \neq
0$) for the isotropic ratio $J^\prime/J \approx 1$ (see, e.g.,
Ref.~\onlinecite{Naef1999}) as obtained for the strongly anisotropic
$J^\prime/J \approx 0.5$ that we have inferred, but only if the Cu$^{+2}$
$g$-factor is strongly reduced from $\gtrsim 2$ to a value which 
has been
determined, and which we have confirmed, to be
about~1.4.\cite{Sandvik1996}  We are not aware of any case of either a
Cu$^{+2}$ oxide compound, including the layered cuprate superconductors and
parent compounds,\cite{Johnston1997} or a Cu$^{+2}$ defect in any oxide
system, in which the $g$-factor of the $S = 1/2$ Cu$^{+2}$ ion, whether
determined by ESR or inferred indirectly, is less than~2 (see, e.g.,
Table~\ref{TabCuGs}, the discussion in Sec.~\ref{SecExpDatModIntro} and
Ref.~\onlinecite{Johnston1997}).  Therefore we consider the fit for
$J^\prime/J \approx 1$ and $g \approx 1.4$ to be unphysical.  As we have
shown in Fig.~\ref{Fig26}, if a physically reasonable
$g$-value is used in the fit, a very poor fit by the isolated two-leg
ladder model to the experimental $\chi(T)$ data is obtained 
\newpage
\begin{figure}
\epsfxsize=3in
\centerline{\epsfbox{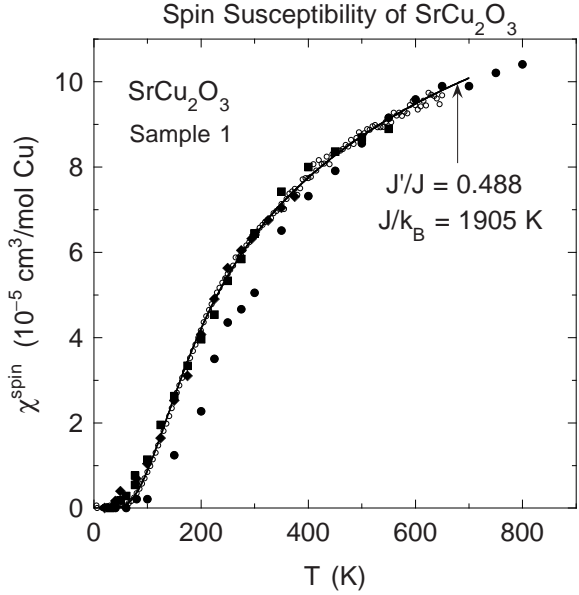}}
\vglue 0.1in
\caption{Magnetic spin susceptibility $\chi^{\rm spin}$ versus temperature
$T$ for SrCu$_2$O$_3$ sample~1 ($\circ$) derived by subtracting a constant
term and a Curie-Weiss impurity term from the
observed\protect\cite{Azuma1994} $\chi(T)$ data.  The solid curve is the
$\chi^{\rm spin}(T)$ derived from a fit to the data by a model of isolated
two-leg ladders with the intraladder exchange constants $J$ and~$J^\prime$
listed in the figure.  The filled symbols are the Knight shift $K(T)$ NMR
data of Imai {\it et al.}\protect\cite{Imai1998} for 
$^{17}$O in the rungs of the two-leg ladders in
La$_6$Ca$_8$Cu$_{24}$O$_{41}$ (circles), Sr$_{14}$Cu$_{24}$O$_{41}$
(squares) and Sr$_{11}$Ca$_3$Cu$_{24}$O$_{41}$ (diamonds), scaled to the
$\chi^{\rm spin}(T)$ data.  In the case of La$_6$Ca$_8$Cu$_{24}$O$_{41}$,
the $K(T)$ data were scaled to match the $\chi^{\rm spin}(T)$ data at the
highest temperatures; a match does not occur over any appreciable
temperature range.}
\label{Fig50}
\end{figure}
\noindent even for the
small increase of $J^\prime/J$ from~0.5 to only 0.7.

The spin susceptibility $\chi^{\rm spin}(T)$ derived for SrCu$_2$O$_3$ from
our fits to the experimental $\chi(T)$ data using Eq.~(\ref{EqChiExp:a}) is
nearly the same irrespective of the model used for $\chi^{\rm spin}(T)$,
because the $\chi_0$, $C_{\rm imp}$ and $\theta$ parameters are nearly the
same for the best fits of the various models to the data.  The $\chi^{\rm
spin}(T)$ derived from the data for sample~1, obtained using the optimum
isolated ladder fit in Fig.~\ref{Fig26}, is shown as the
open circles in Fig.~\ref{Fig50}, together with the
$\chi^{\rm spin}(T)$ fit (solid curve).  \mbox{Imai {\it et
al.}}\cite{Imai1998} have measured the paramagnetic (``Knight'') shift
$K(T)$ for $^{17}$O in the rungs of the Cu$_2$O$_3$ two-leg ladders in
Sr$_{14}$Cu$_{24}$O$_{41}$, Sr$_{11}$Ca$_3$Cu$_{24}$O$_{41}$ and
La$_6$Ca$_8$Cu$_{24}$O$_{41}$, with the field applied perpendicular to the
rung axis.  The relationship between
$K(T)$ and $\chi^{\rm spin}(T)$ is written as\cite{Imai1998}
\begin{mathletters}
\label{EqKChi:all}
\begin{equation} K(T) = K^{\rm orb} + K^{\rm spin}(T)~,
\label{EqKChi:a}
\end{equation} where
\begin{equation} K^{\rm spin}(T) = 2 F\, {\chi^{\rm spin}(T)\over N_{\rm
A}\mu_{\rm B}}~,
\label{EqKChi:b}
\end{equation}
\end{mathletters}
$K^{\rm orb}$ is the nominally anisotropic and $T$-independent orbital shift
arising from the anisotropic orbital Van Vleck susceptibility,
$K^{\rm spin}(T)$ is the contribution to $K$ from hyperfine coupling to the
spin susceptibility of a Cu spin and should be isotropic apart from the
small anisotropy in the $g$ factor, the prefactor ``2'' in
Eq.~(\ref{EqKChi:b}) comes from the two Cu neighbors of each rung O atom,
$F$ is the hyperfine coupling constant of an $^{17}$O nucleus to each Cu
spin and $\chi^{\rm spin}$ is in units of cm$^3$/mol\,Cu.  They determined
the orbital shifts to be
$K^{\rm orb} = -0.02(2)$\%, $-0.02(2)$\% and 0.04(4)\% for
Sr$_{14}$Cu$_{24}$O$_{41}$, Sr$_{11}$Ca$_3$Cu$_{24}$O$_{41}$ and
La$_6$Ca$_8$Cu$_{24}$O$_{41}$, respectively, assuming that $\chi^{\rm
spin}(T = 0) = 0$.  We scaled their derived $K^{\rm spin}(T)$ data in
Fig.~1(a) of Ref.~\onlinecite{Imai1998} to our experimental $\chi^{\rm
spin}(T)$ data in Fig.~\ref{Fig50}, as shown by the filled
symbols in Fig.~\ref{Fig50}.  If the $\chi^{\rm spin}(T)$ is
assumed to be the same as in SrCu$_2$O$_3$, Eq.~(\ref{EqKChi:b}) yields the
hyperfine coupling constants $F = 46$\,kOe and~55\,kOe for
Sr$_{14}$Cu$_{24}$O$_{41}$ and Sr$_{11}$Ca$_3$Cu$_{24}$O$_{41}$,
respectively, which are similar to the preliminary estimate of $F$ in
Ref.~14 of Imai {\it et al.}\cite{Imai1998}  The $K^{\rm spin}(T)$ data for
Sr$_{14}$Cu$_{24}$O$_{41}$ and  Sr$_{11}$Ca$_3$Cu$_{24}$O$_{41}$, in which
the doped-hole concentrations per Cu atom in the two-leg ladders are
estimated to be $p \sim 0.06$ and~0.12, respectively,\cite{Imai1998} scale
very well with the $\chi^{\rm spin}(T)$ for (undoped)  SrCu$_2$O$_3$,
indicating that the shapes of $\chi^{\rm spin}(T)$ of the two-leg ladders
in Sr$_{14}$Cu$_{24}$O$_{41}$ and Sr$_{11}$Ca$_3$Cu$_{24}$O$_{41}$ are each
about the same as in SrCu$_2$O$_3$ (of course, this does not mean that the
magnitudes of $\chi^{\rm spin}(T)$ of all three compounds are the same). 
Remarkably and inexplicably, however, Fig.~\ref{Fig50} shows
that the $K^{\rm spin}(T)$ data for La$_6$Ca$_8$Cu$_{24}$O$_{41}$ cannot be
scaled to be in agreement with the
$\chi^{\rm spin}(T)$ for SrCu$_2$O$_3$ over any appreciable temperature
range; this observation was also previously made by Naef and
Wang.\cite{Naef1999}  Of the three 14-24-41 compounds,
La$_6$Ca$_8$Cu$_{24}$O$_{41}$ is the one for which
$K(T)$ would have been expected to scale best with $\chi^{\rm spin}(T)$ of
SrCu$_2$O$_3$ because the Cu$_2$O$_3$ two-leg ladders in both of these two
compounds are presumably undoped.

Our modeling of $\chi(T)$ for the three-leg ladder compound
Sr$_2$Cu$_3$O$_5$ by QMC simulations for isolated three-leg ladders with
spatially anisotropic exchange yielded intraladder exchange constants
similar to those in SrCu$_2$O$_3$.  However, we reiterate that in this fit
we implicitly assumed that the exchange coupling along the inner leg is the
same as along the outer two legs of the ladder, which is not necessarily
the case (see below).  \mbox{Thurber {\it et al.}}\ have concluded from
$^{63}$Cu NMR measurements on Sr$_2$Cu$_3$O$_5$ that a crossover of the
instantaneous two-spin AF correlation length $\xi$ from a 1D behavior ($\xi
\sim 1/T$) to an anisotropic 2D behavior ($\xi \sim {\rm e}^{2\pi\rho_{\rm
s}/T}$) occurs upon cooling below $\sim 300$\,K.\cite{Thurber1999}  To
investigate how this effect might quantitatively affect our fitted exchange
constants, we increased the low-$T$ limit of the fit from 100\,K to 300\,K
and found no change in the fitted exchange constants, to within their
respective error bars, from those obtained for the low-$T$ fit limit of
100\,K\@.

The intraladder exchange constants in the two-leg ladder compound
LaCuO$_{2.5}$ are found to be $J/k_{\rm B} \sim 1700$\,K and $J^\prime/J
\sim 0.5$, which are similar to those we obtained for the above strontium
cuprate ladders, even though the interladder exchange paths and types are
qualitatively different from those in the latter two  compounds.  The
interladder coupling is found to be $J^{\rm 3D}/J\approx -0.05$, which is
close to our theoretical value at the QCP, $J^{\rm 3D}_{\rm QCP}/J =
-0.036(1)$ for $J^\prime/J = 0.5$, and is on the AF-ordered side of the QCP
with no spin gap, as expected since LaCuO$_{2.5}$ exhibits long-range AF
ordering at 110--125\,K.\cite{Matsumoto1996,Kadono1996}  The $\chi(T)$ data
for sample~1, but only up to $\sim 200$\,K, were previously fitted by
Troyer, Zhitomirsky and Ueda\cite{Troyer1997} assuming $g = 2$, $J^\prime/J
= 1$ and a $T^2$ dependence of $\chi^{\rm spin}$, yielding $J/k_{\rm B} =
1340$(150)\,K\@.  Thus our results show that when the full data sets for
samples~1 and~2 are fitted by accurate QMC simulations, the fitted exchange
constants are quite different from these earlier values.  Our experimental
$\chi(T)$ data are in agreement with the previous theoretical conclusion
that ``due to the dominance of quantum fluctuations in this nearly critical
system no anomaly can be observed at the N\'eel
temperature.''\cite{Troyer1997}

It is perhaps significant that the average intraladder exchange constant
$(2J + J^\prime)/(3k_{\rm B}) \sim 1500$\,K that we find in SrCu$_2$O$_3$,
Sr$_2$Cu$_3$O$_5$ and LaCuO$_{2.5}$ is about the same as in the layered
cuprate superconductor parent compounds with spatially isotropic exchange
within the CuO$_2$ planes.\cite{Johnston1997}  

In contrast to the range of intraladder exchange constant anisotropy
$J^\prime/J \approx 0.5$--0.7 in Table~\ref{TabJSumm} for the cuprate
ladder compounds, we find  that CaV$_2$O$_5$ is essentially a dimer
compound (or perhaps a spin-ladder compound in the strong
interchain-coupling limit
\mbox{$J^\prime/J \gg 1$}).  The  AF intradimer exchange constant is
$J^\prime/k_{\rm B} = 669(3)$\,K, in agreement with the $\chi(T)$ analysis
of Onoda and Nishiguchi\cite{Onoda1996} who obtained $J^\prime/k_{\rm B} =
660$\,K\@.  These values are both larger than the spin gaps 464\,K and
616\,K inferred by Iwase {\it et al.}\cite{Iwase1996} from $^{51}$V NMR
paramagnetic shift and nuclear spin-lattice relaxation rate versus
temperature measurements, respectively, suggesting the presence of
nonnegligible interdimer interactions.  Because these interdimer
interactions are so weak, we could not determine them  unambiguously.  We
conclude however that they must be included in order to obtain the best fit
to the $\chi(T)$ data: $J/J^\prime,\ J^{\prime\prime}/J^\prime$ and
$J^{\prime\prime\prime}/J^\prime$ $\sim -0.05$ to~0.1, subject to the
constraint that $(J + J^{\prime\prime} + J^{\prime\prime\prime})/J^\prime
\approx 0.1$.  Our value for the rung coupling constant $J^\prime$ is about
10\% larger than obtained from LDA+U calculations by Korotin {\it et
al},\cite{Korotin1999} and our estimated value of the sum $(J + 
J^{\prime\prime} + J^{\prime\prime\prime})/J^\prime$ is a little smaller
than the value $(J +  J^{\prime\prime})/J^\prime = 0.15$ from these
calculations (see Table~\ref{tab:calcJ}).  However, we did not include a
frustrating AF second-neighbor diagonal intraladder interaction $J^{\rm
diag}$ in our QMC simulations of $\chi(T)$ (due to the ``negative sign
problem''), which Korotin {\it et al.}\ find to be $J^{\rm diag}/J^\prime =
0.033$.  Our exchange constants for CaV$_2$O$_5$ strongly disagree with
those estimated using ``empirical laws'' (involving only the
nearest-neighbor V-V distance and  V-O-V bond angle)\cite{Harrison1980} by
Millet {\it et al.},\cite{Millet1998} who obtained
$J/J^\prime = 0.80$, $J^{\prime\prime}/J^\prime = 0.23$ and
$J^\prime/k_{\rm B} = 730$\,K, but who also include the caveat that these
estimates are not expected to be very accurate.

Our $\chi(T)$ fits for CaV$_2$O$_5$ do not rule out a spin-freezing
transition at $\sim 50$\,K as inferred from $\mu$SR and
$\chi(T)$ measurements by Luke {\it et al.}\cite{Luke1998}  In fact, close
inspection of our $\chi(T)$ data revealed a small but clear cusp at 44\,K,
qualitatively consistent with their $\chi(T)$ data. In analogy with
Sr(Cu$_{1-x}$Zn$_x$)$_2$O$_3$ discussed in the Introduction, the spin-gap
phase is evidently very delicate and the measurements of Luke {\it et al}.\
indicate that the otherwise singlet spin liquid ground state in
CaV$_2$O$_5$ can be significantly perturbed or destroyed by a relatively
small concentration of defects.

The exchange constants in MgV$_2$O$_5$ were obtained by fitting the
experimental $\chi(T)$ data for two polycrystalline samples using our
theoretical results for $\chi^{\rm spin}(T)$ of $S = 1/2$ anisotropic
two-leg Heisenberg ladders which are coupled together using the molecular
field approximation.  The average intraladder exchange constants we
obtained for the two samples, $J^\prime/J = 0.46(13)$ and $J/k_{\rm B} =
151(9)$\,K, are rather close to those in Table~\ref{tab:calcJ} predicted by
LDA+U calculations.\cite{Korotin1999}  LDA band structure and hopping
integral calculations by \mbox{Korotin {\it et al.}}\ indicate that the
strikingly different exchange interactions and spin gaps in CaV$_2$O$_5$ and
MgV$_2$O$_5$ arise from the stronger tilting of the VO$_5$ pyramids in
MgV$_2$O$_5$ compared to CaV$_2$O$_5$.\cite{Korotin2000}  There is a rather
large variability in $\chi(T)$ between various polycrystalline samples of
MgV$_2$O$_5$.  A more definitive analysis of the magnetic interactions in
this compound will hopefully become possible when
$\chi(T)$ data for single crystals become available.

The similarity between the $J^\prime/J\ (\sim 0.5$--0.7) ratios in
SrCu$_2$O$_3$, Sr$_2$Cu$_3$O$_5$ and LaCuO$_{2.5}$ demonstrates that the
spatial anisotropy in these nearest-neighbor \mbox{intraladder} exchange
constants, deduced on the basis of bilinear exchange only, is an intrinsic
property of cuprate two- and three-leg ladders, irrespective of how they
are (weakly) coupled to each other.  Since the  nearest-neighbor exchange
constants in the 2D CuO$_2$ square lattice are by symmetry necessarily
spatially isotropic, this result indicates that higher order exchange paths
than the nearest-neighbor Cu-O-Cu exchange path are present which are
important to determining the magnetic properties.  Thus, as discussed in
the Introduction, the (effective) nearest-neighbor exchange constants in
different compounds, obtained assuming bilinear exchange only, are not
simply determined by, e.g., the distance between nearest-neighbor
transition metal $R$ ions and the
$R$-O-$R$ bond angle as has often been assumed (see also the discussion of
the exchange constants in CaV$_2$O$_5$ above).  This inference is
consistent with theoretical studies of the four-spin cyclic exchange
interaction around a Cu$_4$ plaquette in cuprate spin ladders and in
high-$T_{\rm c}$ layered cuprate superconductor parent compounds, to be
discussed below.

Since the initial work in 1996 indicating that the \mbox{intraladder}
bilinear exchange constants in SrCu$_2$O$_3$ are spatially strongly
anisotropic,\cite{Johnston1996} a great deal of experimental and theoretical
research of various kinds in addition to that reported here has been done
which further quantifies the spatial anisotropy of the intraladder exchange
constants in two-leg Cu$_2$O$_3$ ladders, which we now discuss.

On the experimental side, Imai {\it et al.}\ have carried out $^{17}$O NMR
investigations of ${\rm Sr_{14}Cu_{24}O_{41}}$ and La$_6$Ca$_8{\rm
Cu_{24}O_{41}}$ single crystals and inferred from the ratio between the
spin component of the Knight shift for O in the rungs to that in the legs
of the Cu$_2$O$_3$ two-leg ladders that the anisotropy in the intraladder
exchange constants is $J^\prime/J
\sim 0.5$ in both compounds.\cite{Imai1998}  Then by comparing their
results with results which they obtained for the square lattice
antiferromagnet ${\rm Sr_2CuO_2Cl_2}$, for which $J$ is known
well,\cite{Johnston1997} they deduced
$J^\prime/k_{\rm B} = 950(300)$\,K for the exchange coupling along the
ladder rungs.\cite{Imai1998}  These exchange constants are in good
agreement with our results for SrCu$_2$O$_3$.

Eccleston {\it et al.}\cite{Eccleston1998} and Katano {\it et
al.}\cite{Katano1999} both fitted their magnetic inelastic neutron
scattering data for the two-leg ladders in ${\rm Sr_{14}Cu_{24}O_{41}}$ and
${\rm Sr_{2.5}Ca_{11.5}Cu_{24}O_{41}}$ single crystals, respectively, using
an assumed one-magnon dispersion relation $E(k_y) = \big[{E^{\rm
max}}^2\sin^2(k_ya) + \Delta^2\big]^{1/2}$, which is quite different from
our accurate one-magnon dispersion relations in Fig.~\ref{Fig22}.  These
two groups respectively obtained $\Delta/k_{\rm B} = 377(1)$\,K and
372(35)\,K, $E^{\rm max}/k_{\rm B} = 2245(28)$\,K and 1830(200)\,K, and
$E^{\rm max}/\Delta = 5.95$ and 4.9.  Using their $\Delta$ and $E^{\rm
max}/\Delta$ values, the $\Delta/J$ versus $J^\prime/J$ in
Eq.~(\ref{EqDelta0}) (Ref.~\onlinecite{Johnston1996}) and $E^{\rm
max}/\Delta$ versus  $J^\prime/J$ (cf.\ Fig.~\ref{Fig23}) obtained from the
dispersion relations of Barnes and Riera\cite{Barnes1994} for the two-leg
ladder, they estimated that $J^\prime/J = 0.55$ and~0.7(2) and $J/k_{\rm B}
= 1510$\,K and 1040(170)\,K, respectively.  However, Eccleston {\it et
al.}\ noted that ``the low intensity of the signal away from the
antiferromagnetic zone center means that ($E^{\rm max}$) is not well
defined.''   In addition, we have shown in Sec.~\ref{SecDisprsnRlns} that 
two-magnon scattering contributions to the neutron scattering function are
important near the maximum in the one-magnon dispersion relation and must
therefore be accounted for when extracting the value of $E^{\rm max}$ from
inelastic neutron scattering data.  Our fit to our $\chi(T)$ data for
SrCu$_2$O$_3$ using Eccleston {\it et al.}s' parameters and assuming $g =
2.1$ yielded an unacceptably poor fit.  A much better fit was obtained by
allowing $J$ and $J^\prime$ to vary, but still subject [in
Eq.~(\ref{EqDelta1})] to the constraint that $\Delta$ is given by the
accurate and reliable neutron scattering result $\Delta/k_{\rm B} = 377$\,K,
which gave $J^\prime/J = 0.47$ and $J/k_{\rm B} = 1880$\,K\@.  These values
are identical within the errors with our independently determined
intraladder exchange constants for SrCu$_2$O$_3$ in Table~\ref{TabJSumm}
and close to our LDA+U calculation predictions. 

\mbox{Regnault {\it et al.}}\cite{Regnault1999} analyzed their inelastic
neutron scattering data for the two-leg ladders in a large single crystal
of ${\rm Sr_{14}Cu_{24}O_{41}}$ somewhat differently.  They assumed a
dispersion relation parallel to the ladders given by $E(k_y) = \big[(\pi
J/2)^2\sin^2(k_ya) +
\Delta^2\big]^{1/2}$ and obtained $\Delta/k_{\rm B} = 370$\,K, $J^\prime/J
\sim 0.50$ and $J/k_{\rm B} \sim 1860$\,K from fits to their data.  These
exchange constants are essentially identical with our values obtained from
modeling
$\chi(T)$ for SrCu$_2$O$_3$.  They also inferred that the interladder
coupling is ``extremely weak'', consistent with the very small interladder
exchange constants in \mbox{Table~\ref{tab:calcJ}} that we infer from our
LDA+U calculations and with the very small value in Table~\ref{TabJSumm} of
the interladder exchange coupling perpendicular to the trellis layers
determined from our fits to our experimental $\chi(T)$ data.

Sugai {\it et al.}\ have carried out polarized micro-Raman scattering
experiments on single domains in polycrystalline
LaCuO$_{2.5}$.\cite{Sugai1999}  They inferred from the energies of the
broad two-magnon peaks in the spectra that $J/k_{\rm B} = 1456$\,K and that
the intraladder exchange is nearly isotropic, with $J^\prime/J = 0.946$. 
Our derived anisotropy is much stronger than this.  They also estimated
from similar measurements on ${\rm La_6Ca_8Cu_{24}O_{41}}$ and ${\rm
Sr_{14}Cu_{24}O_{41}}$ that similarly small anisotropies $J^\prime/J =
0.95$ and~1 occur in the two-leg ladders in these two compounds,
respectively, which are in strong disagreement with the anisotropies
inferred from the above NMR and neutron scattering studies on these
compounds, respectively.

On the theoretical side, large-scale quantum Monte Carlo simulations of
both the uniform and staggered susceptibilities of site-depleted two-leg
ladders coupled in a mean-field approximation were performed by
\mbox{Miyazaki {\it et al.}}\ for ladders with spatially {\it isotropic}
exchange.\cite{Miyazaki1998}  The magnetic phase
diagram\cite{Azuma1998,Azuma1997,Fujiwara1998} of
Sr(Cu$_{1-x}$Zn$_x$)$_2$O$_3$ was qualitatively reproduced, but the optimum
doping level for maximum $T_{\rm N}$ was found to be $\sim 10$--12\% and
the AF ordering persisted to above 20\% doping, contrary to the
experimental values of 4\% and
$\sim 10$\%, respectively.  Theoretical analyses of the site-diluted two-leg
ladder with spatially {\it anisotropic} exchange by \mbox{Laukamp {\it et
al.}}\ indicated that this long-range order can arise at the temperatures
and compositions found experimentally if $J^\prime/J =  0.5$ but not if
$J^\prime/J = 1$ or~5.8.\cite{Laukamp1998}

Greven and Birgeneau found from Monte Carlo simulations that the observed
$T_{\rm N}(x)$ in Sr(Cu$_{1-x}$Zn$_x$)$_2$O$_3$ is consistent with the
site-diluted ladders having \mbox{$J^\prime/J \lesssim 0.5$}, in agreement
with our results for SrCu$_2$O$_3$, and a constant correlation length
$\xi/a = 18(2)$ over the experimental doping range.\cite{Greven1998}  They
suggested that the interladder exchange coupling in the direction
perpendicular to the trellis layers is similar to that between the CuO$_2$
bilayers in YBa$_2$Cu$_3$O$_{6.2}$, i.e.\ $J^{\prime\prime\prime}/J \approx
0.05$, whereas we find a somewhat smaller value $J^{\prime\prime\prime}/J  =
0.01(1)$.  We found theoretically that the QCP for stacked ladders with
$J^\prime/J = 0.5$ occurs at an interladder coupling given by
$J_{\rm QCP}^{\prime\prime\prime}/J = 0.048(2)$, which increases to
$\approx 0.16$ for $J^\prime/J = 1$.  The fact that SrCu$_2$O$_3$ is on the
gapped spin liquid side of the QCP requires that $J^{\prime\prime\prime}$
in this compound satisfy $0 \leq J^{\prime\prime\prime}/J < J_{\rm
QCP}^{\prime\prime\prime}/J$.

Azzouz, Dumoulin and Benyoussef have carried out so-called bond-mean-field
theory calculations of the $\chi^{\rm spin}(T)$ and the NMR nuclear-spin
lattice relaxation rate 1/$T_1(T)$ for both isolated and coupled two-leg
ladders with spatially anisotropic exchange.\cite{Azzouz1997}  For
SrCu$_2$O$_3$, they obtained
$J/k_{\rm B} = 850$\,K and $J^\prime/J = 0.67$ by fitting the $^{63}$Cu
1/$T_1(T)$ measurements of Azuma {\it et al.}\cite{Azuma1994} from $\approx
100$ to~200\,K by their theoretical prediction (the theory did not fit the
data at higher temperatures).

Using a transfer-matrix density-matrix renormalization group method, Naef
and Wang recently computed the NMR $1/T_1(T)$ for $^{17}$O in the rungs and
for $^{63}$Cu in isolated two-leg ladders with both spatially isotropic
($J^\prime/J = 1,\ J/k_{\rm B} = 1300$\,K) and anisotropic ($J^\prime/J =
0.6,\ J/k_{\rm B} = 2000$\,K) exchange and compared the results with
corresponding experimental NMR data for ${\rm SrCu_2O_3}$ and ${\rm
La_6Ca_8Cu_{24}O_{41}}$.\cite{Naef1999}  The $^{63}(1/T_1)(T)$ calculation
reproduced the experimental $^{63}(1/T_1)(T)$ data quite well from 210\,K
to 720\,K assuming $J^\prime/J = 1$, whereas the temperature dependence for
$J^\prime/J = 0.6$ disagreed strongly with that of the experimental data. 
However, poor agreement between the theory and experiment for
$^{17}(1/T_1)(T)$ was obtained for both values of $J^\prime/J$, although the
calculation for $J^\prime/J = 1$ was closer to the experimental data than
that for $J^\prime/J = 0.6$.  The authors thus could not reach a firm
conclusion about the value of $J^\prime/J$, and suggested that inclusion of
additional exchange interactions into the Hamiltonian and/or a
re-evaluation of the hyperfine coupling tensor may be necessary to bring
the theory into agreement with experiment.

An electronic structure calculation in the local density approximation was
carried out by M\"uller {\it et al.}\ for SrCu$_2$O$_3$.\cite{Muller1998}
The hopping matrix elements estimated by fitting the bands were quite
anisotropic, with a rung-to-leg matrix element ratio of $\approx 0.7$, from
which $J^\prime/J < 1$ was expected.

Estimates of the intraladder exchange constants $J$ and $J^\prime$ and the
interladder exchange constant $J^{\prime\prime}$ within a trellis layer of
SrCu$_2$O$_3$ were obtained by de~Graaf {\it et al.}\ using {\it ab initio}
quantum chemical calculations for Cu$_2$O$_6$, Cu$_2$O$_7$ and
Cu$_4$O$_{10}$ cluster segments of the layer.\cite{deGraaf1999}  They
determined the values
$J/k_{\rm B} = 1810$\,K, $J^\prime/J = 0.894$ and $J^{\prime\prime}/J =
-0.080$.  The interladder exchange was thus found to be ferromagnetic as
expected from the nearly 90$^\circ$ Cu-O-Cu bond angle between Cu atoms in
adjacent ladders (see Table~\ref{TabBonds}), with a value in approximate
agreement with the initial estimate $J^{\prime\prime}/J = -0.1$ to~$-0.2$
by Rice, Gopalan and Sigrist in 1993.\cite{Rice1993}  The value of
$J^{\prime}/J$ obtained by de~Graaf {\it et al.}\ is much more isotropic
than the value we obtained from LDA+U calculations in
Table~\ref{tab:calcJ}.  de~Graaf {\it et al.}\ suggested that neglect of
the frustrating trellis-layer interladder coupling $J^{\prime\prime}$ in
the previous modeling of experimental data for SrCu$_2$O$_3$ may be the
reason that values
$J^\prime/J \sim 0.5$ were obtained rather than $J^\prime/J \approx 1$.  We
have demonstrated here, however, that a ferromagnetic interladder coupling
$J^{\prime\prime}/J = -0.1$ has no detectable influence on the value of
$J^\prime/J$ obtained from modeling $\chi(T)$ data for SrCu$_2$O$_3$.

Mizuno, Tohyama and Maekawa studied the microscopic origin of the 
superexchange in both 1D and 2D cuprates.\cite{Mizuno1998}  They found that
the hopping matrix elements between Cu 3$d$ and O 2$p$ orbitals and between
O 2$p$ orbitals depend strongly on the Madelung potential, which is a
function of the dimensionality of the spin-lattice and details of the
crystal structure.  Their calculated exchange constants show qualitatively
the same variations in magnitude and types of spatial anisotropies as
observed for the 1D chain, two-leg ladder and 2D layer cuprates discussed
in the present paper.  For quantitative comparisons, however, they pointed
out that next-nearest-neighbor exchange and the four-spin ring exchange
interaction should be included in the calculations.  In a study of the
magnetic excitation spectra of two-leg ladders, Brehmer {\it et
al.}\cite{Brehmer1998} concluded that ``a moderate amount of ring exchange
reduces the spin gap substantially and makes equal bilinear exchange on
legs and rungs consistent with experimentally observed (magnetic inelastic
neutron scattering) spectra.''  Unfortunately, we were not able to carry
out QMC simulations of $\chi^*(t)$ for Hamiltonians containing such cyclic
four-spin exchange interactions to test this idea, due to a severe
``negative sign'' problem which occurs in the simulations.

Subsequently, Mizuno, Tohyama and Maekawa exactly diagonalized the 
$d$-$p$ model Hamiltonian for a Cu$_6$O$_{17}$ cluster in a two-leg ladder
in SrCu$_2$O$_3$ using open boundary conditions, and then mapped the
eigenenergies onto those of the Heisenberg model and thereby determined the
bilinear $J$,
$J^\prime$ and $J^{\rm diag}$ exchange constants and the cyclic four-spin
exchange interaction $J^{\rm cyc}$.\cite{Mizuno1999}  The bilinear part of
the spin Hamiltonian is the same as given in Eq.~(\ref{EqHeisHam}) and the
cyclic part is\cite{Mizuno1999}
\begin{eqnarray} {\cal H}^{\rm cyc} &=& J^{\rm cyc} \sum_{\rm plaquettes}
4[(\bbox{S}_1\cdot\bbox{S}_2)(\bbox{S}_3\cdot\bbox{S}_4)\nonumber\\
&+&(\bbox{S}_1\cdot\bbox{S}_4)(\bbox{S}_2\cdot\bbox{S}_3) -
(\bbox{S}_1\cdot\bbox{S}_3)(\bbox{S}_2\cdot\bbox{S}_4)]\nonumber\\
&+&\bbox{S}_1\cdot\bbox{S}_2 + \bbox{S}_2\cdot\bbox{S}_3 + 
\bbox{S}_3\cdot\bbox{S}_4 + \bbox{S}_4\cdot\bbox{S}_1\nonumber\\
&+&\bbox{S}_1\cdot\bbox{S}_3 + \bbox{S}_2\cdot\bbox{S}_4 + {1\over 4}~,
\label{EqHcyc}
\end{eqnarray} 
where the sum is over all Cu$_4$ plaquettes on the two-leg
ladder, labeled Cu$_1$-Cu$_2$-Cu$_3$-Cu$_4$ around a \mbox{plaquette}. 
They calculated $J/k_{\rm B} = 2260(60)$\,K, $J^\prime/J = 0.77(12)$,
$J^{\rm diag}/J = 0.015(10)$ and $J^{\rm cyc}/J = 0.092(13)$ for
SrCu$_2$O$_3$.  Comparison of these exchange constants with our
experimentally-derived ones indicates that the cyclic 
exchange interaction
increases $J$ and reduces the anisotropy between $J^\prime$ and~$J$ from the
(effective) values that are obtained assuming only bilinear exchange
interactions.  The $J$ and $J^\prime/J$ values are both larger than our
values in Table~\ref{tab:calcJ} obtained from our LDA$+$U calculations. 
Their $J^{\rm diag}/J$ is antiferromagnetic rather than ferromagnetic as we
found from our LDA$+$U calculations, with a magnitude roughly an order of
magnitude smaller than our (small) value.

Mizuno, Tohyama and Maekawa also calculated $\chi^*(t)$ for the
\mbox{$2\times 8$} ladder by exact diagonalization of their spin
Hamiltonian.\cite{Mizuno1999}  They found that $\chi^*(t)$ is very
sensitive to the presence and value of $J^{\rm cyc}$ for a given set of
bilinear exchange constants.  Using a set of values of the exchange
constants \ within \ their \ range \ determined \ for \ SrCu$_2$O$_3$ 
\begin{figure}
\epsfxsize=3in
\centerline{\epsfbox{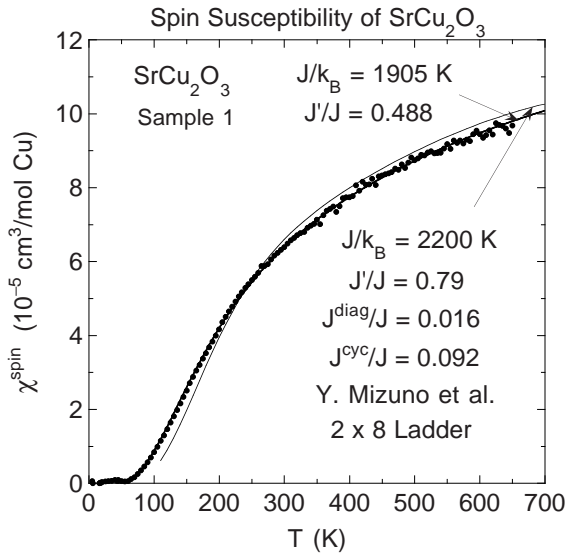}}
\vglue 0.15in
\caption{Magnetic spin susceptibility $\chi^{\rm spin}$ versus temperature
$T$ for SrCu$_2$O$_3$ sample~1 ($\bullet$) derived by subtracting a
constant term and a Curie-Weiss impurity term from the
observed\protect\cite{Azuma1994} $\chi(T)$ data.  The solid curve best
fitting the data is the theoretical $\chi^{\rm spin}(T)$ for isolated
two-leg ladders with the \mbox{intraladder} exchange constants $J$
and~$J^\prime$ listed.  The other solid curve is the first-principles
calculation of Mizuno, \mbox{Tohyama} and Maekawa for a spin-1/2
$2\times 8$ ladder, obtained using a spin Hamiltonian derived for
SrCu$_2$O$_3$ which also includes a next-nearest neighbor intraladder
diagonal interaction $J^{\rm diag}$ and a four-spin cyclic exchange
interaction $J^{\rm cyc}$ with the values listed.\protect\cite{Mizuno1999} 
Both calculations of
$\chi^{\rm spin}(T)$ assume
$g = 2.1$.}
\label{Fig51}
\end{figure}
\vglue0.08in
\noindent cited above, they
found that their calculation of $\chi^{\rm spin}(T)$ for $g = 2$ is in good
agreement with the experimental $\chi(T)$ data\cite{Azuma1994} for our
SrCu$_2$O$_3$ sample~1.\cite{Mizuno1999}  We find even better agreement
with the experimentally determined $\chi^{\rm spin}(T)$, particularly for
the higher temperatures at which the calculations of Mizuno, Tohyama and
Maekawa are expected to be most accurate, using a slightly larger $g$ value
when computing $\chi^{\rm spin}(T)$ from their $\chi^*(t)$ calculation, as
shown in Fig.~\ref{Fig51} for $g = 2.1$, where our $\chi^{\rm
spin}(T)$ fit for isolated anisotropic ladders with parameters in
Eq.~(\ref{EqSCOIsoLadPars}) is shown for comparison.  The good agreement
with the experimental data of both theoretical $\chi^{\rm spin}(T)$
calculations, derived respectively from different spin Hamiltonians which
cannot be mapped onto each other, shows that a good fit of a theoretical
$\chi^{\rm spin}(T)$ to experimentally-derived
$\chi^{\rm spin}(T)$ data can demonstrate consistency of a given spin
Hamiltonian with experimental data, but cannot prove the uniqueness of that
Hamiltonian, as discussed for (VO)$_2$P$_2$O$_7$ in the Introduction.

From the calculations of Mizuno, Tohyama and Maekawa of $\chi^{\rm
spin}(T)$ for the two-leg ladder compound SrCu$_2$O$_3$ which fit the
experimental data well as just discussed,\cite{Mizuno1999} the largest
exchange constant is along the ladder legs (chains) with a value $J/k_{\rm
B} \approx 2300$\,K, which is about 50\% larger than in the layered
cuprates and also larger than we infer for this compound from our
theoretical fits to the experimental $\chi(T)$ data, both determined
assuming the presence of only bilinear exchange interactions.  Their
proposed cyclic exchange interaction is not significant in linear chain
compounds.\cite{Mizuno1999b}  Therefore, \mbox{indirect} support for the
importance of the cyclic exchange interaction in the cuprate ladder
compounds is that a similarly large value $J/k_{\rm B}\approx 2200$\,K has
been inferred from experimental $\chi(T)$ data for the linear chain
compound Sr$_2$CuO$_3$ as discussed in the Introduction, which is also
consistent with their theoretical predictions.\cite{Mizuno1999b}

It is useful to point out here that the physics of $S = 1/2$ two-leg ladders
with a spin gap is not modified at low temperatures $k_{\rm B}T \ll
\min(J,J^\prime)$ by the presence of a four-spin cyclic exchange term.  As
long as there is a finite spin gap, the effective field theory is an $O(3)$
nonlinear sigma model, characterized completely by two parameters, the spin
gap (magnon mass gap) $\Delta$ and the spin wave velocity $c$, even if a
substantial four-spin cyclic exchange term turns out to be present.  The
low-$T$ magnetic properties are completely determined by these
two parameters.  Fits to experimental $\chi(T)$ data at low $T$ can in
principle directly determine  the value of $\Delta$, but in practice this
is often made difficult by a strong contribution from a Curie-Weiss
impurity term at low $T$\@. The spin gap and the velocity $c$ can in
principle be obtained from QMC $\chi(T)$ simulations and/or exact
diagonalization calculations for Heisenberg ladders using calculated (e.g.\
from LDA+U) and/or experimentally determined exchange constants (where a
4-spin exchange interaction could be included).  The Heisenberg ladder
model studied here can thus in any case be viewed as an effective model for
the low-temperature behavior of gapped spin ladders.  At higher
temperatures, deviations from the predictions of the sigma model are
implicitly contained in the effective exchange constants determined by
fitting various models to experimental data. This explains why many
different spin Hamiltonians can fit the same set of experimental $\chi(T)$
data for a given compound containing spin ladders with a mass gap.  Of
course, the values of the exchange constants determined from such fits
depend on the Hamiltonian assumed.

The physical consequences of an additional four-spin cyclic exchange
interaction for several other experimentally observed properties of cuprate
spin ladders as well as layered cuprates\cite{Imada1990} have also been
investigated.  For {\it two-leg ladders}, Matsuda {\it et al.}\ found
theoretically that the one-magnon dispersion relation $E(k)$ along the
ladders is not very sensitive to the presence of the cyclic exchange
interaction.\cite{Matsuda1999}  Their data for $E(k)$ of the Cu$_2$O$_3$
two-leg ladders in La$_6$Ca$_8$Cu$_{24}$O$_{41}$, as determined from
inelastic neutron scattering measurements on four aligned single crystals at
$T =  20$\,K, are consistent with the presence of this
interaction,\cite{Matsuda1999}  but the resolutions and accuracies of the
$E(k)$ data were not high enough to be able to discriminate between the
validities of the bilinear exchange Heisenberg model and one with an
additional cyclic exchange term.  On the other hand, Mizuno, Tohyama and
Maekawa calculated that even a weak cyclic exchange interaction in the
{\it layered cuprates} strongly influences the dispersion, but not the
intensity, of magnetic excitation spectra.\cite{Mizuno1999b}  Sakai and
Hasegawa have found theoretically that a plateau occurs for two-leg ladders
at $T = 0$ in the magnetization vs magnetic field at a magnetic moment 1/2
that of full saturation if $J^{\rm cyc}/J$ is larger than a critical value  
0.05(4).\cite{Sakai1998}  Roger and Delrieu,\cite{Roger1989} Honda, Kuramoto
and Watanabe\cite{Honda1993} and Eroles {\it et al.}\cite{Eroles1999} have
discussed the influence of the cyclic exchange interaction on the predicted
magnetic Raman scattering spectrum for insulating layered cuprates. 
Lorenzana, Eroles and Sorella have concluded that this interaction must be
included in the Hamiltonian in order to quantitatively describe the infrared
optical absorption spectra due to phonon-assisted multimagnon excitations
observed in the layered cuprate AF insulators La$_2$CuO$_4$,
Sr$_2$CuO$_2$Cl$_2$ and YBa$_2$Cu$_3$O$_6$.\cite{Lorenzana1999}  The
influence of four-spin exchange on the properties of the $S = 1/2$ 2D
triangular lattice antiferromagnet have also been studied.\cite{LiMing1999}

The four-spin cyclic exchange interaction may be important to include in
theory for and in the quantitative interpretation of, e.g., dilution
experiments on ladder compounds such as the system
Sr(Cu$_{1-x}$Zn$_x$)$_2$O$_3$, in which the three Cu spins around an
isolated nonmagnetic Zn impurity (acting as a spin vacancy) are members of
only one Cu$_4$ plaquette, whereas Cu spins in the bulk are members of
two.  Similarly, the spin interactions and correlations between Cu spins
along the central leg of the three-leg ladder compound Sr$_2$Cu$_3$O$_5$
may be significantly different than along the outer two legs, even apart
from effects expected from nearest-neighbor interactions and the different
numbers of Cu nearest-neighbors of Cu spins in the central and outer legs,
because each spin in the central leg is a member of four Cu$_4$ plaquettes,
whereas each spin in the outer two legs is a member of only two.  Since the
nuclear-spin lattice relaxation rate $1/T_1$ is a direct measure of the
imaginary part of the local spin susceptibility at low frequency and is not
yet fully understood in the cuprate spin ladder compounds as discussed
above, it will be interesting to compare experimental data for $1/T_1$ of
both O and Cu in the cuprate spin ladders with corresponding calculations
of this quantity in which the four-spin interaction is included.
\newpage
In conclusion, we have demonstrated that when the magnetic spin
susceptibilities of cuprate spin ladder compounds are analyzed in terms of
the bilinear Heisenberg exchange model, strong anisotropy between the
intraladder exchange constants in the legs and rungs is consistently found,
which confirms and extends the conclusion of Ref.~\onlinecite{Johnston1996}
for SrCu$_2$O$_3$.  This anisotropy strongly violates the conventional
empirical rules for  nearest-neighbor superexchange interactions in
oxides.  It is not yet clear to us whether the nearest-neighbor
exchange is really as anisotropic as we deduce, which however is
corroborated by a number of calculations (including our own LDA+U
calculations) and experimental inferences enumerated above, or whether the
values we derive are actually ``effective'' values which indirectly
incorporate the effects of additional terms in the spin Hamiltonian as
indicated by other calculations and experiments.  Analyses of the
temperature-dependent NMR Knight shift and $1/T_1$ in terms of the
nearest-neighbor Heisenberg model have respectively yielded contradictory
results regarding the anisotropy, and the analyses of Raman scattering
experiments have indicated a much smaller anisotropy than we have
inferred.  Much work remains to be done to establish a spin Hamiltonian and
calculational procedures which can self-consistently describe the
$\chi(T)$, NMR, optical and inelastic neutron scattering  measurements
probing the magnetism of the cuprate spin ladder materials.  If additional
terms are in fact present in the Hamiltonian which are important to
determining the magnetic properties, the evidence to date suggests that
higher-order superexchange processes are likely candidates.  In the case of
the magnetic spin susceptibility studied here, the presence of a
four-spin cyclic exchange interaction can exert a strong influence on the
strengths of the (effective) nearest-neighbor exchange interactions
inferred by analyzing experimental $\chi(T)$ data assuming only bilinear
exchange interactions, as we have discussed.  Such four-spin cyclic exchange
processes may also strongly contribute to various properties of the layered
cuprate high-$T_{\rm c}$ superconductors and undoped parent compounds as
indicated by recent work on their optical properties.  It would be very
interesting to determine how this interaction influences the calculated
superconducting correlations in doped spin ladder compounds and also in the
doped layered cuprate high-$T_{\rm c}$ materials.
\newpage
\acknowledgments

We are grateful to S.~Eggert, B.~Frischmuth and M.~Greven for sending us
their theoretical numerical results for $\chi^*(t)$ of the Heisenberg chain,
$\chi^*(t)$ of isolated $n$-leg ladders, and $\chi^*(t)$ and the spin gap
versus $J^\prime/J$ for isolated two-leg ladders, respectively, to Y.~Mizuno
for sending calculations of $\chi^*(t)$ for the two-leg ladder including the
influence of the four-spin cyclic exchange interaction prior to
publication,\cite{Mizuno1999} to T. Imai for sending us the experimental
$^{17}$O Knight shift data\cite{Imai1998} for three
(Sr,Ca,La)$_{14}$Cu$_{24}$O$_{41}$ compounds, and to R.~K.~Kremer,
S.~Maekawa, Y.~Mizuno and T.~M.~Rice for helpful discussions and
correspondence.  Ames Laboratory is operated for the U.S. Department of
Energy by Iowa State University under Contract No.\ W-7405-Eng-82.  The
work at Ames was supported by the Director for Energy Research, Office of
Basic Energy Sciences.  The QMC program was written in C++ using a
parallelizing Monte Carlo library developed by one of the
authors.\cite{Troyer1998}  The QMC simulations were performed on the
Hitachi SR2201 massively parallel computer of the University of Tokyo and
of the Center for Promotion of Computational Science and Engineering of the
Japan Atomic Energy Research Institute.  The LDA+U work (M.K. and V.A.) was
supported by the Russian Foundation for Basic Research (grant
RFFI-98-02-17275).
\vglue0.8in
\appendix
\section*{QMC $\bbox{\chi^*(\lowercase{t})}$ Simulation Fits}

In the following weighted fits to the QMC $\chi^*(t)$ simulation data, the
quality of a fit to a data set is expressed as the statistical $\chi^2$ per
degree of freedom, defined as $\chi^2/{\rm DOF} \equiv (N_{\rm p} -
P)^{-1}\sum_{i=1}^{N_{\rm p}} w_i(\chi^*_i - \chi^{*,{\rm fit}}_i)^2$,
where $N_{\rm p}$ is the number of data points in the data set,
$P$ is the number of fitting parameters, the weighting function $w_i =
1/\sigma_i^2$, and $\sigma_i$ is the estimated error for the $i^{\rm th}$
data point.  An additional measure of the quality of a fit is the absolute
rms deviation $\sigma_{\rm rms}$ of the fit from the data.  The fits were
carried out on Macintosh \mbox{PowerPC} G3 computers (233 and 400\,MHz)
using the software \mbox{Mathematica~3.0}.

\subsection{Isolated Ladders, Isolated Ladders with Ferromagnetic Diagonal
Coupling, Stacked Ladders and 3D-Coupled Ladders with
$\bbox{J^\prime/J \leq 1}$ and Spin Gaps}

All of these QMC data were fitted by a single multidimensional function
with ${\cal P}^{(6)}_{(6)}$ in Eqs.~(\ref{EqChiFit:all}), where
\newpage
\begin{eqnarray} N_n = N_{n0} &+& N_{1n1} \Big({J^\prime\over J}\Big) +
N_{1n2}
\Big({J^\prime\over J}\Big)^2 + N_{1n3}
\Big({J^\prime\over J}\Big)^3\nonumber\\ &+& N_{2n1} \Big({J^{\rm
diag}\over J}\Big) + N_{2n2}
\Big({J^{\rm diag}\over J}\Big)^2\nonumber\\
 &+& N_{2n3} \Big({J^\prime\over J}\Big)\Big({J^{\rm diag}\over
J}\Big)\nonumber\\ &+& N_{3n1}\Big({J_{0.5}^{\prime\prime\prime}\over
J}\Big) + N_{3n1}\Big({J_{0.5}^{\prime\prime\prime}\over
J}\Big)^2\nonumber\\  &+& N_{4n1}\Big({J_1^{\prime\prime\prime}\over
J}\Big)\nonumber + N_{4n2}\Big({J_1^{\prime\prime\prime}\over
J}\Big)^2\nonumber\\  &+& N_{5n1}\Big({J_{0.5}^{\rm 3D}\over J}\Big) +
N_{5n2}\Big({J_{0.5}^{\rm 3D}\over J}\Big)^2\nonumber\\  &+&
N_{6n1}\Big({J_{1}^{\rm 3D}\over J}\Big) + N_{6n3}\Big({J_{1}^{\rm 3D}\over
J}\Big)^3~,\ (n = 1-6)\nonumber\\ D_n = D_{n0} &+& D_{1n1}
\Big({J^\prime\over J}\Big) + D_{1n2}
\Big({J^\prime\over J}\Big)^2 + D_{1n3}
\Big({J^\prime\over J}\Big)^3\nonumber\\ &+& D_{2n1} \Big({J^{\rm
diag}\over J}\Big) + D_{2n2}
\Big({J^{\rm diag}\over J}\Big)^2\nonumber\\
 &+& D_{2n3} \Big({J^\prime\over J}\Big)\Big({J^{\rm diag}\over
J}\Big)\nonumber\\ &+& D_{3n1}\Big({J_{0.5}^{\prime\prime\prime}\over
J}\Big) + D_{3n2}\Big({J_{0.5}^{\prime\prime\prime}\over
J}\Big)^2\nonumber\\   &+& D_{4n1}\Big({J_1^{\prime\prime\prime}\over
J}\Big) + D_{4n2}\Big({J_1^{\prime\prime\prime}\over J}\Big)^2\nonumber\\
&+& D_{5n1}\Big({J_{0.5}^{\rm 3D}\over J}\Big) + D_{5n2}\Big({J_{0.5}^{\rm
3D}\over J}\Big)^2\nonumber\\ &+& D_{6n1}\Big({J_{1}^{\rm 3D}\over J}\Big) +
D_{6n3}\Big({J_{1}^{\rm 3D}\over J}\Big)^3~,\ (n = 1-6) \nonumber\\ 
\Delta^*_{\rm fit} = \Delta^*_0  &+&
\Delta_{32}\Big({J_{0.5}^{\prime\prime\prime}\over J}\Big)^2 +
\Delta_{42}\Big({J_1^{\prime\prime\prime}\over J}\Big)^2\nonumber\\  &+&
\Delta_{52}\Big({J_{0.5}^{\rm 3D}\over J}\Big)^2 +
\Delta_{62}\Big({J_{1}^{\rm 3D}\over J}\Big)^2~,\label{EqNDisoLadJ'<=J}
\label{EqIsoLadPars}
\end{eqnarray} with $J^{\rm max} = J$ and $\Delta^*_0 \equiv \Delta_0/J$
from Eq.~(\ref{EqDelta1}).  The notation, e.g.,
$J_1^{\prime\prime\prime}/J$, means that the respective fit parameter
coefficient and fit for
$J^{\prime\prime\prime}/J$ apply only for $J'/J = 1$.  In most of the fits,
the parameters $D_1,\ D_2$ and~$D_3$ were determined from
$N_1,\ N_2,\ N_3$ and $\Delta^*_{\rm fit}$ by the three HTSE constraints in
Eqs.~(\ref{EqD:all}) and were thus not fitted.  For ease of implementing
our fit functions by the reader, we have included the values of the
constrained parameters in the tables of fitted parameters.

\subsubsection{Isolated Ladders with Nearest-Neighbor Couplings}
\label{SecIsoLadJ'<=J}

We first fitted the data for the isolated ladders ($J^{\prime\prime} =
J^{\prime\prime\prime} = J^{\rm diag} = J^{\rm 3D} = 0$) in 
Fig.~\ref{Fig03} and additional data not shown (see figure
caption).  Here $J^{\rm max} = J$.  Obtaining a reliable fit was essential
because the fit is used as a foundation or backbone of the fits to the data
for the cases below incorporating additional exchange interactions.  The
spin gap $\Delta^*_{\rm fit}(J^\prime/J)$ was set identical to
$\Delta_0(J^\prime/J)$ in Eq.~(\ref{EqDelta1}) and thus was not fitted.  
For $J^\prime/J = 0$ for which $\Delta_0(J^\prime/J) = 0$, \ we \ used \ the
\ $\{N_{n0},D_{n0}\}$ \ fit \ (``Fit~1'') 
\begin{figure}
\epsfxsize=3in
\centerline{\epsfbox{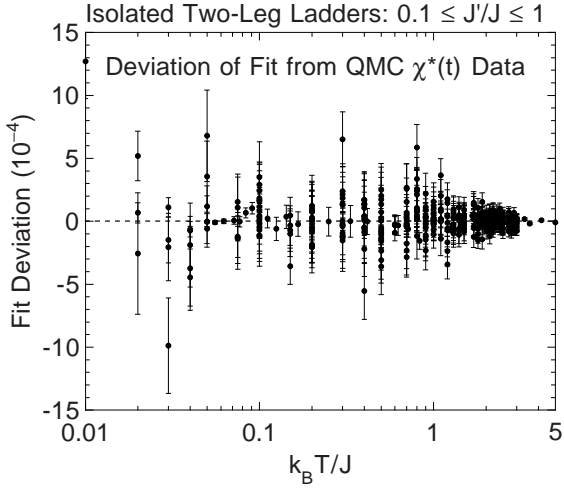}}
\vglue 0.1in
\caption{Semilog plot versus temperature $T$ of the deviation of the fit
from the 378 QMC $\chi^*(t)$ data points for $S = 1/2$ spatially anisotropic
Heisenberg ladders with $0.1 \leq J^\prime/J \leq 1$.  The error bars are
the estimated accuracies of the QMC data.  The fit function is given in
Eqs.~(\protect\ref{EqChiFit:all}) and~(\protect\ref{EqIsoLadPars}) with
parameters in Table~\protect\ref{TabIsoLadJ'/J<=1Pars}.}
\label{Fig52}
\end{figure}
\noindent  parameter set in ${\cal
P}^{(5)}_{(6)}$ in Eqs.~(\ref{EqChiFit:all}) which was obtained for the
range $0.01\leq t\leq 5$ in Ref.~\onlinecite{Johnston1999} for the uniform
chain.  In addition, we separately fitted the 30 QMC data points of
Frischmuth et al.\cite{Frischmuth1996} for $J^\prime/J = 1$ using ${\cal
P}^{(6)}_{(6)}$ in Eqs.~(\ref{EqChiFit:all}); we found that using ${\cal
P}^{(5)}_{(6)}$ resulted in a $\{N_{1n}(J^\prime/J = 1),\ D_{1n}(J^\prime/J
= 1)\}$ parameter set with several very large parameters ($\gg 1$) which
would not allow  accurate high temperature extrapolations.  We thereby
obtained a
$\{N_{1n}(J^\prime/J = 1),\ D_{1n}(J^\prime/J = 1)\}$ nine-parameter set for
$J^\prime/J = 1$, which yielded $\chi^2/{\rm DOF} = 0.079$ and $\sigma_{\rm
rms} = 4.76\times 10^{-5}$; this fit will also be used as one of the
end-function fits for the range $J^\prime/J \geq 1$ in later sections.  The
fit parameters for the general data set $0.1 < J^\prime/J \leq 0.8$ (348
data points) were constrained by these $J^\prime/J = 0$ and $J^\prime/J =
1$ fit parameters.  The complete parameter set $\{N_{1nj},\ D_{1nj}\}$ for
this range was then determined and is given in
Table~\ref{TabIsoLadJ'/J<=1Pars}, including the constrained parameters
$D_1,\ D_2$ and~$D_3$.  Note that from Eqs.~(\ref{EqD:all}), $D_2$ and
$D_3$ are sixth and ninth order in
$J^\prime/J$, respectively.  The goodnesses of fit for the 348 data points
with $0.1 \leq J^\prime/J \leq 0.8$ were $\chi^2/{\rm DOF} = 1.07$ and
$\sigma_{\rm rms} = 0.000159$, and for the total 378 point data set with
$0.1
\leq J^\prime/J \leq 1$ were $\chi^2/{\rm DOF} = 1.02$ and $\sigma_{\rm
rms} = 0.000154$.  The deviations of the fit from the data are plotted
versus temperature in Fig.~\ref{Fig52}.  The
two-dimensional fit is shown as the set of solid curves in
Fig.~\ref{Fig03}, including an example of an exchange
parameter interpolation $\chi^*(t,J^\prime/J = 0.9)$ for which we did not
obtain QMC simulation data.  The fit can also be interpolated in the range
$0 < J^\prime/J < 0.1$.  Each of the following fits in this section were
done with the parameter set $\{N_{n0},\ N_{1nj},\ D_{n0},\ D_{1nj}\}$ found
for the isolated ladders fixed, and with all other exchange parameters
except the one being fitted set to zero.

\begin{figure}
\epsfxsize=3in
\centerline{\epsfbox{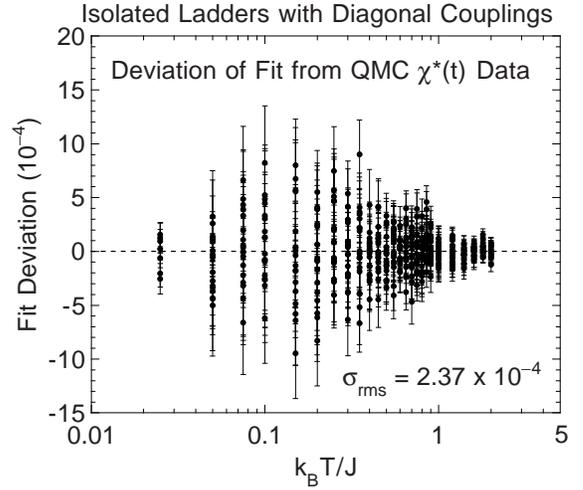}}
\vglue 0.1in
\caption{Semilog plot versus temperature $T$ of the deviation of the fit
from the 457 QMC $\chi^*(t)$ data points for $S = 1/2$ spatially anisotropic
Heisenberg ladders with $0.4 \leq J^\prime/J \leq 0.65$ and with
ferromagnetic diagonal  couplings $J^{\rm diag}/J = -0.05$, $-$0.1
and~$-$0.11.  The error bars are the estimated accuracies of the QMC data. 
The fit function is given in Eqs.~(\protect\ref{EqChiFit:all})
and~(\protect\ref{EqIsoLadPars}) with parameters in
Table~\protect\ref{TabIsoLadJ'/J<=1Pars}.}
\label{Fig53}
\end{figure}

\subsubsection{Isolated Ladders with Ferromagnetic Diagonal Second-Neighbor
Couplings}

For the fit to the 457 QMC data points for the isolated ladders with $0.4
\leq J^\prime/J \leq 0.65$ and with ferromagnetic diagonal couplings
$J^{\rm diag}/J = -0.05$, $-$0.1 and~$-$0.11, we could only use the HTSE
constraints on $D_1$ and $D_2$ in Eqs.~(\ref{EqD:a}) and~(\ref{EqD:b}),
respectively, because for $D_3$ in Eq.~(\ref{EqD:c}) the diagonal coupling
violates a condition for the validity of $d_3$ in  Eq.~(\ref{EqGenHTS3:c})
that no second-neighbor couplings occur.  The fit yielded the parameter set
$\{N_{2nj},\ D_{2nj}\}$ listed in Table~\ref{TabIsoLadJ'/J<=1Pars}.  Because
the $D_{22j}$ parameters are constrained, i.e.\ determined by other
parameters according to Eq.~(\ref{EqD:b}), the $D_{22}$ expansion has nine
terms which are designated in the table by
\begin{eqnarray} D_{22} &=& D_{221}\Big({J^{\rm diag}\over J}\Big) +
D_{222}\Big({J^{\rm diag}\over J}\Big)^2\nonumber\\ &+&
D_{223}\Big({J^\prime\over J}\Big)\Big({J^{\rm diag}\over J}\Big)\nonumber\\
&+& \bigg[D_{224}\Big({J^\prime\over J}\Big)^2 + D_{225}\Big({J^\prime\over
J}\Big)^3 + D_{226}\Big({J^\prime\over J}\Big)^4\bigg]\Big({J^{\rm
diag}\over J}\Big)\nonumber\\ &+&  \bigg[D_{227}\Big({J^\prime\over J}\Big)
+ D_{228}\Big({J^\prime\over J}\Big)^3\bigg]\Big({J^{\rm diag}\over
J}\Big)^2\nonumber\\ &+& D_{229}\Big({J^{\rm diag}\over J}\Big)^3~,
\end{eqnarray} 
\vglue0.05in
\noindent where the first three parameters are the same as defined in
Eq.~(\ref{EqIsoLadPars}).  For this fit $\chi^2/{\rm DOF} = 0.91$ and
$\sigma_{\rm rms} = 0.000237$.  The fit for $J^\prime/J = 0.4$ and~0.6 and
$J^{\rm diag}/J = -0.05$ and $-0.1$ is shown as \ the \ set \ of
\ solid \ curves \ in
\newpage
\noindent Fig.~\ref{Fig04} along with the corresponding data and fit for
$J^{\rm diag}/J = 0$.  The fit deviations for all the $J^{\rm diag}\neq 0$
data are plotted vs temperature in Fig.~\ref{Fig53}.  The
fit should not be extrapolated into the antiferromagnetic diagonal coupling
regime because poles develop in the fit function at low temperatures for
$J^{\rm diag}/J \gtrsim 0.05$.

\subsubsection{Stacked Ladders}

\begin{figure}
\epsfxsize=3in
\centerline{\epsfbox{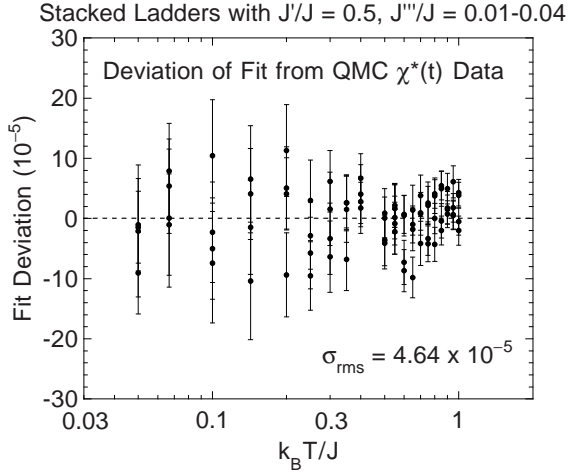}}
\vglue 0.1in
\caption{Semilog plot versus temperature $T$ of the deviation of the fit
from the 84 QMC $\chi^*(t)$ data points for $S = 1/2$ stacked spatially
anisotropic Heisenberg ladders with $J^\prime/J = 0.5$ and  AF interladder
couplings $J^{\prime\prime\prime}/J = 0.01$, 0.02, 0.03 and~0.04.  The
error bars are the estimated accuracies of the QMC data.  The fit function
is given in Eqs.~(\protect\ref{EqChiFit:all})
and~(\protect\ref{EqIsoLadPars}) with parameters in
Table~\protect\ref{TabIsoLadJ'/J<=1Pars}.}
\label{Fig54}
\end{figure}

As shown in Fig.~\ref{Fig11}, the stacked ladders with
$J^\prime/J = 0.5$ have a QCP at $J^{\prime\prime\prime}/J = 0.048(2)$. 
The two regions on either side of the QCP require separate fits.  We fitted
the 84 data points for gapped ladders with $J^{\prime\prime\prime}/J =
0.01$, 
\widetext
\begin{table}
\squeezetable
\caption{Parameters in Eq.~(\protect\ref{EqNDisoLadJ'<=J}) obtained by
fitting quantum Monte Carlo simulations of $\chi^*(t)$ for isolated
ladders, stacked ladders and LaCuO$_{2.5}$-type 3D-coupled ladders with
$J^\prime/J \leq 1$ for parameter regimes which have spin gaps.}
\label{TabIsoLadJ'/J<=1Pars}
\begin{tabular}{lcccccccc}
$n$ & $N_{n0}$ & $N_{1n1}$ & $N_{1n2}$ & $N_{1n3}$ & $D_{n0}$ & $D_{1n1}$ 
& $D_{1n2}$ & $D_{1n3}$ \\
\hline 1 & $-$0.05383784 & $-$0.67282213 & 0.03896299 & 0.01103114 &
0.44616216  & $-$0.82582213 & 0.03896299 & $-$0.08786886 \\ 2 &   
0.09740136  & 0.12334838 & $-$0.0253489 & 0.00655748 & 0.32048245  &
$-$0.40632550 & 0.20252880 & $-$0.03801372 \\
  &                &            & & & & $D_{124}$: 0.07998604  & $D_{125}$:
$-$0.00385344& $D_{126}$: 0.00379963 \\ 3 &    0.01446744  & $-$0.03965984
& $-$0.03120146 & 0.02118588 & 0.13304199  & $-$0.25099527 & 0.11749096 &
$-$0.07871375 \\
  &                &            & & & & $D_{134}$: 0.04106834  & $D_{135}$:
$-$0.01886681 & $D_{136}$: 0.00157755 \\
  &                &            & & & & $D_{137}$: $-$0.00387185  &
$D_{138}$: 0.00019055 & $D_{139}$: $-$0.00010728 \\ 4 & 0.001392519 &
0.006657608 & $-$0.020207553 & 0.008830122 & 0.03718413  & $-$0.10249898 &
0.04316152 & 0.01936105 \\  5 & 0.0001139343 & 0.0001341951 & 0.0016684229
& $-$0.0001396407 & 0.002813608  & 0.000402749 & 0.001958564 &
$-$0.003803837 \\  6 & 0 & 0.0000422531 & $-$0.0001609830 & 0.0001335788 &
0.0002646763  & $-$0.0010424633 & 0.0015813041 & $-$0.0002914845\\
\hline && $N_{2n1}$ & $N_{2n2}$ & $N_{2n3}$ & & $D_{2n1}$ & $D_{2n2}$ &
$D_{2n3}$ \\
\hline 1 && $-$2.812719 & $-$0.227040 & 1.282815 && $-$2.312719 &
$-$0.227040 & 1.282815 \\ 2 && 0.5793014 & $-$0.5930882 & $-$0.1876388 &&
$-$0.853977 & $-$1.862968 & 0.346204 \\
  && & & &&$D_{224}$: $-$0.176789 & $D_{225}$: 0.234243 & $D_{226}$:
$-$0.126870\\
  && & & &&$D_{227}$: 0.676145 & $D_{228}$: 0.022454 & $D_{229}$:
$-$0.113520\\  3 && $-$0.1384016 & 0.5058027 & $-$0.1388999 && $-$0.4919953
& 0.5264363 &0.1538512\\ 4 && $-$0.00509828 & $-$0.05463381 & 0.02008685 &&
$-$0.1013551 & 0.4839774 & 0.2117160 \\  5 && 0.00174376 & $-$0.00553088 &
$-$0.01572268 && $-$0.0154461 & $-$0.1073785 &
$-$0.0566799 \\ 6 && 0.0000469804 & 0.0005268598 & 0.0001876060 &&
$-$0.0016760123 & 0.0064578167 & 0.0039956271 \\
\hline && $N_{3n1}$ & $N_{3n2}$ & $\Delta_{32}$& & $D_{3n1}$ & $D_{3n2}$ &
$D_{3n3}$ \\
\hline 1 && $-$1.666276 & $-$7.682011 & $-$78.10006 & & $-$1.666276 &
70.41805 & \\ 2 &&    2.521203 & $-$17.49621 & & & 1.539638 & $-$18.73787 &
$-$94.92723 \\
  && & & && $D_{324}$: 2449.844 \\ 3 && $-$0.9419028 & 3.589057 & & &
$-$0.06061213 & 9.508413 & 111.9860 \\
  && & & &&$D_{334}$: $-$1562.968 & $D_{335}$: $-$3856.902 & $D_{336}$:
55968.11\\  4 && 0.1787010 & $-$1.092329 & & & 0.2163155 & $-$1.686861 & \\ 
5 && $-$0.02955773 & 0.2617274 & & & $-$0.1110507 & 1.013217 & \\ 6 &&
0.001132174 & $-$0.02132289 & & & 0.004337158 & 0.1318350 & \\
\hline
  && $N_{4n1}$ & $N_{4n2}$ & $\Delta_{42}$ && $D_{4n1}$ & $D_{4n2}$ &
$D_{4n3}$ \\
\hline 1 && 6.148834 & -52.149512 & $-$21.01218 && 6.648834 & $-$31.13733 \\
2 && 0.5547782 & 1.3178812 &&& 1.491021 & $-$17.30110 & 113.6317 \\
  &&           &           &&& $D_{424}$: $-$875.0191 \\ 3 && $-$1.307790 &
7.686068 &&& $-$0.07069018 & 3.291341 & 46.69600 \\
  &&           &          &&& $D_{434}$: $-$281.9628 & $D_{435}$: 919.8816 &
$D_{436}$: $-$9966.124 \\ 4 && 0.5749293 & $-$3.292545 &&& 0.05701761 &
$-$0.8714753 \\ 5 && $-$0.06766142 & 0.3635859 &&& 0.05170433 &
$-$0.2843813 \\ 6 && 0.003379759 & -0.02309552 &&& $-$0.005558845 &
0.1690838 \\
\hline
  && $N_{5n1}$ & $N_{5n2}$ & $\Delta_{52}$ && $D_{5n1}$ & $D_{5n2}$ &
$D_{5n3}$ \\
\hline 1 && $-$17.77588 & $-$183.3631 & $-$145 && $-$17.27587 & $-$38.36313
\\ 2 && $-$0.6504693 & $-$88.97643 &&& $-$8.255295 & $-$168.3606 &
$-$2596.684 \\
  &&              &             &&& $D_{524}$: $-$16075.16 \\ 3 &&
$-$0.4344193 & 13.81917 &&& $-$3.603076 & $-$28.21008 & $-$1226.259 \\
  &&              &          &&& $D_{534}$: $-$24794.62 & $D_{535}$:
$-$194906.5 &
$D_{536}$: $-$1419501 \\ 4 && $-$0.2963010 & $-$9.737364 &&& $-$1.314690 &
$-$8.006486 \\ 5 && 0.0469558 & 0.8909398 &&& 0.2074800 & 5.052001 \\ 6 &&
$-$0.0003390179 & $-$0.007869661 &&& 0.0001023351 & 0.003324225 \\
\hline
  && $N_{6n1}$ & $N_{6n3}$ & $\Delta_{62}$ && $D_{6n1}$ & $D_{6n2}$ &
$D_{6n3}$\\
\hline 1 && 3.113662 & 14.39181 & $-$27.31860 && 3.613662 & 27.31860 &
14.39181 \\ 2 && $-$1.013862 & 7.979673 &&  & $-$0.8306449 & $-$9.900986 &
110.2705 \\
  &&          &           &&&  $D_{624}$: 380.3488 & $D_{625}$: 393.1642 \\
3 && 0.2569465 & $-$9.659867 &&& 0.7269061 & 2.749900 & $-$27.71825 \\
  &&             &          &&& $D_{634}$: $-$110.1823 & $D_{635}$:
1667.584 &
$D_{636}$: 3594.586 \\ 
  &&             &          &&& $D_{637}$: 5370.346 \\ 4 && 0.1382070 &
$-$1.436330 &&& $-$0.09150437 & & 6.226186 \\ 5 && $-$0.01967034 &
$-$0.3392272 &&& 0.09652782 & & $-$2.830070 \\ 6 && $-$0.0005646914 &
0.06430789 &&& $-$0.01310045 & & 1.654887 \\
\end{tabular}
\end{table}
\narrowtext
\newpage
\noindent 0.02, 0.03 and~0.04 and obtained the $\{N_{3nj}$, $D_{3nj}$,
$\Delta_{32}\}$ parameter set listed in Table~\ref{TabIsoLadJ'/J<=1Pars}.
The value of the fitted parameter $\Delta_{32}$ predicts that the spin gap
should decrease to zero at $J^{\prime\prime\prime}_{\rm QCP}/J = 0.052$,
close to the value inferred from Fig.~\ref{Fig11}.  The deviations of the
fit from the data are plotted vs $t$ in Fig.~\ref{Fig54}.  The goodnesses
of fit are
$\chi^2$/DOF = 1.39 and $\sigma_{\rm rms} = 0.0000464$.  The relative rms
deviation is 0.125\%.  In Table~\ref{TabIsoLadJ'/J<=1Pars}, the constrained
parameters $D_{3n}$ with $n = 1$, 2 and~3 are written as power series,
$D_{3n} = \sum_j D_{3nj}(J^{\prime\prime\prime}/J)^j$.  The fit is shown as
the set of solid curves through the data for $J^{\prime\prime\prime}/J =
0.01$--0.04 in Fig.~\ref{Fig09}.  

We fitted the 70 stacked ladder data points for $J^\prime/J = 1$ and
$J^{\prime\prime\prime}/J = 0.05$, 0.1 and 0.15, parameters which are
\newpage
\begin{figure}
\epsfxsize=3in
\centerline{\epsfbox{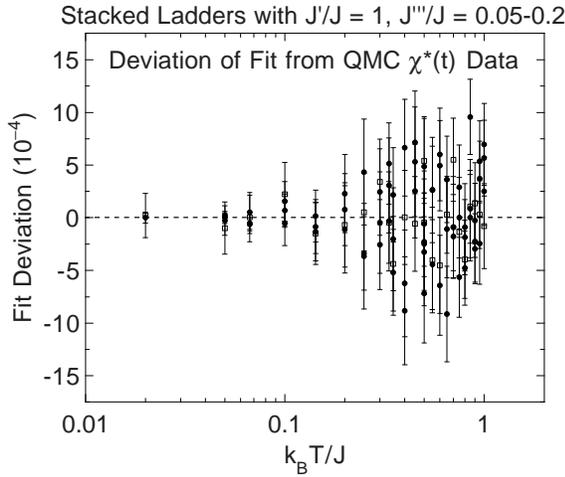}}
\vglue 0.1in
\caption{Semilog plot versus temperature $T$ of the deviation of the fit
from the 70 QMC $\chi^*(t)$ data points for $S = 1/2$ stacked spatially
isotropic Heisenberg ladders with $J^\prime/J = 1$ and  AF interladder
couplings $J^{\prime\prime\prime}/J = 0.05$, 0.1, and~0.15 for which there
is a spin gap ($\bullet$); here $\sigma_{\rm rms} = 0.000384$.  The open
squares are the deviations of a separate fit to the 24 data points for
$J^{\prime\prime\prime}/J = 0.2$ from those data ($\sigma_{\rm rms} =
0.000256$), for which there is no spin gap.  The error bars are the
estimated accuracies of the QMC data.  The fit function for the gapped fit
is given in Eqs.~(\protect\ref{EqChiFit:all})
and~(\protect\ref{EqIsoLadPars}) with parameters in
Table~\protect\ref{TabStackLadJ'/J=0.5Pars}.}
\label{Fig55}
\end{figure}
\noindent 
all in the gapped region with $J^{\prime\prime\prime}/J <
J^{\prime\prime\prime}_{\rm QCP}/J
\approx 0.16$, and obtained $\chi^2/{\rm DOF} = 1.19$ and $\sigma_{\rm rms}
= 0.000384$ for the $\{N_{4nj},\ D_{4nj},\ \Delta_{42}\}$ parameter set
listed in Table~\ref{TabIsoLadJ'/J<=1Pars}.  The two-dimensional fit is
shown as the set of solid curves for these exchange constant combinations in
Fig.~\ref{Fig12}.  The 
\begin{figure}
\epsfxsize=3in
\centerline{\epsfbox{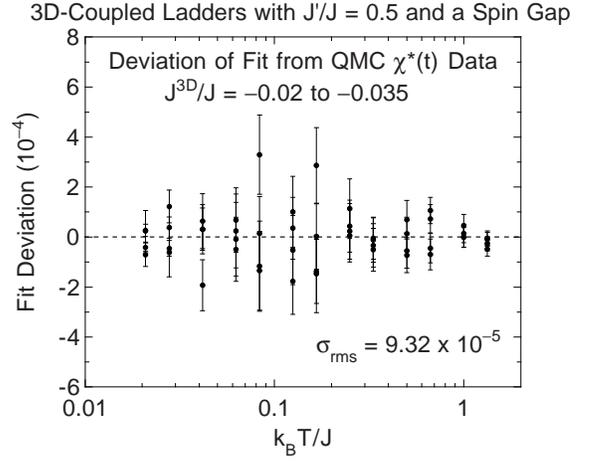}}
\vglue 0.1in
\caption{Semilog plot versus temperature $T$ of the deviation of the fit
from the 52 QMC $\chi^*(t)$ data points for $S = 1/2$ 3D-coupled spatially
anisotropic Heisenberg ladders with $J^\prime/J = 0.5$ and  ferromagnetic
interladder couplings $J^{\rm 3D}/J = -0.02,\ -0.025,\ -0.03$, and~$-0.035$
for which there is a spin gap ($\bullet$).  The error bars are the
estimated accuracies of the QMC data.  The fit function for the gapped fit
is given in Eqs.~(\protect\ref{EqChiFit:all})
and~(\protect\ref{EqIsoLadPars}) with parameters in
Table~\protect\ref{TabIsoLadJ'/J<=1Pars}.}
\label{Fig56}
\end{figure}
\noindent deviations of the fit from the data
are shown as the filled circles in Fig.~\ref{Fig55}. 
The fitted value of
$\Delta_{42}$ predicts that
$J^{\prime\prime\prime}_{\rm QCP}/J = 0.155$, at which the spin gap
vanishes.

\subsubsection{LaCuO$_{2.5}$-type 3D-Coupled Ladders}

\begin{figure}
\epsfxsize=3in
\centerline{\epsfbox{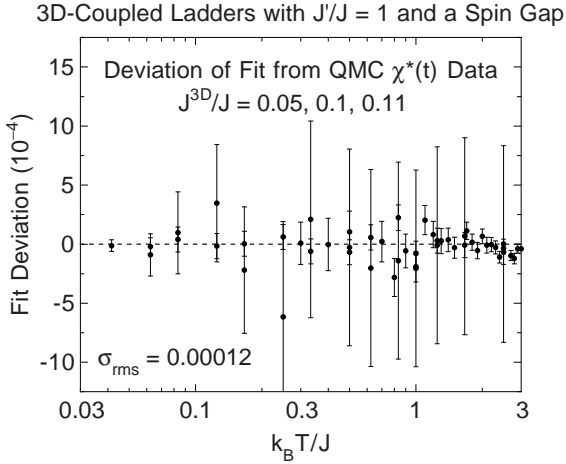}}
\vglue 0.1in
\caption{Semilog plot versus temperature $T$ of the deviation of the fit
from the 84 QMC $\chi^*(t)$ data points for $S = 1/2$ 3D-coupled spatially
anisotropic Heisenberg ladders with $J^\prime/J = 1$ and  AF interladder
couplings $J^{\rm 3D}/J = 0.05$, 0.1  and~0.11 for which there is a spin
gap ($\bullet$).  The error bars are the estimated accuracies of the QMC
data.  The fit function for the gapped fit is given in
Eqs.~(\protect\ref{EqChiFit:all}) and~(\protect\ref{EqIsoLadPars}) with
parameters in Table~\protect\ref{TabIsoLadJ'/J<=1Pars}.}
\label{Fig57}
\end{figure}

For the LaCuO$_{2.5}$-type 3D-coupled ladder data for $J^\prime/J = 0.5$
and $J^{\rm 3D}/J = -0.035,\ -0.03,\ -0.025$ and $-0.02$ (52 data points),
which exhibit a spin-gap, we fixed the parameter $\Delta_{52}=-145$, which
yields a QCP at $J^{\rm 3D}_{\rm QCP}/J = -0.0384$. We obtained a
$\chi^2/{\rm DOF} = 1.44$ and $\sigma_{\rm rms} = 0.0000932$ for the
$\{N_{5nj},\ D_{5nj},\ \Delta_{52}\}$ parameter set listed in
Table~\ref{TabIsoLadJ'/J<=1Pars}.  The fit is shown as the set of solid
curves in Fig.~\ref{Fig14} for these parameter
combinations, and the deviations of the fit from the data are shown in
Fig.~\ref{Fig56}.  Similarly, for the 84 data points for
$J^\prime/J = 1$ and $J^{\rm 3D}/J = 0.05,\ 0.1,\ {\rm and}\
0.11$,\cite{Troyer1997} which also exhibit a spin-gap, we obtained a
$\chi^2/{\rm DOF} = 0.92$ and $\sigma_{\rm rms} = 0.00012$ for the
$\{N_{6nj},\ D_{6nj},\ \Delta_{62}\}$ parameter set listed in
Table~\ref{TabIsoLadJ'/J<=1Pars}.  The fit is shown as the set of solid
curves for the respective exchange constant combinations in
Fig.~\ref{Fig16}, and the fit deviations are plotted vs
temperature in Fig.~\ref{Fig57}.

\subsection{Trellis Layers with $\bbox{J^\prime/J \leq 1}$}
\vglue0.17in
The 136 trellis layer QMC data points with $J^{\prime\prime} = -0.2,\
-0.1$, 0.1 and~0.2 for $J^\prime/J = 0.5$ and~1 in
Fig.~\ref{Fig06} were computed using QMC simulations by
Miyahara~{\it et al.}\cite{Miyahara1998}  \,They \,also \,computed \,data
\,for \,$J^{\prime\prime} = -0.5$ \,and~\,0.5
\noindent  which are not shown in the figure.  Due
to the ``negative sign problem'' arising from the geometric frustration in
the interactions between spins on adjacent ladders, accurate
$\chi^*(t)$ data could not be obtained to low temperatures.  In addition,
for the trellis layer the spin gap is expected to be nearly independent of
the interladder coupling.  For both of these reasons, we fitted the data by
expressions having the form of the modified MFT expression in
Eqs.~(\ref{EqMFTGenf:all}).  For the case of trellis layers with
$J^\prime/J \leq 1$, one has $J^{\rm max} = J$ and we write the function
$f(J_{ij},t)$ in Eq.~(\ref{EqMFTGenf:b}) in the general form
\begin{eqnarray} f\Big({J^\prime\over J}&,&{J^{\prime\prime}\over J},t\Big)
= 2 c_{11}
\Big({J^{\prime\prime}\over J}\Big) + c_{12} \Big({J^{\prime\prime}\over
J}\Big)^2 + c_{13} \Big({J^{\prime\prime}\over J}\Big)^3\nonumber\\ 
&+&{c_{21} \Big({J^{\prime\prime}\over J}\Big) + c_{22} \Big({J^\prime\over
J}\Big) \Big({J^{\prime\prime}\over J}\Big) + c_{23}
\Big({J^{\prime\prime}\over J}\Big)^2\over t} \nonumber\\ &+& {-{9\over 8}
c_{33} \Big({J^{\prime\prime}\over J}\Big)^2 + c_{34}
\Big({J^\prime\over J}\Big) \Big({J^{\prime\prime}\over J}\Big)^2 + c_{35}
\Big({J^{\prime\prime}\over J}\Big)^3\over 3 t^2} \nonumber\\ &+& { c_{43}
\Big({J^{\prime\prime}\over J}\Big)^2 + c_{44} \Big({J^\prime\over J}\Big)
\Big({J^{\prime\prime}\over J}\Big)^2 + c_{45}
\Big({J^{\prime\prime}\over J}\Big)^3\over t^3}~,
\label{EqTrelLayf}
\end{eqnarray} in which the $c_{ij}$ parameters may be varied and used as
fitting parameters.  For the exact HTSE to ${\cal O}(1/t^2)$ on the
right-hand-side of Eq.~(\ref{EqTrelLayf}), one would have
$c_{11}=c_{23}=c_{33}=c_{35}=1$ with the remaining $c_{ij}$ parameters
being zero.  In their fits for $-0.5\leq J^{\prime\prime}/J \leq 0.5$ and
$J^\prime/J = 0.5$ and~1 in the temperature range \mbox{$1\leq t \leq 1.5$},
\mbox{Miyahara {\it et al.}}\cite{Miyahara1998} set
$c_{11} = 1$ and then obtained $c_{12} = 0.3436$.  Since the correct Weiss
temperature in the Curie-Weiss law requires $c_{11}=1$ {\it and} $c_{12} =
0$, this parametrization does not give the correct Curie-Weiss behavior 
but does yield the correct Curie law in the limit of high temperatures.

In our fits to the QMC data in Fig.~\ref{Fig06}, various
combinations of nonzero $c_{ij}$ parameters in Eq.~(\ref{EqTrelLayf}) were
tried.  The lowest-order, zero-parameter MFT fit with
$c_{11}=1$ and with the remaining $c_{ij}$ parameters being zero yielded
the fit shown as the set of solid curves in
Fig.~\ref{Fig06}.  For this  ``fit'' to all the data points,
the goodnesses of fit were $\chi^2$/DOF = 98.7 and $\sigma_{\rm rms} =
0.0020$.  Excluding the two data points with
$J^\prime/J = 0.5, J^{\prime\prime}/J = \pm 0.1$ at $t = 0.1$ gave a much
lower
$\sigma_{\rm rms} = 0.00080$.  This is the fit function we use to fit our
experimental $\chi(T)$ data for SrCu$_2$O$_3$.  We also tried various
combinations of up to four $c_{ij}$ fitting parameters in fitting the data
for
$t\geq 0.25$, with or without one or  more of the constraints above
associated with the HTSE, but the $\sigma_{\rm rms}$ of the fit could not
be significantly improved compared to that of the above MFT ``fit'' with
zero fitting parameters.  In the attempts with one or more nonzero
$c_{4j}$, the fits diverged significantly from the trend of the data at low
temperatures because of the lack of enough data points at low temperatures
to constrain these parameters.

\subsection{Stacked Ladders with No Spin Gap}
\begin{figure}
\epsfxsize=3in
\centerline{\epsfbox{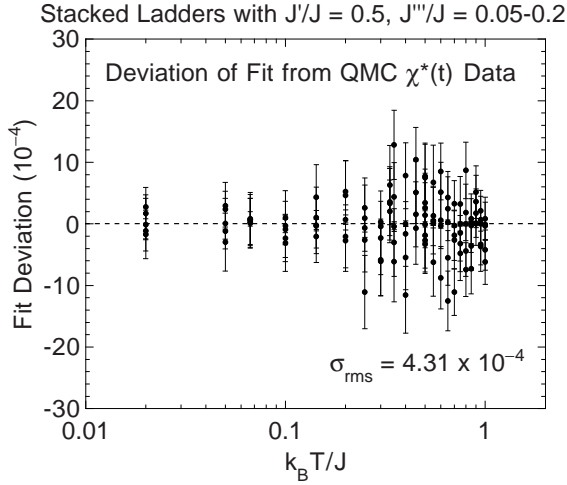}}
\vglue 0.1in
\caption{Semilog plot versus temperature $T$ of the deviation of the fit
from the 118 QMC $\chi^*(t)$ data points for $S = 1/2$ stacked spatially
anisotropic Heisenberg ladders with $J^\prime/J = 0.5$ and  AF interladder
couplings $J^{\prime\prime\prime}/J = 0.05$, 0.1, 0.15 and~0.2 for which
there is no spin gap.  The error bars are the estimated accuracies of the
QMC data.  The fit function is given in Eqs.~(\protect\ref{EqChiFit:all})
and~(\protect\ref{EqNDStakLadJ'/J=0.5}) with parameters in
Table~\protect\ref{TabStackLadJ'/J=0.5Pars}.}
\label{Fig58}
\end{figure}

\vglue0.07in
Our 118 QMC data points for the stacked ladders with $J^\prime/J = 0.5$ and
$J^{\prime\prime\prime}/J = 0.05$, 0.1, 0.15 and~0.2 are on the side of the
QCP at
$J^{\prime\prime\prime}_{\rm QCP}/J = 0.048(2)$ with no spin gap and were
fitted by the  expression with $J^{\rm max} = J$, ${\cal P}^{(5)}_{(6)}$ and
$\Delta_{\rm fit}^* = 0$ in Eqs.~(\ref{EqChiFit:all}), with 
\begin{eqnarray} N_n = N_{n0} &+& N_{n1}\Big({J^{\prime\prime\prime}\over
J}\Big) + N_{n2}
\Big({J^{\prime\prime\prime}\over J}\Big)^2~,\ (n = 1-5)\nonumber\\ D_n =
D_{n0} &+& D_{n1} \Big({J^{\prime\prime\prime}\over J}\Big) + D_{n2}
\Big({J^{\prime\prime\prime}\over J}\Big)^2.\ (n =
1-6)\label{EqNDStakLadJ'/J=0.5}
\end{eqnarray} The $D_1,\ D_2$ and $D_3$ series coefficients were not
fitted but were determined from the $N_1,\ N_2$ and $N_3$ series according
to the three HTSE constraints in Eqs.~(\ref{EqD:all}).  The low-$T$
expansion is of the correct form for a 2D quantum critical point, $\chi^*
\propto t$;\cite{Chubukov1994,Chubukov1993,Sandvik1994,Elstner1998} on the
side of the QCP with no spin gap, we find that $\chi^*$ continues to be
linear in
$t$ at low $t$, as is also predicted.  The parameters of the fit are given
in Table~\ref{TabStackLadJ'/J=0.5Pars}, for which $\chi^2/{\rm DOF} = 1.09$
and $\sigma_{\rm rms} = 0.000431$ were obtained.  The fit is shown as the
set of four solid curves for $J^{\prime\prime\prime}/J = 0.05$, 0.1, 0.15
and~0.2 in Fig.~\ref{Fig09}.  Note that the fits for all
four $J^{\prime\prime\prime}/J$ values cross at the temperature $t \approx
0.16$; i.e., at this temperature the $\chi^*$ is nearly independent of
$J^{\prime\prime\prime}/J$ in the gapless regime.  The deviations of the
fit from the data are plotted vs $t$ in
Fig.~\ref{Fig58}.

Our 24 QMC data points for the stacked ladders with $J^\prime/J = 1$ and
$J^{\prime\prime\prime}/J = 0.2$ are also on the side of the QCP at
$J^{\prime\prime\prime}_{\rm QCP}/J \approx 0.16$ with no spin gap and were
therefore also fitted by the eight-parameter expression in
Eq.~(\ref{EqNDStakLadJ'/J=0.5}) with \ \mbox{$J^{\rm max} = J$}, \ ${\cal
P}^{(5)}_{(6)}$ \ and \ $\Delta_{\rm fit}^* = 0$ \,in
\,Eqs.~(\ref{EqChiFit:all}), \,with 
\noindent parameters shown in Table~\ref{TabStackLadJ'/J=1NoGapPars}.  The
three
$D_1,\ D_2$ and $D_3$ parameters were not fitted but were determined from
the
$N_1,\ N_2$ and $N_3$ fitting parameters as constrained by the HTSE\@.  The
fit is shown as the solid curve in Fig.~\ref{Fig12} and the
deviations of the fit from the data as the open squares in
Fig.~\ref{Fig55}.  The goodnesses of fit are
$\chi^2/$DOF = 0.60 and $\sigma_{\rm rms} = 0.000256$.

\subsection{LaCuO$_{2.5}$-Type 3D Interladder Couplings \protect\\ with
$\bbox{J^\prime/J = 0.5}$ 0.6, 0.7, 0.8, 0.9 \protect\\ and 1.0 and with No
Spin Gap}
\label{SecLaCuO2.5NoGap}

The LaCuO$_{2.5}$-Type 3D interladder couplings $-0.1 \leq J^{\rm 3D}/J \leq
-0.035$ and presumably $0.05
\leq J^{\rm 3D}/J \leq 0.2$ with $J^\prime/J = 0.5$, and $0.12 \leq
J^{\prime\prime}/J \leq 0.2$ with $J^\prime/J = 1$,\cite{Troyer1997} are on
the side of the QCP with no spin gap and a finite $\chi^*(t = 0)$.  We estimate that our QMC data with $J^{\rm 3D}/J = 0.05$
and $J^\prime/J = 0.6$ and with $J^{\rm 3D}/J = 0.1$, 0.15 and~0.2 and
$J^\prime/J = 0.6$, 0.7, 0.8 and~0.9 are all in the gapless regime.   
These data total 564 data points.  
\widetext
\begin{table}
\caption{Parameters in Eq.~(\protect\ref{EqNDStakLadJ'/J=0.5}) obtained by
fitting quantum Monte Carlo simulations of $\chi^*(t)$ for stacked ladders
with
$J^\prime/J = 0.5$ which have no spin gap.}
\label{TabStackLadJ'/J=0.5Pars}
\begin{tabular}{lcccccc}
$n$ & $N_{n0}$ & $N_{n1}$ & $N_{n2}$ & $D_{n0}$ & $D_{n1}$ & $D_{n2}$ \\
\hline 1 & $-$0.07395387 & $-$1.1046032 & $-$7.7000424 & 0.5510461 &
$-$0.6046032 &
$-$7.7000424 \\  2 & 0.17151127 & $-$1.16738784 & 5.24878372 & 0.4065401 &
$-$1.8947418 & 0.1339557 \\
  &            &               &            &        & $D_{23}$:
$-$3.8500212 \\ 3 & $-$0.00486493 & 0.02453618 & $-$0.53626512 & 0.1700718
& $-$0.9299952 &
$-$0.0235946 \\
  &               &            &            &        & $D_{33}$: 2.4315744 &
$D_{34}$: $-$1.9250106 \\ 4 & 0.001860855 & $-$0.019272190 & 0.082869123 &
0.05484099 &
$-$0.53559069 & 0.82169392 \\  5 & $-$0.0003207081 & 0.0074168695 &
$-$0.0175909734 & 0.001093210 & $-$0.051478243 & 0.223840382 \\ 6 &&&&
$-$0.000527212 & 0.025391112 & $-$0.063286649
\end{tabular}
\end{table}
\narrowtext

One expects that $\chi^*
\propto t^2$ at the 3D quantum critical point, and for further increases in
$|J^{\rm 3D}/J|$ one expects $\chi^* = \chi^*(0) + b t^2$ with a finite
$\chi^*(0)$.\cite{Normand1997,Troyer1997} Thus we fitted all of the above
QMC data using $\Delta^*_{\rm fit} = 0$ and using ${\cal P}^{(5)}_{(6)}$ in
Eqs.~(\ref{EqChiFit:all}), with
\begin{eqnarray} N_n = N_{n0} &+& N_{n1}\Big({J^\prime\over J}\Big) +
N_{n2}\Big({J^\prime\over J}\Big)^2\nonumber\\ &+& N_{n3}
\Big({J^\prime\over J}\Big)\Big({J^{\rm 3D}\over J}\Big) + N_{n4}
\Big({J^{\rm 3D}\over J}\Big)^2\nonumber\\ &+& N_{n5}\Big({J^\prime\over
J}\Big)\Big({J^{\rm 3D}\over J}\Big)^2~,~~(n = 1-3,\ 5)\nonumber\\
\label{EqNDLCO0.5}\\ D_n = D_{n0} &+& D_{n1}\Big({J^\prime\over J}\Big) +
D_{n2} \Big({J^\prime\over J}\Big)^2\nonumber\\ &+& D_{n3}
\Big({J^\prime\over J}\Big)\Big({J^{\rm 3D}\over J}\Big) + D_{n4}
\Big({J^{\rm 3D}\over J}\Big)^2\nonumber\\ &+& D_{n5}\Big({J^\prime\over
J}\Big)\Big({J^{\rm 3D}\over J}\Big)^2~,~~(n = 1-4,\ 6)\nonumber
\end{eqnarray} The three HTSE constraints in Eqs.~(\ref{EqD:all}) were
enforced, so the series for $D_n$ with $n = 1$, 2 and 3 are determined from
$N_1$, $N_2$ and/or
$N_3$ series, and Eqs.~(\ref{EqD:all}) with $\Delta^* = 0$ yield the
$D_n$ series terms in Eq.~(\ref{EqNDLCO0.5}) plus the respective additional
terms
\vglue1.1in
\begin{table}
\caption{Parameters in Eq.~(\protect\ref{EqNDStakLadJ'/J=0.5}) obtained by
fitting quantum Monte Carlo simulations of $\chi^*(t)$ for stacked ladders
with
$J^\prime/J = 1$ which have no spin gap.}
\label{TabStackLadJ'/J=1NoGapPars}
\begin{tabular}{ldd}
$n$ & $N_{n0}$ & $D_{n0}$ \\
\hline 1 & $-$0.6740395 & 0.1759605 \\ 2 & 0.3915161 & 0.2035826 \\ 3 &
0.007222304 & 0.2061725 \\ 4 & 0.01429191 & 0.1497920 \\ 5 & 0.005028154 &
$-$0.01677125 \\ 6 & 0 & 0.05307760 \\
\end{tabular}
\end{table}
\newpage
\begin{figure}
\epsfxsize=3in
\centerline{\epsfbox{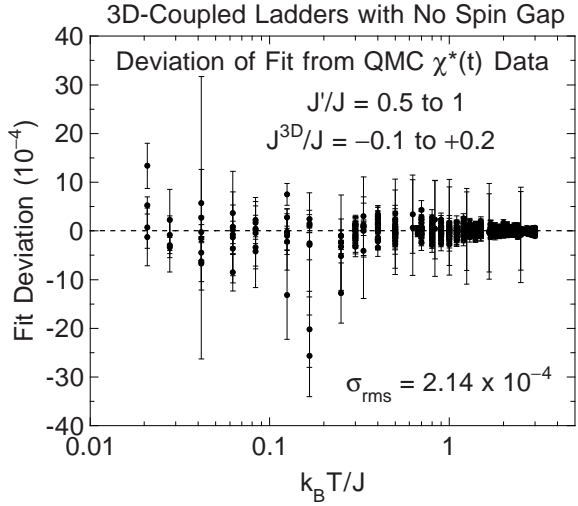}}
\vglue 0.1in
\caption{Semilog plot versus temperature $T$ of the deviation of the fit
from the 564 QMC $\chi^*(t)$ data points for $S = 1/2$ 3D-coupled spatially
anisotropic Heisenberg ladders with $J^\prime/J = 0.5$ and $J^{\rm 3D}/J =
-0.1,\ -0.08,\ -0.06,\ -0.04$ and~0.05, $J^\prime/J = 1$ and $J^{\rm 3D}/J =
0.12$, and $J^\prime/J = 0.5$, 0.6, 0.7, 0.8 and~0.9 and $J^{\rm 3D}/J =
0.1$, 0.15 and 0.2 ($\bullet$), for which there is no spin gap.  The error
bars are the estimated accuracies of the QMC data.  The fit function for the
gapped fit is given in Eqs.~(\protect\ref{EqChiFit:all})
and~(\protect\ref{EqNDLCO0.5}) with parameters in
Table~\protect\ref{TabLCOPars}.}
\label{Fig59}
\end{figure}

\begin{eqnarray} D_{\rm 1\,add} &=& D_{16}\Big({J^{\rm 3D}\over
J}\Big)~,\nonumber\\
\nonumber\\ D_{\rm 2\,add} &=& D_{26}\Big({J^{\rm 3D}\over J}\Big) +
D_{27}\Big({J^\prime\over J}\Big)^2\Big({J^{\rm 3D}\over J}\Big)\nonumber\\
&+& D_{28}\Big({J^{\rm 3D}\over J}\Big)^3 + D_{29}\Big({J^\prime\over
J}\Big)\Big({J^{\rm 3D}\over J}\Big)^3\nonumber\\ &+&
D_{210}\Big({J^\prime\over J}\Big)^2\Big({J^{\rm 3D}\over
J}\Big)^2~,\nonumber\\
\nonumber\\ D_{\rm 3\,add} &=&  D_{36}\Big({J^{\rm 3D}\over J}\Big) +
D_{37}\Big({J^\prime\over J}\Big)^2\Big({J^{\rm 3D}\over J}\Big)\nonumber\\
&+& D_{38}\Big({J^{\rm 3D}\over J}\Big)^3 +  D_{39}\Big({J^\prime\over
J}\Big)\Big({J^{\rm 3D}\over J}\Big)^3\nonumber\\ &+&
D_{310}\Big({J^\prime\over J}\Big)^2\Big({J^{\rm 3D}\over J}\Big)^2 +
D_{311}\Big({J^\prime\over J}\Big)^3 \nonumber\\  &+&
D_{312}\Big({J^\prime\over J}\Big)^3
\Big({J^{\rm 3D}\over J}\Big) + D_{313}\Big({J^{\rm 3D}\over
J}\Big)^4\nonumber\\ &+& D_{314}\Big({J^\prime\over J}\Big)^3\Big({J^{\rm
3D}\over J}\Big)^2 + D_{315}\Big({J^\prime\over J}\Big)\Big({J^{\rm
3D}\over J}\Big)^4.
\label{EqNDLCOD1D2D3}
\end{eqnarray} This fit function satisfies the low-temperature limit
prediction $\chi(T) = A + B T^2$.  The 32 parameters of the fit in
Eqs.~(\ref{EqNDLCO0.5}) and the series in Eqs.~(\ref{EqNDLCOD1D2D3}) for
the constrained parameters are given in Table~\ref{TabLCOPars}, for which
$\chi^2/{\rm DOF} = 0.95$ and $\sigma_{\rm rms} = 0.000214$ were obtained. 
The three-dimensional fit is shown as the sets of solid curves in
Fig.~\ref{Fig18} and for the gapless parameter regimes in
Figs.~\ref{Fig14} and~\ref{Fig16}.  The
deviations of the fit from the data are plotted vs temperature in
Fig.~\ref{Fig59}.

\subsection{Isolated Ladders, Trellis Layers and Stacked Ladders with
$\bbox{J^\prime/J \geq 1}$}
\vglue0.3in
Since these spin lattices for all the exchange constant combinations with
$J^\prime/J \geq 1$ that we simulated have a spin gap, we were able to fit
all the QMC  data with the same multidimensional function, using ${\cal
P}^{(6)}_{(6)}$ in Eqs.~(\ref{EqChiFit:all}), with
\begin{eqnarray} N_n = N_{n0} &+& N_{1n1}\Big({J\over J^\prime}\Big) +
N_{1n2}
\Big({J\over J^\prime}\Big)^2 + N_{1n3} \Big({J\over
J^\prime}\Big)^3\nonumber\\  &+& N_{1n4} \Big({J\over J^\prime}\Big)^4 +
N_{2n1}\Big({J^{\prime\prime}\over J^\prime}\Big) +
N_{2n2}\Big({J^{\prime\prime}\over J^\prime}\Big)^2\nonumber\\ &+&
N_{3n1}\Big({J^{\prime\prime\prime}\over J^\prime}\Big) +
N_{3n2}\Big({J^{\prime\prime\prime}\over J^\prime}\Big)^2\nonumber\\ &+&
N_{3n3}\Big({J\over J^\prime}\Big)\Big({J^{\prime\prime\prime}\over
J^\prime}\Big)~,\ (n = 1-6)\nonumber\\ D_n = D_{n0} &+& D_{1n1}
\Big({J\over J^\prime}\Big) + D_{1n2}
\Big({J\over J^\prime}\Big)^2 + D_{1n3} \Big({J\over
J^\prime}\Big)^3\nonumber\\ &+& D_{1n4} \Big({J\over J^\prime}\Big)^4 +
D_{2n1}
\Big({J^{\prime\prime}\over J^\prime}\Big) + D_{2n2}
\Big({J^{\prime\prime}\over J^\prime}\Big)^2\nonumber\\ &+&
D_{3n1}\Big({J^{\prime\prime\prime}\over J^\prime}\Big) +
D_{3n2}\Big({J^{\prime\prime\prime}\over J^\prime}\Big)^2\nonumber\\ &+&
D_{3n3} \Big({J\over J^\prime}\Big)\Big({J^{\prime\prime\prime}\over
J^\prime}\Big)~,\ (n = 1-6)\nonumber\\   {\Delta_{\rm fit}\over J^\prime} =
1 &+&
\Delta_{11}\Big({J\over J^\prime}\Big) + \Delta_{12}\Big({J\over
J^\prime}\Big)^2+ \Delta_{13}\Big({J\over J^\prime}\Big)^3~,
\label{EqNDisoLadJ'>=J}
\end{eqnarray} 
\vglue0.28in
\noindent with $J^{\rm max} = J^\prime$.  All of the fits in this
section utilized the three respective HTSE constraints on $D_1,\ D_2$ and
$D_3$ in Eqs.~(\ref{EqD:all}).  The seven $N_{n0}$ and $D_{n0}$ fitting
coefficients were first determined to high accuracy by fitting to the exact
expression for the $S = 1/2$ dimer given in Eq.~(\ref{EqChiDimer}), for
which $J/J^\prime = 0$ and $\Delta/J^\prime = 1$.  We were able to fit a
498-point double-precision representation of this function for
$0.02 \leq t \leq 4.99$ to a variance of $1.2\times 10^{-16}$.  The fit
parameters were further constrained by requiring that the $\{N_n,\ D_n,\
\Delta_{\rm fit}\}$ values for $J/J^\prime = 1$ be identical with those
found in Sec.~\ref{SecIsoLadJ'<=J} above for $J^\prime/J =1$ from the fit
to the QMC data for $J^\prime/J \leq 1$.
  
\begin{figure}
\epsfxsize=3in
\centerline{\epsfbox{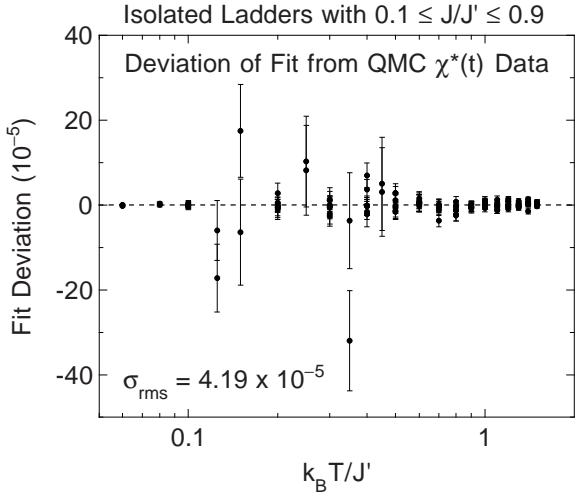}}
\vglue 0.1in
\caption{Semilog plot versus temperature $T$ of the deviation of the fit
from the 119 QMC $\chi^*(t)$ data points for $S = 1/2$ spatially anisotropic
Heisenberg ladders with $0.1 \leq J/J^\prime \leq 0.9$.  The error bars are
the estimated accuracies of the QMC data.  The fit function is given in
Eqs.~(\protect\ref{EqChiFit:all}) and~(\protect\ref{EqNDisoLadJ'>=J}) with
parameters in Table~\protect\ref{TabIsoLadJ'/J>=1Pars}.}
\label{Fig60}
\end{figure}

\begin{figure}
\epsfxsize=3.3in
\centerline{\epsfbox{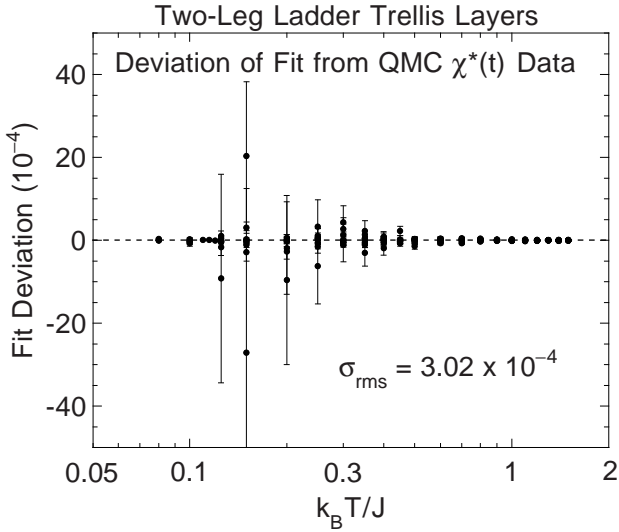}}
\vglue 0.1in
\caption{Semilog plot versus temperature $T$ of the deviation of the fit
from the 162 QMC $\chi^*(t)$ data points for $S = 1/2$ spatially
anisotropic Heisenberg two-leg ladder trellis layers with intraladder
couplings $J/J^\prime = 0.1$ and~0.2, and interladder intralayer couplings
$J^{\prime\prime}/J^\prime = -0.2,\ -0.1$, 0.1 and~0.2 ($\bullet$), for
which there is a spin gap.  The error bars are the estimated accuracies of
the QMC data.  The fit function for the QMC data is given in
Eqs.~(\protect\ref{EqChiFit:all}) and~(\protect\ref{EqNDisoLadJ'>=J}) with
parameters in Table~\protect\ref{TabIsoLadJ'/J>=1Pars}.}
\label{Fig61}
\end{figure}

\begin{figure}
\epsfxsize=3in
\centerline{\epsfbox{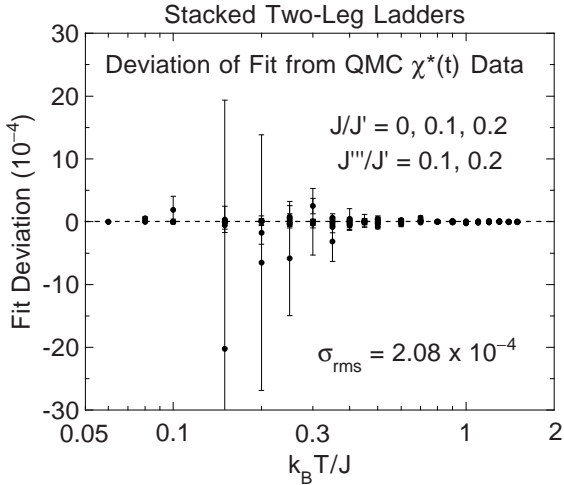}}
\vglue 0.1in
\caption{Semilog plot versus temperature $T$ of the deviation of the
29-parameter three-dimensional fit from the 119 QMC $\chi^*(t)$ data points
for stacked $S = 1/2$ spatially anisotropic Heisenberg two-leg ladders with
intraladder couplings $J/J^\prime = 0,\ 0.1$ and~0.2, and interladder
couplings $J^{\prime\prime\prime}/J^\prime = 0.1$ and~0.2 ($\bullet$), for
which there is a spin gap.  The error bars are the estimated accuracies of
the QMC data.  The fit function for the QMC data is given in
Eqs.~(\protect\ref{EqChiFit:all}) and~(\protect\ref{EqNDisoLadJ'>=J}) with
parameters in Table~\protect\ref{TabIsoLadJ'/J>=1Pars}.}
\label{Fig62}
\end{figure}

The 29 $N_{1nj},\ D_{1nj}$ and~$\Delta_{1j}$ coefficient fitting parameters
for the isolated ladders were determined by fitting to the 119 QMC
simulation data points for $J/J^\prime = 0.1$, 0.2, 0.3, 0.4, 0.5, 0.7
and~0.9  and the results are given in Table~\ref{TabIsoLadJ'/J>=1Pars},
yielding a $\chi^2/{\rm DOF} = 1.24$ and
$\sigma_{\rm rms} = 4.19\times 10^{-5}$ for the fit to those data.  The fit
is shown as the set of solid curves in Fig.~\ref{Fig05},
including two curves showing exchange parameter example interpolations for
$\chi^*(t,J/J^\prime = 0.6,\ 0.8)$ for which we did not obtain QMC
simulation data.  The fit deviations from all the fitted data are plotted
vs temperature in Fig.~\ref{Fig60}.  In the following fits, the
$\{N_{n0},\ D_{n0},\ N_{1nj},\ D_{1nj},\ \Delta_{1j}\}$ set of parameters
for the isolated ladders was held fixed. 
\newpage
The 18-parameter three-dimensional fit to the 162 QMC trellis layer
$\chi^*(t)$ data points with intraladder couplings $J/J^\prime = 0.1$
and~0.2, each with interladder couplings $J^{\prime\prime}/J = -0.2,\
-0.1$, 0.1 and~0.2,\cite{Miyahara1998} yielded $\chi^2/{\rm DOF} = 1.17$ and
$\sigma_{\rm rms} = 0.000302$.  As in the fit to the trellis layer QMC data
for $J^\prime/J \leq 1$ above, the data could be fitted well assuming that
the spin gap is independent of the interladder coupling.  The set of
$\{N_{2nj},\ D_{2nj}\}$ parameters obtained, including the constrained
$D_{21},\ D_{22}$ and $D_{23}$ series, is given in
Table~\ref{TabIsoLadJ'/J>=1Pars} and the fit is plotted as the set of solid
curves in Figs.~\ref{Fig07}
and~\ref{Fig08}, respectively. 
The fit deviations are
plotted vs temperature in Fig.~\ref{Fig61}.  The series for
$D_{2n}$ with $n = 2$ and~3 have the following terms in addition to those in
Eq.~(\ref{EqNDisoLadJ'>=J})
\widetext
\begin{table}
\caption{Parameters in Eq.~(\protect\ref{EqNDLCO0.5}) obtained by fitting
quantum Monte Carlo simulations of $\chi^*(t)$ for LaCuO$_{2.5}$-type 3D
coupled $S = 1/2$ Heisenberg ladders with no spin gap.}
\label{TabLCOPars}
\begin{tabular}{lcccccc}
$n$ & $N_{n0}$ & $N_{n1}$ & $N_{n2}$ & $N_{n3}$  & $N_{n4}$ & $N_{n5}$ \\
\hline 1 & $-$0.4102143 & 0.6183598 & 0 & 0.1499184 & $-$16.56455 &
3.798390 \\ 2 & 0.2411046 & $-$0.2644301 & 0 & 0.2756975 & $-$0.03651210 &
1.150476 \\  3 &
$-$0.0318917 & 0.09842195 & $-$0.05803464 & $-$0.01381445 & $-$0.5306434 &
0.4387261
\\   5 & 0.0008320664 & $-$0.002259251 & 0.001012209 & 0 & 0.03673690 &
$-$0.005869753 \\ 
\hline
$n$ & $D_{n0}$ & $D_{n1}$ & $D_{n2}$ & $D_{n3}$  & $D_{n4}$ & $D_{n5}$ \\
\hline  1 & 0.08978571 & 0.8683598 & 0 & 0.1499184 & $-$16.56455 & 3.798390
\\
  &            & $D_{16}$: 0.5 \\  2 & 0.2859975 & $-$0.05780379 &
0.2795899 & 0.6598366 & $-$8.068789 & $-$1.016509 \\ 
  && $D_{26}$: $-$0.2051071 & $D_{27}$: 0.03747960 & $D_{28}$: $-$8.282277 &
$D_{29}$: 1.899195 & $D_{210}$: 0.9495975 \\  3 & 0.06944034 & 0.1810730 &
$-$0.1754189 & 0.02929887 & $-$4.792592 & 2.246872 \\ 
  && $D_{36}$: 0.1205523 & $D_{37}$: 0.06892439 & $D_{38}$: 0.06507728 &
$D_{39}$: 0.6127174 & $D_{310}$: $-$1.782950 \\
  && $D_{311}$: 0.1189616 & $D_{312}$: 0.01873980 & $D_{313}$: $-$4.141139 &
$D_{314}$:0.4747988  & $D_{315}$: 0.9495975 \\   4 & 0.01957895 &
0.03025742 & $-$0.007991254 & 0.1063007 & $-$1.631987 & 0.3758439 \\ 6 &
$-$0.0002842768 & 0.001377244 & $-$0.001162417 & 0 & 0.09843710 & 0.0408774
\\ 
\end{tabular}
\end{table}
 \narrowtext
\begin{eqnarray} D_{2n\,{\rm add}} &=& D_{2n3} \Big({J\over
J^\prime}\Big)^5 + D_{2n4} \Big({J\over J^\prime}\Big)^6 + D_{2n5}
\Big({J\over J^\prime}\Big)^7\nonumber\\
 &+& \Big({J^{\prime\prime}\over J^\prime}\Big)\bigg[D_{2n6}\Big({J\over
J^\prime}\Big) + D_{2n7}
\Big({J\over J^\prime}\Big)^2\nonumber\\
 &+& D_{2n8}\Big({J\over J^\prime}\Big)^3 + D_{2n9}
\Big({J\over J^\prime}\Big)^4\bigg]\nonumber\\ &+&
\Big({J^{\prime\prime}\over J^\prime}\Big)^2\bigg[D_{2n10} \Big({J\over
J^\prime}\Big) + D_{2n11}
\Big({J\over J^\prime}\Big)^2 \nonumber\\ &+& D_{2n12} \Big({J\over
J^\prime}\Big)^3\bigg] + D_{2n13}\Big({J^{\prime\prime}\over
J^\prime}\Big)^3\nonumber\\
 &+& D_{2n14} \Big({J\over J^\prime}\Big)^8 + D_{2n15}
\Big({J\over J^\prime}\Big)^9\nonumber\\
 &+& D_{2n16}\Big({J\over J^\prime}\Big)^{10} \nonumber\\ &+&
\Big({J^{\prime\prime}\over J^\prime}\Big)\bigg[D_{2n17} \Big({J\over
J^\prime}\Big)^5 + D_{2n18} \Big({J\over J^\prime}\Big)^6\nonumber\\
 &+& D_{2n19}\Big({J\over J^\prime}\Big)^7\bigg]\nonumber\\
 &+&\Big({J^{\prime\prime}\over J^\prime}\Big)^2\bigg[D_{2n20}\Big({J\over
J^\prime}\Big)^4 + D_{2n21}\Big({J\over J^\prime}\Big)^5\nonumber\\
 &+& D_{2n22} \Big({J\over J^\prime}\Big)^6\bigg]\nonumber\\
 &+& \Big({J^{\prime\prime}\over J^\prime}\Big)^3\bigg[D_{2n23} \Big({J\over
J^\prime}\Big) + D_{2n24}\Big({J\over J^\prime}\Big)^2\nonumber\\
 &+& D_{2n25}\Big({J\over J^\prime}\Big)^3\bigg] + D_{2n26}
\Big({J^{\prime\prime}\over J^\prime}\Big)^4~.
\end{eqnarray}
\vglue0.09in
The 27-parameter three-dimensional fit to the 119 stacked ladder data
points for
$J^\prime/J = 0,$ 0.1 and~0.2 and $J^{\prime\prime\prime}/J = 0.1$ and~0.2
yielded
$\chi^2/{\rm DOF} = 0.81$ and $\sigma_{\rm rms} = 0.000208$, and is plotted
as the set of solid curves in Fig.~\ref{Fig13}.  The fit
deviations are plotted vs temperature in Fig.~\ref{Fig62}. 
The set of $\{N_{3nj},\ D_{3nj}\}$ parameters are given 
\newpage
\widetext
\begin{table}
\squeezetable
\caption{Parameters in Eq.~(\protect\ref{EqNDisoLadJ'>=J}) obtained by
fitting quantum Monte Carlo simulations of $\chi^*(t)$ for isolated $S =
1/2$ Heisenberg ladders, trellis layers and stacked ladders with
$J^\prime/J \geq 1$, all of which have spin gaps.}
\label{TabIsoLadJ'/J>=1Pars}
\begin{tabular}{lcccccc}
$n$ & $N_{n0}$ & $N_{1n1}$ & $N_{1n2}$ & $N_{1n3}$ & $N_{1n4}$ \\
\hline 1 & 0.6342799 & $-$0.4689967 & $-$0.1224498 & $-$0.6316720 &
$-$0.08782728 \\ 2 & 0.1877696 & $-$0.1498959 & $-$0.4760102 & 0.2714945 &
0.3686003 \\ 3 & 0.03360362 & $-$0.01319703 & $-$0.3269535 & 0.7854194 &
$-$0.5140804 \\ 4 & 0.003861107 & $-$0.01530859 & 0.1567169 & $-$0.2790342
& 0.1304374 \\ 5 & 0.001821501 & $-$0.02615567 & 0.02974644 & $-$0.05230788
& $-$0.06695909 \\ 6 & 0            & $-$0.0002501523 & 0.001069419 &
$-$0.001893068 & 0.001088651 \\
\hline
 & $D_{n0}$ & $D_{1n1}$ & $D_{1n2}$ & $D_{1n3}$ & $D_{1n4}$ \\
\hline 1 & $-$0.1157201 & 1.493088 & $-$1.5046567 & $-$0.2134494 &
$-$0.08782728 \\ 2 & 0.08705969 & $-$0.1502010 & 0.9054526 & $-$1.607161 &
0.9440189 \\
  &  & $D_{125}$: 0.07149545 & $D_{126}$: $-$0.05532895 & $D_{127}$:
$-$0.03673135 \\ 3 & 0.005631367 & 0.07738460 & $-$0.1639982 & 0.3932678 &
$-$0.7370737 \\
  &  & $D_{135}$: 0.6755368 & $D_{136}$: $-$0.2865834 & $D_{137}$:
$-$0.1009833 \\
  &  & $D_{138}$: 0.07759403 & $D_{139}$: 0.007719311 & $D_{1310}$:
$-$0.007680941\\ 4 & 0.001040887 & 0.01252745 & 0.1183833 & $-$0.2857871 &
0.1510432 \\  5 & 0.00006832857 & 0.004243732 & $-$0.03901711 & 0.1055626 &
$-$0.06948651 \\  6 & 0            & $-$0.0001868979 & 0.009010690 &
$-$0.019630625 & 0.01131886 \\
\hline & & $\Delta_{11}$ & $\Delta_{12}$ & $\Delta_{13}$ \\
\hline & & $-$1.462084 & 1.382207 & $-$0.4182226 \\
\hline
 & $N_{2n1}$ & $N_{2n2}$ & $D_{2n1}$ & $D_{2n2}$ & $D_{2n3}$ & $D_{2n4}$ \\
\hline 1 & $-$0.1760291 & 0.0004871780 & 0.3239709 & 0.0004871780 \\ 2 &
$-$0.1153545 & $-$0.1173427 & $-$0.1661927 & 0.04427737 & 0.07149545 &
$-$0.0553289 \\ 
  &&& $D_{225}$: $-$0.03673135 & $D_{226}$: 0.1511598 & $D_{227}$:
$-$0.5090197 &
$D_{228}$: $-$0.1803441 \\
  &&& $D_{229}$: $-$0.0439136 & $D_{2210}$: 0.0009558843 & $D_{2211}$:
$-$0.0006733808 &$D_{2212}$: 0.0002037488 \\
  &              &             & $D_{2213}$: 0.0002435890 \\ 3 &
$-$0.02937641 & $-$0.0613319 & 0.01787339 & $-$0.03423491 & 0.6755368 &
$-$0.2865834
\\ 
  &&& $D_{235}$: $-$0.1009833 & $D_{236}$: $-$0.05309999 & $D_{237}$:
$-$0.1163412 &
$D_{238}$: 0.08362761 \\
  &&& $D_{239}$: 0.2237513 & $D_{2310}$: $-$0.2051769 & $D_{2311}$:
$-$0.09081426 &$D_{2312}$: $-$0.1407214 \\
  &&& $D_{2313}$: $-$0.01958888 & $D_{2314}$: 0.07759403 & $D_{2315}$:
0.007719311 & $D_{2316}$: $-$0.007680941 \\
  &&& $D_{2317}$: 0.1594617 & $D_{2318}$: $-$0.04305911 & $D_{2319}$:
$-$0.01836568 &$D_{2320}$: $-$0.02109167 \\
  &&& $D_{2321}$: $-$0.0002816231 & $D_{2322}$: 0.00004260618 & $D_{2323}$:
0.0003561476 &$D_{2324}$: $-$0.0003366904 \\
  &&& $D_{2325}$: 0.0001018744 & $D_{2326}$: 0.0001217945 \\ 4 &
$-$0.0004271643 & 0.01065699 & $-$0.002927824 & $-$0.01012968 \\  5 &
0.001469929 & $-$0.001090253 & 0.0001835023 & $-$0.002659887 \\ 6 &
$-$0.0001162990 & 0.001595892 & 0.000004275399 & 0.0004027895 \\
\hline
& $N_{3n1}$ &
$N_{3n2}$ & $N_{3n3}$ & $D_{3n1}$ & $D_{3n2}$ & $D_{3n3}$ \\
\hline 1 & $-$0.1054832 & 0.9135603 & 0.4642452 & 0.3945168 & 0.9135603 &
0.4642452 \\ 2 & $-$0.04904426 & 0.4481929 & $-$0.02425882 & $-$0.1527919 &
$-$0.03971898 &
$-$0.08286582 \\  &&&& $D_{324}$: 0.07149545 & $D_{325}$: $-$0.05532895 &
$D_{326}$: $-$0.03673135 \\ &&&& $D_{327}$: 0.3043595 & $D_{328}$:
$-$0.7925231 & $D_{329}$: 0.1502442 \\ &&&& $D_{3210}$: 2.024605 &
$D_{3211}$: $-$1.262729 & $D_{3212}$: 0.3820716 \\ &&&& $D_{3213}$:
0.4567802 \\ 3 & $-$0.01992624 & 0.2093460 & $-$0.2041661 & 0.004045611 &
0.1525759 &
$-$0.04750486 \\  &&&& $D_{334}$: 0.6755368 & $D_{335}$: $-$0.2865834 &
$D_{336}$: $-$0.1009833 \\ &&&& $D_{337}$: $-$0.7790473 & $D_{338}$:
1.364354 & $D_{339}$: $-$1.065771 \\ &&&& $D_{3310}$: $-$0.6522574 &
$D_{3311}$: 2.236366 & $D_{3312}$: $-$2.972952 \\ &&&& $D_{3313}$:
$-$0.1757212 & $D_{3314}$: 0.07759403 & $D_{3315}$: 0.007719311 \\ &&&&
$D_{3316}$: $-$0.007680941 & $D_{3317}$: 0.9431046 & $D_{3318}$:
$-$0.3052558 \\ &&&& $D_{3319}$: 0.02223492 & $D_{3320}$: 1.697456 &
$D_{3321}$: $-$0.5281020 \\ &&&& $D_{3322}$: 0.07989548 & $D_{3323}$:
0.7839124 & $D_{3324}$: $-$0.6313647 \\ &&&& $D_{3325}$: 0.1910358 &
$D_{3326}$: 0.22839009 \\ 4 & $-$0.003145802 & 0.02093948 & 0.079086430 &
0.006791683 & $-$0.1052993 & 0.09720203 \\   5 & 0.007236379 &
$-$0.04068458 & $-$0.04175775 & $-$0.0005392557 & 0.01629503 &
$-$0.03986072 \\ 6 & $-$0.0007577840 & 0.008442882 & 0.004628364 &
0.00003040315 & $-$0.0006839610 & 0.005075806  
\end{tabular}
\end{table}
\narrowtext
\newpage
\mbox{ }
\newpage
\noindent in Table~\ref{TabIsoLadJ'/J>=1Pars}.  The constrained series for
$D_{2n}$ with
$n = 2$ and~3 have the following terms in addition to those in
Eq.~(\ref{EqNDisoLadJ'>=J})
\begin{eqnarray} D_{3n\,{\rm add}} &=& D_{3n4} \Big({J\over
J^\prime}\Big)^5 + D_{3n5} \Big({J\over J^\prime}\Big)^6 + D_{3n6}
\Big({J\over J^\prime}\Big)^7\nonumber\\
 &+& \Big({J^{\prime\prime\prime}\over J^\prime}\Big)\bigg[D_{3n7}
\Big({J\over J^\prime}\Big)^2\nonumber\\
 &+& D_{3n8}\Big({J\over J^\prime}\Big)^3 + D_{3n9}
\Big({J\over J^\prime}\Big)^4\bigg]\nonumber\\ &+&
\Big({J^{\prime\prime\prime}\over J^\prime}\Big)^2\bigg[D_{3n10}
\Big({J\over J^\prime}\Big) + D_{3n11}
\Big({J\over J^\prime}\Big)^2 \nonumber\\ &+& D_{3n12} \Big({J\over
J^\prime}\Big)^3\bigg] + D_{3n13}\Big({J^{\prime\prime\prime}\over
J^\prime}\Big)^3\nonumber\\
 &+& D_{3n14} \Big({J\over J^\prime}\Big)^8 + D_{3n15}
\Big({J\over J^\prime}\Big)^9\nonumber\\
 &+& D_{3n16}\Big({J\over J^\prime}\Big)^{10} \nonumber\\ &+&
\Big({J^{\prime\prime\prime}\over J^\prime}\Big)\bigg[D_{3n17} \Big({J\over
J^\prime}\Big)^5 + D_{3n18} \Big({J\over J^\prime}\Big)^6\nonumber\\
 &+& D_{3n19}\Big({J\over J^\prime}\Big)^7\bigg]\nonumber\\
 &+&\Big({J^{\prime\prime\prime}\over
J^\prime}\Big)^2\bigg[D_{3n20}\Big({J\over J^\prime}\Big)^4 +
D_{3n21}\Big({J\over J^\prime}\Big)^5\nonumber\\
 &+& D_{3n22} \Big({J\over J^\prime}\Big)^6\bigg]\nonumber\\
 &+& \Big({J^{\prime\prime\prime}\over J^\prime}\Big)^3\bigg[D_{3n23}
\Big({J\over J^\prime}\Big) + D_{3n24}\Big({J\over
J^\prime}\Big)^2\nonumber\\
 &+& D_{3n25}\Big({J\over J^\prime}\Big)^3\bigg] + D_{3n26}
\Big({J^{\prime\prime\prime}\over J^\prime}\Big)^4~.
\end{eqnarray}

\subsection{$\bbox{S=1/2}$ Three- and Five-Leg Ladders}

In these $n$-leg ladders, the spin gap $\Delta = 0$.  The spins are not all
equivalent, since the outer leg spins have three nearest neighbors whereas
the inner leg spins have four.  Therefore, only the Curie-Weiss terms to
${\cal O}(1/t^2)$ in the HTSE in Eqs.~(\ref{EqGenHTS2})
and~(\ref{EqGenHTS3}) can be utilized to constrain the fits.  Frischmuth
{\it et al.}\cite{Frischmuth1996} have carried out QMC simulations of
$\chi^*(t)$ for ladders with $n = 1$--6 and spatially isotropic exchange
and also for $n = 3$ with anisotropic exchange.

For the three-leg ($n = 3$) ladders with anisotropic exchange, we fitted
the 126 QMC $\chi^*(t)$ data points\cite{Frischmuth1996} for $J^\prime/J =
0.4,\ 0.5,\ 0.6\ 0.7$ and~1 by Eqs.~(\ref{EqChiFit:all}) using $J^{\rm max}
= J$, $\Delta = 0$, the constrained $D_1$ given by Eq.~(\ref{EqD:a}), $4
d_1$ given by the average value $(2/3)(3 + 2J^\prime/J)$, and  ${\cal
P}^{(4)}_{(5)}$ in Eq.~(\ref{EqChiFit:all}).  The error bars for
$J^\prime/J = 1$ (which were not available to us) were estimated from those
of the other data sets.  The $N_n$ and
$D_n$ parameters are written as power series in $J^\prime/J$ according to
\begin{eqnarray} N_n = N_{n0} &+& N_{1n}\Big({J^{\prime}\over J}\Big) +
N_{2n}
\Big({J^{\prime}\over J}\Big)^2~,\ (n = 1-4)\nonumber\\ D_n = D_{n0} &+&
D_{1n} \Big({J^{\prime}\over J}\Big) + D_{2n}
\Big({J^{\prime}\over J}\Big)^2~.\ (n = 1-5)\label{Eq3-LegLadFit}
\end{eqnarray} 
\begin{figure}
\epsfxsize=3in
\centerline{\epsfbox{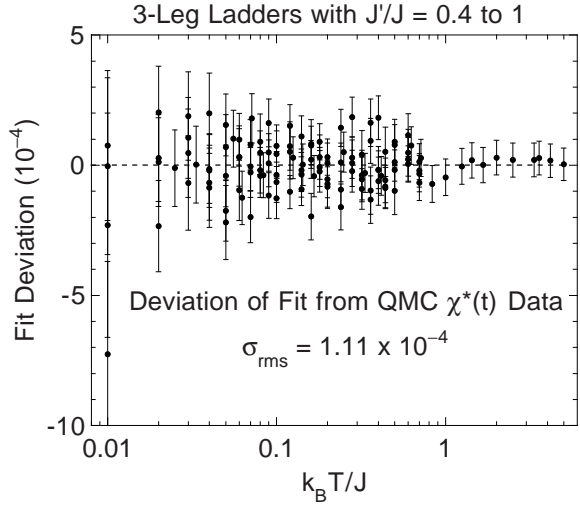}}
\vglue 0.1in
\caption{Semilog plot versus temperature $T$ of the deviation of the fit
from the 126 QMC $\chi^*(t)$ data points for $S = 1/2$ spatially
anisotropic three-leg Heisenberg ladders with $J^\prime/J = 0.5$, 0.5, 0.6,
0.7 and~1 ($\bullet$), for which there is no spin gap.  The error bars are
the estimated accuracies of the QMC data.  The fit function for the QMC
data is given in Eqs.~(\protect\ref{EqChiFit:all})
and~(\protect\ref{Eq3-LegLadFit}) with parameters in
Table~\protect\ref{Tab3-LegLadPars}.}
\label{Fig63}
\end{figure}

\noindent The $\{N_{n},D_{n}\}$ series coefficients are given in
Table~\ref{Tab3-LegLadPars}.  The two-dimensional fit is shown as the set
of solid curves through the data in Fig.~\ref{Fig19}, where
extrapolations to $t = 2$ and interpolation curves for $J^\prime/J = 0.8$
and~0.9 are also shown.  The deviations of the fit from all the data are
plotted vs temperature in Fig.~\ref{Fig63}.  The qualities
of the fit are $\chi^2$/DOF = 1.08, $\sigma_{\rm rms} = 0.000111$, and the
relative rms deviation is 0.227\%.

We also obtained an unweighted fit to the $\chi^*(t)$ data of Frischmuth
{\it et al.}\cite{Frischmuth1996} for the isotropic five-leg ladder (for
which  the error bars were not available to us) using ${\cal
P}^{(3)}_{(4)}$ in Eq.~(\ref{EqChiFit:all}), $\Delta = 0$, $D_1$ given by
Eq.~(\ref{EqD:a}) and $4 d_1$ given by the average coordination number $4 -
(2/n) = 3.6$,\cite{Johnston1996,Johnston1997} with parameters
\begin{eqnarray} N_1 &=&	-0.2732853~,~~~N_2 =	0.09333487~,\nonumber\\ N_3
&=&	0.006660300~,~~~D_1 =	0.6267147~,\nonumber\\ D_2 &=&	0.3077097~,~~~D_3
=	0.04438012~,\nonumber\\ D_4 &=&	0.07488932~,~~~{\chi^2\over{\rm DOF}} =
5.6\times 10^{-9}~.
\label{Eq5-LegPars}
\end{eqnarray}

\mbox{ }

\mbox{ }

\widetext
\begin{table}
\caption{Parameters in Eq.~(\protect\ref{Eq3-LegLadFit}) obtained by
fitting quantum Monte Carlo simulations of $\chi^*(t)$ for three-leg
ladders\protect\cite{Frischmuth1996} with $J^\prime/J = 0.4,\ 0.5,\ 0.6,\
0.7$ and~1.0 which have no spin gap.}
\label{Tab3-LegLadPars}
\begin{tabular}{lcccccc}
$n$ & $N_{n0}$ & $N_{n1}$ & $N_{n2}$ & $D_{n0}$ & $D_{n1}$ & $D_{n2}$ \\
\hline 1 & 0.5016864 & $-$0.8090646 & 0.1273943 & 1.001686 & $-$0.4757312 &
0.1273943 \\ 2 & 0.3820011 & $-$0.8079494 & 0.4709254 & 0.4967571 &
0.1455232 &
$-$0.3643260\\ 3 & $-$0.05586766 & 0.2124001 & $-$0.1450983 & 0.9862886 &
$-$2.353475 & 1.5390197 \\ 4 & 0.0008251446 & $-$0.007962278 & 0.01431864 &
$-$0.1471295 & 0.7458894 & $-$0.6521486 \\ 5 & & & & 0.004095701 &
$-$0.04832765 & 0.09224402 
\end{tabular}
\end{table}
\narrowtext


\end{document}